\documentclass[12pt,a4paper]{report}
\usepackage[paperwidth=21cm, paperheight=31cm]{geometry}
\usepackage[T1]{fontenc}
\usepackage[latin1]{inputenc}
\usepackage{graphicx}
\usepackage{amsmath}
\usepackage{amssymb}
\usepackage[nottoc]{tocbibind}

\usepackage{makeidx}
\makeindex

\newtheorem{theorem}{Theorem}
\newtheorem{corollary}{Corollary}

\DeclareMathOperator{\diag}{diag}
\DeclareMathOperator{\tr}{tr}

\newcommand{\bl}{\boldsymbol}
\newcommand{\ph}{\phantom}

\newcommand{\lef}{\left(}
\newcommand{\rig}{\right)}
\newcommand{\lan}{\langle}
\newcommand{\ran}{\rangle}

\makeindex
\begin{document}

\begin{titlepage}
\begin{center}
\begin{figure}[h]
	\centering
		\includegraphics[width=12cm]{{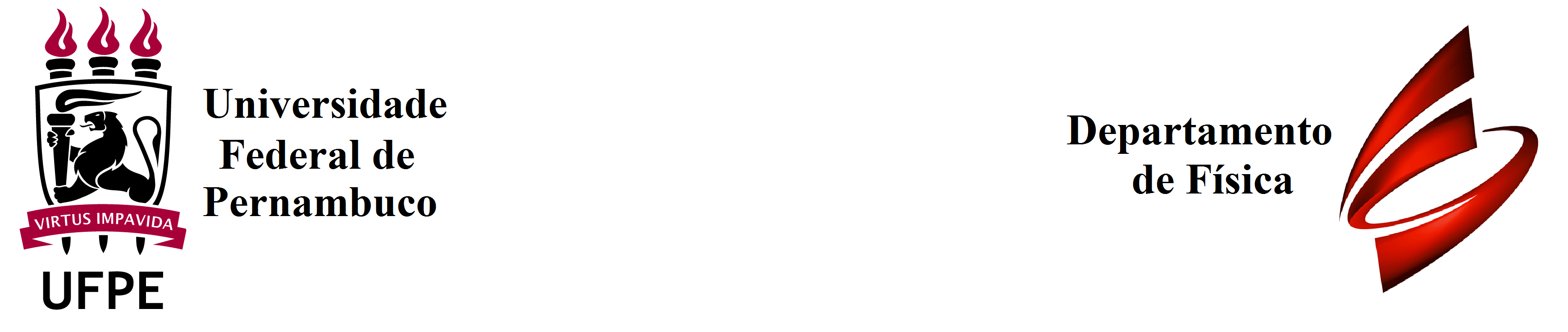}}
\end{figure}
\end{center}
\vspace{2cm}
\begin{center}
\large{\textbf{\textsc{On the Pursuit of Generalizations for the Petrov Classification and the Goldberg-Sachs Theorem}}}
\end{center}
\vspace{0.4cm}
\begin{center}
  \Large{\textbf{Carlos Batista}}
\end{center}
\vspace{3cm}
\begin{center}
  Doctoral Thesis \\
  Universidade Federal de Pernambuco, Departamento de F\'{\i}sica\\
  Supervisor: Bruno Geraldo Carneiro da Cunha\\
  Brazil - November - 2013
\end{center}

\vspace{3cm}

\begin{center}
\begin{figure}[h]
	\centering
		\includegraphics[width=11cm]{{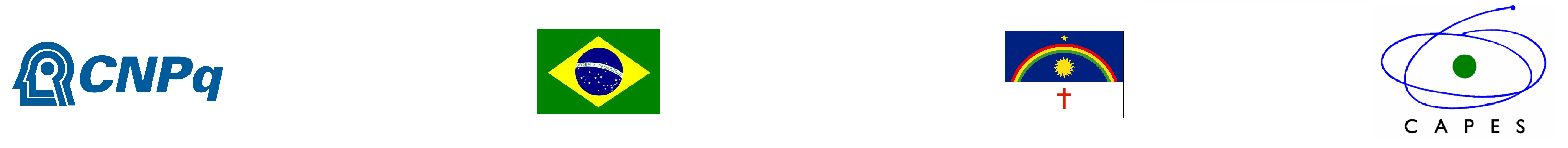}}
\end{figure}
\end{center}
\end{titlepage}


\begin{titlepage}

\quad
\vspace{2cm}

\small{
\begin{flushright}
\parbox{3.4in}{Thesis presented to the graduation program of the Physics Department of Universidade Federal de Pernambuco as part of the duties to obtain the degree of Doctor of Philosophy in Physics.}
\end{flushright}}
\normalsize

\vspace{3cm}

\small
\begin{flushleft}
  \textbf{Examining Board:}\\
  Prof. Amilcar Rabelo de Queiroz (IF-UNB, Brazil)\\
  Prof. Antônio Murilo Santos Macêdo (DF-UFPE, Brazil)\\
  Prof. Bruno Geraldo Carneiro da Cunha (DF-UFPE, Brazil)\\
  Prof. Fernando Roberto de Luna Parisio Filho (DF-UFPE, Brazil)\\
  Prof. Jorge Antonio Zanelli Iglesias (CECs, Chile)
\end{flushleft}

\vfill\begin{center}
\begin{figure}[h]
	\centering
		\includegraphics[width=12cm]{{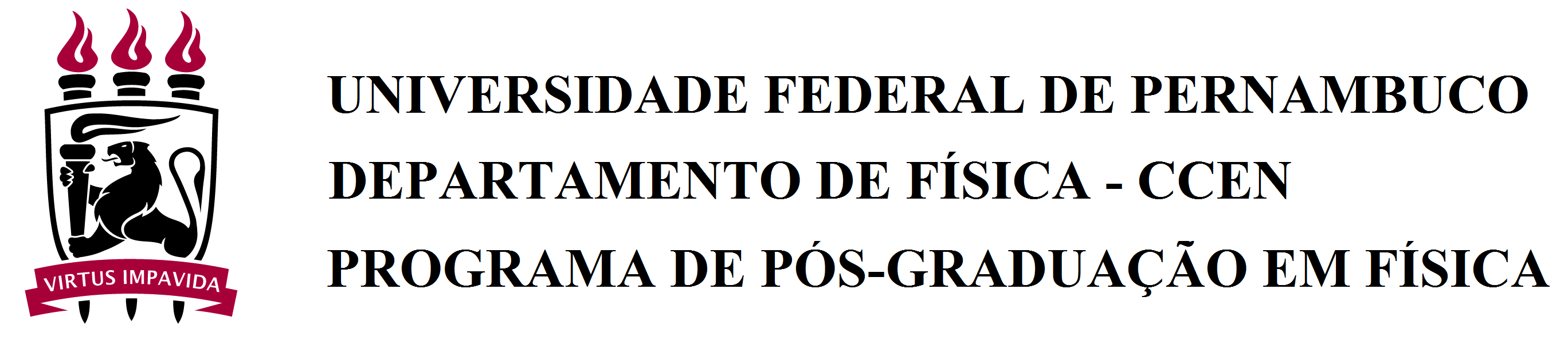}}
\end{figure}
\end{center}

\end{titlepage}


\begin{titlepage}

\begin{center}
  \LARGE{\textbf{Abstract}}\\
  \end{center}
\normalsize
  \vspace{0.3cm}
  \quad

The Petrov classification is an important algebraic classification for the Weyl tensor valid in 4-dimensional space-times. In this thesis such classification is generalized to manifolds of arbitrary dimension and signature. This is accomplished by interpreting the Weyl tensor as a linear operator on the bundle of $p$-forms, for any $p$, and computing the Jordan canonical form of this operator. Throughout this work the spaces are assumed to be complexified, so that different signatures correspond to different reality conditions, providing a unified treatment. A higher-dimensional generalization of the so-called self-dual manifolds is also investigated.

The most important result related to the Petrov classification is the Goldberg-Sachs theorem. Here are presented two partial generalizations of such theorem valid in even-dimensional manifolds. One of these generalizations states that certain algebraic constraints on the Weyl ``operator'' imply the existence of an integrable maximally isotropic distribution. The other version of the generalized Goldberg-Sachs theorem states that these algebraic constraints imply the existence of a null congruence whose optical scalars obey special restrictions.

On the pursuit of these results the spinorial formalism in 6 dimensions was developed from the very beginning, using group representation theory. Since the spinors are full of geometric significance and are suitable tools to deal with isotropic structures, it should not come as a surprise that they provide a fruitful framework to investigate the issues treated on this thesis. In particular, the generalizations of the Goldberg-Sachs theorem acquire an elegant form in terms of the pure spinors.\\
\\
\\
\\
\small{\textbf{Keywords:} General relativity, Weyl tensor, Petrov classification, Integrability, Isotropic distributions, Goldberg-Sachs theorem, Spinors, Clifford algebra.}
\normalsize

\end{titlepage}


\begin{titlepage}

\large{\textbf{This thesis is based on the following published articles:}}
\\
\\
\\
\\
\small
$\bullet$\;Carlos Batista, \textit{Weyl tensor classification in four-dimensional manifolds of all signatures}, General Relativity and Gravitation \textbf{45} (2013),
    785. \\
\\
\\
\\
$\bullet$\;Carlos Batista, \textit{A generalization of the Goldberg-Sachs theorem and its consequences}, General Relativity and Gravitation \textbf{45} (2013), 1411.\\
    \\
\\
\\
$\bullet$\;Carlos Batista and Bruno G. Carneiro da Cunha, \textit{Spinors and the Weyl tensor classification in six dimensions}, Journal of Mathematical Physics \textbf{54} (2013),
    052502. \\
    \\
\\
\\
$\bullet$\;Carlos Batista, \textit{On the Weyl tensor classification in all dimensions and its relation with integrability properties}, Journal of Mathematical Physics
    \textbf{54} (2013), 042502.

\normalsize

\end{titlepage}


\begin{titlepage}

\begin{center}
 \LARGE{\textbf{Acknowledgments}}\\
     \end{center}
   \normalsize
  \vspace{0.3cm}
  \quad

In order for such a long work, lasting almost five years, to succeed it is unavoidable to have the aid and the support of a lot of people. In this section I would like to sincerely thank to everybody that contributed in some way to my doctoral course.

I want to acknowledge my supervisor, Bruno Geraldo Carneiro da Cunha, for the sensitivity in suggesting a research project that fully matches my professional tastes. I also thank for all advise he gave me during our frequent meetings. It is inspiring to be supervised by such a wise scientist as Professor Bruno. Finally I thank for the freedom and the continued support he provided me, so that I could follow my own track. I take the chance to acknowledge all other Professors from UFPE that contributed to my education, particularly the Professors Ant\^{o}nio Murilo, S\'{e}rgio Coutinho, Henrique Ara\'{u}jo and Liliana Gheorghe, whose knowledge and commitment have inspired me.

In the same vein, I thank to all the mates as well as to the staff of the physics department. Specially, I thank to my doctorate fellow F\'{a}bio Novaes Santos for all the times he patiently helped me, thank you very much. I also would like to mention my friends Carolina Cerqueira, Danilo Pinheiro, Diego Leite and Rafael Alves, who contributed for a more pleasant environment in the physics department. I acknowledge the really qualified and efficient work of the graduation secretary  Alexsandra Melo as well as the friendship and support of the under-graduation secretary Paula Franssinete.

Finally, and most importantly, I would like to thank for the unconditional support of all my family. Particularly, I thank to my mother, Ana L\'{u}cia, and to my sister, Nat\'{a}lia Augusta, for always encouraging me to study, since my childhood, as well as stimulating my vocation. I also thank to my parents in law, Guilherme e L\'{u}cia Helena, for taking responsibility on the construction of my house, what allowed me to proceed using my whole time to study.  To conclude, I want to effusively and repeatedly thank to my wife, Juliana. In addition for her being my major inspiration, she supports me and encourages me like no one else. There are no words to say how much I am glad for having her besides me. I love you, my wife!!

During my Ph.D. I received financial support from CAPES (Coordena\c{c}\~{a}o de Aperfei\c{c}oamento de Pessoal de N\'{\i}vel Superior) and CNPq (Conselho Nacional de Desenvolvimento Cient\'{\i}fico e Tecnol\'{o}gico). It is worth mentioning that I really appreciate doing what I enjoy, study physics and mathematics, on my own country and still be paid for this. I will do my best, as I always tried to, in order for my work as a researcher and as a Professor, in the future, to return this investment.

\end{titlepage}


\begin{titlepage}

\quad
\vspace{5cm}
\begin{quote}
  \emph{ ``The black holes of nature are the most perfect macroscopic objects there are in the universe: the only elements in their construction are our concepts of space and time. And since the general theory of relativity provides only a single unique family of solutions for their descriptions, they are the simplest objects as well.''}\\
  \begin{flushright}
\small{Subrahmanyan Chandrasekhar $\ph{aAAA}$\\
(The Mathematical Theory of Black Holes)}
 \end{flushright}

\end{quote}
\normalsize

\end{titlepage}


\pagenumbering{gobble}{\tableofcontents}

\newpage

\pagenumbering{arabic}
\setcounter{page}{10}

\setcounter{chapter}{-1}


\chapter{Motivation and Outline}

The so called Petrov classification is an algebraic classification for the Weyl tensor of a 4-dimensional curved space-time that played a prominent role in the development of general relativity. Particularly, it helped on the search of exact solutions for Einstein's equation, the most relevant example being the Kerr metric. Furthermore, such classification contributed for the physical understanding of gravitational radiation. There are several theorems concerning this classification, they associate the Petrov type of the Weyl tensor with physical and geometric properties of the space-time. Probably the most important of these theorems is the Goldberg-Sachs theorem, which states that in vacuum the Weyl tensor is algebraically special if, and only if, the space-time admits a shear-free congruence of null geodesics. It was because of this theorem that Kinnersley was able to find all type $D$ vacuum solutions for Einstein's equation, an impressive result given that such equation is highly non-linear.

Since the Petrov classification and the Goldberg-Sachs theorem have been of major importance for the study of 4-dimensional Lorentzian spaces, it is quite natural trying to generalize these results to manifolds of arbitrary dimension and signature. This is the goal of the present thesis. In what follows the Petrov classification will be extended to all dimensions and signatures in a geometrical approach. Moreover, there will be presented few generalizations of the Goldberg-Sachs theorem valid in even-dimensional spaces. The relevance of this work is enforced by the increasing significance of higher-dimensional manifolds in physics and mathematics.

This thesis was split in two parts. The part \ref{Part-Rev} shows the classical results concerning the Petrov classification and its associated theorems, while part \ref{Part-Orig} presents the work developed by the present author during the doctoral course. In chapter \ref{Chap. GR} the basic tools of general relativity and differential geometry necessary for the understanding of this thesis are reviewed. It is shown that gravity manifests itself as the curvature of the space-time and it is briefly discussed the relevance of higher-dimensional manifolds. Chapter \ref{Chap. Pet} shows six different routes to define the Petrov classification. In addition, the so called principal null directions are interpreted from the physical and geometrical points of view. Chapter \ref{Chap. Theorems4} presents some of the most important theorems concerning the Petrov classification, as the Goldberg-Sachs, the Mariot-Robinson and the Peeling theorems. In chapter \ref{Chap. Gen4D} the Petrov classification is generalized to 4-dimensional spaces of arbitrary signature in a unified approach, with each signature being understood as a choice of reality condition on a complex space. Moreover, it is shown that this generalized classification is related to the existence of important geometric structures. Chapter \ref{Chap._Spin6D} develops the spinorial formalism in 6 dimensions with the aim of uncovering results that are hard to perceive by means of the standard vectorial approach. In particular, the spinorial language reveals that the Weyl tensor can be seen as an operator on the space of 3-vectors, which is exploited in order to classify this tensor. It is also proved an elegant partial generalization of the Goldberg-Sachs theorem making use of the concept of pure spinors. An algebraic classification for the Weyl tensor valid in arbitrary dimension and signature is then developed in chapter \ref{Chap. AllD}, where it is also proved two partial generalizations of the Goldberg-Sachs theorem valid in even-dimensional manifolds. Finally, chapter \ref{Chap. Conclusion} discuss the conclusions and perspectives of this work.

Some background material is also presented in the appendices. Appendix \ref{App._Segre} introduces a classical algebraic classification for square matrices called the Segre classification and defines a refinement for it. Such refined classification is used throughout the thesis. Appendix \ref{App._NullTetrad} describes what a null tetrad is. The formal treatment of Clifford algebra and spinors is addressed in appendix \ref{App._Cliff&Spinors}, where some pedagogical examples are also worked out. Finally, appendix \ref{App._Group} introduces and give some examples of the basics concepts on group representation theory.

\part{Review and Classical Results}\label{Part-Rev}

\chapter{Introducing General Relativity}\label{Chap. GR}
Right after Albert Einstein arrived at his special theory of relativity, in 1905, he noticed that the Newtonian theory of gravity needed to be modified. Newton's theory predict that when a gravitational system is perturbed the effect of such perturbation is immediately felt at all points of space, in other words the gravitational interaction propagates with infinite velocity. This, however, is in contradiction with one of the main results of special relativity, that no information can propagate faster than light. Moreover, according to Einstein's results energy and mass are equivalent, which implies that the light must feel the gravitational attraction, in disagreement with the Newtonian gravitational theory.

It took long 10 years for Einstein to establish a relativistic theory of gravitation, the General Theory of Relativity. In spite of the sophisticated mathematical background necessary to understand this theory, it turns out that it has a beautiful geometrical interpretation. According to general relativity,  gravity shows itself as the curvature of the space-time. Such theory has had several experimental confirmations, notably the correct prediction of Mercury's perihelion precession and the light deflection. In particular, it is worth noting that the GPS technology strongly relies on the general theory of relativity.

The aim of the present chapter is to describe the basic tools of general relativity necessary in the rest of the thesis. Readers already familiar with such theory are encouraged to skip this chapter. Throughout this thesis it will be assumed that repeated indices are summed, the so-called Einstein summation convention. The symmetrization and anti-symmetrization of indices are respectively denoted by round and square brackets. So that, for instance, $T_{(\mu\nu)}= \frac{1}{2}(T_{\mu\nu}+ T_{\nu\mu})$ and\label{Symmetrization} $L_{[\mu\nu\rho]}= \frac{1}{6}(L_{\mu\nu\rho}+ L_{\nu\rho\mu}+ L_{\rho\mu\nu}- L_{\nu\mu\rho}- L_{\rho\nu\mu}- L_{\mu\rho\nu})$.

\section{Gravity is Curvature}
According to the special theory of relativity we live in a four-dimensional flat space-time endowed with the metric:
$$ ds^2 \,=\, \eta_{\mu\nu}\,dx^\mu\,dx^\nu \,=\, dt^2 \,-\, dx^2 \,-\, dy^2 \,-\,dz^2\,,  $$
where $\{x^\mu\}=\{t,x,y,z\}$ are cartesian coordinates. Note that if we make a Poincar\'{e} transformation, $x^\mu\mapsto \Lambda^\mu_{\phantom{\mu}\nu}x^\nu + a^\mu$, where $a^\mu$ is constant and $\eta_{\rho\sigma}\Lambda^\rho_{\phantom{\rho}\mu}\Lambda^\sigma_{\phantom{\sigma}\nu}= \eta_{\mu\nu}$, then the metric remains invariant. Physically, performing a Poincar\'{e} transformation means changing from one inertial frame to another, which should not change the Physics. But, in addition to the inertial coordinates we are free to use any coordinate system of our preference. For example, in a particular problem it might be convenient to use spherical coordinates on the space. The procedure of changing coordinates is simple, for example, if $g_{\mu\nu}$ is the metric on the coordinate system $\{x^\mu\}$ then using new coordinates, $\{x'^\mu\}$, we have:
$$ g_{\mu\nu}\,dx^\mu\,dx^\nu \,=\, g_{\mu\nu} \, \left(\frac{\partial x^{\mu}}{\partial x'^{\rho}}dx'^\rho\right)\,\left(\frac{\partial x^{\nu}}{\partial x'^{\sigma}}dx'^\sigma\right) \;\Rightarrow\; g'_{\rho\sigma}\,=\,  \frac{\partial x^{\mu}}{\partial x'^{\rho}}\frac{\partial x^{\nu}}{\partial x'^{\sigma}}\,g_{\mu\nu}\,.$$
Where $g'_{\rho\sigma}$ are the components of the metric on the coordinates $\{x'^\mu\}$. In general, if $T^{\mu_1\ldots\mu_p}_{\phantom{\mu_1\ldots\mu_p}\nu_1\ldots\nu_q}$ are the components of a tensor $\bl{T}$ on the coordinate system $\{x^\mu\}$, then its components on the coordinates $\{x'^\mu\}$ are:
 \begin{equation}\label{Tensor_Transf.}
 T'^{\mu_1\ldots\mu_p}_{\phantom{\mu_1\ldots\mu_p}\nu_1\ldots\nu_q} \,=\, \left(\frac{\partial x'^{\mu_1}}{\partial x^{\rho_1}}\ldots\frac{\partial x'^{\mu_p}}{\partial x^{\rho_p}}\right)\,  \left(\frac{\partial x^{\sigma_1}}{\partial x'^{\nu_1}}\ldots\frac{\partial x^{\sigma_q}}{\partial x'^{\nu_q}}\right)\, T^{\rho_1\ldots\rho_p}_{\phantom{\rho_1\ldots\rho_p}\sigma_1\ldots\sigma_q}\,.
 \end{equation}

So far so good. But there is one important thing whose transformation under coordinate changes is non trivial, the derivative. Let $V^\mu$ be the components of a vector on the coordinate system $\{x^\mu\}$. Then it is a simple matter to prove that $\partial_\nu V^\mu$ does not transform as a tensor under a general coordinate change. Nevertheless, after some algebra, it can be proved that defining
\begin{equation}\label{chr}
  \Gamma^\mu_{\nu\rho} \,\equiv\, \frac{1}{2}\,g^{\mu\sigma}\left( \partial_\nu g_{\rho\sigma} + \partial_\rho g_{\nu\sigma} - \partial_\sigma g_{\nu\rho} \right)\,,
\end{equation}
with $g^{\mu\nu}$ being the inverse of $g_{\mu\nu}$ and $\partial_\nu$ being the partial derivative\label{PartialD} with respect to the coordinate $x^\nu$, then the combination
\begin{gather}\label{Cov.Deriv}
 \nabla_{\nu}\,V^\mu \,\equiv\, \frac{\partial V^\mu}{\partial x^\nu} \,+\, \Gamma^\mu_{\nu\rho}\,V^\rho \,=\, \partial_\nu\, V^\mu \,+\, \Gamma^\mu_{\nu\rho}\,V^\rho
\end{gather}
does transform as a tensor. The object $\Gamma^\mu_{\nu\rho}$, called Christoffel symbol (it is not a tensor), serves to correct the non-tensorial character of the partial derivative. The operator $\nabla_{\nu}$ is called the covariant derivative, it has the remarkable property that when acting on a tensor it yields another tensor. Its action on a general tensor is, for example,
\begin{equation}\label{Cov.Deriv.Tens}
 \nabla_{\nu}\,T^{\mu_1}_{\phantom{\mu_1}\mu_2\mu_3}\,=\, \partial_\nu\,T^{\mu_1}_{\phantom{\mu_1}\mu_2\mu_3} \,+\,
 \Gamma^{\mu_1}_{\nu\sigma}\,T^{\sigma}_{\phantom{\sigma}\mu_2\mu_3} - \Gamma^{\sigma}_{\nu\mu_2}\, T^{\mu_1}_{\phantom{\mu_1}\sigma\mu_3} - \Gamma^{\sigma}_{\nu\mu_3}\, T^{\mu_1}_{\phantom{\mu_1}\mu_2\sigma}\,.
\end{equation}
Using this formula it is straightforward to prove that $\nabla_\rho g_{\mu\nu}= 0$, so the metric is covariantly constant. Since coordinates are physically  meaningless we should always work with tensorial objects, because they are invariant under coordinate changes. Therefore, we should only use covariant derivatives instead of partial derivatives. Although they seem awkward, the covariant derivatives are, actually, quite common. For instance, in 3-dimensional calculus it is well-known that the divergence of a vector field in spherical coordinates looks different than in cartesian coordinates, this happens because we are implicitly using the covariant derivative.

Now comes a puzzle. From the physical point of view one might expect that no reference frame is better than another, all of them are equally arbitrary. In particular, the concept of acceleration is relative, since according to the classical Einstein's mental experiment (\textit{Gedankenexperiment}) gravity and acceleration are locally indistinguishable, the so-called equivalence principle. In spite of this, Minkowski space-time has an infinite class of privileged frames, the cartesian frames (also called inertial frames). From the geometrical point of view these frames are special because the Christoffel symbols, $\Gamma^\mu_{\nu\sigma}$, vanish identically in all points. But, as just advocated, the existence of these preferred frames is not a reasonable assumption. Therefore, we conclude that the space-time should not admit the existence of a frame such that $\Gamma^\mu_{\nu\sigma}$ vanishes in all points. Geometrically this implies that the space-time is curved! Somebody could argue that the inertial frames represent non-accelerated observers and, therefore, may exist. But our universe is full of mass everywhere, which implies that the gravitational field is omnipresent. Using then the equivalence principle we conclude that all objects are accelerated, so that it is nonsense to admit the existence of globally non-accelerated frames. Now we might wonder ourselves: If the space-time is not flat then why has special relativity been so successful? The reason is that in every point of a curved space-time we can always choose a reference frame such that $g_{\mu\nu}=\eta_{\mu\nu}$ and $\Gamma^\mu_{\nu\sigma}= 0$ at this point. Hence, special relativity is always valid locally.

Another natural question that emerges is: What causes space-time bending? Let us try to answer this. In special relativity a free particle moves on straight lines, which are the geodesics of flat space-time. Analogously, on a curved space-time the free particles shall move along the geodesics. Thus, no matter the peculiarities of a particle, if it is free it will follow the geodesic path compatible with its initial conditions of position and velocity. This resembles gravity, which, due to the equality of the inertial and gravitational masses, is such that all particles with the same initial condition follow the same trajectory. For example, a canon-ball and a feather both acquire the same acceleration under the gravitational field. Therefore, it is reasonable to say that the gravity bends the space-time. There is another path which leads us to the same conclusion. In line with Einstein's elevator experiment, gravity is locally equivalent to acceleration. Now suppose we are in a reference frame such that $\Gamma^\mu_{\nu\sigma}= 0$, then if this referential is accelerated it is simple matter to verify that the Christoffel symbol will be different from zero. Thus acceleration is related to the non-vanishing of $\Gamma^\mu_{\nu\sigma}$. Furthermore, the lack of a coordinate system such that $\Gamma^\mu_{\nu\sigma}= 0$ in all points of the space-time implies that the space-time is curved. So that we arrive at the following relations:
$$ \textrm{Gravity} \;\;\longleftrightarrow\;\; \textrm{Acceleration} \;\;\longleftrightarrow\;\; \Gamma^\mu_{\nu\sigma}\neq 0 \;\;\longleftrightarrow\;\; \textrm{Curvature}\,,$$
which again leads us to the conclusion that gravity causes the curvature of the space-time. This is the main content of the General Theory of Relativity.

In the standard model of particles the fundamental forces of nature are transmitted by bosons: photons carry the electromagnetic force, $W$ and $Z$ bosons communicate the weak interaction and gluons transmit the strong nuclear force. In the same vein, the gravitational interaction might be carried by a boson, dubbed the graviton\index{Graviton}. Indeed, heuristically speaking, since the emission of a particle of non-integer spin changes the total angular momentum of the system\footnote{For instance, suppose that a particle has integer spin and then emits a fermion. So, by the rule of angular momenta addition (see eq. (\ref{WignerEck}) in appendix \ref{App._Group}), it follows that its angular momentum after the emission is a superposition of non-integer values. Therefore it must have changed.} it follows that interactions carried by fermions are generally incompatible with the existence of static forces \cite{Feynman-Grav}. Now comes the question: What are the mass and the spin of the graviton? Since the gravitational force has a long range (energy goes as $1/r$) it follows that the mass must be zero, just as the mass of the photon. Moreover, since the graviton is a boson its spin must be integer. One can prove that it must be different from zero, since a scalar theory of gravitation predicts that the light is not affected by gravity \cite{Gravitation}, which contradicts the experiments and the fact that energy and mass are equivalent. The spin should also be different from one, since the interaction carried by a massless particle of spin one is the electromagnetic force which can be both attractive and repulsive, whereas gravity only attracts. It turns out that the graviton has spin 2. Indeed, in \cite{Feynman-Grav} it is shown how to start from the theory of a massless spin 2 particle on flat space-time and arrive at the general theory of relativity. For a wonderful introductory course in general relativity see \cite{Hartle_Book}. More advanced texts are available at \cite{Wald,FrolovBook}. Historical remarks and interesting philosophical thoughts can be found in \cite{Rovelli_Book}.


\section{Riemannian Geometry, the Formalism of \\Curved Spaces}\label{Sec._RiemannianGeom}
In order to make calculations on general relativity it is of fundamental importance to get acquainted with the tools of Riemannian geometry. The intent of the present section is to briefly introduce the bare minimum concepts on such subject necessary for the understanding of this thesis.

Roughly, an $n$-dimensional manifold $M$ is a smooth space such that locally it looks like $\mathbb{R}^n$. For example, the 2-sphere is a 2-dimensional manifold, since it is smooth and if we look very close to some patch of the spherical surface it will look like a flat plane (the Earth surface is round, but for its inhabitants it, locally, looks like a plane). More precisely, a manifold of dimension $n$ is a topological set such that the neighborhood of each point can be mapped into a patch of $\mathbb{R}^n$ by a coordinate system in a way that the overlapping neighborhoods are consistently joined \cite{Wald,Schutz-book}. Now imagine curves passing through a point $p$ belonging to the surface of the 2-sphere. The possible directions that these curves can take generate a plane, called the tangent space of $p$. Generally, associated to each point $p\in M$ of an $n$-dimensional manifold we have a vector space of dimension $n$, denoted by $T_pM$ and called the tangent space of $p$. A vector field $\bl{V}$ is then a map that associates to every point of the manifold a vector belonging to its tangent space. The union of the tangent spaces of all points of a manifold $M$ is called the tangent bundle and denoted by $TM$. A vector field is just an element of the tangent bundle.

Now, suppose that we introduce a coordinate system $\{x^\mu\}$ in the neighborhood of $p\in M$ and let $\bl{V}$ be a vector field in this neighborhood. Denoting by $V^\mu$ the components of $\bl{V}$ on such coordinate system then it is convenient to use the following abstract notation:
$$  \boldsymbol{V} \,=\, V^\mu \, \frac{\partial\,}{\partial x^\mu} \,\equiv\, V^\mu \,\partial_\mu\,.   $$
This is useful because when we make a coordinate transformation, $x^\mu\mapsto x'^\mu$, and use the chain rule to transform the partial derivative we find that the components of the vector field change just as displayed in (\ref{Tensor_Transf.}). Therefore, the vector fields on a manifold can be interpreted as differential operators that act on the space of functions over the manifold. Furthermore, the partial derivatives $\{\partial_\mu\}$ provide a basis for the tangent space at each point, forming the so-called coordinate frame. For example, on the 2-sphere we can say that $\{\partial_\theta, \partial_\phi\}$ is a coordinate frame, where $\theta$ is the polar angle while $\phi$ denotes the azimuthal angle.

A metric $\bl{g}$\label{Metric} is a symmetric non-degenerate map that act on two vector fields and gives a function over the manifold. In this thesis it will always be assumed that the manifold is endowed with a metric, hence the pair $(M,\bl{g})$ will sometimes be called the manifold. In particular, note that the Minkowski manifold is $(\mathbb{R}^4,\eta_{\mu\nu})$. The components of the metric on a coordinate frame are denoted by $g_{\mu\nu}= \bl{g}(\partial_\mu,\partial_\nu)$.   By conveniently choosing a coordinate frame, we can always manage to put the matrix $g_{\mu\nu}$ in a diagonal form such that all slots are $\pm1$ at some arbitrary point $p\in M$, $g_{\mu\nu}\mapsto g'_{\mu\nu}= \diag(1,1,\ldots,-1,-1,\ldots)$. The modulus of the metric trace when it is in such diagonal form is called the signature of the metric and denoted by $s$, $s= |\Sigma_\mu\, g'_{\mu\mu}|$\index{Signature}. Denoting by $n$ the dimension of the manifold then if $s=n$ the metric is said to be Euclidean, for $s=(n-2)$ the signature is Lorentzian and if $s=0$ the metric is said to have split signature. In Riemannian geometry it is customary to low and raise indices using the metric, $g_{\mu\nu}$, and its inverse, $g^{\mu\nu}$.

The partial derivative of a scalar function, $\partial_\mu f\equiv \nabla_\mu f$, is a tensor. But, as discussed in the preceding section, when acting on tensors this partial derivative must be replaced by the covariant derivative, defined on equations (\ref{chr}) and (\ref{Cov.Deriv.Tens}). In the formal jargon, this tensorial derivative is called a connection. Particularly, the connection defined by (\ref{chr}) and (\ref{Cov.Deriv.Tens}) is named the Levi-Civita connection.
The covariant derivative share many properties with the usual partial derivative, it is linear and obey the Leibniz rule. However, these two derivatives also have a big difference: while the partial derivatives always commute, the covariant derivatives generally do not. More precisely it is straightforward to prove that:
\begin{gather}
  (\nabla_\mu \nabla_\nu - \nabla_\nu \nabla_\mu)\,V^\rho \,=\, R^\rho_{\phantom{\rho}\sigma\mu\nu}\,V^\sigma\,, \label{Cov.Deriv.Commut}\\
  R^\rho_{\phantom{\rho}\sigma\mu\nu} \,\equiv\, \partial_\mu\Gamma^\rho_{\sigma\nu} - \partial_\nu\Gamma^\rho_{\sigma\mu} + \Gamma^\rho_{\kappa\mu}\Gamma^\kappa_{\sigma\nu} - \Gamma^\rho_{\kappa\nu}\Gamma^\kappa_{\sigma\mu}\,.\label{Riemann tensor}
\end{gather}
The object $R^\rho_{\phantom{\rho}\sigma\mu\nu}$ is called the Riemann tensor. Although its definition was made in terms of the non-tensorial Christoffel symbols, $R^\rho_{\phantom{\rho}\sigma\mu\nu}$ is indeed a tensor, as the left hand side of equation (\ref{Cov.Deriv.Commut}) is a tensor. The Riemann tensor is also called the curvature tensor, because it measures the curvature of the manifold\footnote{Actually it measures the curvature of the tangent bundle.}. In particular, a manifold is flat if, and only if, the Riemann tensor vanishes. Defining
$R_{\rho\sigma\mu\nu}= g_{\rho\kappa}R^\kappa_{\phantom{\kappa}\sigma\mu\nu}$ then, after some algebra, it is possible to prove that this tensor has the following symmetries.
\begin{equation}\label{Riemann-Symm}
  R_{\rho\sigma\mu\nu}= R_{[\rho\sigma][\mu\nu]} \,\,;\; R_{\rho\sigma\mu\nu}=R_{\mu\nu\rho\sigma} \,\,;\; R_{\rho[\sigma\mu\nu]}=0 \,\,;\; \nabla_{[\kappa} R_{\rho\sigma]\mu\nu}=0
\end{equation}
Particularly, the last two symmetries above are called Bianchi identities. There are other important tensors that are constructed out of the Riemann curvature tensor:
\begin{gather*}\label{WeylTensor}
  R_{\mu\nu}\,\equiv\, R^\rho_{\phantom{\rho}\mu\rho\nu} \quad;\quad R\,\equiv\, g^{\mu\nu}R_{\mu\nu} \,=\, R^{\nu}_{\phantom{\nu}\nu} \\
   C_{\rho\sigma\mu\nu} \equiv R_{\rho\sigma\mu\nu} - \frac{2}{n-2} \left( g_{\rho[\mu}R_{\nu]\sigma}- g_{\sigma[\mu}R_{\nu]\rho} \right) + \frac{2}{(n-1)(n-2)} R \, g_{\rho[\mu}g_{\nu]\sigma}\,.
\end{gather*}
These tensors are respectively called Ricci tensor, Ricci scalar and Weyl tensor\index{Weyl tensor}. The Ricci tensor is symmetric, while the Weyl tensor has all the symmetries of equation (\ref{Riemann-Symm}) except for the last one, the differential Bianchi identity. The Weyl tensor will be of central importance in this piece of work, since the main goal of this thesis is to define an algebraic classification for this tensor and relate such classification with integrability properties. The Weyl tensor has two landmarks: it is traceless, $C^\rho_{\phantom{\rho}\sigma\rho\nu}= 0$, and it is invariant under conformal transformations, \textit{i.e.}, if we transform the metric as $g_{\mu\nu}\mapsto \Omega^2g_{\mu\nu}$ then the tensor $C^\rho_{\phantom{\rho}\sigma\mu\nu}$ remains invariant.

\section{Geodesics}
Given two points $p_1$ and $p_2$ on a manifold $(M,\bl{g})$, the trajectory of minimum length connecting these points is called a geodesic. If $x^\mu(\tau)$ is a curve joining these points, with $x^\mu(\tau_i)= p_i$, then its length is given by:
$$ \Delta(\tau_1,\tau_2) \,=\, \int_{\tau_1}^{\tau_2}\,\sqrt{g_{\mu\nu}\,\frac{dx^\mu}{d\tau}\frac{dx^\nu}{d\tau}}\,\,d\tau\,. $$
Note that $\Delta$ is invariant under the change of parametrization of the curve. Let us exploit this freedom adopting the arc length, $s(\tau)\equiv \Delta(\tau_1,\tau)$, as the curve parameter. Then performing a standard variational calculation we find that the curve of minimum length connecting $p_1$ and $p_2$ satisfies the following differential equation known as the geodesic equation:
\begin{equation}\label{Geodesic Eq.}
  \frac{d^2x^\rho}{ds^2} \,+\, \Gamma^\rho_{\mu\nu} \frac{dx^\mu}{ds}\frac{dx^\nu}{ds} \,=\,0
\end{equation}
Note that using cartesian coordinates on the Minkowski space we have that $\Gamma^\rho_{\mu\nu}= 0$, so that eq. (\ref{Geodesic Eq.}) implies that the geodesics of flat space are straight lines, as it should be. Using equations (\ref{Cov.Deriv}) and (\ref{Geodesic Eq.}) we find that the geodesic equation can be elegantly expressed by:
\begin{equation}\label{Geodesic Eq. Cov}
 T^\mu\,\nabla_{\mu}\, T^\nu \,=\,0\,, \quad  T^\mu\equiv \frac{dx^\mu}{ds}\,.
\end{equation}
Note that the vector field $T^\mu$ is tangent to the curve. If instead of the arc length parameter, $s$, we have used another parameter $\tau$, we would have found the equation $N^\mu\nabla_{\mu}N^\nu = fN^\nu$, where $N^\mu\equiv\frac{dx^\mu}{d\tau}$ and $f$ is some function. The parameters $\tau'$ such that $f=0$ are called affine parameters. It is simple matter to verify that the affine parameters are all linearly related to the arc length, $\tau'=a\,s+b$ with $a\neq0$ and $b$ being constants. Physically, the arc length $s$ of a time-like curve (geodesic or not) represents the proper time of the observer following this curve. In general relativity, free massive particles follow time-like geodesics, whereas free massless particles describe null geodesics. It is worth remarking that  here a particle is said to be free when the only force acting on it is the gravitational force.

\begin{figure}[h]
	\centering
		\includegraphics[width=10cm]{{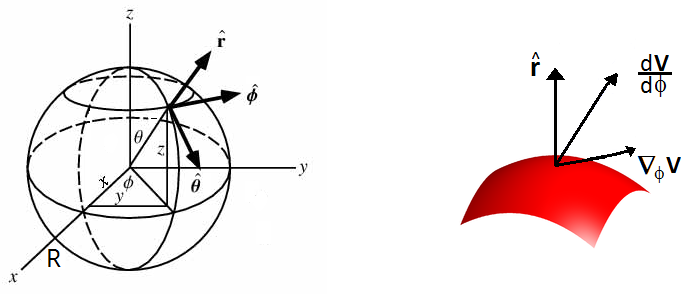}}
	\caption{\footnotesize{Sphere embedded in the 3-dimensional Euclidean space. The vector fields $\hat{\bl{\theta}}$ and $\hat{\bl{\phi}}$ are tangent to the sphere. On the right hand side it is illustrated that the covariant derivative of a vector field tangent to the sphere is the projection of the ordinary derivative onto the plane tangent to the spherical surface. }}
	\label{Fig. Sphere}
\end{figure}

In order to gain some intuition on the formalism introduced so far, let us go back to the example of the $2$-sphere. Let $S$ be a sphere of radius $r$ embedded on the 3-dimensional Euclidean space $\mathbb{R}^3$, as depicted in figure \ref{Fig. Sphere}. The metric of the 3-dimensional space is $ds^2 = dx^2 + dy^2 + dz^2$. Then, the points on the sphere can be locally labeled by the coordinates $\theta$ and $\phi$ related to the cartesian coordinates by $x= r\sin\theta\cos\phi$,   $y= r\sin\theta\sin\phi$ and  $z= r\cos\theta$. Inserting these expressions in the 3-dimensional metric and assuming that $r$ is constant we are led to the metric of the $2$-sphere, $ds^2= r^2d\theta^2 + r^2 \sin^2\theta\, d\phi^2$. Once we have this metric we can compute its associated curvature by means of equation (\ref{Riemann tensor}). In particular, the Ricci scalar is found to be $R= 2/r^2$. So, the bigger the radius the smaller the curvature. Now, let $\bl{V}$ be a vector field tangent to the sphere, $\bl{V}\cdot \hat{\bl{r}}= 0$. Where the dot denotes the inner product of $\mathbb{R}^3$. Then, the covariant derivative of $\bl{V}$ along some curve tangent to the sphere is just the projection of the ordinary derivative of $\bl{V}$ along this curve onto the tangent planes of the sphere, see figure \ref{Fig. Sphere}. For instance, the covariant derivative of $\bl{V}$ along the great circle $\theta= \frac{\pi}{2}$ is $\nabla_{\phi}\bl{V} = \frac{d\bl{V}}{d\phi} - (\hat{\bl{r}}\cdot \frac{d\bl{V}}{d\phi}) \hat{\bl{r}}$. Particularly, one can prove that $\nabla_{\phi} \hat{\bl{\phi}}= 0$, which implies that such great circle is a geodesic curve. In general, all great circles of the $2$-sphere are geodesic curves.

%
%

\section{Symmetries and Conserved Quantities}\label{Sec._Symmetries}

Suppose that a space-time is symmetric on the direction of the coordinate vector $\boldsymbol{K}= \partial_1$, \textit{i.e.}, it looks the same irrespective of the value of the coordinate $x^1$. This implies that in this coordinate system we have $\partial_1\,g_{\mu\nu}= 0$. Then, using the fact that $K^\mu= \delta^\mu_1$ and the expression for the Christoffel symbol in terms of the derivatives of the metric, we easily find that:
\begin{equation}\label{KillingEq}
 \nabla_\mu\,K_{\nu} \,=\, \frac{1}{2}\,\left( \partial_\mu \,g_{\nu 1} \,-\, \partial_\nu\,g_{\mu 1}  \right) \;\Rightarrow\;\;  \nabla_\mu\,K_{\nu} \,+\, \nabla_\nu\,K_{\mu}\,=\,0\,.
\end{equation}
Conversely, if a vector field $\boldsymbol{K}$ satisfies $\nabla_{(\mu}K_{\nu)}= 0$ then it is simple matter to prove that on a coordinate system in which $\bl{K}$ is a coordinate vector the relation $K^\mu\partial_\mu g_{\rho\sigma}= 0$ holds. The equation $\nabla_{(\mu}K_{\nu)}= 0$ is the so-called Killing equation and the vector field $\boldsymbol{K}$ is called a Killing vector field\index{Killing vector}. In general the symmetries of a space-time are not obvious from the expression of the metric. For example, the Minkowski space-time has 10 independent Killing vector fields, although only 4 symmetries are obvious from the usual expression of this metric. That is the reason why the Killing vectors are so important, they characterize the symmetries of a manifold without explicitly using coordinates.

From the Noether theorem it is known that continuous symmetries are associated to conserved charges. So the Killing vector fields must be related to conserved quantities. Indeed, if $\boldsymbol{K}$ is a Killing vector and $\bl{T}$ is the affinely parameterized vector field tangent to a geodesic curve then the scalar $T^\mu K_\mu$ is constant along such geodesic, $T^\nu\nabla_\nu(T^\mu K_\mu)= T^\nu T^\mu \nabla_{(\nu} K_{\mu)}= 0$. Physically, this means that along free-falling orbits the component of the  momentum along the direction of a Killing vector is conserved. The use of these conserved quantities are generally quite helpful to find the solutions of the geodesic equation. For instance, since the Schwarzschild space-time has 4 independent Killing vectors it follows that the geodesic trajectories can be found without solving the geodesic equation. But, in addition to the Killing vectors, there are other tensors associated with the symmetries of a manifold. For example, let $K_{\nu_1\nu_2\ldots\nu_p}$ be a completely symmetric tensor obeying to the equation
$$  \nabla_{(\mu}\, K_{\nu_1\nu_2\ldots\nu_p)} \,=\, 0\,,  $$
then the scalar $K_{\nu_1\ldots\nu_p}T^{\nu_1}\ldots T^{\nu_q}$ is conserved along the geodesic generated by $\boldsymbol{T}$. The tensor $K_{\nu_1\nu_2\ldots\nu_p}$ is called a Killing tensor of order $p$.

Another important class of tensors associated to symmetries is formed by the Killing-Yano\index{Killing-Yano tensor} (KY) tensors. These are skew-symmetric tensors, $Y_{\nu_1\nu_2\ldots\nu_p}= Y_{[\nu_1\nu_2\ldots\nu_p]}$, that obey to the equation $\nabla_\mu Y_{\nu_1\ldots\nu_p} + \nabla_{\nu_1} Y_{\mu\ldots\nu_p}= 0$. If $T^\mu$ generates an affinely parameterized geodesic then $Y_{\nu_1\nu_2\ldots\nu_p}T^{\nu_p}$ is covariantly constant along the geodesic. Note also that if $Y_{\mu\nu}$ is a Killing-Yano tensor then $K_{\mu\nu}= Y_{\mu}^{\phantom{\mu}\rho}Y_{\rho\nu}$ is a Killing tensor of order two. Although we can always construct Killing tensors out of KY tensors, not all Killing tensors are made from KY tensors \cite{Collinson}. For more details about KY tensors see \cite{FrolovBook}.

There are also tensors associated to scalars conserved only along null geodesics. A totally symmetric tensor $\bl{L}$ is said to be a conformal Killing tensor (CKT) when the equation $\nabla_{(\nu}L_{\mu_1\ldots\mu_p)}= g_{(\nu\mu_1}A_{\mu_2\ldots\mu_p)}$ holds for some tensor $\boldsymbol{A}$. If $\boldsymbol{L}$ is a CKT of order $p$ and $\bl{l}$ is tangent to an affinely parameterized null geodesic then the scalar $L_{\mu_1\ldots\mu_p}l^{\mu_1}\ldots l^{\mu_p}$ is constant along such geodesic. It is not so hard to prove that if $\bl{K}$ is a Killing tensor on the manifold $(M,\boldsymbol{g})$ then $L_{\mu_1\ldots\mu_p}= \Omega^{2p}\,K_{\mu_1\ldots\mu_p}$ is a CKT of the manifold $(M,\tilde{\boldsymbol{g}})$ with $\tilde{g}_{\mu\nu}= \Omega^2\,g_{\mu\nu}$. In the same vein, we say that a completely skew-symmetric tensor $\bl{Z}$ is a conformal Killing-Yano (CKY) tensor if it satisfies the equation $\nabla_{(\nu}Z_{\mu_1)\mu_2\ldots\mu_p}= g_{\nu[\mu_1}H_{\mu_2\ldots\mu_p]} + g_{\mu_1[\nu}H_{\mu_2\ldots\mu_p]}$ for some tensor $\boldsymbol{H}$ \cite{FrolovBook}.

Generally it is highly non-trivial to guess whether a manifold possess a Killing tensor, a KY tensor as well as its conformal versions. Therefore, such tensors are said to represent hidden symmetries\index{Hidden symmetries}. Since the Kerr metric has just 2 independent Killing vectors it is not possible to find the geodesic trajectories using only these symmetries. But, in 1968, B. Carter was able to discover another conserved quantity that enabled him to solve the geodesic equation \cite{Carter-Kerr}. Two years later Walker and Penrose demonstrated that this ``new'' conserved scalar is associated to a Killing tensor of order two \cite{Walk.Penr.}. Thereafter it has been proved that this Killing tensor is the ``square'' of a KY tensor \cite{Collinson}.

\section{Einstein's Equation}

Hopefully we already convinced ourselves that the gravitational field is represented by the metric, $g_{\mu\nu}$, of a curved manifold $(M,\boldsymbol{g})$. But we do not know yet how to find this metric given the distribution of masses throughout the space-time. For example, in the Newtonian theory the gravitational field is represented by a scalar, the gravitational potential $\phi$, whose equation of motion is $\nabla^2\phi= 4\pi G\varrho$, where $G$ is the gravitational constant and $\varrho$ is the mass density. Analogously, we need to find the equation of motion for the metric $g_{\mu\nu}$. It can already be expected that, differently from the Newtonian theory, the source of gravity is not just the mass density, but the energy content as a whole, since in relativity mass and energy are equivalent.

A wise path to find the correct field equation satisfied by $g_{\mu\nu}$ is to guess a reasonable action representing the gravitational field and its interaction with the other fields. Let us start analyzing how the metric couples to the matter fields. Well, this is simple: given the action of a field in special relativity we just need to replace the Minkowski metric by $\boldsymbol{g}$ and substitute the partial derivatives by covariant derivatives. There is, however, an important detail missing. In order for the action to look the same in any coordinate system we must impose for it to be a scalar. It is simple matter to prove that the volume element of space-time $d^4x= dx^0dx^1dx^2dx^3$ is not invariant under coordinate transformations. This can be fixed by taking $\sqrt{|g|}d^4x$ as the volume element, with $g$ being the determinant of $g_{\mu\nu}$. Regarding the action of the gravitational field, the simplest non-trivial scalar that can be constructed out of the metric is the Ricci scalar $R$, defined in section \ref{Sec._RiemannianGeom}. Therefore we find that a reasonable action is:
\begin{equation}\label{ActionRG}
  S \,=\, \frac{1}{16\pi G}\int R\,\sqrt{|g|}d^4x \,+\,  \int \mathcal{L}_m(\varphi_i,\nabla_{\mu}\varphi_i,g_{\mu\nu})\, \sqrt{|g|}d^4x\,.
\end{equation}
Where $\mathcal{L}_m$ is the Lagrangian density of the matter fields $\varphi_i$. Then, using the least action principle, we can prove that the equation of motion for the field $g_{\mu\nu}$ is given by the so-called Einstein's equation \cite{FrolovBook}:
\begin{equation}\label{Eisntein.Eq}
  R_{\mu\nu} \,-\, \frac{1}{2}R\,g_{\mu\nu} \,=\, 8\pi G\,T_{\mu\nu}\;;\quad T^{\mu\nu}\,\equiv\, \frac{2}{\sqrt{|g|}}\,\frac{\delta S_m}{\delta g_{\mu\nu}}\,.
\end{equation}
The symmetric tensor $T_{\mu\nu}$ is the energy-momentum tensor of the matter fields. Particularly, in vacuum we have $T_{\mu\nu}=0$. Einstein's equation matches  the geometry of the space-time, on the left hand side, to the energy content, on the right hand side. Note that this equation is highly non-linear, since the Ricci tensor and the Ricci scalar depends on the square of the metric as well as on the inverse of the metric. This non-linearity can be easily grasped using physical intuition. Since the graviton carries energy it produces gravity, which then interact with this graviton and so on. In other words, the graviton interacts with itself. This differs from classical electrodynamics, where the photon has zero electric charge and, therefore, generates no electromagnetic field.

As a simple and important example let us work out the case where just the electromagnetic field is present. In relativistic theory this field is represented by a co-vector $A_{\mu}$, the vector potential. From this field one can construct the skew-symmetric tensor $F_{\mu\nu}= \nabla_{\mu}A_{\nu}- \nabla_{\nu}A_{\mu}$. The action of the electromagnetic field is given by:
\begin{equation}\label{Action_EM}
 S_{em} \,=\, -\frac{1}{16\pi}\, \int g^{\mu\rho}g^{\nu\sigma} F_{\mu\nu}F_{\rho\sigma} \, \sqrt{|g|}d^4x\,.
\end{equation}
Taking the functional derivative of this action with respect to the metric yields the following energy-momentum tensor for the electromagnetic field:
\begin{equation}\label{Tab-EM}
 \mathcal{T}_{\mu\nu} \,=\, \frac{1}{4\pi} \, \left(  F_{\mu\sigma} F_{\nu}^{\phantom{\nu}\sigma} - \frac{1}{4}\, g_{\mu\nu} F^{\rho\sigma}F_{\rho\sigma} \right)\,.
\end{equation}
Furthermore, computing the functional derivative of the action (\ref{Action_EM}) with respect to $A_{\mu}$ and equating to zero yields $\nabla^\nu F_{\mu\nu}= 0$, which is equivalent to Maxwell's equations without sources. The set of equations $R_{\mu\nu} - \frac{1}{2}Rg_{\mu\nu}= 8\pi G\mathcal{T}_{\mu\nu}$, $\nabla^\nu F_{\mu\nu}= 0$ and $F_{\mu\nu}= 2\nabla_{[\mu}A_{\nu]}$ is called Einstein-Maxwell's equations.

In this section we have considered that the gravitational Lagrangian is given by the Ricci scalar $R$, which yields Einstein's theory. Although general relativity has had several experimental confirmations it is expected that for really intense gravitational fields this Lagrangian shall be corrected by higher order terms, such as $R^2$, $R^{\mu\nu\rho\sigma}R_{\mu\nu\rho\sigma}$, $\partial_\mu R \,\partial^\mu R$ and so on. Indeed, string theory predicts that the gravitational action contains terms of all orders on the curvature. In this picture the Einstein-Hilbert action, $S= \frac{1}{16\pi G}\int R\sqrt{|g|}d^nx$, is just a weak field approximation for the complete action.

\section{Differential Forms}\label{Sec. DiffForms}
Just as in section \ref{Sec._RiemannianGeom} it was valuable to say that the tangent space is spanned by the differential operators $\partial_\mu$, it is also fruitful to assume that the dual of this space, the space of linear functionals on $T_p M$, is generated by the differentials $dx^\mu$. Thus if $A_{\mu}$ are the components of a co-vector field in the coordinates $\{x^\mu\}$, then we shall represent the abstract tensor $\boldsymbol{A}$ as follows:
$$  \boldsymbol{A} \,=\, A_{\mu}\,dx^\mu \,.$$
With such definition it follows that $A_{\mu}$ will properly transform under coordinate changes, see eq. (\ref{Tensor_Transf.}). Therefore, an arbitrary tensor $\boldsymbol{T}$ has the following abstract representation:
$$ \boldsymbol{T} \,=\,  T^{\mu_1\ldots\mu_p}_{\phantom{\mu_1\ldots\mu_p}\nu_1\ldots\nu_q}\, \partial_{\mu_1}\otimes\ldots\otimes\partial_{\mu_p}\otimes dx^{\nu_1}\otimes\ldots \otimes dx^{\nu_q}\,.  $$
Since formally $dx^\mu$ is a linear functional on the space of vector fields, its action on a vector field gives a scalar. Such action is defined by $dx^\mu(\partial_\nu)= \delta^\mu_{\,\nu}$, so that if $\boldsymbol{A}$ is co-vector and $\boldsymbol{V}$ is a vector then $\boldsymbol{A}(\boldsymbol{V})=A_{\mu}V^{\mu}$.

A particularly relevant class of tensors are the so-called differential forms, which are tensors with all indices down and totally skew-symmetric. For instance,  $F_{\mu_1\ldots\mu_p}= F_{[\mu_1\ldots\mu_p]}$ is called a $p$-form and the vectorial space generated by all $p$-forms at some point $x\in M$ is denoted by $\wedge^pM|_x$. A fundamental operation when dealing with forms is the exterior product, whose definition is:
\begin{equation*}
  \boldsymbol{F}\wedge \boldsymbol{H} \,=\, \frac{(p+q)!}{p!\,q!}\, F_{[\mu_1\ldots\mu_p}\,H_{\nu_1\ldots\nu_q]}\,dx^{\mu_1}\otimes\ldots \otimes dx^{\mu_p}\otimes dx^{\nu_1}\otimes\ldots \otimes dx^{\nu_q}\,.
\end{equation*}
Where $\boldsymbol{F}$ is a $p$-form and $\boldsymbol{H}$ is a $q$-form, so that their exterior product yields a $(p+q)$-form. As an example note that the following relation holds:
\begin{align*}
 dx^1\wedge dx^2\wedge dx^3 \,=\,&( dx^1\otimes dx^2 \otimes dx^3 + dx^2\otimes dx^3 \otimes dx^1 + dx^3\otimes dx^1 \otimes dx^2 + \\
   \,-&\phantom{(} dx^2\otimes dx^1 \otimes dx^3  -  dx^3\otimes dx^2 \otimes dx^1 - dx^1\otimes dx^3 \otimes dx^2 \,)\,.
\end{align*}
In $n$ dimensions the set $\{1,dx^{\mu_1}, dx^{\mu_1}\wedge dx^{\mu_2},\ldots,dx^1\wedge\ldots\wedge dx^n\}$, which contains $2^n$ elements, forms a basis for the space of differential forms, called exterior bundle. In particular, a general $p$-form $\boldsymbol{F}$ can be written as:
$$  \boldsymbol{F} \,=\, \frac{1}{p!} \, F_{\mu_1\ldots\mu_p} \, dx^{\mu_1}\wedge dx^{\mu_2}\wedge\ldots\wedge dx^{\mu_p}\,. $$
A $p$-form is called simple\index{Simple form} when it can be expressed as the exterior product of $p$ 1-forms. For instance, every $n$-form is simple.

Another important operation involving differential forms is the interior product\index{Interior product}\label{InteriorProduct}, which essentially is the contraction of a differential form $\bl{F}$ with a vector field $\bl{V}$ yielding another form $\bl{H}\equiv \bl{V}\lrcorner \bl{F}$. If $\bl{F}$ is a $p$-form then the interior product of $\bl{V}$ and $\bl{F}$ is the $(p-1)$-form defined by $H_{\mu_2\ldots\mu_p}\equiv V^{\mu_1}F_{\mu_1\mu_2\ldots\mu_p}$. When $\bl{V}\lrcorner \bl{F}= 0$ we say that the differential form $\bl{F}$ annihilates $\bl{V}$.

Suppose that $(M,\bl{g})$ is an $n$-dimensional manifold. Then we can introduce the so-called Levi-Civita symbol $\varepsilon_{\mu_1\ldots\mu_n}$, defined as the unique object, up to a sign, that is totally skew-symmetric and normalized as $\varepsilon_{12\ldots n}= \pm1$. Although this symbol is not a tensor we can use it to define the important tensor $\boldsymbol{\epsilon}$ called the volume-form\index{Volume-form}\label{VolumeForm} and defined by \cite{SeanCarrol}:
$$ \epsilon_{\mu_1\ldots\mu_n} \,\equiv\, \sqrt{|g|}\,\varepsilon_{\mu_1\ldots\mu_n}\;\Rightarrow\quad \boldsymbol{\epsilon}\,=\, \sqrt{|g|}\,dx^{1}\wedge \ldots\wedge dx^{n}\,, $$
where $g$ denotes the determinant of the matrix $g_{\mu\nu}$. After some algebra it can be proved that this tensor obeys to the following useful identity \cite{SeanCarrol}:
\begin{equation}\label{EE_DELTA}
  \epsilon^{\mu_1\ldots\mu_p\,\nu_{p+1}\ldots\nu_n}\,\epsilon_{\mu_1\ldots\mu_p\,\sigma_{p+1}\ldots\sigma_n}\;=\;p!(n-p)!\, (-1)^{\frac{n-s}{2}} \delta_{\sigma_{p+1}}^{\;[\nu_{p+1}}\ldots\delta_{\sigma_{n}}^{\;\nu_{n}]}\,.
\end{equation}
Where $s$ is the signature of the metric. Moreover, the volume-form can be used to define an important operation called Hodge dual\index{Hodge dual}. The Hodge dual of a $p$-form $\bl{F}$ is a $(n-p)$-form denoted by $\star\bl{F}$ and defined by\label{HodgeDual}:
 \begin{equation}\label{HodgeDual-Def}
 \left(\star F\right)_{\mu_1\ldots\mu_{n-p}}\;=\; \frac{1}{p!} \,\epsilon^{\nu_1\ldots\nu_{p}}_{\phantom{\nu_1\ldots\nu_{p}}\mu_1\ldots\mu_{n-p}} \, F_{\nu_1\ldots\nu_p}\,.
 \end{equation}

%

Finally, the last relevant operation on the space of forms is the exterior differentiation, $d$. This differential operation maps $p$-forms into $(p+1)$-forms as follows:
$$ d\boldsymbol{F} = \frac{1}{p!}\, \partial_\nu F_{\mu_1\ldots\mu_p}\, dx^{\nu}\wedge dx^{\mu_1}\wedge\ldots\wedge dx^{\mu_p}\,.$$
Although we have used the partial derivative, we could have used the covariant derivative and the result would be the same, because of the symmetry $\Gamma^\rho_{\mu\nu}= \Gamma^\rho_{\nu\mu}$ of the Christoffel symbol. Therefore, the term on the right hand side of the above equation is indeed a tensor. A remarkable property of the exterior derivative is that its square is zero, $d(d\boldsymbol{F})= 0$, which stems from the commutativity of the partial derivatives.

As an application of this formalism note that the source-free Maxwell's equations can be elegantly expressed in terms of differential forms. The vector potential $A_{\mu}$ is a 1-form, $\boldsymbol{A}= A_{\mu}dx^\mu$. The field strength, $F_{\mu\nu}\equiv \nabla_{\mu}A_{\nu}- \nabla_{\nu}A_{\mu}$, is nothing more than the exterior derivative of $\boldsymbol{A}$, $\boldsymbol{F}= d\boldsymbol{A}$. In particular, this implies that $d\boldsymbol{F}= 0$. The missing equation is $\nabla^\nu F_{\mu\nu}= 0$, which can be proved to be equivalent to $d(\star \boldsymbol{F})= 0$.  Hence, in the absence of sources, the electromagnetic field is represented by a 2-form, $\boldsymbol{F}$, obeying the equations $d\boldsymbol{F}= 0$ and $d(\star \boldsymbol{F})= 0$.

\section{Cartan's Structure Equations}\label{Sec. Cartan}
Up to now we have adopted the coordinate frames $\{\partial_\mu\}$ and $\{dx^\mu\}$ as bases for the tangent space and for its dual respectively. Often it is convenient to use a non-coordinate frame $\{\boldsymbol{e}_a= e_a^{\phantom{a}\mu} \partial_\mu\}$, where the index $a$ is not a vectorial index, but rather a label for the $n$ vector fields composing the frame. Associated to this non-coordinate vector frame is the so-called dual frame $\{\boldsymbol{e}^a= e^a_{\phantom{a}\mu} dx^\mu\}$, defined to be such that $\boldsymbol{e}^a(\boldsymbol{e}_b)= \delta^a_{\,b}$. Given a tensor, say $T^\mu_{\phantom{\mu}\nu}$, its components in the frame $\{\boldsymbol{e}_a\}$ are defined by $T^a_{\phantom{a}b}\equiv T^\mu_{\phantom{\mu}\nu} e^a_{\phantom{a}\mu}e_b^{\phantom{b}\nu}$. In particular, note that $g_{ab}= \boldsymbol{g}(\boldsymbol{e}_a,\boldsymbol{e}_b)$. Once fixed the frame $\{\boldsymbol{e}_a\}$, let us define the set of $n^2$ connection 1-forms $\boldsymbol{\omega}^a_{\phantom{a}b}$ by the following relation\label{Connection}:
\begin{equation}\label{w-connect-1-form}
 V^\mu\nabla_\mu \boldsymbol{e}^a  \,=\, -\, \boldsymbol{\omega}^a_{\phantom{a}b}(\boldsymbol{V})\,\boldsymbol{e}^b\,, \quad  \forall\;\textrm{ vector field }\;\bl{V}\,.
\end{equation}
Then expanding $\boldsymbol{e}^a$ in a coordinate frame and using equation (\ref{Riemann tensor}) we can, after some algebra, prove the following identities \cite{chandrasekhar}:
\begin{equation}\label{Cartan Struc}
  d\boldsymbol{e}^a + \boldsymbol{\omega}^a_{\phantom{a}b}\wedge \boldsymbol{e}^b \,=\,0 \quad;\quad \frac{1}{2}R^a_{\phantom{a}bcd} \, \boldsymbol{e}^c\wedge\boldsymbol{e}^d \,=\, d\boldsymbol{\omega}^a_{\phantom{a}b} + \boldsymbol{\omega}^a_{\phantom{a}c}\wedge \boldsymbol{\omega}^c_{\phantom{c}b}\,.
\end{equation}
Where $R^a_{\phantom{a}bcd}$ are the components of the Riemann tensor with respect to the frame $\{\boldsymbol{e}_a\}$. These equations are known as the Cartan structure equations\index{Cartan structure equation}. Moreover, defining the scalars $\omega_{ab}^{\phantom{ab}c}\equiv \boldsymbol{\omega}^c_{\phantom{c}b}(\boldsymbol{e}_a)$ we can easily prove that $\nabla_a \boldsymbol{e}_b = \omega_{ab}^{\phantom{ab}c}\boldsymbol{e}_c$.

Sometimes it is of particular help to work with frames such that $g_{ab}$ is a constant scalar. In this case the components of the connection 1-forms obey to the constraint $\omega_{abc}= -\omega_{acb}$, where $\omega_{abc}\equiv \omega_{ab}^{\phantom{ab}d}\,g_{dc}$. Indeed, using the fact that the metric is covariantly constant along with the Leibniz rule yield:
$$ 0\,=\,  \nabla_c\,\left[\,\bl{g}(\bl{e}_a,\bl{e}_b)\,\right]\,=\, \bl{g}(\nabla_c\bl{e}_a,\bl{e}_b) + \bl{g}(\bl{e}_a,\nabla_c\bl{e}_b) \,=\, \omega_{ca}^{\phantom{ca}d}\,g_{db} + \omega_{cb}^{\phantom{cb}d}\,g_{ad}\,.$$

Just as the language of differential forms provides an elegant and fruitful way to deal with Maxwell's equations, Cartan's structure equations do the same in Riemannian geometry. Particularly, equation (\ref{Cartan Struc}) gives, in general, the quicker way to compute the Riemann tensor of a manifold. For applications and geometrical insights on the meaning of these equations see \cite{Gravitation}.

From the physical point of view, the relevance of Cartan's structure equations stems from its relation with the formulation of general relativity as a gauge theory. It is well-known that, except for gravity, the fundamental interactions of nature are currently described by gauge theories, more precisely Yang-Mills theories. Although not widely advertised, it turns out that general relativity can also be cast in the language of gauge theories\footnote{Actually, the most simple gauge formulation of gravity, called Einstein-Cartan theory, is equivalent to general relativity just in the absence of spin. In the presence of matter with spin the former theory allows a non-zero torsion \cite{GaugeTheories}.}. In this approach the gauge group of gravity is the group of Lorentz transformations, $SO(3,1)$\index{SO(3,1)} \cite{GaugeTheories}. Indeed, those acquainted with the formalism of non-abelian gauge theory will recognize the second identity of (\ref{Cartan Struc}) as the equation defining the curvature associated  to the connection $\bl{\omega}^a_{\ph{a}b}$.

\section{Distributions and Integrability}\label{Sec. Integrab}\index{Distribution}

Let $(M,\bl{g})$ be an $n$-dimensional manifold, then a $q$-dimensional distribution in $M$ is a smooth map that associates to every point $p\in M$ a vector subspace of dimension $q$, $\Delta_p \subset T_pM$. We say that the set of vector fields $\{\bl{V}_i\}$ generates this distribution when they span the vector subspace $\Delta_p$ for every point $p\in M$. For instance, a non-vanishing vector field generates a 1-dimensional distribution. We say that a distribution of dimension $q$ is integrable when there exists a smooth family of submanifolds of $M$ such that the tangent spaces of these submanifolds are $\Delta_p$. This means that locally $M$ admits coordinates $\{x^1,\ldots,x^q,y^1,\ldots,y^{n-q}\}$ such that the vector fields $\{\partial_{x^i}\}$ generate $\Delta_p$. In this case the family of submanifolds is given by the hyper-surfaces of constant $y^\alpha$.

Given a set of $q$ vector fields $\{\bl{V}_i\}$ that are linearly independent at every point then it generates a $q$-dimensional distribution denoted by $Span\{\bl{V}_i\}$\label{Span}. One might then wonder, how can we know if such distribution is integrable? Before answering this question it is important to introduce the Lie bracket\index{Lie bracket}. If $\bl{V}$ and $\bl{Z}$ are vector fields then their Lie bracket is another vector field defined by:
$$ [\bl{V},\bl{Z}] \,\equiv\, V^\mu\nabla_\mu \bl{Z} - Z^\mu\nabla_\mu \bl{V} \,=\, \left( V^\mu\partial_\mu\,Z^\nu - Z^\mu\partial_\mu\,V^\nu  \right) \partial_\nu \,. $$

\begin{figure}[h]
	\centering
		\includegraphics[width=5cm]{{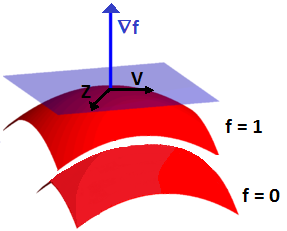}}
	\caption{\footnotesize{The space is foliated by the surfaces of constant $f$. The vector field $\bl{\nabla}f$ is orthogonal to the leafs of the foliation, while $\bl{V}$ and $\bl{Z}$ are tangent.}}
	\label{Fig. foliation}
\end{figure}

As a warming exercise let us work out an example on the $n$-dimensional Euclidian space, $(\mathbb{R}^n,\delta_{\mu\nu})$. Let $f(\bl{r})$ be some function on this manifold, then generally the surfaces of constant $f$ foliate the space, with the leafs being orthogonal to $\bl{\nabla}f$ as depicted in figure \ref{Fig. foliation}. Therefore, if $\bl{V}$ is some vector field tangent to the foliating surfaces then $\bl{V}\cdot\bl{\nabla}f= 0$. Differentiating this last equation we get
$$ \partial_\mu(\bl{V}\cdot\bl{\nabla}f)\,=\, 0\;\;\Rightarrow\;\; (\partial_\mu V^\nu)\,\partial_\nu f \,+\, V^\nu \, \partial_\mu\partial_\nu f \,=\, 0 \,.$$
Therefore, if $\bl{Z}$ is another vector field tangent to the leafs of constant $f$ then
$$ [\bl{V},\bl{Z}]\cdot\bl{\nabla}f = \left( V^\mu\partial_\mu Z^\nu - Z^\mu\partial_\mu V^\nu \right)\,\partial_\nu f  =
 - V^\mu Z^\nu \partial_\mu\partial_\nu f + Z^\mu V^\nu \partial_\mu\partial_\nu f   = 0\,.$$
This means that the Lie bracket of two vector fields tangent to the foliating surfaces yield another vector field tangent to these surfaces. Now let $\bl{\theta}\neq 0$ be a 1-form proportional to $df$, $\bl{\theta}= h\,df$. Then note that a vector field $\bl{V}$ is tangent to the leafs of constant $f$ if, and only if, $\bl{\theta}(\bl{V})= 0$. In addition, note that $d\bl{\theta}\wedge\bl{\theta}= 0$ and that $d(\frac{1}{h}\bl{\theta})= 0$.

The results obtained in the preceding paragraph are just a special case of a well-known theorem called the Frobenius theorem\index{Frobenius theorem}, which states that the distribution generated by the vector fields $\{\bl{V}_i\}$ is integrable if, and only if, there exists a set of functions $C_{ij}^k$ such that $[\bl{V}_i,\bl{V}_j]= C_{ij}^k\,\bl{V}_k$. In other words, this distribution is integrable if, and only if, the vector fields $\bl{V}_i$ form a closed algebra under the Lie brackets \cite{Frankel}.

The Frobenius theorem can be presented in a ``dual'' version, in terms of differential forms. Let $\{\bl{V}_i\}$ be a set of $q$ vector fields generating a $q$-dimensional distribution. Then we can complete this set with more $(n-q)$ vector fields, $\{\bl{U}_\alpha\}$, so that $\{\bl{V}_i,\bl{U}_\alpha\}$ spans the tangent space at every point. Associated to this frame is a dual frame of 1-forms $\{\bl{\omega}^i, \bl{\theta}^\alpha\}$ such that $\bl{\omega}^i(\bl{V}_j)= \delta^i_{\,j}$, $\bl{\omega}^i(\bl{U}_\alpha)= 0$, $\bl{\theta}^\alpha(\bl{V}_i)= 0$ and $\bl{\theta}^\alpha(\bl{U}_\beta)= \delta^\alpha_{\,\beta}$. Note that a  vector field is tangent to the distribution if, and only if, it is annihilated by all the $(n-q)$ 1-forms $\bl{\theta}^\alpha$. The dual version of the Frobenius theorem then states that the distribution generated by $\{\bl{V}_i\}$ is integrable if, and only if,
\begin{equation}\label{Froben. Theo}
  d\bl{\theta}^\alpha\wedge\bl{\theta}^1\wedge\bl{\theta}^2\wedge\ldots \wedge\bl{\theta}^{(n-q)}\,=\,0\quad \;\forall\;\;\alpha\in\{1,\ldots,(n-q)\} \,.
\end{equation}
Defining $\bl{\Theta}\equiv \bl{\theta}^1\wedge\ldots \wedge\bl{\theta}^{(n-q)}$, then note that a vector field $\bl{V}$ is tangent to the distribution generated by $\{\bl{V}_i\}$ if, and only if, $\bl{V}\lrcorner\bl{\Theta}= 0$. Now suppose that there exists a non-zero function $h$ such that $d(h\bl{\Theta})= 0$, then expanding this equation and taking the wedge product with $\bl{\theta}^\alpha$ we arrive at the equation (\ref{Froben. Theo}). Conversely, if the distribution generated by $\{\bl{V}_i\}$ is integrable then, by definition, one can introduce coordinates $\{x^1,\ldots,x^q,y^1,\ldots,y^{n-q}\}$ such that the vector fields $\{\partial_{x^i}\}$ generate this distribution. Since $dy^\alpha(\partial_{x^i})= 0$, it follows that $\bl{\Theta}= \frac{1}{h}(dy^1\wedge\ldots \wedge dy^{n-q})$ for some non-vanishing function $h$, which implies that $d(h\bl{\Theta})= 0$. We proved, therefore, that the distribution annihilated by  $\bl{\Theta}$ is integrable if, and only if, there exists some non-zero function $h$ such that $d(h\bl{\Theta})= 0$. Equivalently, it can be stated that the distribution annihilated by a simple form $\bl{\Theta}$ is integrable if, and only if, there exists a 1-form $\bl{\varphi}$ such that $d\bl{\Theta}= \bl{\varphi}\wedge\bl{\Theta}$.

The integrability of distributions plays an important role in Caratheodory's formulation of thermodynamics. In his formalism, the equilibrium states of a thermodynamical system form a differentiable manifold $\mathcal{M}$. In such a manifold it is defined a global function $U$, the internal energy, and two $1$-forms, $\bl{W}$ and $\bl{Q}$, representing the work done and the received heat, respectively. The first law of thermodynamics is then written as $dU= \bl{Q} - \bl{W}$. A curve in this manifold is called adiabatic if its tangent vector field is annihilated by $\bl{Q}$. According to Caratheodory, the second law of thermodynamics says that in the neighborhood of every point $x\in \mathcal{M}$ there are points $y$ such that there is no adiabatic curve joining $x$ to $y$. He was able to prove that this formulation of the second law guarantees that the distribution annihilated by $\bl{Q}$ is integrable. Particularly, this implies that there exist functions $T$ and $S$ such that $\bl{Q}= T dS$. Physically, these functions are the temperature, $T$, and the entropy, $S$. For more details see \cite{Frankel} and references therein.

\section{Higher-Dimensional Spaces}\label{Sec.HighD}
Einstein's general relativity postulates that we live in a 4-dimensional Lorentzian manifold, which means that the space-time has 3 spatial dimensions
and one time dimension. There are, however, some theories claiming that our space-time can have more spatial dimensions. Particularly, in order to provide a consistent quantum theory, superstring theory requires the space-time dimension to be 10 or 11 \cite{BeckerB}. Which justifies the study of higher-dimensional general relativity.

One might wonder: If these extra dimensions exist then why they have not been perceived yet? A reasonable reason is that these dimensions can be highly wrapped. For example, if we look at a long pipe that is far from us it will appear that it is just a one-dimensional line. But as we get closer and closer to the pipe we will note that it is actually a cylinder, which has two dimensions. An instructive example for understanding the role played by a curled dimension is to solve Schr\"{o}dinger equation for a particle of mass $m$ inside an infinite well. Let the space be 2-dimensional with one of the dimensions being a circle of radius $\textsf{R}$ while the other dimension is open and has an infinite well of size $\textsf{L}$, then the energy spectrum of this system is easily proved to be \cite{Zwiebach}:
$$ E_{p,q} \,=\, \frac{\hbar^2\pi^2}{2m} \,\lef\frac{p}{\textsf{L}} \rig^2 \,+\, \frac{\hbar^2}{2m}\,\lef \frac{q-1}{\textsf{R}}\rig^2 \,, \quad p,q\in\{1,2,3,\ldots\}\,. $$
The first term on the right hand side of this equation is just the regular spectrum of a 1-dimensional infinite well of size $\textsf{L}$, while the second term is the contribution from the extra dimension. Then note that if $\textsf{R}$ is very small, $\textsf{R}\ll \textsf{L}$, then it will be necessary a lot of energy to excite the modes with quantum number $q$. Thus in the limit $\textsf{R}\rightarrow0$ the system will remain in a state with $q=1$, which implies that we retrieve the spectrum of a 1-dimensional well. Thus if the extra dimensions are very tiny the only hope to detect them is through very energetic experiments\footnote{In closed string theory a new phenomenon emerges. Since strings can wrap around a curled dimension there exist winding modes that need little energy to be excited when $\textsf{R}$ is much smaller than the Planck length. Furthermore, due to a symmetry called $T$-duality, in closed string theory very small radius turns out to be equivalent to very large radius.}. Indeed, currently the LHC\footnote{LHC is the abbreviation for Large Hadron Collider, the most energetic particle accelerator in the world.} is probing the existence of extra dimensions.

In addition to the possibility of our universe having extra dimensions and to the obvious mathematical relevance, the study of higher-dimensional curved spaces has other applications. For example, in classical mechanics the phase space of a system is a $2p$-dimensional manifold endowed with a symplectic structure, where $p$ is the number of degrees of freedom \cite{Arnold}. As a consequence, higher-dimensional spaces are also of interest to thermodynamics and statistical mechanics.

It is needless to explain the physical relevance of the Lorentzian signature. But it is worth highlighting that other signatures are also important in physics, let alone in mathematics. Spaces with split signature are of relevance for the theory of integrable systems, Yang-Mills fields and for twistor theory \cite{MasonBook}. Moreover, the Euclidean signature emerges when we make a Wick rotation on the time coordinate in order to make path integrals convergent. The
Euclidean curved spaces are sometimes called gravitational instantons, although it is more appropriate to define a gravitational instanton as a complete 4-dimensional Ricci-flat Euclidean manifold that is asymptotically-flat and whose Weyl tensor is self-dual \cite{Eguchi}. Analogously to the instantons solutions of Yang-Mills theory, gravitational instantons provide a dominant contribution to Feynman path integral, justifying its physical interest \cite{Eguchi}. Non-Lorentzian signatures are also of relevance for string theory.

Given the importance of these topics, the present thesis will investigate some properties of higher-dimensional curved spaces of arbitrary signature. The path adopted here is to work with complexified manifolds so that the results can be carried to any signature by judiciously choosing a reality condition \cite{Trautman}. The technique of using complexified geometry with the aim of extracting results for real spaces can be fruitful and enlightening, an approach that was advocated by McIntosh and Hickman in a series of papers \cite{McIntosh}, where 4-dimensional general relativity was explored using complexified manifolds.

\chapter{Petrov Classification, Six Different Approaches}\label{Chap. Pet}

The Petrov classification is an algebraic classification for the curvature, more precisely for the Weyl tensor, valid in 4-dimensional Lorentzian manifolds. It has been of invaluable relevance for the development of general relativity, in particular it played a prominent role on the discovery of Kerr metric \cite{Kerr}, which is probably the most important solution of general relativity. Furthermore, guided by such classification and a theorem due to Goldberg and Sachs \cite{Goldberg-Sachs}, Kinnersley was able to find all type $D$ vacuum solutions \cite{typeD}, a really impressive accomplishment since Einstein's equation is non-linear. Moreover, this classification contributed for the study of gravitational radiation \cite{Bel,Pirani}, the peeling theorem being one remarkable example \cite{Peeling-Penrose}.

Such classification was created by the Russian mathematician Alexei Zinovievich Petrov in 1954\footnote{Petrov obtained this classification in a previous article published  in 1951 but, as himself acknowledges in \cite{Petrov}, the proofs in this first work were not precise.} \cite{Petrov} with the intent of classifying Einstein space-times. A. Z. Petrov has worked on differential geometry and general relativity, and he has been one of the most important scientists responsible for the spread of Einstein's gravitational theory inside the Soviet Union\footnote{A short biography of A. Z. Petrov can be found in Kazan University's website \cite{Petrov_Biography}.}. In particular, around 1960 he has written a really remarkable book on general relativity that certainly has been of great relevance for the dissemination of this theory on such an isolated nation \cite{Petrov_Book}.

In its original form, this classification consisted only of three types, $I$, $II$ and $III$. Few years later, in 1960, Roger Penrose developed spinorial techniques to general relativity and, as a consequence, has found out that these types could be further refined, adding the types $D$ and $N$ to the classification \cite{Penrose-Spinors}. It is worth mentioning that by the same time Robert Debever and Louis Bel arrived at such refinement by a different path \cite{Bel,Debever}, in particular they have developed an alternative approach to define the Petrov types, the so-called Bel-Debever criteria.

The route adopted by A. Z. Petrov to arrive at his classification amounts to reinterpreting the Weyl tensor as an operator acting on the space of bivectors. As time passed by, several other methods to attack such classification were developed. Since these approaches look very different from each other, it comes as a surprise that all of them are equivalent. The intent of the present chapter is to describe six different ways to attain this classification. As one of the goals of this thesis is to describe an appropriate generalization for the Petrov classification valid in dimensions greater than four, the analysis of these different approaches proves to be important because in higher dimensions many of these methods are not equivalent anymore. Therefore, in order to find a suitable higher-dimensional generalization for the Petrov classification it is helpful to investigate the benefits and flaws of each method in 4 dimensions.

Throughout this chapter it will be assumed that the space-time is a 4-dimensional manifold endowed with a metric of Lorentzian signature, $(M,\bl{g})$. Furthermore, the tangent bundle is assumed to be endowed with the Levi-Civita connection, hence the curvature referred here is with respect to this connection. All calculations are assumed to be local, in a neighborhood of an arbitrary point $p\in M$.

\section{Weyl Tensor as an Operator on the Bivector Space}\label{Petrov_Biv}
In this section the so-called bivector approach will be used to define the Petrov classification. To this end the results of appendices \ref{App._Segre} and \ref{App._NullTetrad} will be necessary, so that the reader is advised to take a look at these appendices before proceeding.

The Weyl tensor\index{Weyl tensor} is the trace-less part of the Riemann tensor, it has the following symmetries (see section \ref{Sec._RiemannianGeom}):
\begin{equation}\label{weylsymm}
 C_{\mu\nu\rho\sigma} = C_{[\mu\nu][\rho\sigma]} = C_{\rho\sigma\mu\nu}\; ;\; \;
 C^\mu_{\phantom{\mu}\nu\mu\sigma} = 0 \; ; \;\; C_{\mu[\nu\rho\sigma]} = 0\,.
 \end{equation}
Skew-symmetric tensors of rank 2 are called a bivectors, $B_{\mu\nu}=B_{[\mu\nu]}$. Since the Weyl tensor is anti-symmetric in the first and second pairs of indices, it follows that this tensor can be interpreted as a linear operator that maps bivectors into bivectors in the following way:
\begin{equation}\label{C_Petrov}
   B_{\mu\nu} \mapsto T_{\mu\nu} = C_{\mu\nu\rho\sigma}B^{\rho\sigma} \;,\; \textrm{where}\; \;B_{\mu\nu}=B_{[\mu\nu]} \; ,T_{\mu\nu}=T_{[\mu\nu]}\,.
 \end{equation}
Studying the possible eigenbivectors of this operator we arrive at the Petrov classification, actually this was the original path taken by A. Z. Petrov \cite{Petrov}. In order to enlighten the analysis it is important to review some properties of bivectors in four dimensions. Let us denote the volume-form of the 4-dimensional Lorentzian manifold $(M, g_{\mu\nu})$ by $\epsilon_{\mu\nu\rho\sigma}$. This is a totally anti-symmetric tensor, $\epsilon_{\mu\nu\rho\sigma} = \epsilon_{[\mu\nu\rho\sigma]}$, whose non-zero components in an orthonormal frame are $\pm1$. It is well-known that it satisfies the following identity \cite{SeanCarrol}:
%
%
\begin{equation}\label{epsiepsi}
\epsilon^{\mu_1\mu_2\nu_1\nu_2}\, \epsilon_{\mu_1\mu_2\sigma_1\sigma_2} = - 2 \,\left(\,\delta_{\sigma_1}^{\nu_1}\,\delta_{\sigma_2}^{\nu_2} \,-\, \delta_{\sigma_2}^{\nu_1}\,\delta_{\sigma_1}^{\nu_2} \,\right)\,.
\end{equation}
By means of the volume-form we can define the Hodge dual\index{Hodge dual} operation that maps bivectors into bivectors. The dual of the bivector $\bl{B}$ is defined by
\begin{equation}\label{dualofbivector}
    \lef\star B\rig_{\mu\nu} \,\equiv\, \frac{1}{2} \epsilon_{\mu\nu\rho\sigma}B^{\rho\sigma}\,.
\end{equation}
Let us denote by $\mathfrak{B}_{\mathbb{C}}$ the complexification of the bivector bundle.
Using equation (\ref{epsiepsi}) it is easy matter to see that the double dual of a bivector is it negative, $[\star(\star B)]_{\mu\nu} = -B_{\mu\nu}$. This implies that the 6-dimensional space $\mathfrak{B}_{\mathbb{C}}$ can be split into the direct sum of the two 3-dimensional eigenspaces of the dual operation.
\begin{equation}\label{oplusc}
\mathfrak{B}_{\mathbb{C}} = \mathfrak{D} \oplus \mathfrak{\overline{D}}
\end{equation}
$$ \mathfrak{D} = \{Z_{\mu\nu} \in  \mathfrak{B}_{\mathbb{C}}\, | \, \lef\star Z\rig_{\mu\nu}= i Z_{\mu\nu}\} \; ; \;
 \mathfrak{\overline{D}} = \{ Y_{\mu\nu} \in  \mathfrak{B}_{\mathbb{C}} \,|\, \lef\star Y\rig_{\mu\nu}= -i Y_{\mu\nu}\} $$
The elements of $\mathfrak{D}$ are called self-dual bivectors\index{Self-dual bivector}, whereas a bivector belonging to $\mathfrak{\overline{D}}$ is dubbed anti-self-dual. By means of the volume-form it is also possible to split the Weyl tensor into a sum of the dual part, $C^+$, and the anti-dual part, $C^-$:
\begin{equation}\label{cpartition}
C_{\mu\nu\rho\sigma} = C^+_{\mu\nu\rho\sigma} + C^-_{\mu\nu\rho\sigma}\; \; ; \;\;C^{\pm}_{\mu\nu\rho\sigma} \equiv \frac{1}{2} \left(C_{\mu\nu\rho\sigma} \mp \frac{i}{2}\,  C_{\mu\nu}^{\phantom{\mu\nu}\alpha\beta} \epsilon_{\alpha\beta\rho\sigma}\right)\,.
\end{equation}
It is then immediate to verify the following relations:
$$ C^+_{\mu\nu\rho\sigma}\,Y^{\rho\sigma}\,=\,0\quad \forall\; \bl{Y}\in\overline{\mathfrak{D}} \quad;\quad C^-_{\mu\nu\rho\sigma}\,Z^{\rho\sigma}\,=\,0\quad \forall\;\bl{Z}\in \mathfrak{D}\,. $$
This means that in order to analyse the action of Weyl tensor on $\mathfrak{B}_{\mathbb{C}}$ it is sufficient to study the action of $C^+$ in $\mathfrak{D}$ and the action of $C^-$ in $\overline{\mathfrak{D}}$. However, by the definition on eq. (\ref{cpartition}), $C^-$ is the complex conjugate of $C^+$, so that it is enough to study just the operator $C^+:\,\mathfrak{D}\rightarrow\mathfrak{D}$. Since this operator is trace-less and act on a 3-dimensional space it follows that it can have the following algebraic types according to the refined Segre classification\index{Segre classification} (see appendix \ref{App._Segre}):
\\
\begin{equation}\label{C^+ types}
\left\{
  \begin{array}{ll}
    \textbf{Type O}&\rightarrow \; C^+\,=\,0  \\
    \textbf{Type I}&\rightarrow \; C^+\, \textrm{ is type } \; [1,1,1|\,] \,\textrm{ or } [1,1|1] \\
    \textbf{Type D}&\rightarrow  \; C^+\, \textrm{ is type } \; [(1,1),1|\,]  \\
    \textbf{Type II}&\rightarrow \; C^+\, \textrm{ is type } \; [2,1|\,]  \\
   \textbf{Type N}&\rightarrow \; C^+\, \textrm{ is type } \; [\,|2,1] \\
    \textbf{Type III}&\rightarrow \; C^+\, \textrm{ is type } \; [\,|\,3]\,.
  \end{array}
\right.
\end{equation}
\\

These are the so-called Petrov types. Therefore, in order to determine the Petrov classification of the Weyl tensor using this approach we must follow four steps: 1) Choose a basis for the space of self-dual bivectors $\mathfrak{D}$; 2) Calculate the action of the operator defined by (\ref{C_Petrov}) in this basis in order to find a $3\times3$ matrix representation for $C^+$; 3) Find the eigenvalues and eigenvectors of this matrix; 4) Use this eigenvalue structure to determine the algebraic type of such matrix according to the refined Segre classification (appendix \ref{App._Segre}) and after this use equation (\ref{C^+ types}).

With the aim of making connection with the forthcoming sections, let us follow some of these steps explicitly. Once introduced a null tetrad frame $\{\bl{l},\bl{n},\bl{m},\overline{\bl{m}}\}$ (see appendix \ref{App._NullTetrad}), the ten independent components of the Weyl tensor can be written in terms of five complex scalars\label{WeylScalars1}:
 \begin{equation}\label{weylscalars}
 \Psi_0 \equiv C_{\mu\nu\rho\sigma}l^\mu m^\nu l^\rho m^\sigma \; ; \; \Psi_1 \equiv C_{\mu\nu\rho\sigma}l^\mu n^\nu l^\rho m^\sigma \; ; \; \Psi_2 \equiv C_{\mu\nu\rho\sigma}l^\mu m^\nu \overline{m}^\rho n^\sigma  $$
$$\Psi_3 \equiv C_{\mu\nu\rho\sigma}l^\mu n^\nu \overline{m}^\rho n^\sigma \; ; \;\Psi_4 \equiv C_{\mu\nu\rho\sigma}n^\mu \overline{m}^\nu n^\rho \overline{m}^\sigma\,.
 \end{equation}
 These are the so-called Weyl scalars. A basis to the space of self-dual bivectors, $\mathfrak{D}$, is given by:
\begin{equation}\label{Z}
    Z^{1}_{\mu\nu} = 2\,l_{[\mu}m_{\nu]} \; ; \; Z^{2}_{\mu\nu} = 2\,\overline{m}_{[\mu}n_{\nu]} \; ; \; Z^{3}_{\mu\nu} = 2\,n_{[\mu}l_{\nu]} + 2\,m_{[\mu}\overline{m}_{\nu]}
\end{equation}
In this basis the representation of operator $C^+: \,\mathfrak{D}\rightarrow\mathfrak{D}$ is
\begin{equation}\label{Cplusmatrix}
[C^+] = 2\left[
                                                                             \begin{array}{ccc}
                                                                               \Psi_2 & \Psi_4 & -2\Psi_3 \\
                                                                               \Psi_0& \Psi_2 & -2\Psi_1 \\
                                                                               \Psi_1 & \Psi_3 & -2\Psi_2 \\
                                                                             \end{array}
                                                                           \right]
\end{equation}
Note that this matrix has vanishing trace, as claimed above equation (\ref{C^+ types}). Thus, in order to get the Petrov type of the Weyl tensor we just have to calculate the Weyl scalars, using eq. (\ref{weylscalars}), plug them on the above matrix and investigate the algebraic type of such matrix.

When the Weyl tensor is type $I$ it is said to be algebraically general, otherwise it is called algebraically special. If the Weyl tensor is type O in all points we say that the space-time is conformally flat, which means there exists a coordinate system such that $g_{\mu\nu}=\Omega^2\eta_{\mu\nu}$. Note that the Petrov classification is local, so that the type of the Weyl tensor can vary from point to point on space-time. In spite of this it is interesting that the majority of the exact solutions has the same Petrov type in all points of the manifold. For instance, all known black holes are type $D$ and the plane gravitational waves are type $N$.

As pointed out at the beginning of this chapter, when Petrov classification first emerged only three types were defined, known as types I, II and III \cite{Pirani,Petrov}. With the contributions of Penrose, Debever and Bel these types were refined as depicted below.
$$ \textrm{I} -\textrm{Refinement}-^\nearrow_\searrow\begin{array}{l}
                   I \\
                   D
                 \end{array}
                 \quad;\quad \textrm{II}  -\textrm{Refinement}-^\nearrow_\searrow \begin{array}{c}
                   II \\
                   N
                 \end{array}  \quad;\quad  \textrm{III} \longrightarrow  III
 $$
Indeed, from the definition of Petrov types presented on equation (\ref{C^+ types}) it is already clear that the type $D$ can be seen as special case of the type $I$, while type $N$ is a specialization of type $II$\footnote{ It is worth mentioning that in ref. \cite{Bel} L. Bel has used a different convention, denoting the types $I$, $D$, $II$, $III$ and $N$ by $I$, $II_a$, $II_b$, $III_a$ and $III_b$ respectively.}.

More details about the bivector method will be given in chapter \ref{Chap. Gen4D}, where this approach will be used to classify the Weyl tensor in any signature, see also \cite{art1}. In particular, chapter \ref{Chap. Gen4D} advocates that the bivector approach is endowed with an enlightening geometrical significance. A careful investigation of the bivector method in higher dimensions was performed in \cite{BivectHighDim}.

\section{Annihilating Weyl Scalars}\label{Sec._Annihil_Scalars}
In this section a different characterization of the Petrov types will be presented. In this approach the different types are featured by the possibility of annihilating some Weyl tensor components using a suitable choice of basis. As a warming up example let us investigate the type $D$. According to eq. (\ref{C^+ types}), in this case the algebraic type of $C^+$ is $[(1,1),1|\,]$, which means that such operator can be put on the diagonal form $\diag(\lambda,\lambda,\lambda')$. But since $\tr(C^+)=0$, we must have $\lambda'= -2\lambda$, hence $C^+ = \diag(\lambda,\lambda,-2\lambda)$. Now, looking at  eq. (\ref{Cplusmatrix}) we see that this is compatible with the Weyl scalars $\Psi_0,\Psi_1,\Psi_3$ and $\Psi_4$ being all zero. In general, each Petrov type enables one to find a suitable basis where some Weyl scalars can be made to vanish.

The Lorentz transformations at point $p\in M$ is the set of linear transformations on tangent space, $T_pM$, which preserves the inner products. These transformations can be obtained by a composition of the following three simple operations in a null tetrad frame $\{\bl{l},\bl{n},\bl{m},\overline{\bl{m}}\}$: \\ \\
\textbf{(i) Lorentz Boost}
\begin{equation}\label{BOOST}
\bl{l}\rightarrow\lambda \bl{l} \; ;\; \; \bl{n}\rightarrow\lambda^{-1}\bl{n} \; ;\; \; \bl{m}\rightarrow e^{i\theta}\bl{m} \; ;\; \; \overline{\bl{m}}\rightarrow e^{-i\theta}\overline{\bl{m}}
\end{equation}
\textbf{(ii) Null Rotation Around $\boldsymbol{l}$}
 \begin{equation}\label{NullRotl}
 \bl{l}\rightarrow \bl{l} \, ; \;\, \bl{n}\rightarrow \bl{n} + w \bl{m} + \overline{w} \,\overline{\bl{m}} + |w|^2 \bl{l} \, ; \;\, \bl{m}\rightarrow \bl{m} + \overline{w} \bl{l} \, ; \;\, \overline{\bl{m}} \rightarrow \overline{\bl{m}} + w\bl{l}
 \end{equation}
\textbf{(iii) Null Rotation Around $\boldsymbol{n}$}
\begin{equation}\label{NullRotn}
\bl{l}\rightarrow \bl{l} + \overline{z} \bl{m} + z \overline{\bl{m}} + |z|^2 \bl{n} \, ; \;\, \bl{n}\rightarrow \bl{n} \, ; \;\, \bl{m} \rightarrow \bl{m} + z \bl{n} \, ; \;\, \overline{\bl{m}} \rightarrow \overline{\bl{m}} + \overline{z}\bl{n}.
\end{equation}
Where $\lambda$ and $\theta$ are real numbers while $z$ and $w$ are complex, composing a total of six real parameters. This should be expected from the fact that the Lorentz group\index{SO(3,1)}, in a 4-dimensional space-time, has 6 dimensions. In order to verify that these transformations do indeed preserve the inner products, note that the metric  $g_{\mu\nu}=2l_{(\mu}n_{\nu)} - 2m_{(\mu}\overline{m}_{\nu)}$ remains invariant by them.

Now let us try to annihilate the maximum number of Weyl scalars by transforming the null tetrad under the Lorentz group. After performing a null rotation around $\bl{n}$ the Weyl scalars change as follows:
\begin{align}
  \nonumber   \Psi_0\rightarrow \Psi'_0(z) \,=\, \Psi_0 \,+\, 4\,z\,\Psi_1\,& +\, 6\,z^2\,\Psi_2 \,+\, 4\,z^3\,\Psi_3 \,+\, z^4\,\Psi_4\; ; \\
  \Psi_1\rightarrow \Psi'_1(z) \,=\, \frac{1}{4}\frac{d}{dz}\Psi'_0(z) \;\;\; ;&\;\; \;  \Psi_2\rightarrow \Psi'_2(z) = \frac{1}{3}\frac{d}{dz}\Psi'_1(z)\; ;\label{Weylscalartransf}\\
  \nonumber \Psi_3\rightarrow \Psi'_3(z) \,=\, \frac{1}{2}\frac{d}{dz}\Psi'_2(z) \;\; ;& \;\;\Psi_4\rightarrow \Psi'_4(z) = \frac{d}{dz}\Psi'_3(z) = \Psi_4\,,
\end{align}
which can be proved using equations (\ref{weylscalars}) and (\ref{NullRotn}). Now if we set $\Psi'_0=0$ we will have a fourth order polynomial in $z$ equal to zero\footnote{Here it is being assumed that $\Psi_4\neq0$, which is always allowed if the Weyl tensor does not vanish identically. Indeed, if the Weyl tensor is non-zero and $\Psi_4=0$ then by means of a null rotation around $\bl{l}$ we can easily make $\Psi_4\neq0$.}. Thus, in general we have four distinct values of the parameter $z$ which accomplish this, call these values $\{z_1, z_2, z_3, z_4\}$. Then the Petrov types can be defined  as follows:
\begin{equation}\label{Types roots}
\left\{
  \begin{array}{ll}
    \textbf{Type O}&\rightarrow \; \textrm{Weyl tensor is zero}  \\
    \textbf{Type I}&\rightarrow \; \textrm{All roots are different} \\
    \textbf{Type D}&\rightarrow  \; \textrm{Two pairs of roots coincide},\,z_1= z_2\neq z_3= z_4 \\
    \textbf{Type II}&\rightarrow \; \textrm{Two roots coincide}, \,z_1= z_2\neq z_3\neq z_4\neq z_1  \\
   \textbf{Type III}&\rightarrow \; \textrm{Three roots coincide}, \, z_1= z_2= z_3\neq z_4\\
    \textbf{Type N}&\rightarrow \; \textrm{All roots coincide}, \, z_1= z_2= z_3= z_4\,.
  \end{array}
\right.
\end{equation}
These four roots define four Lorentz transformations. By means of eq. (\ref{NullRotn}) such transformations lead to four privileged null vector fields $\bl{l}'_i$, which are the ones obtained by performing these transformations on the vector field $\bl{l}$ of the original null tetrad:
\begin{equation}\label{PNDvanishing}
  \bl{l}\,\rightarrow \,\, \bl{l}'_{i} \,=\,  \bl{l} + \overline{z_i}\, \bl{m} + z_i \,\overline{\bl{m}} + |z_i|^2\, \bl{n} \; , \quad i\in\{1,2,3,4\}\,.
  \end{equation}
These real null directions are called the principal null directions (PNDs)\index{Principal null directions(PND)} of the Weyl tensor. Moreover, when $z_i$ is a degenerated root the PND $\bl{l}'_{i}$ is said to be a repeated PND\footnote{The concept of repeated PND can also be extracted from the bivector formalism of section \ref{Petrov_Biv}, as proved on reference \cite{art2}.}. When $z_i$ is a root of order $q$, we say that the associated PND has degeneracy $q$.  By the above definition of Petrov classification it then follows that the Petrov type $I$ admits four distinct PNDs; in type $D$ there are two pairs of repeated PNDs; in type $II$ there exists three distinct PNDs, one being repeated; in type $III$ we have two PNDs, one of which is repeated with triple degeneracy; in type $N$ there is only one PND, this PND in repeated and has degree of degeneracy four.

In type $I$ once we set $\Psi'_0= 0$, by making $z=z_i$, the other Weyl scalars are all different from zero, as can be seen from equations  (\ref{Weylscalartransf}) and (\ref{Types roots}). Then performing a null rotation around $\bl{l}$, which makes $\Psi'_\alpha\rightarrow\Psi''_\alpha$, it is possible to make $\Psi''_4$ vanish while keeping $\Psi''_0= 0$, no other scalars can be made to vanish. Thus in type $I$ the Weyl scalars $\Psi_0$ and $\Psi_4$ can always be made to vanish by a judicious choice of null tetrad. As a further example let us treat the type $D$. In the type $D$ setting $z=z_1$ it follows from equations  (\ref{Weylscalartransf}) and (\ref{Types roots}) that $\Psi'_0 = \Psi'_1 = 0$. After this we can perform a null rotation around $\bl{l}$ in order to set $\Psi''_3 = \Psi''_4 = 0$ while keeping $\Psi''_0 = \Psi''_1 = 0$. The table below sums up what can be accomplished using this kind of procedure.
\begin{table}[!htbp]
\begin{center}
\begin{tabular}{c||c||c}
  \hline
  \hline
  Type $O$ $-$ All  &  Type $II$ $-$ $\Psi_0,\Psi_1,\Psi_4$ & Type $D$ $-$ $\Psi_0,\Psi_1,\Psi_3,\Psi_4$\\
   Type $I$ $-$ $\Psi_0,\Psi_4$    &   Type $III$ $-$ $\Psi_0,\Psi_1,\Psi_2,\Psi_4 $ & Type $N$ $-$ $\Psi_0,\Psi_1,\Psi_2,\Psi_3$ \\
  \hline
  \hline
\end{tabular}\caption{\footnotesize{Weyl scalars that can be made to vanish, by a suitable choice of basis, on each Petrov type.}}\label{Tab._vanishScalars}
\end{center}
\end{table}

Although the definition of the Petrov types given in the present section looks completely different from the one given in section (\ref{Petrov_Biv}) it is not hard to prove that they are actually equivalent. As an example let us work out the type $N$ case. According to the table \ref{Tab._vanishScalars}, if the Weyl tensor is type $N$ it follows that it is possible to find a null tetrad on which the only non-vanishing Weyl scalar is $\Psi_4$. In this basis eq. (\ref{Cplusmatrix}) yield that $C^+$ has the following matrix representation:
$$ C_{N} \,=\, 2\left[
                                                                             \begin{array}{ccc}
                                                                               0 & \Psi_4 & 0 \\
                                                                               0& 0 & 0 \\
                                                                               0 & 0& 0 \\
                                                                             \end{array}
                                                                           \right]\,.
                                                                          $$
Along with appendix \ref{App._Segre} this means that the algebraic type of the operator $C_N$ is $[\,|2,1]$, which perfectly matches the definition of eq. (\ref{C^+ types}). More details about the approach adopted in this section can be found in \cite{chandrasekhar}.


\section{Boost Weight}
In this section the boost transformations\index{Boost weight}, eq. (\ref{BOOST}), will be used to provide another form of expressing the Petrov types. In order to accomplish this we first need to see how the Weyl scalars behave under Lorentz boosts. Inserting eq. (\ref{BOOST}) into the definition of the Weyl scalars, eq. (\ref{weylscalars}), we easily find the following transformation:
\begin{equation}\label{Boostweylscalars}
    \Psi_\alpha\,\longrightarrow\, \gamma^{(2-\alpha)}\,\Psi_\alpha\ \; \; , \quad\, \gamma\equiv e^{i\theta}\,\lambda\; ,\, \; \alpha\in\{0,1,2,3,4\}\,.
\end{equation}
In jargon we say that the Weyl scalar $\Psi_\alpha$ has boost weight $\mathfrak{b}=(2-\alpha)$. Note, particularly, that the maximum boost weight (b.w.) that a component of the Weyl tensor can have is $\mathfrak{b} = 2$, while the minimum is $\mathfrak{b} = -2$.

Given the components of the Weyl tensor on a particular basis, we shall denote by $\mathfrak{b}_+$ the b.w. of the non-vanishing Weyl scalar with maximum boost weight. Analogously, $\mathfrak{b}_-$ denotes the b.w. of the non-vanishing Weyl tensor component with minimum boost weight. For instance, using eq. (\ref{Boostweylscalars}) and table (\ref{Tab._vanishScalars}) we see that if the Weyl tensor is type $III$ then it is possible to find a null frame in which $\mathfrak{b}_+ = -1$. In general we can define the Petrov types using this kind of reasoning, the bottom line is summarized below:
\begin{equation}\label{Types boost}
\left\{
  \begin{array}{ll}
     \textbf{Type I}&\rightarrow \; \textrm{There is a frame in which}\; \mathfrak{b}_+ = + 1 \\
    \textbf{Type II}&\rightarrow  \; \textrm{There is a frame in which} \; \mathfrak{b}_+ = \,0\\
    \textbf{Type III}&\rightarrow \; \textrm{There is a frame in which} \; \mathfrak{b}_+ = -1  \\
   \textbf{Type N}&\rightarrow \; \textrm{There is a frame in which} \; \mathfrak{b}_+ = -2 \\
   \textbf{Type D}&\rightarrow \; \textrm{There is a frame in which} \; \mathfrak{b}_+ = \; \mathfrak{b}_- = 0\\
   \textbf{Type O}&\rightarrow \; \textrm{Weyl tensor vanishes identically} \,.
  \end{array}
\right.
\end{equation}
On the boost weight approach the different Petrov types have a hierarchy: The type $I$ is the most general, type $II$ is a special case of the type $I$, type $III$ is a special case of type $II$ and the type $N$ is a special case of type $III$. The type $D$ is also a special case of type $II$, in this type all non-vanishing components of the Weyl tensor have zero boost weight.

A classification for the Weyl tensor using the boost weight method can be naturally generalized to higher dimensions, which yields the so-called
CMPP\index{CMPP classification} classification \cite{CMPP}. The CMPP classification has been intensively investigated in the last ten years, see, for example, \cite{Rev. Coley, Rev. Ortaggio} and references therein.

\section{Bel-Debever and Principal Null Directions}\label{Sec._Bel-Deb}\index{Principal null directions(PND)}
Few years after the release of Petrov's original article defining his classification, Bel and Debever have, independently, found an equivalent, but quite different, way to define the Petrov types \cite{Bel,Debever}. On such approach the Petrov types are defined in terms of algebraic conditions involving the Weyl tensor and the principal null directions defined in section \ref{Sec._Annihil_Scalars}.

Since the null tetrad frame at a point $p\in M$ forms a local basis for the tangent space $T_pM$, it follows that the Weyl tensor can be expanded in terms of the tensorial product of this basis. Because of the symmetries of this tensor, eq. (\ref{weylsymm}), it follows that the expansion shall be expressed in terms of the following kind of combination:
$$\langle e,v,u,t\rangle_{\mu\nu\rho\sigma} \,\equiv\, 4\,e_{[\mu} v_{\nu]}\, u_{[\rho}t_{\sigma]} \,+\, 4\,u_{[\mu} t_{\nu]} \,e_{[\rho}v_{\sigma]}\,. $$
Once introduced a null tetrad $\{\bl{l},\bl{n},\bl{m},\overline{\bl{m}}\}$, the Weyl tensor can be written as the following expansion:
 \begin{gather}
   \nonumber C_{\mu\nu\rho\sigma} = \Big{\{}\,\frac{1}{2}(\Psi_2+ \overline{\Psi}_2)\big{[}\langle l,n,l,n \rangle + \langle m,\overline{m},m,\overline{m} \rangle\big{]}  + \Psi_0\langle n,\overline{m},n,\overline{m}\rangle + \\
   +  \Psi_4\langle l,m,l,m\rangle -\Psi_2\langle l,m,n,\overline{m}\rangle - \frac{1}{2}(\Psi_2- \overline{\Psi}_2)\langle l,n,m,\overline{m} \rangle + \label{C_abcd expansion} \\ \nonumber +  \Psi_1\big{[}\langle l,n,n,\overline{m}\rangle + \langle n,\overline{m},\overline{m},m \rangle\big{]}
   + \Psi_3\big{[}\langle l,m,m,\overline{m}\rangle - \langle l,n,l,m \rangle\big{]}+\; c.c. \,\Big{\}}_{\mu\nu\rho\sigma}\,.
 \end{gather}
Where $c.c.$ denotes the complex conjugate of all previous terms inside the curly bracket. In particular, note that the right hand side of the above equation is real and has the symmetries of the Weyl tensor. We can verify that such expansion is indeed correct by contracting equation (\ref{C_abcd expansion}) with the null frame and checking that equation (\ref{weylscalars}) is satisfied. Now, contracting equation (\ref{C_abcd expansion}) with $l^\nu l^\rho$ yield:
$$  C_{\mu\nu\rho\sigma}l^\nu l^\rho= \left[\Psi_1(l_\mu \overline{m}_\sigma+\overline{m}_\mu l_\sigma) + c.c. \right] -  2\left(\Psi_0\overline{m}_\mu \overline{m}_\sigma + c.c. \right) - 2\left(\Psi_2+ \overline{\Psi}_2\right)l_\mu l_\sigma\,.
 $$
The above expression, in turn, immediately implies the following identities:
\begin{equation}\label{Clll}
    l_{[\alpha}C_{\mu]\nu\rho\sigma}l^\nu l^\rho \,=\, \left(\Psi_1 \,l_{[\alpha}\overline{m}_{\mu]} l_\sigma + c.c. \right) \,-\, 2\,\left(\Psi_0 \, l_{[\alpha}\overline{m}_{\mu]} \overline{m}_\sigma + c.c. \right)\,,
\end{equation}
\begin{equation}\label{Cllll}
    l_{[\alpha}C_{\mu]\nu\rho[\sigma}l_{\beta]}l^\nu l^\rho \,=\, -2\,\left(\Psi_0 \, l_{[\alpha}\overline{m}_{\mu]} \overline{m}_{[\sigma}l_{\beta]}  + c.c.\right)\,.
 \end{equation}
From which we conclude that the combination on the left hand side of eq. (\ref{Cllll}) vanishes if, and only if, $\Psi_0 = 0$. Hence, by the definition given in section \ref{Sec._Annihil_Scalars}, it follows that $\bl{l}$ is a principal null direction if, and only if, $l_{[\alpha}C_{\mu]\nu\rho[\sigma}l_{\beta]}l^\nu l^\rho = 0 $. Analogously, eq. (\ref{Clll}) and the definition below eq. (\ref{PNDvanishing}) imply that $\bl{l}$ is a repeated PND if, and only if, $l_{[\alpha}C_{\mu]\nu\rho\sigma}l^\nu l^\rho = 0$. In the same vein, the following relations can be proved:
\begin{align*}
  \Psi_0=\Psi_1=\Psi_2=0 \;\;\Leftrightarrow\;\;& C_{\mu\nu\rho[\sigma}l_{\alpha]}l^\rho \,=\,0 \\
  \Psi_0=\Psi_1=\Psi_2=\Psi_3=0 \;\;\Leftrightarrow\;\;& C_{\mu\nu\rho\sigma}l^\rho\,=\,0
\end{align*}
Using these results and table \ref{Tab._vanishScalars} it is then simple matter to arrive at the following alternative definition for the Petrov types:
\begin{equation*}\index{Bel-Debever criteria}
\left\{
  \begin{array}{ll}
     \textbf{Type I}&\rightarrow \; \textrm{ exists $\bl{l}$ such that } l_{[\alpha}C_{\mu]\nu\rho[\sigma}l_{\beta]}l^\nu l^\rho =0  \\
    \textbf{Type II}&\rightarrow  \; \textrm{ exists $\bl{l}$ such that } l_{[\alpha}C_{\mu]\nu\rho\sigma}l^\nu l^\rho =0\\
    \textbf{Type III}&\rightarrow \; \textrm{ exists $\bl{l}$ such that } C_{\mu\nu\rho[\sigma}l_{\alpha]}l^\rho = 0  \\
   \textbf{Type N}&\rightarrow \; \textrm{ exists $\bl{l}$ such that } C_{\mu\nu\rho\sigma}l^\rho =0 \\
   \textbf{Type D}&\rightarrow \; \textrm{ exist $\bl{l},\bl{n}$ such that } l_{[\alpha}C_{\mu]\nu\rho\sigma}l^\nu l^\rho =0 = n_{[\alpha}C_{\mu]\nu\rho\sigma}n^\nu n^\rho\\
   \textbf{Type O}&\rightarrow \;\textrm{ exist $\bl{l},\bl{n}$ such that } C_{\mu\nu\rho\sigma}l^\rho=0 =C_{\mu\nu\rho\sigma}n^\rho\,.
  \end{array}
\right.
\end{equation*}
Where it was assumed that $\bl{l}$ and $\bl{n}$ are real null vectors such that $l^\mu\,n_\mu = 1$. On such definition it is assumed that the Petrov types obey the same hierarchy of the preceding section:
 $$O \,\subset\, N \,\subset\, III \,\subset\, II\,\subset\, I \;\textrm{ and }\; O \,\subset\, D\,\subset\, II\,.$$
These algebraic constraints involving the Weyl tensor and null directions are called Bel-Debever conditions. In reference \cite{Bel-Deb.Higher} these conditions were investigated in higher-dimensional space-times and  connections with the CMPP classification were made.

\section{Spinors, Penrose's  Method}\label{Sec.Spinor4d}\index{Spinor}
In this section we will take advantage of the spinorial formalism in order to describe the Petrov classification, an approach introduced by R. Penrose \cite{Penrose-Spinors}. Here it will be assumed that the reader is already familiar with the spinor calculus in 4-dimensional general relativity.  For those not acquainted with this language, a short course is available in \cite{AdvancedGR}. For a more thorough treatment with diverse applications \cite{Penrose} is recommended. Appendix \ref{App._Cliff&Spinors} of the present thesis provides the general formalism of spinors in arbitrary dimensions.

On the spinorial formalism of 4-dimensional Lorentzian manifolds we have two types of indices, the ones associated with Weyl spinors of positive chirality, $A,B,C,...\in\{1,2\}$, and the ones related to semi-spinors of negative chirality, $\dot{A},\dot{B},\dot{C},...\in\{1,2\}$. It is also worth mentioning that the complex conjugation changes the chirality of the spinorial indices. In this language a vectorial index is equivalent to the ``product'' of two spinorial indices, one of positive chirality and one of negative chirality:
$$ V_\mu \,\sim\, V_{A\dot{A}}\,.  $$
The spaces of semi-spinors are endowed with skew-symmetric metrics $\varepsilon_{AB}=\varepsilon_{[AB]}$ and $\overline{\varepsilon}_{\dot{A}\dot{B}}= \overline{\varepsilon}_{[\dot{A}\dot{B}]}$. This anti-symmetry implies, for instance, that $\zeta^A\zeta_A = \zeta^A\varepsilon_{AB}\zeta^B = 0$ for every spinor $\bl{\zeta}$. These spinorial metrics are related to the space-time metric by the relation $g_{\mu\nu}\sim \varepsilon_{AB} \overline{\varepsilon}_{\dot{A}\dot{B}}$. In this formalism the Weyl tensor\index{Weyl tensor} is represented by
\begin{equation}\label{WeylSpinor4d}
C_{\mu\nu\rho\sigma} \,\sim\, \left(\, \Psi_{ABCD}\,\overline{\varepsilon}_{\dot{A}\dot{B}}\overline{\varepsilon}_{\dot{C}\dot{D}} \,+\, c.c.\, \right)\,.
\end{equation}
Where $\Psi$ is a completely symmetric object, $\Psi_{ABCD} = \Psi_{(ABCD)}$, and $c.c.$ denotes the complex conjugate of the previous terms inside the bracket. Since $\bl{\varepsilon}$ carry the degrees of freedom of the space-time metric, it follows that the degrees of freedom of the Weyl tensor are entirely contained on $\bl{\Psi}$. Therefore, classify the Weyl tensor is then equivalent to classify $\bl{\Psi}$.

It is a well-known result in this formalism that every object with completely symmetric chiral indices, $S_{A_1A_2\ldots A_p} = S_{(A_1A_2\ldots A_p)}$, can be decomposed as a symmetrized direct product of spinors, $S_{A_1A_2\ldots A_p} = \zeta^1_{(A_1}\zeta^2_{A_2}\ldots\zeta^p_{A_p)}$ \cite{AdvancedGR}. Particularly, we can always find spinors $\bl{\zeta},\bl{\theta},\bl{\xi}$ and $\bl{\chi}$ such that
\begin{equation}\label{WeylSpinorDecomp}
    \Psi_{ABCD} \,=\, \zeta_{(A}\,\theta_{B}\,\xi_{C}\,\chi_{D)}\,.
\end{equation}
We can then easily classify the Weyl tensor according to the possibility of the spinors $\bl{\zeta},\bl{\theta},\bl{\xi}$ and $\bl{\chi}$ being proportional to each other. Denoting de proportionality of the spinors by ``$\leftrightarrow$'' and the non-proportionality by ``$\nleftrightarrow$'', we shall define:
\begin{equation}\label{Petrv_Spinors}
\left\{
  \begin{array}{ll}
     \textbf{Type I}&\rightarrow \;  \bl{\zeta},\bl{\theta},\bl{\xi} \textrm{ and } \bl{\chi } \textrm{ are non-propotional to each other}\\
    \textbf{Type II}&\rightarrow  \; \textrm{One pair coincide, } \bl{\zeta}\leftrightarrow\bl{\theta}\nleftrightarrow\xi\nleftrightarrow\bl{\chi}\nleftrightarrow\bl{\zeta}\\
    \textbf{Type III}&\rightarrow \; \textrm{Three spinors coincidence, } \bl{\zeta}\leftrightarrow\bl{\theta}\leftrightarrow\bl{\xi}\nleftrightarrow\bl{\chi} \\
   \textbf{Type D}&\rightarrow \; \textrm{Two pairs coincide, } \bl{\zeta}\leftrightarrow\bl{\theta}\nleftrightarrow\bl{\xi}\leftrightarrow\bl{\chi}\\
   \textbf{Type N}&\rightarrow \; \textrm{All spinors coincide, } \bl{\zeta}\leftrightarrow\bl{\theta}\leftrightarrow\bl{\xi}\leftrightarrow\bl{\chi} \\
   \textbf{Type O}&\rightarrow \;  \bl{\zeta}\,=\, \bl{\theta}\,=\, \bl{\xi}\,=\, \bl{\chi } \,=\, 0 \,.
  \end{array}
\right.
\end{equation}
The spinors that appear on the decomposition of $\bl{\Psi}$ are called the principal spinors of the Weyl tensor, since they are intimately related to the principal null directions. Indeed, the real null vectors generated by these spinors,
$$ l_1^{\,\mu}\,\sim \, \zeta^A\overline{\zeta}^{\dot{A}} \;\; ; \;\;  l_2^{\,\mu}\,\sim \,  \theta^A\overline{\theta}^{\dot{A}} \;\; ; \;\; l_3^{\,\mu}\,\sim \,  \xi^A\overline{\xi}^{\dot{A}} \;\; ;\; \;l_4^{\,\mu} \,\sim \, \chi^A\overline{\chi}^{\dot{A}}\,, $$
point in the principal null directions of the Weyl tensor. Hence, the coincidence of the principal spinors is equivalent to coincidence of PNDs, which makes a bridge between the spinorial approach to the Petrov classification and the approach adopted in section \ref{Sec._Annihil_Scalars}.

The spinorial formalism allows us to see quite neatly which Weyl scalars can be made to vanish by a suitable choice of null tetrad frame on each Petrov type. If $\{o_{_A},\iota_{_A}\}$ forms a spin frame, $o_{_A} \iota^{A} =1$, then we can use them to build a null tetrad frame, as shown in appendix \ref{App._NullTetrad}. So using equations (\ref{weylscalars}) and (\ref{NulltetradSPIN}) we can prove that the Weyl scalars are given by:
\begin{gather}
 \nonumber \Psi_0 =  \Psi_{ABCD}o^Ao^Bo^Co^D \; ;\; \Psi_1 =  \Psi_{ABCD}o^Ao^Bo^C\iota^D \; ;\; \Psi_2 =  \Psi_{ABCD}o^Ao^B\iota^C\iota^D \\
 \Psi_3 =  \Psi_{ABCD}o^A\iota^B\iota^C\iota^D \; ;\; \Psi_4 =  \Psi_{ABCD}\iota^A\iota^B\iota^C\iota^D\,. \label{WeylSca_Spin}
\end{gather}
Thus, for example, if the Weyl tensor is type $D$ according to eq. (\ref{Petrv_Spinors}) then there exists non-zero spinors $\bl{\zeta}$ and $\bl{\xi}$ such that $\Psi_{ABCD} = \zeta_{(A}\zeta_{B}\xi_{C}\xi_{D)}$. Since $\bl{\zeta}\nleftrightarrow \bl{\xi}$ it follows that $\zeta_A\xi^A = w\neq0$. Therefore, setting $o_{_A} = \zeta_{_A}$ and $\iota_{_A} = w^{-1}\xi_{_A}$ it follows that $\{\bl{o},\bl{\iota}\}$ forms a spin frame. Then using equation (\ref{WeylSca_Spin}) we easily find that in this frame $\Psi_0 = \Psi_1 = \Psi_3 = \Psi_4 = 0$, which agrees with table \ref{Tab._vanishScalars}. By means of the same reasoning it is straightforward to work out the other types and verify that the definitions of the Petrov types presented on (\ref{Petrv_Spinors}) perfectly matches the table \ref{Tab._vanishScalars}.

In the same vein, the bivector method of section \ref{Petrov_Biv} can be easily understood on the spinorial formalism. In the spinorial language a self-dual bivector\index{Self-dual bivector} is represented by a symmetric spinor $\phi^{AB}=\phi^{(AB)}$, so that the map $C^+$ is represented by $\phi_{AB}\mapsto\phi'_{AB} = \Psi_{ABCD} \phi^{CD}$. Thus, for example, if the Weyl tensor is type $N$ then we can find a spin frame $\{\bl{o},\bl{\iota}\}$ such that $\Psi_{ABCD} = o_{_A}o_{_B}o_{_C}o_{_D}$. Then defining $\phi_1^{AB}=o^Ao^B$, $\phi_2^{AB}=o^{(A}\iota^{B)}$ and $\phi_3^{AB}=\iota^A\iota^B$, it follows that the action of $C^+$ in this basis of self-dual bivectors yields $\phi'_1=0$, $\phi'_2=0$ and $\phi'_3=\phi_1$, which agrees with equation (\ref{C^+ types}).


\section{Clifford Algebra}\index{Clifford algebra}
In this section the formalism of Clifford algebra will be used to describe another form to arrive at the Petrov classification. For those not acquainted with the tools of geometric algebra, appendix \ref{App._Cliff&Spinors} introduces the necessary background. Let $\{\hat{\bl{e}}_0,\hat{\bl{e}}_1,\hat{\bl{e}}_2,\hat{\bl{e}}_3,\}$ be a local orthonormal frame on a 4-dimensional Lorentzian manifold $(M,\bl{g})$, $$\frac{1}{2} \lef \hat{\bl{e}}_a\hat{\bl{e}}_b \,+\, \hat{\bl{e}}_b\hat{\bl{e}}_a \rig\,=\,\bl{g}(\hat{\bl{e}}_a,\hat{\bl{e}}_b)\,=\, \eta_{ab}\,=\, \diag(1,-1,-1,-1)\,.$$
Denoting by $\eta^{ab}$ the inverse matrix of $\eta_{ab}$, we shall define $\hat{\bl{e}}^a=\eta^{ab}\hat{\bl{e}}_b$. Let us denote the space spanned by the bivector fields by $\Gamma(\wedge^2 M)$. Then, in the formalism of geometric calculus \cite{Lasenby,Hestenes} the Weyl tensor  is a linear operator on the space of bivectors, $\mathcal{C}: \Gamma(\wedge^2 M) \rightarrow \Gamma(\wedge^2 M)$, whose action is\footnote{All results in this thesis are local, so that it is always being assumed that we are in the neighborhood of some point. Thus, formally, instead of $\Gamma(\wedge^2 M)$ we should have written $\Gamma(\wedge^2 M)|_{N_x}$, which is the restriction of the space of sections of the bivector bundle to some neighborhood $N_x$ of a point $x\in M$. So we are choosing a particular local trivialization of the bivector bundle.}
\begin{equation}\label{C_Clifford}
  \mathcal{C}(\bl{V}\wedge \bl{U}) \,=\,  V^a\, U^b \, C_{abcd}\, \hat{\bl{e}}^c\wedge \hat{\bl{e}}^d\,,
  \end{equation}
where $C_{abcd}$ are the components of the Weyl tensor on the frame $\{\hat{\bl{e}}_a\}$. In the above equation $\bl{V}\wedge \bl{U}$ means the anti-symmetrized part of the Clifford product of $\bl{V}$ and $\bl{U}$, $\bl{V}\wedge \bl{U}= \frac{1}{2}(\bl{V}\bl{U}- \bl{U}\bl{V})$. Then using (\ref{C_Clifford}) and equation (\ref{abc}) of appendix \ref{App._Cliff&Spinors} we find:
\begin{align}
 \nonumber \hat{\bl{e}}^a\,\mathcal{C}(\hat{\bl{e}}_a\wedge \hat{\bl{e}}_b) \,&=\, C_{abcd}\,\hat{\bl{e}}^a\, (\hat{\bl{e}}^c\wedge \hat{\bl{e}}^d) = C_{abcd}\,\hat{\bl{e}}^a\, \frac{1}{2}(\hat{\bl{e}}^c \hat{\bl{e}}^d - \hat{\bl{e}}^d \hat{\bl{e}}^c)  \\
  \nonumber \,&=\,C_{abcd} \, \frac{1}{2} (2\,\eta^{ac}\,\hat{\bl{e}}^d - 2\,\eta^{ad}\,\hat{\bl{e}}^c + 2 \,\hat{\bl{e}}^a\wedge\hat{\bl{e}}^c\wedge\hat{\bl{e}}^d) \\
  \,&=\, 2\,C_{\phantom{c}bcd}^{c}\, \hat{\bl{e}}^d \,-\, C_{b[acd]}\, \hat{\bl{e}}^a\wedge\hat{\bl{e}}^c\wedge\hat{\bl{e}}^d \label{Weyl_Id_Cliff}
\end{align}
Equation (\ref{Weyl_Id_Cliff}) makes clear that on the Clifford algebra formalism the single equation $\hat{\bl{e}}^a \mathcal{C}(\hat{\bl{e}}_a\wedge \hat{\bl{e}}_b) =0$ is equivalent to the trace-less property and the Bianchi identity satisfied by the Weyl tensor. There are two other symmetries satisfied by this tensor, see (\ref{weylsymm}), which are the anti-symmetry on the first and second pairs of indices, $C_{abcd}=C_{[ab][cd]}$ and the symmetry by the exchange of these pairs, $C_{abcd}=C_{cdab}$. But the latter symmetry can be derived from the Bianchi identity, while the former is encapsulated in the present formalism by the fact that the operator $\mathcal{C}$ maps bivectors into bivectors. Thus we conclude that on the Clifford algebra approach all the symmetries of the Weyl tensor are encoded in the following relations:
\begin{equation}\label{WeylSym_Cliff}
 \mathcal{C}:\, \Gamma(\wedge^2 M) \rightarrow \Gamma(\wedge^2 M) \quad ; \quad  \hat{\bl{e}}^a\,\mathcal{C}(\hat{\bl{e}}_a\wedge \hat{\bl{e}}_b) \,=\,0  \,.
\end{equation}

Before proceeding let us define the following bivectors:
$$ \bl{\sigma}_i\,=\, \hat{\bl{e}}_0 \wedge \hat{\bl{e}}_i \;\,;\quad \bl{I}\bl{\sigma}_i \,=\, \frac{1}{2} \epsilon^{ijk}\, \hat{\bl{e}}_j \wedge \hat{\bl{e}}_k $$
Where $i,j,k$ are indices that run from 1 to 3, $\epsilon^{ijk}$ is a totally anti-symmetric object with $\epsilon^{123}=1$ and $\bl{I}=\hat{\bl{e}}_0\hat{\bl{e}}_1\hat{\bl{e}}_2\hat{\bl{e}}_3$ is the pseudo-scalar\index{Pseudo-scalar} defined on appendix \ref{App._Cliff&Spinors}. In particular, using these definitions and the Bianchi identity it is not difficult to prove that the following equation holds:
\begin{equation}\label{C(D_i)}
  \mathcal{C}(\bl{\sigma}_i) \,=\, -2 \left[\, C_{0i0j} + \bl{I}\, C_{0kli}\epsilon^{klj}  \,\right]\,\bl{\sigma}_j
\end{equation}
Also, expanding equation (\ref{WeylSym_Cliff}) we find the following explicit relations:
\begin{gather}
 \nonumber \bl{\sigma}_1\,\mathcal{C}(\bl{\sigma}_1) \,+\, \bl{\sigma}_2\,\mathcal{C}(\bl{\sigma}_2) \,+\, \bl{\sigma}_3\,\mathcal{C}(\bl{\sigma}_3) \,=\, 0  \\
  \nonumber \bl{\sigma}_1\,\mathcal{C}(\bl{\sigma}_1) \,=\, \bl{I}\bl{\sigma}_2\,\mathcal{C}(\bl{I}\bl{\sigma}_2) \,+\, \bl{I}\bl{\sigma}_3\,\mathcal{C}(\bl{I}\bl{\sigma}_3) \\
     \bl{\sigma}_2\,\mathcal{C}(\bl{\sigma}_2) \,=\, \bl{I}\bl{\sigma}_1\,\mathcal{C}(\bl{I}\bl{\sigma}_1) \,+\, \bl{I}\bl{\sigma}_3\,\mathcal{C}(\bl{I}\bl{\sigma}_3) \label{Explic.Cliff} \\
  \nonumber \bl{\sigma}_3\,\mathcal{C}(\bl{\sigma}_3) \,=\, \bl{I}\bl{\sigma}_1\,\mathcal{C}(\bl{I}\bl{\sigma}_1) \,+\, \bl{I}\bl{\sigma}_2\,\mathcal{C}(\bl{I}\bl{\sigma}_2)
\end{gather}
Summing the last three relations above and then using the first one, we find $\sum_i \bl{I}\bl{\sigma}_i\,\mathcal{C}(\bl{I}\bl{\sigma}_i)=0$. Then using this identity on the last three relations of (\ref{Explic.Cliff}) we conclude that $\mathcal{C}(\bl{I}\bl{\sigma}_i)=\bl{I}\mathcal{C}(\bl{\sigma}_i)$. By means of  this and the identity $\bl{I}^2=-1$ we also find that $\mathcal{C}(\bl{I}\,\bl{I}\bl{\sigma}_i)= \bl{I}\mathcal{C}(\bl{I}\bl{\sigma}_i)$. Since $\{\bl{\sigma}_i,\bl{I}\bl{\sigma}_i\}$ is a basis for the bivector space it follows that in general
\begin{equation}\label{C(IB)=IC(B)}
  \mathcal{C}(\bl{I}\bl{B}) \,=\, \bl{I}\,\mathcal{C}(\bl{B}) \quad\; \forall\;\bl{B}\in \Gamma(\wedge^2 M)\,.
\end{equation}

Now recall from appendix \ref{App._Cliff&Spinors} that the pseudo-scalar $\bl{I}$ commutes with the elements of even order, in particular it commutes with all bivectors. Moreover, equation (\ref{C(IB)=IC(B)}) guarantees that $\bl{I}$ commutes with the Weyl operator. Therefore, when dealing with the Weyl operator acting on the bivector space we can treat the $\bl{I}$ as if it were a scalar. Furthermore, since $\bl{I}^2=-1$ we can pretend that $\bl{I}$ is the imaginary unit, $\bl{I}\sim i=\sqrt{-1}$, so that we can reinterpret the operator $\mathcal{C}$ as an operator on the complexification of the real space generated by $\{\bl{\sigma}_1,\bl{\sigma}_2,\bl{\sigma}_3\}$. With these conventions the equation (\ref{C(D_i)}) can be written as\footnote{A similar phenomenon happens on the Clifford algebra of the space $\mathbb{R}^3$. In this case the pseudo-scalar commutes with all elements of the algebra and obeys to the relation $\bl{I}^2=-1$, so that it can actually be interpreted as the imaginary unit, $\bl{I}\sim i=\sqrt{-1}$. This is the geometric explanation of why the complex numbers are so useful when dealing with rotations in 3 dimensions.}:
\begin{equation}\label{C_ij_Cliff.}
  \mathcal{C}(\bl{\sigma}_i)\,=\, \mathcal{C}_{ij}\,\bl{\sigma}_j\quad;\; \mathcal{C}_{ij}\sim -2\lef C_{0i0j} + i\, C_{0kli}\epsilon^{klj}\rig
\end{equation}
Now we can easily define a classification for the Weyl tensor. Using equation (\ref{C_ij_Cliff.}) and the symmetries of the Weyl tensor it is trivial to prove that this matrix is  trace-less, $\mathcal{C}_{ii}=0$. Therefore, the possible algebraic types for the operator $\mathcal{C}$ are the same as the ones listed on eq. (\ref{C^+ types}).

Note that this classification is, in principle, different from the one shown on subsection \ref{Petrov_Biv}. While the latter uses the space of self-dual bivectors to define a 3-dimensional operator, the operator introduced in the present subsection acts on the space generated by $\{\bl{\sigma}_1,\bl{\sigma}_2,\bl{\sigma}_3\}$, which is not the space of self-dual bivectors. The remarkable thing is that these two classifications turns out to be equivalent. This can be seen by noting that to every eigen-bivector of $\mathcal{C}$ we can associate a self-dual bivector that is eigen-bivector of $C^+$ with the ``same'' eigenvalue. Indeed, if $\bl{B}$ is an eigen-bivector of the operator $\mathcal{C}$ on the Clifford algebra approach then $\mathcal{C}(\bl{B})=(\lambda_1+\bl{I}\lambda_2)\bl{B}$, where $\lambda_1$ and $\lambda_2$ are real numbers. Then using equation (\ref{Hodge_Cliff}) of appendix \ref{App._Cliff&Spinors} we see that $\bl{B}_+=(1-i\bl{I})\bl{B}$ is a self-dual bivector. Moreover, we can use equation (\ref{C(IB)=IC(B)}) to prove that $\mathcal{C}(\bl{B}_+)=(\lambda_1+i\lambda_2)\bl{B}_+$. To finish the proof just note that the Weyl operator defined on (\ref{C_Clifford}) agrees with the definition of the section \ref{Petrov_Biv}, see equation (\ref{C_Petrov}). Hence we have that $C^+(\bl{B}_+)=(\lambda_1+i\lambda_2)\bl{B}_+$. More details about this method can be found in \cite{Hestenes,Sobcz_Cliff}. In particular, reference \cite{Hestenes} has exploited the Clifford algebra formalism to find canonical forms for the Weyl operator for each algebraic type. As an aside, it is worth mentioning that the whole formalism of general relativity can be translated to the Clifford algebra language with some advantages \cite{Cliff_GR}.

\section{Interpreting the PNDs}\label{Sec._PNDinterp.}\index{Principal null directions(PND)}
In the previous sections it has been proved that every space-time with non-vanishing Weyl tensor admits some privileged null directions, four at most, called the principal null directions (PNDs). In the present section we will investigate the role played by these directions both from the geometrical and physical points of view.

According to \cite{ShildBook,Bel-Deb.Higher}, in 1922 \'{E}lie Cartan has pointed out that the Weyl tensor of a general 4-dimensional space-time defined four distinguished null directions endowed with some invariance properties under the parallel transport over infinitesimal closed loops. It turns out that these directions were the principal null directions of the Weyl tensor, in spite of Petrov's article defining his classification have appeared three decades later. Suppose that a vector $\bl{v}$  belonging to the tangent space at a point $p\in M$ is parallel transported along an infinitesimal parallelogram with sides generated by $\bl{t}_1$ and  $\bl{t}_2$, as illustrated on the figure below.
\begin{figure}[h]
	\centering
		\includegraphics[width=3.5cm]{{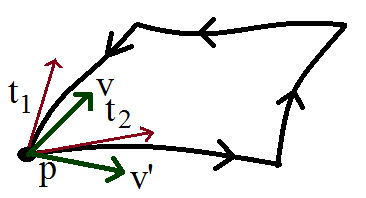}}
	\label{fig:Specialization Diagrams}
\end{figure}\\
It is a well-known result of Riemannian geometry that the change on the vector $\bl{v}$ caused by the parallel transport over the loop is given by
\begin{equation}\label{paralleltransport}
    \delta v^\mu \equiv v'^\mu - v^\mu = -\epsilon\, R^\mu_{\phantom{\mu}\nu\rho\sigma}\,v^\nu \,t_1^\rho \,t_2^\sigma\,.
\end{equation}
Where $\bl{v}'$ is the vector after the parallel transport and $\epsilon$ is proportional to the area of the parallelogram. In vacuum, as henceforth assumed in this section, Einstein's equation implies that the Riemann tensor is equal to the Weyl tensor. So that in this case one can substitute $R^\mu_{\phantom{\mu}\nu\rho\sigma}$ by $C^\mu_{\phantom{\mu}\nu\rho\sigma}$ in equation (\ref{paralleltransport}). Now let us search for null directions that are preserved by this kind of parallel transport.

Let $\bl{v}=\bl{l}$ be a PND and $\bl{n}$ a null vector such that $l^\mu n_\mu= 1$. Then, from section \ref{Sec._Bel-Deb}, we have that $l_{[\alpha}C_{\mu]\nu\rho[\sigma}l_{\beta]}l^\nu l^\rho =0$. Contracting this equation with $t_2^\mu n^\beta$ we easily find that $C^\sigma_{\phantom{\sigma}\nu\rho\mu}l^\nu l^\rho t_2^\mu \propto l^\sigma$ for any $\bl{t}_2$ orthogonal to $\bl{l}$. Thus PNDs are the null directions with the property of being invariant by the parallel transport around infinitesimal parallelograms generated by the PND itself and any direction orthogonal to it. In the same vein, if $\bl{l}$ is a repeated principal null direction then $l_{[\alpha}C_{\mu]\nu\rho\sigma}l^\nu l^\rho =0$. Contracting this last equation with $t_2^\sigma n^\alpha$ we find that $\delta l^\mu \propto l^\mu$ for any parallelogram such that one of the sides is generated by $\bl{l}$. If $\bl{l}$ is a triply degenerated PND then $C_{\mu\nu\rho[\sigma}l_{\alpha]}l^\rho = 0$, which by contraction with $t_1^\mu t_2^\nu n^\alpha$ yield that $\delta l^\mu \propto l^\mu$ for any parallelogram. Finally, if $\bl{l}$ is a PND with degree of degeneracy four then $C_{\mu\nu\rho\sigma}l^\sigma=0$, so that $\delta l^\mu=0$ for any parallelogram. Table \ref{Tab.PND} summarizes these geometric properties of the PNDs.
\begin{table}
\begin{center}
\begin{tabular}{cccc}
  \hline
  \hline
  $q\,=\,1$ & $q\,=\,2$ & $q\,=\,3$ & $q\,=\,4$ \\
  \hline
  $\bl{t}_1\,=\,\bl{l}$  & $\bl{t}_1\,=\,\bl{l}$   &  $\bl{t}_1$ \small{arbitrary}  & $\bl{t}_1$ \small{arbitrary} \\
  $\;t_2^\mu \,l_\mu=0$\; & \; $\bl{t}_2$ \small{arbitrary} \;& \;$\bl{t}_2$ \small{arbitrary}\; &\; $\bl{t}_2$ \small{arbitrary}\;\\
  $\delta l^\mu \propto l^\mu$ & $\delta l^\mu \propto l^\mu$ & $\delta l^\mu \propto l^\mu$ & $\delta l^\mu =0$ \\
    \hline
    \hline
      \end{tabular}
\caption{\footnotesize{Invariance of the PNDs under parallel transport over an infinitesimal parallelogram with sides generated by $\bl{t}_1$ and $\bl{t}_2$. In the first row $q$ denotes the degeneracy of the PND $\bl{l}$.}}\label{Tab.PND}
\end{center}
\end{table}

In ref. \cite{HallPND} it was shown another geometric interpretation for the principal null directions. Glossing over the subtleties, it was proved there that a null direction is a PND when the Riemannian curvature of a 2-space generated by this null direction and a space-like vector field $\bl{t}$ is independent of $\bl{t}$.

One of the first physicists to investigate the physical meaning of the Petrov types was F. Pirani. In ref. \cite{Pirani} he has tried to find a plausible definition of gravitational radiation by comparing with the electromagnetic case. In this article it has been shown that the energy-momentum tensor associated with electromagnetic radiation admits no time-like eigenvector and one null eigenvector at most, this null vector turned out to point in the direction of the radiation propagation. Searching for an analogous condition in general relativity Pirani investigated the eigenbivectors of Riemann tensor. The intersection of the planes generated by such eigenbivectors defined what he called Riemann principal directions (RPDs), which are not the PNDs, as they are not necessarily null. But it turns out that the null Riemann principal directions are repeated PNDs. Thus, mimicking the electromagnetic case, Pirani arrived at the conclusion that if a space-time admits a time-like RPD then no gravitational radiation should be present. Along with the results of Bel \cite{Bel}, this means that no gravitational radiation is allowed on Petrov types $I$ and $D$, which is reasonable since all static space-times are either type $I$ or $D$. Pirani and Bel interpreted the repeated PNDs of types $II$, $III$ and $N$ as the direction of the gravitational radiation propagation \cite{Bel,Pirani}.

In order to understand the physical meaning of the PNDs, the analogy between the electromagnetic theory and general relativity was also exploited by other physicists. In \cite{Bel_tensor,Bel} L. Bel has introduced a tensor of rank four that is quadratic on the Riemann tensor and that in vacuum has properties that perfectly mimics the electromagnetic energy-momentum tensor. Such tensor is now called the Bel-Robinson tensor \cite{Penrose}. Then Debever proved that in vacuum this tensor is completely determined by the principal null directions of the Weyl tensor \cite{Debever}, a result that can be easily verified using the spinorial formalism. In ref. \cite{Penrose-Spinors}, Penrose has argued that the PNDs are related to the gravitational energy density, enforcing and complementing Debever's results. Penrose also concluded that pure gravitational radiation should be present only in type $N$ space-times, since only in this case the Weyl tensor satisfies the massless wave-equation.

Finally, according to the Goldberg-Sachs theorem, the repeated PNDs in vacuum are tangent to a congruence of null geodesics that is shear-free. This celebrated theorem is behind the integrability of Einstein's equation for space-times of type $D$ \cite{typeD}. This important result will be deeply exploited on the forthcoming chapters. One of the goals of this thesis is to prove a suitable generalization of this theorem valid in higher dimensions, which will be accomplished in chapters \ref{Chap._Spin6D} and \ref{Chap. AllD}.

\section{Examples}
\textbf{1)} \textbf{Schwarzschild space-time}\\
Schwarzschild space-time is the unique spherically-symmetric solution of Einstein's equation in vacuum. In a static and spherically symmetric coordinate system its metric is given by
$$ds^2 = f^2\,dt^2 -  f^{-2}\, dr^2 - r^2(d\theta^2 + \sin^2\theta\, d\varphi^2)\,,\;\; f^2 = 1-\frac{2M}{r}  \,.$$
A suitable orthonormal frame and a suitable null tetrad are then,
\begin{gather*}
  \hat{\bl{e}}_{0} = f^{-1}\,\partial_t \; ; \;\; \hat{\bl{e}}_{1} =f\,\partial_r \; ;\; \; \hat{\bl{e}}_{2} =\frac{1}{r}\partial_\theta  \; ; \; \; \hat{\bl{e}}_{3}=\frac{1}{r\sin\theta}\partial_\varphi\,;\,\textrm{ and }  \\
  \bl{l}=\frac{1}{\sqrt{2}}(\hat{\bl{e}}_0+\hat{\bl{e}}_1) \, ; \; \bl{n} =\frac{1}{\sqrt{2}}(\hat{\bl{e}}_0-\hat{\bl{e}}_1) \, ; \; \bl{m} =\frac{1}{\sqrt{2}}(\hat{\bl{e}}_2+i\hat{\bl{e}}_3) \, ; \; \overline{\bl{m}} =\frac{1}{\sqrt{2}}(\hat{\bl{e}}_2-i\hat{\bl{e}}_3)\,.
\end{gather*}
Since the vector field $\partial_t= f\hat{\bl{e}}_{0}$ is a time-like hyper-surface orthogonal Killing vector field\index{Killing vector}, the space-time is called static. In other words this means that the above metric is invariant by the transformations $t\rightarrow-t$ and $t\rightarrow t + \epsilon$, where $\epsilon$ is a constant. Such symmetries imply that the Weyl tensor cannot be of Petrov types $II$, $III$ or $N$. For instance, if some static space-time were type $N$ it would have just one PND, $\bl{l}=\hat{\bl{e}}_{0}+ \hat{\bl{e}}$ where $\hat{\bl{e}}$ is some space-like vector of unit norm. But using the symmetry $t\rightarrow-t$ we conclude that the null vector $\bl{l}'= -\hat{\bl{e}}_{0}+ \hat{\bl{e}}$ should also be a PND, which contradicts the type $N$ hypothesis. Thus the Schwarzschild solution must be either type $I$ or $D$. Indeed, calculating the Weyl scalars, by means of (\ref{weylscalars}), on the above null frame we get:
$$\Psi_0 = \Psi_1 = \Psi_3 = \Psi_4 = 0\;\; ;\;\; \Psi_2 = \frac{M}{r^3}\,.$$
Then, thanks to table \ref{Tab._vanishScalars}, we conclude that the Schwarzschild space-time has Petrov type $D$, with $\bl{l}$ and $\bl{n}$ being repeated PNDs. Actually, it can be proved that the whole family of Kerr-Newman solutions is type $D$.\\
\\
%
%
\textbf{2)} \textbf{Plane Gravitational Waves}\\
Physically, plane waves are characterized by the existence of plane wave-fronts (equipotentials) orthogonal to the direction of propagation. Since the graviton is a massless particle, it follows that the gravitational field propagates along a null direction $\bl{l}$. In order for all the points on a wave-front remain on the same phase as propagation occurs, the null vector field $\bl{l}$ must be covariantly constant throughout the space-time. In particular, this implies that  $\bl{l}$ remains unchanged by parallel transport, which according to table \ref{Tab.PND} implies that the space-time must be type $N$ if vacuum is assumed. Therefore, a manifold that represents the propagation of plane gravitational waves might be type $N$. Indeed, if a space-time admits a covariantly constant null vector $\bl{l}$ then its metric must be of the following form \cite{Stephani,Exact_Sol}:
$$ ds^2 \,=\, 2dudr \,+\, 2H(u,x,y)du^2 \,-\, dx^2\,-\, dy^2\,,$$
where $\bl{l}=\partial_r$. A manifold with such metric is called a $pp$-wave space-time. Choosing the other vectors of the null tetrad to be $\bl{n}=\partial_u-H\partial_r$ and $\bl{m}=\frac{1}{\sqrt{2}}(\partial_x + i \partial_y)$ it follows that all the Weyl scalars vanish except for $\Psi_4\propto (\partial_w\partial_wH)$, where $w$ is a complex coordinate defined by $w=x+iy$. This implies that in points of space-time where $\partial_w\partial_wH\neq0$ the Weyl tensor is type $N$ with PND given by $\bl{l}=\partial_r$. Note that in general this $pp$-wave metric is not a vacuum solution, since its Ricci tensor generally does not vanish,  $R_{\mu\nu}\propto (\partial_{\overline{w}}\partial_w H) l_\mu l_\nu$. In order to gain some insight on the meaning of the these coordinates, note that in the limit $H\rightarrow0$ the above metric is just the Minkowski metric with $u= \frac{1}{\sqrt{2}}(t+z)$ and $r= \frac{1}{\sqrt{2}}(t-z)$, where the frame $\{\partial_t, \partial_x, \partial_y, \partial_z\}$ is a global inertial frame on the Minkowski space-time. The plane wave space-time is of great relevance for the quantum theory of gravity because all its curvature invariants vanish \cite{Pravda_VSI}, so that the quantum corrections for the Einstein-Hilbert action do not contribute \cite{pp_QuantGrav}. There is also an interesting article by Penrose proving that all space-times in a certain limit are $pp$-wave \cite{Penrose_pp}.

The $pp$-wave solution provides an illustration that the Petrov type can vary from point to point on the manifold, it is local classification. For instance, if $H= (x^2+ y^2)^2= ww\bar{w}\bar{w}$ then the only non-vanishing Weyl scalar is $\Psi_4\propto \bar{w}\bar{w}$. Therefore, in this case the Petrov classification is type $O$ at the points satisfying $(x^2+ y^2)= 0$ and type $N$ outside the 2-dimensional time-like surface $(x^2+ y^2)= 0$.
\\
\\
\textbf{3)} \textbf{Cosmological Model (FLRW)}
\\
Astronomical observations reveal that on large scales (above $10^{24}m$) the universe looks homogeneous and isotropic on the spatial sections. This leads us to the so-called FLRW cosmological model, whose metric is of the following form \cite{Landau_Field}:
$$ ds^2 = dt^2 - R^2(t) \left[\frac{dr^2}{1-\kappa r^2}+r^2(d\theta^2 + \sin^2\theta\, d\varphi^2)\right]\; ;\; \kappa=0,\pm1\,.  $$
The metric inside the square bracket is the general metric of a 3-dimensional homogeneous and isotropic space, the case $\kappa=0$ being the flat space, $\kappa= 1$ being the 3-sphere and $\kappa= -1$ is the hyperbolic 3-space. Now let us see that the Petrov classification of such metric must be type $O$. Suppose, by contradiction, that the Petrov type is different from $O$ at some point. Then at this point the Weyl tensor would admit at least one and at most four PNDs. If $\bl{l}$ is a PND then, as it is a null vector, it must be of the form $\bl{l}=\lambda(\partial_t+\hat{\bl{e}})$, where $\hat{\bl{e}}$ is a unit space-like vector and $\lambda\neq 0$ is a real scalar. But this distinguishes a privileged spatial direction, the one tangent to  $\hat{\bl{e}}$, which contradicts the isotropy assumption. Homogeneity then guarantees that the same is true on the other points of space. Thus we conclude that the FLRW space-time is type $O$. Indeed, it is not so hard to verify that the Weyl tensor of this metric vanishes.

\section{Other Classifications}
In this chapter it was shown that a space-time can be classified using the Petrov type of the Weyl tensor. In the next chapter it will be presented several important theorems involving the Petrov classification, confirming its usefulness. But this is not the only form to classify a manifold at all. In this section three other noteworthy methods to classify a space-time will be presented.

In section \ref{Sec._Symmetries} it was said that the symmetries of a manifold are represented by the Killing vectors. These vector fields have an important property, the Lie bracket of any two Killing vectors is another Killing vector. Therefore, the Killing vectors of a manifold generate a Lie group known as the group of motions of the space-time. For instance, the group of motions of the flat space-time is the Poincar\'{e} group. We can, thus, classify the space-times according to the group of motions. For details and applications see \cite{Stephani,Petrov_Book}.

Let $\bl{v}$ be a vector belonging to the tangent space at a point $p\in M$ of the 4-dimensional space-time $(M,\bl{g})$. Then if we perform the parallel transport of such vector along a closed loop then the final result will be another vector $\bl{v}'$. It is easy to see that $\bl{v}'$ is related to $\bl{v}$ by a linear transformation. The group formed by all such transformations, for all closed loops, is called the Holonomy group of $p$ and denoted by $H_p$. Since the metric is covariantly constant it follows that $H_p\subset O(1,3)$. Moreover, the holonomy group is the same at all points of a connected domain \cite{Nakahara}, so the holonomy provides a global classification for the space-times. Connections between the Petrov classification and holonomy groups were studied in \cite{Hall-Schell}.

Just as the Weyl tensor provides a map of bivectors into bivectors, the Ricci tensor can be seen as an operator on the tangent space whose action is defined by $V^\mu \mapsto V'^\mu= R^\mu_{\phantom{\mu}\nu}V^\nu$. Such operator can be algebraically classified by means of the refined Segre classification (appendix \ref{App._Segre}), yielding another independent way to classify the curvature of a manifold. For instance, in the $pp$-wave space-time (see the preceding section) the Ricci tensor has the form $R_{\mu\nu}= \lambda l_\mu l_\nu$ with $\bl{l}$ being a null vector field. In this case, if $\lambda\neq 0$ the algebraic type of the Ricci tensor is $[\,|1,1,2]$. Since Einstein's equation (\ref{Eisntein.Eq}) connects the Ricci tensor to the energy-momentum tensor it turns out that classify one of these tensors is tantamount to classify the other. Because of the latter fact it follows that the so-called energy conditions impose restrictions over such algebraic classification. For example, the type $[1,3|\,]$ is not compatible with the dominant energy condition.  The classification of the Ricci tensor is of particular help when the Weyl tensor vanishes, since in this case the curvature is entirely determined by the former tensor. More about this classification is available in \cite{Stephani}. In the forthcoming chapters we will be interested in the vacuum case, $R_{\mu\nu}= 0$, so that the classification of the Ricci tensor will play no role.

\chapter{Some Theorems on Petrov Types}\label{Chap. Theorems4}
One could devise a lot of different forms to classify the curvature of a space-time, but certainly many of them will be of little help both for the Physical understanding and for solving equations. The major relevance of the Petrov classification does not come from the algebraic classification in itself, but from its connection with Physics and, above all, with geometry. The Physical content behind this classification is mainly based on the interpretation of the principal null directions, discussed in section \ref{Sec._PNDinterp.}. Regarding the geometric content there exist several theorems relating the Petrov classification with geometric restrictions on the space-time. The intent of the present chapter is to show some of the most important theorems along this line.

As a warming up for what comes, let us consider an example showing that it is quite natural that algebraic restrictions on the curvature yield geometric constraints on the space-time and \textit{vice versa}. Let $(M,\bl{g})$ be a 4-dimensional space-time containing a covariantly constant vector field, $\nabla_\mu\,K_\nu= 0$. Then, using equation (\ref{Cov.Deriv.Commut}) we arrive at the following consequence:
\begin{equation}\label{Const.Vector}
 R_{\mu\nu\rho}^{\phantom{\mu\nu\rho}\sigma}\,K_\sigma \,=\, \left(\nabla_\mu\nabla_\nu - \nabla_\nu\nabla_\mu\right)\,K_\rho  \,=\,0\,.
\end{equation}
Conversely, if $R_{\mu\nu\rho}^{\phantom{\mu\nu\rho}\sigma}K_\sigma=0$ then $K_\mu$ must be a multiple of a covariantly constant vector field. Thus we obtained a connection between an algebraic condition, $R_{\mu\nu\rho}^{\phantom{\mu\nu\rho}\sigma}K_\sigma=0$, and a geometric restriction, the constancy of $\bl{K}$. In particular, if $\bl{K}$ is null then equation (\ref{Const.Vector}) implies that Petrov classification is type $N$. Note also that some geometric constraints are quite severe. For instance, if the space-time admits four constant vector fields that are linearly independent at every point then eq. (\ref{Const.Vector}) implies that $ R_{\mu\nu\rho}^{\phantom{\mu\nu\rho}\sigma}= 0$, \textit{i.e.}, the manifold is flat.

%
%

\section{Shear, Twist and Expansion}
Before proceeding to the theorems on Petrov types it is important to introduce the geodesic congruences, which is the aim of this section. In particular, it will be shown the physical interpretation of the expansion, shear and twist parameters. This will be of great relevance for the forthcoming sections.

\begin{figure}[h]
	\centering
		\includegraphics[width=5cm]{{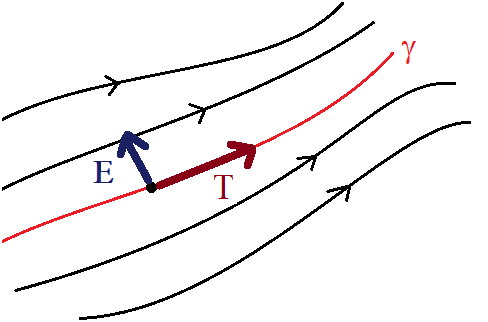}}
	\caption{\footnotesize{A congruence of geodesics, $\bl{T}$ is the tangent vector field and $\bl{E}$ measures the relative deviation of the geodesics.}}
	\label{Fig_Congruence}
\end{figure}

Let $(M,\bl{g})$ be a $4$-dimensional Lorentzian manifold and $N_p\subset M$ be the neighborhood of some point $p\in M$. A congruence of geodesics in $N_p$ is a family of geodesics such that at each point of $N_p$ passes one, and just one, of these geodesics. Such congruence defines a vector field $T^\mu$ that is tangent to the geodesics and affinely parameterized, $T^\mu\nabla_\mu T^\nu = 0$. Now, suppose that the congruence is time-like and that its tangent vector field is normalized so that $T^\mu T_{\mu}= 1$. It is possible to study how the geodesics on the congruence move relative to each other by introducing a set of $3$ vector fields $E_i^\mu$  called deviation vector fields. These vector fields are orthogonal to the direction of propagation and they connect a fiducial geodesic $\gamma$ on the congruence to the neighbors geodesics, as depicted on the figure \ref{Fig_Congruence}.
The vector fields $E_i^\mu$ are assumed to commute with $T^\mu$, so that a suitable coordinate system can be introduced, with the affine parameters of the geodesics, $\tau$, being one of the coordinates. Therefore we have $[\bl{E}_i,\bl{T}]= E_i^\mu \nabla_\mu \bl{T} - T^\mu \nabla_\mu \bl{E}_i = 0$. Then the relative movements of the geodesics on the congruence are measured by the variation of $\bl{E}_i$ along the geodesics:
\begin{equation}\label{dE/dtau}
  \frac{dE_i^\nu}{d\tau} \,=\, T^\mu \nabla_\mu E_i^\nu  \,=\, E_i^\mu \nabla_\mu T^\nu \,=\, M^{\nu\mu}\,E_{i\,\mu}\,\,,\;\, M^{\nu\mu}= \nabla^\mu T^\nu\,.
\end{equation}
The geodesic character of $\bl{T}$ and the constancy of its norm easily implies that $M_{\mu\nu}T^\nu= 0$ and $T^\mu M_{\mu\nu}= 0$. Denoting by $P_{\mu\nu} = g_{\mu\nu} - T_\mu T_\nu$ the projection operator on the space generated by $\{\bl{E}_i\}$, we can split the tensor $M_{\mu\nu}$ into its irreducible parts: the trace, $\theta=M^\mu_{\phantom{\mu}\mu}$, the traceless symmetric part, $\sigma_{\mu\nu}= M_{(\mu\nu)} - \frac{1}{3}\theta P_{\mu\nu}$ and the skew-symmetric part, $\omega_{\mu\nu}= M_{[\mu\nu]}$. These three parts of the tensor $\bl{M}$ are named the expansion, the shear\index{Shear}\label{Shear1} and the twist, respectively. In order to understand the origin of these names let us work out a simple example.

%
%

\begin{figure}[h]
	\centering
		\includegraphics[width=11cm]{{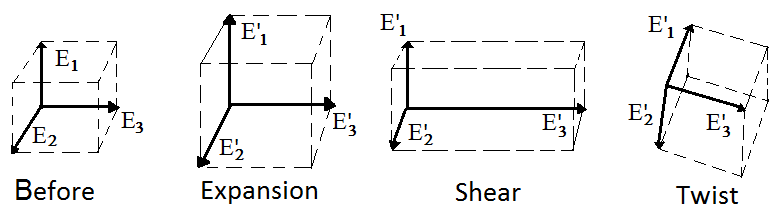}}
	\caption{\footnotesize{The illustration on the left side shows a unit cube before the infinitesimal evolution. Then the next 3 pictures display the changes caused by an expansion, a shear and a twist, respectively. The shear and the twist keep the volume invariant.}}
	\label{Sher_Twist_Exp}
\end{figure}

Suppose that the vectors on the 3-dimensional Euclidian space, $(\mathbb{R}^3,\delta_{ij})$, obey the equation of motion $ \frac{d\hat{\bl{E}}}{dt} = \mathbf{M}\,\hat{\bl{E}}$, where $\mathbf{M}$ is a $3\times3$ matrix. Now let us split this matrix as the sum of its trace, the trace-less symmetric part and the skew-symmetric part, $\mathbf{M} =  \frac{1}{3}\theta\mathbf{1} + \boldsymbol{\sigma}  +  \boldsymbol{\omega}$. Then plugging this into the equation of motion and assuming that $\delta t$ is an infinitesimal time interval, we get:
\begin{equation}\label{E'evlou}
   \hat{\bl{E}'}\,\equiv\, \hat{\bl{E}}(t + \delta t) \,=\,  \hat{\bl{E}}(t) \,+\, \delta t\, \left[\frac{1}{3}\,\theta\,\mathbf{1} \,+\, \boldsymbol{\sigma}  \,+\,  \boldsymbol{\omega}\right]\,\hat{\bl{E}}(t)\,.
  \end{equation}
Now we shall analyse the individual effect of each of the terms inside the square bracket on the above equation. Let $\{\hat{\bl{E}}_1,\hat{\bl{E}}_2,\hat{\bl{E}}_3\}$ be a cartesian frame, $\hat{\bl{E}}_i\cdot\hat{\bl{E}}_j= \delta_{ij}$, so that these vectors generate a cube of unit volume, see figure \ref{Sher_Twist_Exp}. Thus if $\boldsymbol{\sigma} = \boldsymbol{\omega} = 0$ then eq. (\ref{E'evlou}) implies that the infinitesimal evolution of these vectors is $\hat{\bl{E}'}_i = (1 + \frac{1}{3}\delta t\theta) \hat{\bl{E}}_i$. This says that the cube generated by the vectors $\{\hat{\bl{E}}_i\}$ is expanded by the same amount on all sides, so that its shape is kept invariant while its volume get multiplied by $(1 + \delta t\theta)$. Therefore, it is appropriate to call $\theta$ the expansion parameter.

Suppose now that both $\theta$ and $\boldsymbol{\omega}$ vanish. Since $\boldsymbol{\sigma}$ is a symmetric real matrix then it is always possible to choose an orthonormal frame in which it takes the diagonal form. Let us suppose that we are already on this frame, $\boldsymbol{\sigma}= \diag(\lambda_1, \lambda_2, \lambda_3)$. Then eq. (\ref{E'evlou}) yield $\hat{\bl{E}'}_i = (1 + \delta t\lambda_i) \hat{\bl{E}}_i$, \textit{i.e.}, the sides of the cube changes their length by different amounts but keep the direction fixed. It is simple matter to verify that after the infinitesimal evolution the volume changes by $\delta t(\lambda_1+  \lambda_2 + \lambda_3)$, which is zero since the trace of $\boldsymbol{\sigma}$ vanishes. Thus it is reasonable to call $\boldsymbol{\sigma}$ the shear.

Finally, setting $\theta$ and $\boldsymbol{\sigma}$ equal to zero and using the matrix $\boldsymbol{\omega}$ define the vector $\hat{\bl{\omega}} \equiv (\omega_{32}, \omega_{13}, \omega_{21})$. Then a simple algebra reveals that eq. (\ref{E'evlou}) yield $\hat{\bl{E}'}_i = \hat{\bl{E}}_i + \delta t\,\hat{\bl{\omega}}\times\hat{\bl{E}}_i$, where ``$\times$'' denotes the vectorial product of $\mathbb{R}^3$. This implies that the frame vectors are all infinitesimally rotated around the vector $\hat{\bl{\omega}}$ by the angle $\delta t|\hat{\bl{\omega}}|$, which justifies calling $\boldsymbol{\omega}$ the twist. Since this is a rotation it follows that the volume of the cube does not change. Figure \ref{Sher_Twist_Exp} depicts the action of the expansion, the shear and the twist.
\\

To analyze the relative movements of a congruence of null geodesics is a bit trickier. The problem is that in this case the space orthogonal to the geodesics also contains the vectors tangent to the congruence, as a null vector is orthogonal to itself. Therefore, we must ignore the part of the orthogonal space that is tangent to the null geodesics and work in an effective 2-dimensional space-like subspace. Let $\bl{l}$ be a vector field tangent to a congruence of null geodesics affinely parameterized. Thus introducing a frame $\{\bl{l},\bl{n},\hat{\bl{e}}_1,\hat{\bl{e}}_2\}$ such that the non-zero inner products are $l^\mu n_\mu=1$ and $\hat{e}_i^\mu\hat{e}_{j\,\mu}=-\delta_{ij}$, then the space of effective deviation vectors is generated by $\{\hat{\bl{e}}_i\}$. So that equation (\ref{dE/dtau}) yields:
\begin{equation}\label{de/dtau}
  \frac{d\hat{\bl{e}}_i}{d\tau} \,=\, \hat{e}_i^{\,\mu}\,\nabla_{\mu}\,\bl{l} \,\equiv\, \alpha_i\,\bl{l} \,+\, \beta_i\,\bl{n} \,+\, N_{ij}\,\hat{\bl{e}}_{j}\, \;\Rightarrow\quad\frac{d\hat{\bl{e}}_i}{d\tau} \,\sim \,N_{ij}\,\hat{\bl{e}}_{j}\,.
\end{equation}
Where the symbol ``$\sim$'' means equal except for terms proportional to $\bl{l}$ and it was used the fact that $\beta_i =0$, once $l^\mu l_\mu=0$. Thus on a null congruence we say that the expansion, shear and twist are respectively given by the trace, the trace-less symmetric part and the skew-symmetric part of the $2\times2$ matrix $N_{ij}$. By means of equation (\ref{de/dtau}) we see that the matrix $\mathbf{N}$ is defined by, $N_{ij} = -\bl{g}(\nabla_{\hat{\bl{e}}_i}\bl{l},\hat{\bl{e}}_j)$. We can encapsulate the four real components of the matrix $\mathbf{N}$ on the following three parameters called the optical scalars\index{Optical scalars} of the null congruence:
\begin{gather*}
  \theta\,\equiv\, \frac{1}{2}\left( N_{11} + N_{22} \right)\,;\;\, \omega\,\equiv\, \frac{1}{2}\left( N_{21} - N_{12} \right)\,;\\
  \sigma \,\equiv\, -\frac{1}{2}\left[ (N_{11} - N_{22}) +i(N_{12} + N_{21}) \right]\,.
\end{gather*}
The real scalars $\theta$ and $\omega$ are respectively called expansion and twist, while the complex scalar $\sigma$ is the shear\index{Shear} of the null geodesic congruence. Using these definitions it is possible to split the matrix $\mathbf{N}$ as the sum of its trace, its symmetric and trace-less part and its skew-symmetric part as follows:
$$  \mathbf{N} \,=\,  \theta\left[
                                     \begin{array}{cc}
                                       1 & 0 \\
                                       0 & 1 \\
                                     \end{array}
                                   \right] \,+\,  \frac{1}{2}\left[
                                     \begin{array}{cc}
                                       -(\sigma+\overline{\sigma}) & i(\sigma-\overline{\sigma}) \\
                                       i(\sigma-\overline{\sigma}) & (\sigma+\overline{\sigma}) \\
                                     \end{array}
                                   \right] \,+\, \omega\left[
                                     \begin{array}{cc}
                                       0 & -1 \\
                                       1 & 0\\
                                     \end{array} \right] \,.$$
Now it is useful to introduce the complex vector $\bl{m} = \frac{1}{\sqrt{2}}(\hat{\bl{e}}_1+i\,\hat{\bl{e}}_2)$, so that $\{\bl{l},\bl{n},\bl{m},\overline{\bl{m}}\}$ forms a null tetrad frame (appendix \ref{App._NullTetrad}). Then using the definitions of $\bl{m}$ and $\mathbf{N}$ it is straightforward to prove the following relations:
\begin{equation}\label{Shear}
\bl{g}(m^\mu\,\nabla_{\mu}\bl{l},\bl{m}) \,=\, \sigma     \;;\quad \bl{g}(m^\mu\,\nabla_{\mu}\bl{l},\overline{\bl{m}}) \,=\,  -(\theta \,+\, i \omega)\,.
\end{equation}
These are useful expressions that will be adopted as the definitions for the optical scalars of a null geodesic congruence in a 4-dimensional space-time.

Some important classes of space-times are defined by means of the optical scalars\index{Optical scalars}. In any dimension the Kundt class of space-times is defined as the one possessing a congruence of null geodesics that is shear-free ($\sigma= 0$), twist-free ($\omega= 0$) and with vanishing expansion ($\theta= 0$), $pp$-wave being the most important member of this class \cite{Rev. Ortaggio,Exact_Sol,pp-waveHD}. The Robinson-Trautman space-times are defined, in any dimension, as the ones containing a congruence of null geodesics that is shear-free, twist-free but with non-zero expansion, the Schwarzschild solution being one important example \cite{Exact_Sol,M. Ortaggio-Robinson-Trautman}. As a final comment it is worth mentioning that a congruence of null orbits is hypersurface-orthogonal ($l_{[\mu}\nabla_\nu l_{\rho]}=0$) if, and only if, the orbits are geodesic and twist-free \cite{M. Ortaggio-Robinson-Trautman}. Now we are ready to go on and study the theorems on the Petrov classification.

%
%

\section{Goldberg-Sachs}\label{Sec. GS4D1}\index{Goldberg-Sachs theorem}
The so-called Goldberg-Sachs (GS) theorem is the most important theorem about the Petrov classification. It was first proved by J. Goldberg and R. Sachs \cite{Goldberg-Sachs} and its mathematical formulation is the following:
\begin{theorem}\label{Theo_Gold-Sachs}
In a non-flat vacuum space-time (vanishing Ricci tensor and non-zero Riemann tensor) the Weyl scalars $\Psi_0$ and $\Psi_1$ vanish simultaneously if, and only if, the null vector field $\bl{l}$ is geodesic and shear-free.
\end{theorem}
Where in the above theorem it was used the notation introduced in section \ref{Petrov_Biv}. A relatively compact proof of this theorem can be found in ref.  \cite{chandrasekhar}. According to section \ref{Sec._Bel-Deb} the condition $\Psi_0= \Psi_1= 0$ is equivalent to the relation $l_{[\alpha}C_{\mu]\nu\rho\sigma}l^\nu l^\rho =0$, which means that $\bl{l}$ is a repeated principal null direction\index{Principal null directions(PND)}. An equivalent form of stating this theorem is saying that in vacuum a null vector field is geodesic and shear-free if, and only if, it points in a repeated PND. In particular, algebraically special vacuum space-times must admit a shear-free congruence of null geodesics.

A particularly interesting situation occurs in vacuum solutions of Petrov type $D$. Since in this case the Weyl tensor admits two repeated PNDs (section \ref{Sec._Annihil_Scalars}) it follows that there exist two independent null geodesic congruences that are shear-free. This apparently inconsequential geometric restriction has enabled the complete integration of Einstein's field equation \cite{typeD}, \textit{i.e.}, all type $D$ vacuum solutions were analytically found. In addition, the Goldberg-Sachs theorem has also played a prominent role on the original derivation of Kerr solution \cite{Kerr}. Interestingly, all known black-holes are of type $D$.

Let us suppose that a conformal transformation is made on the space-time, $(M,\bl{g})\mapsto (M,\tilde{\bl{g}}= \Omega^2\bl{g})$. Then if $\{\bl{l}, \bl{n}, \bl{m}, \overline{\bl{m}}\}$ is a null tetrad frame in $(M,\bl{g})$ then $\{\widetilde{\bl{l}}= \bl{l}, \widetilde{\bl{n}}= \Omega^{-2}\bl{n}, \tilde{\bl{m}}= \Omega^{-1}\bl{m}, \tilde{\overline{\bl{m}}}= \Omega^{-1}\overline{\bl{m}}\}$ will be a null tetrad on $(M,\tilde{\bl{g}})$. Then defining $V_\mu\equiv \partial_\mu\ln\Omega$ and working out the transformation of the Christoffel symbol it is a simple matter to prove the following relation:
$$ \tilde{\nabla}_\mu\, \tilde{l}^\nu \,=\, \nabla_\mu\,l^\nu \,+\,\delta^\nu_{\,\mu}\,l^\rho V_\rho  \,+\,V_\mu\,l^\nu  \,-\, l_\mu\,g^{\nu\rho}\,V_\rho  \,. $$
From which we immediately see that if $\bl{l}$ is geodesic in $(M,\bl{g})$ so will be $\tilde{\bl{l}}$  in $(M,\tilde{\bl{g}})$, although not affinely parameterized in general. Moreover, using equation (\ref{Shear}) we find that $\sigma= 0$ if, and only if, $\tilde{\sigma}= 0$. Therefore, on null congruences the geodesic shear-free condition is invariant under conformal transformations. Since the Weyl tensor is also invariant under these transformations we conclude that there exists a kind of asymmetry on the GS theorem as stated above, as the vacuum condition is not invariant under conformal transformations.  Noting this, I. Robinson and A. Schild have been able to generalize the GS theorem to conformally Ricci-flat space-times \cite{GS-Conformal}.

Fourteen years after the appearance of the GS theorem, J. Plebla\'{n}ski and S. Hacyan noticed that in vacuum the existence of a null congruence that is geodesic and shear-free is equivalent to the existence of two integrable distributions\index{Distribution} of isotropic planes \cite{Plebanski2}. This is of great geometric relevance and will be exploited on the next chapter in order to generalize the GS theorem to 4-dimensional manifolds of all signatures.

Since non-linear equations are hard to deal with, sometimes it is useful to linearize Einstein's equation in order to study some properties of general relativity. But it is very important to keep in mind that many features of the linearized model are not carried to the complete theory. Particularly, in ref. \cite{GS_Linear} it was proved that the Goldberg-Sachs theorem is not valid in linearized gravity. The proof consisted of presenting explicit examples of linearized space-times admitting a null vector field that is geodesic and shear-free but is not a repeated PND on the linearized theory.

Since the GS theorem proved to be of great relevance to 4-dimensional general relativity, recently a lot of effort has been made in order to generalize this theorem to higher dimensions. But this task is not trivial at all. For instance, in \cite{FrolovMyers5D} it was proved that in 5 dimensions a repeated PND (according to Bel-Debever criteria) is not necessarily shear-free. Indeed, the shear-free condition turns out to be quite restrictive in dimensions greater than 4. A suitable higher-dimensional generalization of the PNDs are the so-called Weyl aligned null directions (WANDs) \cite{CMPP}. Although the WANDs share many properties with the 4-dimensional PNDs there are also some important differences. For example, while in four dimensions a non-zero Weyl tensor admits at least one and at most four PNDs, in higher dimensions a non-vanishing Weyl tensor may admit from zero up to infinitely many WANDs \cite{GS-GeodesicPart}. Some progress towards a higher-dimensional generalization of the GS theorem was already accomplished using this formalism \cite{GS-GeodesicPart,GS-5D,GS-HighD,Rev. Ortaggio}. In particular it was proved that every space-time admitting a repeated WAND has at least one repeated WAND that is geodesic. Moreover, in chapter \ref{Chap. AllD} it will be presented a particular generalization of this theorem valid in even dimensions.

The equivalence between the geodesic and shear-free condition and the integrability of null planes provides another path to generalize the GS theorem. A partial generalization of the Goldberg-Sachs theorem using this method has been accomplished in 2011 by Taghavi-Chabert \cite{HigherGSisotropic1,HigherGSisotropic2}. He has proved that in a Ricci-flat manifold of dimension $d=2n+\epsilon$, with $\epsilon=0,1$, if the Weyl tensor is algebraically special but generic otherwise then the manifold admits an integrable $n$-dimensional isotropic distribution. Such generalisation will be exploited and reinterpreted in chapters \ref{Chap._Spin6D} and \ref{Chap. AllD}.

\section{Mariot-Robinson}\label{Sec._Mariot4D}\index{Mariot-Robinson theorem}
We call $F_{\mu\nu}=F_{[\mu\nu]}\neq0$ a null bivector\index{Null bivector} when $F^{\mu\nu}F_{\mu\nu}=0= F^{\mu\nu}\,\star F_{\mu\nu}$, where $\star \bl{F}$ is the Hodge dual of $\bl{F}$, defined on equation (\ref{dualofbivector}). It can be proved that $\bl{F}$ is a real null bivector if, and only if, there exists some null vector $\bl{l}$ and a space-like vector $\bl{e}$ such that:
$$  F_{\mu\nu} \,=\, 2\,l_{[\mu}\,e_{\nu]}\;;\quad l^\mu\,e_\mu\,=\,0\,. $$
The null vector $\bl{l}$ is then called the principal null vector of $\bl{F}$. Up to a multiplicative constant, $\bl{l}$ is the unique vector that simultaneously obeys to the algebraic relations $F_{\mu\nu}\,l^\nu= 0$ and $F_{[\mu\nu}\,l_{\rho]}= 0$. The Mariot-Robinson theorem is then given by \cite{Robinson}:
\begin{theorem}\label{Theo_MariotRob}
A 4-dimensional Lorentzian manifold admits a null bivector obeying to the source-free Maxwell's equations if, and only if, the principal null vector of such  bivector generates a null congruence that is geodesic and shear-free.
\end{theorem}
A simple proof of this theorem using spinors is given in \cite{AdvancedGR}. More explicitly, such theorem guarantees that if $F_{\mu\nu}=l_\mu e_\nu- e_\mu l_\nu$ obeys the equations $\nabla^\mu F_{\mu\nu}= 0$ and $\nabla^\mu\,(\star F)_{\mu\nu}= 0$ then the null vector field $\bl{l}$ must be geodesic and shear-free. Conversely, if $\bl{l}$ generates a null congruence of shear-free geodesics then one can always find a space-like vector field $\bl{e}$ such that $F_{\mu\nu}=l_\mu e_\nu- e_\mu l_\nu$ obeys the equations $\nabla^\mu F_{\mu\nu}= 0$ and $\nabla^\mu \,(\star F)_{\mu\nu}= 0$. Using this result and the Goldberg-Sachs theorem we immediately arrive at the following interesting consequence:
\begin{corollary}
A vacuum space-time is algebraically special according to the Petrov classification if, and only if, it admits a null bivector obeying to source-free Maxwell's equations.
\end{corollary}
In this corollary the Maxwell field, $\bl{F}$, was assumed to be a test field, which means that its energy was assumed to be low enough to be neglected on Einstein's equation, so that the space-time can be assumed to be vacuum. But, actually, this corollary remains valid if we also consider that the electromagnetic field distorts the space-time, \textit{i.e}, if the metric obeys the equation $R_{\mu\nu} -\frac{1}{2}Rg_{\mu\nu}= 8\pi G\, T_{\mu\nu}$, where $T_{\mu\nu}$ is the energy-momentum tensor of the electromagnetic field $\bl{F}$.

Physically, a null Maxwell field represents electromagnetic radiation. Suppose that $\{\hat{\bl{e}}_t,\hat{\bl{e}}_x,\hat{\bl{e}}_y,\hat{\bl{e}}_z\}$ is a Lorentz frame, then a plane electromagnetic wave of frequency $\omega$ propagating on the direction $\hat{\bl{e}}_z$ is generated by the electric field $\mathbf{E}=E_0\cos[\omega(z-t)]\,\hat{\bl{e}}_x$ and the magnetic field $\mathbf{B}=E_0\cos[\omega(z-t)]\,\hat{\bl{e}}_y$. Indeed, it is simple matter to verify that these fields are solutions of the Maxwell's equations without sources. The field $\bl{F}$ associated to such electric and magnetic fields is $F_{\mu\nu} = 2\,l_{[\mu}e_{\nu]}$, with $\bl{l} = (\hat{\bl{e}}_t + \hat{\bl{e}}_z)$ and $\bl{e}=-E_0\cos[\omega(z-t)]\,\hat{\bl{e}}_x$, which is a null
bivector\index{Null bivector}. The energy-momentum tensor of such field is given by $T_{\mu\nu} = \frac{e^\rho e_\rho}{4\pi}l_\mu l_\nu$.

Given the null field $F_{\mu\nu} = 2\,l_{[\mu}e_{\nu]}$ then the bivectors $\bl{F}^\pm= (\bl{F} \pm i \star\bl{F})$ are given by $F^+_{\mu\nu}=2\,l_{[\mu}m_{\nu]}$ and $F^-_{\mu\nu}=2\,l_{[\mu}\overline{m}_{\nu]}$, where $\bl{m}$ is a complex null vector field orthogonal to $\bl{l}$. In section \ref{Sec. GS4D1} it was commented that the existence of a shear-free congruence of null geodesics  is equivalent to the existence of two integrable distributions\index{Distribution} of isotropic planes\index{Isotropic}. Therefore, the Mariot-Robinson theorem guarantees that the existence of a null solution for the source-free Maxwell's equations is equivalent to the existence of two integrable distributions of isotropic planes. These distributions are the ones generated by $\{\bl{l},\bl{m}\}$ and $\{\bl{l},\overline{\bl{m}}\}$.

By means of the language of isotropic distributions, the Mariot-Robinson theorem admits a generalization valid in all even dimensions and all signatures. In \cite{HughstMason} the proof was made using spinors, while in \cite{art4} a simplified proof using just tensors is presented. This generalized version of the Mariot-Robinson theorem will be discussed in chapter \ref{Chap. AllD}.

\section{Peeling Property}\index{Peeling property}

In this section it will be shown that the Weyl tensor of an asymptotically flat space-time has a really simple fall off behaviour near the null infinity. But before enunciating this beautiful result it is necessary to introduce the concept of asymptotic flatness. By an asymptotically flat space-time it is meant one that looks like Minkowski space-time as we approach the infinity. But in order to extract any mathematical consequence of this hypothesis it is necessary to make a rigorous definition of what ``looks like Minkowski'' means. This is a bit complicated since coordinates are meaningless in general relativity, so that it is not reasonable to say that the metric of an asymptotically flat space-time must approach the Minkowski metric as the spatial coordinates go to infinity.

In order to avoid taking coordinates to infinity it is interesting to perform a conformal transformation, $g_{\mu\nu}\mapsto \widetilde{g}_{\mu\nu}= \Omega^2g_{\mu\nu}$, that brings the points from the infinity of an asymptotically flat space-time to a finite distance. Thus although $\int ds = \int \sqrt{g_{\mu\nu}dx^\mu dx^\nu}$ goes to infinity as $x^\mu \rightarrow \infty$ we can manage to make $\int d\tilde{s} = \int \Omega\sqrt{g_{\mu\nu}dx^\mu dx^\nu}$ finite by properly making $\Omega \rightarrow 0$ as $x^\mu \rightarrow \infty$. So that the infinity of the space-time $(M,\bl{g})$ is represented by the boundary $\Omega=0$ on the space-time $(M,\widetilde{\bl{g}})$. Using this reasoning a space-time $(M,\bl{g})$ is said to be asymptotically flat when there exists another space-time $(\widetilde{M},\widetilde{\bl{g}})$, called the non-physical space-time, such that: (1) $M\subset \widetilde{M}$ and $\widetilde{M}$ has a boundary given by $\Omega= 0$ that represents the null infinity of $(M,\bl{g})$; (2) $\widetilde{g}_{\mu\nu}= \Omega^2g_{\mu\nu}$ and $\partial_\mu\Omega\neq 0$ on the boundary $\Omega= 0$; (3) The Ricci tensor of $(M,\bl{g})$ vanishes on the neighborhood of $\Omega= 0$. For details and motivation of this definition see \cite{AdvancedGR,Peeling-Penrose,Wald}.

Since we have some freedom on the definition of $\Omega$, we can choose it to be the affine parameter of a null geodesic on $(\widetilde{M},\widetilde{\bl{g}})$,  let $\tilde{\bl{l}}= \frac{d\;}{d\Omega}$ be the tangent to this geodesic. Such geodesic then defines another null geodesic on $(M,\bl{g})$ whose tangent shall be denoted by $\bl{l}= \frac{d\,}{dr}$. Imposing $r$ to be an affine parameter we find that $r=-\Omega^{-1}$, so that $l^\mu=\Omega^2\widetilde{l}^\mu$. The non-physical manifold, $(\widetilde{M},\widetilde{\bl{g}})$, and the vector $\tilde{n}_\mu = \partial_\mu\Omega$ are assumed to be completely regular on the boundary $\Omega= 0$. Using this and the transformation rule of the Ricci scalar under conformal transformations we find that the vector field $\tilde{\bl{n}}$ becomes null, according to $\tilde{\bl{g}}$, as we approach the boundary of $\widetilde{M}$. Note also that $\tilde{l}^\mu\tilde{n}_\mu = 1$, hence we can find a complex vector $\tilde{\bl{m}}$ so that, near the boundary, $\{\tilde{\bl{l}},\tilde{\bl{n}},\tilde{\bl{m}},\overline{\tilde{\bl{m}}}\}$ forms a null tetrad of $(\widetilde{M},\widetilde{\bl{g}})$. Since $\bl{l}=\Omega^2\widetilde{\bl{l}}$ and $\bl{g}=\Omega^{-2}\widetilde{\bl{g}}$ we find that the corresponding null tetrad of $(M,\bl{g})$ is such that $\bl{n}=\tilde{\bl{n}}$ and $\bl{m}=\Omega\tilde{\bl{m}}$.

\begin{figure}
	\centering
		\includegraphics[width=5cm]{{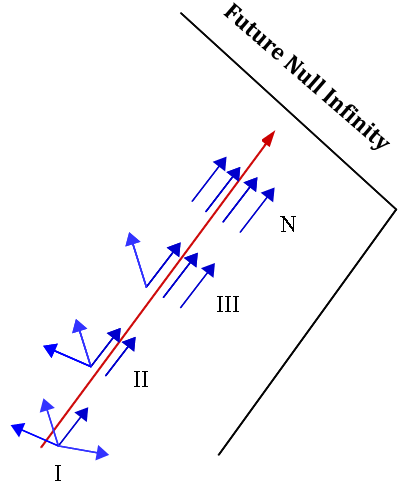}}
	\caption{\footnotesize{According to the peeling theorem, as we approach the null infinity of an asymptotically flat space-time the Petrov type of the Weyl tensor becomes increasingly special. The blue arrows represent the principal null directions of the Weyl tensor, while the red axis represents the null direction along which null infinity is approached.}}
	\label{Fig.Peeling}
\end{figure}

Since $(\widetilde{M},\widetilde{\bl{g}})$ is regular at $\Omega=0$ it is expected that the Weyl scalars of the non-physical space-time are all non-vanishing and of the same order on the boundary. However, it can be proved that the Weyl scalars of $(\widetilde{M},\widetilde{\bl{g}})$ are generally of order $\Omega$ \cite{AdvancedGR}, $\widetilde{\Psi}_\alpha\sim O(\Omega)$. Using this fact along with equation (\ref{weylscalars}) and the transformation of the null tetrad frame, we can easily find the behaviour of the Weyl scalars of $(M,\bl{g})$. For example,
\begin{align*}
  \Psi_0 \,=&\, C_{\mu\nu\rho\sigma}l^\mu m^\nu l^\rho m^\sigma \,=\, \Omega^{-2}\,\widetilde{C}_{\mu\nu\rho\sigma}  l^\mu m^\nu l^\rho m^\sigma \\
  =&\, \Omega^{-2}\,\widetilde{C}_{\mu\nu\rho\sigma}  \,\Omega^2\tilde{l}^\mu \, \Omega \tilde{m}^\nu\, \Omega^2 \tilde{l}^\rho \,\Omega \tilde{m}^\sigma  \,=\, \Omega^4\, \widetilde{\Psi}_0  \,\sim\, O(\Omega^5)\,.
\end{align*}
Where it was used the fact that $C^\mu_{\phantom{\mu}\nu\rho\sigma}$ is invariant by conformal transformations, which implies that $C_{\mu\nu\rho\sigma} = g_{\mu\kappa}C^\kappa_{\phantom{\kappa}\nu\rho\sigma} = \Omega^{-2}\widetilde{C}_{\mu\nu\rho\sigma}$. In general the following behaviour is found:
$$  \Psi_0 \,\sim\, O(\Omega^5) \;,\;\, \Psi_1 \,\sim\, O(\Omega^4) \;,\;\, \Psi_2 \,\sim\, O(\Omega^3) \;,\;\, \Psi_3 \,\sim\, O(\Omega^2) \;,\;\, \Psi_4 \,\sim\, O(\Omega)\,.$$
Since $\Omega = -r^{-1}$, the above relations along with table \ref{Tab._vanishScalars} implies the following result known as the peeling theorem \cite{Peeling-Penrose}:
\begin{theorem}\label{Theo_Peeling}\index{Principal null directions(PND)}
Let $(M,\bl{g})$ be an asymptotically flat space-time. Then if we approach the null infinity, $r\rightarrow \infty$, along a null geodesic whose affine parameter is $r$ and whose tangent vector is $\bl{l}$ then the Weyl tensor has the following fall off behaviour:
$$ \bl{C} \,=\, \frac{\bl{C}_N}{r} \,+\, \frac{\bl{C}_{III}}{r^2} \,+\,  \frac{\bl{C}_{II}}{r^3} \,+\, \frac{\bl{C}_{I}}{r^4} \,+\, O(r^{-5})\,.$$
Where the tensors $\bl{C}_N$, $\bl{C}_{III}$, $\bl{C}_{II}$, and $\bl{C}_{I}$ have the symmetries of a Weyl tensor and are respectively of Petrov type $N$, $III$, $II$ and $I$ (or more special). The vector field $\bl{l}$ is a repeated PND of the first three terms of the above expansion and a PND of the tensor $\bl{C}_{I}$ (see figure \ref{Fig.Peeling}).
\end{theorem}

The peeling theorem\index{Peeling property} has been generalized to higher dimensions just quite recently \cite{Peel-Reall}. It was proved that the fall off behaviour of the Weyl tensor in higher dimensions is both qualitatively and quantitatively different from the 4-dimensional case. Indeed, concerning asymptotic infinity the dimension 4 is a very special one, as the definition of asymptotically flat in other dimensions proved to be fairly tricky \cite{Hollands-Asymp.Flat,Asymp.Flat_Tanabi}. The physical justification for a non-trivial definition of asymptotic flatness in higher dimensions comes from the fact that such definition must be stable under small perturbations, it should be compatible with the existence of a generator for the Bondi energy and it might allow the existence of gravitational radiation.

\section{Symmetries}
Given the Petrov type of a space-time occasionally it is possible to say which symmetries the manifold might have and, conversely, given the symmetries of a space-time sometimes we can guess its Petrov classification. The intent of this section is to present some theorems connecting the Petrov classification with the existence of symmetry tensors. One of the first results on these lines was obtained by Kinnersley in \cite{typeD}, where he explicitly found all type $D$ vacuum solutions and, as a bonus, arrived at the following result:
\begin{theorem}
Every type $D$ vacuum space-time admits either 4 or 2 independent Killing vector fields.
\end{theorem}
Another remarkable result about type $D$ solutions was then found by Walker and Penrose in ref. \cite{Walk.Penr.}, where it was proved that these space-times have a hidden symmetry\index{Hidden symmetries}:
\begin{theorem}\label{Theo_typeD_KT}
Every type $D$ vacuum space-time with less than 4 independent Killing vectors admits a non-trivial conformal Killing tensor (CKT) of order two. Furthermore, if the metric is not a C-metric\footnote{This is an important class of type $D$ vacuum solutions representing a pair of Black Holes accelerating away from each other due to structures represented by conical singularities. The C-metric is a generalization of the Schwarzschild solution with one extra parameter in addition to the mass, so that the Schwarzschild metric is a particular member of this class. For a thorough analysis of these metrics see \cite{Exact_Sol}.} then this CKT is, actually, a Killing tensor.
\end{theorem}\index{Killing-Yano tensor}
The second part of the above theorem can be found in \cite{Hugh_Killing,Stephani}. Later, Collinson \cite{Collinson} and Stephani \cite{Steph_KY} investigated whether these Killing tensors can be constructed out of Killing-Yano tensors (see section \ref{Sec._Symmetries}), arriving at the following result:
\begin{theorem}
Every type $D$ vacuum space-time possessing a non-trivial Killing tensor of order two, $K_{\mu\nu}$, also admits a Killing-Yano tensor $Y_{\mu\nu}$ such that $K_{\mu\nu}= Y_{\mu}^{\phantom{\mu}\sigma}Y_{\sigma\nu}$.
\end{theorem}

As defined in section \ref{Sec._Mariot4D}, a real bivector $B_{\mu\nu}$ is called null when it can be written as $B_{\mu\nu}=l_{[\mu}e_{\nu]}$, where $\bl{l}$ is null, $\bl{e}$ is space-like and $l^\mu\,e_\mu= 0$. On the other hand, if $B'_{\mu\nu}$ is a real non-null bivector then it is always possible to arrange a null tetrad frame such that $B'_{\mu\nu}= a\,l_{[\mu}n_{\nu]} + ib\,m_{[\mu}\overline{m}_{\nu]}$, where $a$ and $b$ are real functions (this can be easily seen using spinors). Using this along with the results of \cite{Ibohal-KY} we can state:
\begin{theorem}
A vacuum space-time admitting a null Killing-Yano tensor of order two, $Y_{\mu\nu}=l_{[\mu}e_{\nu]}$, must be of Petrov type $N$ with $\bl{l}$ being the repeated PND. On the other hand, a vacuum space-time admitting a non-null Killing-Yano tensor of order two, $Y'_{\mu\nu}= a\,l_{[\mu}n_{\nu]} + ib\,m_{[\mu}\overline{m}_{\nu]}$, must have type $D$ with $\bl{l}$ and $\bl{n}$ being repeated PNDs.
\end{theorem}
Actually, this theorem remains valid if instead of vacuum we consider electro-vacuum space-times \cite{Ibohal-KY}. For more theorems on the same line see \cite{Stephani} and references therein.

Regarding higher-dimensional space-times, it is appropriate mentioning references \cite{Mason_KY,HigherGSisotropic1} which, inspired by theorem \ref{Theo_typeD_KT}, have suggested that a suitable generalization of the Petrov type $D$ condition for manifolds of dimension $d=2n+\epsilon$, with $\epsilon= 0,1$, should be the existence of $2^n$ integrable maximally isotropic distributions\index{Maximally isotropic}. For interesting results concerning hidden symmetries and Killing-Yano\index{Hidden symmetries} tensors in higher-dimensional black holes see the nice paper \cite{Frolov_KY}.

\part{Original Research}\label{Part-Orig}

\chapter{Generalizing the Petrov Classification and the Goldberg-Sachs Theorem to All Signatures}\label{Chap. Gen4D}
In the previous chapters it was defined the Petrov classification, an algebraic classification for the Weyl tensor valid in 4-dimensional Lorentzian manifolds that is related to very important theorems. In particular, such classification proved to be helpful in the search of new exact solutions to Einstein's equation, a remarkable example being the Kerr metric \cite{Kerr}. The aim of this chapter is to generalize the Petrov classification to 4-dimensional spaces of arbitrary signature. The strategy adopted here is to work with complexified spaces, interpreting the various signatures as different reality conditions. This approach is based on the reference \cite{art1} and yields a unified classification scheme to all signatures. Generalizations of the Petrov classification were already known before the article \cite{art1}: In \cite{Plebanski75} the complex case was treated using spinors, Euclidean manifolds were investigated in \cite{Hacyan,Karlhede}, while the split signature was studied in \cite{Petrov_Book,Law1,Nurwoski2,Law2,ColeyPSEUD}. But none of these previous works attempted to provide a unified classification scheme such that each signature is just a special case of the complex classification.

The Goldberg-Sachs theorem is the most important result on the Petrov classification. Particularly, it enabled the complete integration of Einstein's vacuum equation for type $D$ space-times \cite{typeD}. In ref. \cite{Plebanski2} Pleba\'{n}ski and Hacyan proved a beautiful generalization of this theorem valid in complexified manifolds. They realised that a suitable complex generalization of a shear-free null geodesic congruence is an integrable distribution of isotropic planes. Here such generalized theorem will be used to show that certain algebraic restrictions on the Weyl tensor imply the existence of important geometric structures on 4-dimensional manifolds of any signature, results that were presented on the article \cite{art2}.

\section{Null Frames}\label{Sec. NullFrame4}\index{Null frame}
Before proceeding it is important to establish the notation that will be adopted throughout this chapter. In particular, let us see explicitly how one can use a complexified space in order to obtain results on real manifolds of arbitrary signature. We shall define a null frame on a 4-dimensional manifold as a frame of vector fields $\{\bl{E}_1,\bl{E}_2, \bl{E}_3,\bl{E}_4\}$ such that the only non-zero inner products are:
\begin{equation}\label{NullFrame4D}
  \bl{g}(\bl{E}_1,\bl{E}_3)\,=\,1   \quad \textrm{ and } \quad   \bl{g}(\bl{E}_2,\bl{E}_4)\,=\,-1 \,.
\end{equation}
Particularly, note that all vector fields on this frame are null. Depending on the signature\index{Signature} of the manifold the vectors of a null frame obey to different reality conditions, let us see this explicitly.
\\
\\
\small
\textsl{
$\bullet$ \textbf{Euclidean Signature, $\bl{s=4}$}\\
In such a case, by definition, it is possible to introduce a real frame $\{\hat{\bl{e}}_a\}$ such that $\bl{g}(\hat{\bl{e}}_a,\hat{\bl{e}}_b)= \delta_{ab}$. Thus it is straightforward to see that the following vectors form a null frame:
$$ \bl{E}_1 = \frac{1}{\sqrt{2}}(\hat{\bl{e}}_1 + i \hat{\bl{e}}_3)\,;\; \bl{E}_2 = \frac{i}{\sqrt{2}}(\hat{\bl{e}}_2 + i \hat{\bl{e}}_4)\,;\;  \bl{E}_3 = \frac{1}{\sqrt{2}}(\hat{\bl{e}}_1 - i \hat{\bl{e}}_3)\,;\;  \bl{E}_4 = \frac{i}{\sqrt{2}}(\hat{\bl{e}}_2 - i \hat{\bl{e}}_4)\,.$$
Note that the following reality conditions hold:
\begin{equation}\label{Real.Eucli}
   \bl{E}_3 \,=\, \overline{\bl{E}_1} \quad; \quad    \bl{E}_4 \,=\, -\overline{\bl{E}_2}\,.
\end{equation}
$\bullet$ \textbf{Lorentzian Signature, $\bl{s=2}$}\\
As shown on appendix \ref{App._NullTetrad} in this signature we can introduce a null tetrad $\{\bl{l},\bl{n},\bl{m},\overline{\bl{m}}\}$, which is a frame such that the only non-zero inner products are $l^\mu n_{\mu}= 1$ and $m^\mu \overline{m}_{\mu}= -1$. Therefore, the following vector fields form a null frame:
\begin{equation}\label{NullFrame-Tetrad}
  \bl{E}_1 \,=\,\bl{l}  \;;\quad \bl{E}_2 \,=\, \bl{m} \;;\quad \bl{E}_3 \,=\,\bl{n} \;;\quad  \bl{E}_4  \,=\, \bl{\overline{m}}
\end{equation}
So a null frame is just a null tetrad reordered. Since, by definition, in a null tetrad $\bl{l}$ and $\bl{n}$ are both real, it follows that on Lorentzian case the reality conditions are:
\begin{equation}\label{Real.Loren}
 \bl{E}_1 \,=\,\overline{\bl{E}_1}  \quad; \quad \bl{E}_3 \,=\, \overline{\bl{E}_3}   \quad; \quad \bl{E}_4 \,=\, \overline{\bl{E}_2} \,.
\end{equation}
$\bullet$ \textbf{Split Signature, $\bl{s=0}$}\\
In such signature there exists a real frame $\{\hat{\bl{e}}_a\}$ such that $\bl{g}(\hat{\bl{e}}_a,\hat{\bl{e}}_b)= \diag(1,1,-1,-1)$. Then the following vectors form a null frame:
$$ \bl{E}'_1 = \frac{1}{\sqrt{2}}(\hat{\bl{e}}_1 +  \hat{\bl{e}}_3)\,;\; \bl{E}'_2 = \frac{1}{\sqrt{2}}(\hat{\bl{e}}_4 + \hat{\bl{e}}_2)\,;\; \bl{E}'_3 = \frac{1}{\sqrt{2}}(\hat{\bl{e}}_1 -  \hat{\bl{e}}_3)\,;\; \bl{E}'_4 = \frac{1}{\sqrt{2}}(\hat{\bl{e}}_4 -  \hat{\bl{e}}_2)\,. $$
Note that all vectors on this frame are real:
\begin{equation}\label{Real.Split}
  \bl{E}'_1 \,=\,\overline{\bl{E}'_1}  \quad; \quad \bl{E}'_2 \,=\,\overline{\bl{E}'_2}  \quad; \quad \bl{E}'_3 \,=\,\overline{\bl{E}'_3}  \quad; \quad \bl{E}'_4 \,=\,\overline{\bl{E}'_4}  \,.
\end{equation}
When the metric has split signature it is also possible to introduce a complex null frame. Indeed, note that the vector fields
$$ \bl{E}_1 = \frac{1}{\sqrt{2}}(\hat{\bl{e}}_1 + i \hat{\bl{e}}_2)\,;\; \bl{E}_2 = \frac{1}{\sqrt{2}}(\hat{\bl{e}}_3+ i \hat{\bl{e}}_4)\,;\;  \bl{E}_3 = \frac{1}{\sqrt{2}}(\hat{\bl{e}}_1 - i \hat{\bl{e}}_2)\,;\;  \bl{E}_4 = \frac{1}{\sqrt{2}}(\hat{\bl{e}}_3 - i \hat{\bl{e}}_4)$$
form a null frame. The reality conditions on this frame are $\bl{E}_3= \overline{\bl{E}}_1$ and $\bl{E}_4= \overline{\bl{E}}_2$.}
\\
\normalsize
\\
Therefore, a wise path to obtain results valid in any signature is to assume that the tangent bundle is complexified and when necessary use a suitable reality condition to specify the signature. This can easily be understood as follows: if we work over the complex field the signature is not fixed, because a vector $\hat{\bl{e}}$ whose norm squared is $1$, $\bl{g}(\hat{\bl{e}},\hat{\bl{e}})= 1$, can be multiplied by $i$ and yield a vector whose norm squared is $-1$, so that the apparent signature can be changed.

Once fixed a null frame $\{\bl{E}_a\}$, one can define the dual frame $\{\bl{E}^a\}$, which is a set of 1-forms such that $\bl{E}^a(\bl{E}_b)= \delta^{\,a}_b$ (see section \ref{Sec. Cartan}). By means of eq. (\ref{NullFrame4D}) it is trivial to note that the components of such 1-forms are:
\begin{equation}\label{E^a-E_a}
   E^{1\,\mu} \,=\, E_3^{\phantom{3}\mu}\,\,;\,\; E^{2\,\mu} \,=\, -E_4^{\phantom{4}\mu}\,\,;\,\; E^{3\,\mu} \,=\, E_1^{\phantom{1}\mu}\,\,;\,\; E^{4\,\mu} \,=\, -E_2^{\phantom{2}\mu}\,.
\end{equation}
The dual frame can be used to define the following 2-forms constituting a basis for the space of bivectors:
\small
\begin{gather*}
 \bl{Z}^{1+} \,=\, \bl{E}^4\wedge\bl{E}^3 \,\,;\;\, \bl{Z}^{2+} \,=\, \bl{E}^1\wedge\bl{E}^2 \,\,;\;\,  \bl{Z}^{3+} \,=\, \frac{1}{\sqrt{2}}\lef\bl{E}^1\wedge\bl{E}^3 + \bl{E}^4\wedge\bl{E}^2\rig  \\
  \,\bl{Z}^{1-} \,=\, \bl{E}^2\wedge\bl{E}^3 \,\,;\;\, \bl{Z}^{2-} \,=\, \bl{E}^1\wedge\bl{E}^4 \,\,;\;\,  \bl{Z}^{3-} \,=\, \frac{1}{\sqrt{2}}\lef\bl{E}^1\wedge\bl{E}^3 + \bl{E}^2\wedge\bl{E}^4\rig .
\end{gather*}
\normalsize
By means of eq. (\ref{E^a-E_a}) we see that the components of the 2-form  $\bl{Z}^{1+}$ are $Z^{1+\,\mu\nu}= 2 E_1^{\phantom{1}[\mu}E_2^{\phantom{1}\nu]}$, which sometimes is written as $\bl{Z}^{1+}= \bl{E}_1\wedge\bl{E}_2$. Because of this we say that $\bl{Z}^{1+}$ generates the family of planes spanned by the vector fields $\bl{E}_1$ and $\bl{E}_2$. Note that since $\bl{g}(\bl{E}_1,\bl{E}_1)= \bl{g}(\bl{E}_2,\bl{E}_2)= \bl{g}(\bl{E}_1,\bl{E}_2)= 0$, all vectors tangent to these planes are null. This kind of plane is called totally null or isotropic\index{Isotropic} and $\bl{Z}^{1+}$ is then called a null
bivector\index{Null bivector}. More about isotropic subspaces can be found in \cite{SimpleSpinors}. In the same vein $\bl{Z}^{2+}$, $\bl{Z}^{1-}$ and $\bl{Z}^{2-}$ generate the isotropic planes spanned by $\{\bl{E}_3,\bl{E}_4\}$, $\{\bl{E}_1,\bl{E}_4\}$ and $\{\bl{E}_2,\bl{E}_3\}$ respectively. From now on a bivector $\bl{Z}$ will be called a null bivector when it can be written as $Z^{\mu\nu}= 2l^{[\mu}k^{\nu]}$ with $Span\{\bl{l},\bl{k}\}$  being a distribution of isotropic planes\footnote{Note that in section \ref{Sec._Mariot4D} the definition of a null bivector was broader than this, there a bivector $\bl{B}= \bl{l}\wedge\bl{e}$ with $\bl{e}$ being space-like and orthogonal to the null vector $\bl{l}$ was also called null. But if we are working with arbitrary signature it is more useful to define a null bivector\index{Null bivector} as a simple bivector that generates an isotropic distribution.}.

%
%

Since the determinant of the matrix $g_{ab}= \bl{g}(\bl{E}_a,\bl{E}_b)$ is $g=1$, the components of the volume-form\index{Volume-form} on the null frame $\{\bl{E}_a\}$ are given by
$$ \epsilon_{abcd}\,=\,  \varepsilon_{abcd}\,,\quad \textrm{where}\quad \varepsilon_{abcd}= \varepsilon_{[abcd]} \;\textrm{ and }\; \varepsilon_{1234}\equiv-1\,.  $$
Thus if $\bl{Z}$ is a bivector, $Z_{ab}= Z_{[ab]}$, then its Hodge dual\index{Hodge dual} is given by:
\begin{equation}\label{Hdual-4D}
  \star Z_{cd} \,=\,\frac{1}{2}\, Z^{ab}\, \varepsilon_{abcd}\,.
\end{equation}
With the aim of improving the notation, let us define $\mathcal{H}$ as an operator that acts on the space of bivectors in some open set of the manifold, $\mathcal{H}: \Gamma(\wedge^2 M)\rightarrow\Gamma(\wedge^2 M)$, and implements the Hodge dual map, $\mathcal{H}(\bl{Z})\equiv \star\bl{Z}$. Then using equation (\ref{Hdual-4D}) it is simple matter to verify that $\mathcal{H}^2= \bl{1}$, where $\bl{1}$ is the identity operator. Thus the eigenvalues of $\mathcal{H}$ are $\pm1$ and the bivector space at such neighborhood can be split as the following direct sum\footnote{All results in this thesis are local, so that it is always being assumed that we are in the neighborhood of some point. Thus, formally, instead of $\Gamma(\wedge^2 M)$ we should have written $\Gamma(\wedge^2 M)|_{N_x}$, which is the restriction of the space of sections of the bivector bundle to some neighborhood $N_x$ of a point $x\in M$. So we are choosing a particular local trivialization of the bivector bundle.}\label{P-FormSections}:
$$  \Gamma(\wedge^2 M) \,=\, \Lambda^{2+} \,\oplus\, \Lambda^{2-} \,. $$
Where $\Lambda^{2\pm}$ is spanned by the bivectors with eigenvalue $\pm1$ with respect to $\mathcal{H}$. $\Lambda^{2+}$ is called the space of self-dual bivectors\index{Self-dual bivector}, while $\Lambda^{2-}$ is the space of anti-self-dual 2-forms. It is simple matter to prove that $\Lambda^{2+}$ is generated by $\{\bl{Z}^{i+}\}$, while $\Lambda^{2-}$ is generated by $\{\bl{Z}^{i-}\}$, with $i\in \{1,2,3\}$. For instance, let us prove that $\bl{Z}^{1+}$ is self-dual:
$$ \star Z^{1+}_{\phantom{1+}cd} \,=\, \frac{1}{2}\, Z^{1+}_{\phantom{1+}ab}\, \varepsilon^{ab}_{\phantom{ab}cd} \,=\, \varepsilon^{43}_{\phantom{43}cd} \,=\,  \varepsilon_{12cd} \,=\, -\lef \delta^{\,3}_c\delta^{\,4}_d - \delta^{\,4}_c\delta^{\,3}_d\rig =  Z^{1+}_{\phantom{1+}cd} \,.$$
Particularly, note that every null bivector must be an eigenbivector of the Hodge operator $\mathcal{H}$. It is worth remarking that what we call a self-dual bivector will be an anti-self-dual bivector if we change the sign of the volume-form. So the spaces $\Lambda^{2+}$ and $\Lambda^{2-}$ can be interchanged by a simple change of sign on the volume-form $\bl{\epsilon}$.

It is useful to introduce the following symmetric inner product on the space of bivectors:
$$ \lan\bl{Z},\bl{B}\ran \,\,\equiv\,\, Z_{\mu\nu}\,B^{\mu\nu}\,. $$
It is simple matter to prove that the operator $\mathcal{H}$ is self-adjoint with respect to this inner product, $\lan\bl{Z},\mathcal{H}(\bl{B})\ran =   \lan \mathcal{H}(\bl{Z}),\bl{B}\ran$. In particular this implies that the inner product of a self-dual bivector and an anti-self-dual bivector vanishes. Indeed, the only non-vanishing inner products of the bivector basis introduced above are:
\begin{equation}\label{InnerP-Biv}
  \lan\bl{Z}^{1\pm},\bl{Z}^{2\pm}\ran \,=\, 2  \quad  \textrm{and} \quad   \lan\bl{Z}^{3\pm},\bl{Z}^{3\pm} \ran \,=\, -2 \,.
\end{equation}

\section{Generalized Petrov Classification}
Now let us define an algebraic classification for the Weyl tensor\index{Weyl tensor} valid for any signature and that naturally generalizes the Petrov classification. To this end we shall define the Weyl operator at a point $x\in M$, $\mathcal{C}: \Gamma(\wedge^2 M)\rightarrow\Gamma(\wedge^2 M)$, by the following action:
$$   \bl{Z}\,\longmapsto\,\bl{B} \,=\, \mathcal{C}(\bl{Z})\,,\,\textrm{ with }\,  B_{\mu\nu} \,=\, Z^{\rho\sigma}\,C_{\rho\sigma\mu\nu} \,.  $$
Where $\bl{Z}$ and $\bl{B}$ are bivectors. Note that the operator $\mathcal{C}$ is self-adjoint with respect to the inner product on the space of bivectors, $ \lan\bl{Z},\mathcal{C}(\bl{B}) \ran =   \lan \mathcal{C}(\bl{Z}),\bl{B} \ran$. Now let us prove that the Weyl operator has a fundamental property, it commutes with the Hodge dual operator $\mathcal{H}$:
\begin{gather*}
  [\mathcal{C}\,\mathcal{H} - \mathcal{H}\,\mathcal{C}](\bl{Z})= 0 \quad\forall\,\, \bl{Z} \;\;\Leftrightarrow\;\; C_{\phantom{\rho\sigma}\mu\nu}^{\rho\sigma}\,\epsilon_{\alpha\beta\rho\sigma}= \epsilon_{\rho\sigma\mu\nu}\, C^{\rho\sigma}_{\phantom{\rho\sigma}\alpha\beta} \;\Leftrightarrow\; \\
 \epsilon^{\alpha\beta\kappa\gamma} C_{\phantom{\rho\sigma}\mu\nu}^{\rho\sigma}\,\epsilon_{\alpha\beta\rho\sigma}= \epsilon^{\alpha\beta\kappa\gamma} \epsilon_{\rho\sigma\mu\nu}\, C^{\rho\sigma}_{\phantom{\rho\sigma}\alpha\beta} \;\Leftrightarrow\;\\
 (-1)^{s/2}\,2!\,2! \,C_{\phantom{\rho\sigma}\mu\nu}^{\rho\sigma}\, \delta^{\,[\kappa}_\rho \delta^{\,\gamma]}_\sigma =  (-1)^{s/2} \,4!\, \delta^{\,[\alpha}_\rho\delta^{\,\beta}_\sigma \delta^{\,\kappa}_\mu \delta^{\,\gamma]}_\nu  \, C^{\rho\sigma}_{\phantom{\rho\sigma}\alpha\beta} \;\Leftrightarrow\;\\
 4 \,C_{\phantom{\rho\sigma}\mu\nu}^{\kappa\gamma} = 4 \,\delta^{\,[\alpha}_\mu \delta^{\,\beta]}_\nu \delta^{\,[\kappa}_\rho \delta^{\,\gamma]}_\sigma C^{\rho\sigma}_{\phantom{\rho\sigma}\alpha\beta} = 4 \,C_{\phantom{\mu\nu}\mu\nu}^{\kappa\gamma}\,.
\end{gather*}
Where equations (\ref{EE_DELTA}) and (\ref{weylsymm}) were used. Thus we conclude that the operators $\mathcal{C}$ and $\mathcal{H}$ commute. This implies that the eigenspaces of $\mathcal{H}$ are preserved by the operator $\mathcal{C}$, \textit{i.e.}, if $\bl{Z}^{\pm}\in\Lambda^{2\pm}$ then $\mathcal{C}(\bl{Z}^{\pm})\in\Lambda^{2\pm}$. Thus the operator $\mathcal{C}$ can be written as
$$  \mathcal{C} \,=\, \mathcal{C}^+ \,\oplus\, \mathcal{C}^-\,,  $$
where $\mathcal{C}^\pm$ is the restriction of $\mathcal{C}$ to $\Lambda^{2\pm}$. In other words, the operators $\mathcal{C}^\pm$ act on the 3-dimensional spaces generated by $\{\bl{Z}^{i\pm}\}$. When $\mathcal{C}^-= 0$ the Weyl tensor is said to be self-dual, while if $\mathcal{C}^+= 0$ it is anti-self-dual.

In 4 dimensions the Weyl tensor has 10 independent components, these can be chosen to be the following scalars\label{WeylScalars2}:
\small
\begin{align}\label{WeylScalars4}
  \nonumber \Psi^+_0 \equiv C_{1212} \; ; \; \Psi^+_1 \equiv C_{1312} \; ; \; \Psi^+_2 \equiv C_{1243}\; ; \;  \Psi^+_3 \equiv
C_{1343} \; ; \;&\Psi^+_4 \equiv C_{3434} \\
 \Psi^-_0 \equiv C_{1414} \; ; \; \Psi^-_1 \equiv C_{1314} \; ; \; \Psi^-_2 \equiv C_{1423}\; ; \;  \Psi^-_3 \equiv
C_{1323} \; ; \;&\Psi^-_4 \equiv C_{3232} \,.
\end{align}
\normalsize
Where $C_{abcd}\equiv C_{\mu\nu\rho\sigma}E_a^{\,\,\mu}E_b^{\,\,\nu}E_c^{\,\,\rho}E_d^{\,\,\sigma}$ are the components of the Weyl tensor on the null frame $\{\bl{E}_a\}$. In order to see that these components of the Weyl tensor are indeed independent of each other it is necessary to verify whether the symmetries of the Weyl tensor impose any relation between them. After some straightforward algebra it can be proved that the trace-less condition, $C^a_{\phantom{a}bad}= 0$, and the Bianchi identity, $C_{a[bcd]}= 0$, are equivalent to the following equations:
\begin{eqnarray*}
  \nonumber C_{2123} &=& C_{4143}= C_{1214}= C_{3234} = 0 \; ; \\
  C_{2124}  &=& \Psi^+_1\;;\; C_{4142} = \Psi_1^-\;;\;C_{2324} = \Psi_3^-\;;\;C_{4342} = \Psi^+_3\;; \\
  \nonumber C_{2424}&=& C_{1313} =  \Psi^+_2+\Psi_2^-\; ; \; C_{1324} = \Psi_2^--\Psi^+_2.
\end{eqnarray*}
Which proves that the scalars defined on (\ref{WeylScalars4}) can, indeed, represent the 10 degrees of freedom of the Weyl tensor. These scalars can also be conveniently written as follows:
\small
\begin{gather}
  \nonumber \Psi^{\pm}_0 = \frac{1}{4}  \lan\bl{Z}^{1\pm}, \mathcal{C}(\bl{Z}^{1\pm}) \ran \;;\; \Psi^{\pm}_1 = \frac{-1}{4\sqrt{2}}  \lan\bl{Z}^{1\pm}, \mathcal{C}(\bl{Z}^{3\pm})\ran \\
  \Psi^{\pm}_2 = \frac{1}{4}  \lan\bl{Z}^{1\pm}, \mathcal{C}(\bl{Z}^{2\pm})\ran \,=  \frac{1}{8}  \lan\bl{Z}^{3\pm}, \mathcal{C}(\bl{Z}^{3\pm})\ran \label{Tracefree-Bianchi}\\
  \nonumber \Psi^{\pm}_3 = \frac{-1}{4\sqrt{2}}  \lan\bl{Z}^{2\pm}, \mathcal{C}(\bl{Z}^{3\pm})\ran \;;\; \Psi^{\pm}_4 = \frac{1}{4}  \lan\bl{Z}^{2\pm}, \mathcal{C}(\bl{Z}^{2\pm}) \ran \,.
\end{gather}
\normalsize

By means of equations (\ref{Tracefree-Bianchi}) and (\ref{InnerP-Biv}) it can be easily proved that the matrix representations of the operators $\mathcal{C}^\pm$ on the basis $\{\bl{Z}^{i\pm}\}$ are given by:
\begin{equation}\label{Cmatrix}
\mathcal{C}^\pm \,=\, 2\left[
                                                                             \begin{array}{ccc}
                                                                               \Psi^\pm_2\vspace{0.15cm} & \Psi^\pm_4 & -\sqrt{2}\Psi^\pm_3 \\
                                                                               \vspace{0.15cm}\Psi^\pm_0& \Psi^\pm_2 & -\sqrt{2}\Psi^\pm_1 \\
                                                                               \sqrt{2}\Psi^\pm_1 & \sqrt{2}\Psi^\pm_3 & -2\Psi^\pm_2 \\
                                                                             \end{array}
                                                                           \right].
\end{equation}
Since the operators $\mathcal{C}^+$ and $\mathcal{C}^-$ have vanishing trace it follows that the possible algebraic types of these operators according to the refined Segre classification are the ones listed on equation (\ref{C^+ types}). It is also worth noting that $\mathcal{C}^+$ and $\mathcal{C}^-$ are independent of each other. So, for instance, we might say that the Weyl tensor is type $(I,N)$ when $\mathcal{C}^+$ is type $I$ and $\mathcal{C}^-$ is type $N$. Note that the type $(I,N)$ is intrinsically equivalent to the type $(N,I)$, since the operators $\mathcal{C}^+$ and $\mathcal{C}^-$ are interchanged if we multiply the volume-form by $-1$. So we conclude that on a complexified 4-dimensional manifold the Weyl tensor can have 21 algebraic types \cite{art1}:
\small
\begin{equation}\label{PossibleTypes}
\begin{array}{ccccccc}
  (O,O) & (O,D) &(O,N) &(O,III) &(O,II) &(O,I)& (D,D) \\
  (D,N)& (D,III)& (D,II)& (D,I)& (N,N)& (N,III)& (N,II)\\
 (N,I) &(III,III)& (III,II)& (III,I)& (II,II)& (II,I)& (I,I)
\end{array}
\end{equation}
\normalsize
As proved in ref. \cite{art1}, the same classification can be attained using the boost weight approach\index{Boost weight}. Up to now the metric was not assumed to be real, so that the Weyl tensor is generally complex. But some of these types are forbidden when the metric is real, as we shall see in what follows.

\subsection{Euclidean Signature}
Let us suppose that $\bl{g}$ is a real metric with Euclidean signature. Then the components $C_{\mu\nu\rho\sigma}$ of the Weyl tensor on a real coordinate frame are real. By means of this fact along with equations (\ref{Real.Eucli}) and (\ref{WeylScalars4}), one can easily prove that in this signature the Weyl scalars obey the following reality conditions:
\begin{equation}\label{WeyScEuc}
  \overline{\Psi^{\pm}_0} \,=\, \Psi^{\pm}_4 \;\;;\;\;\; \overline{\Psi^{\pm}_1} \,=\, -\Psi^{\pm}_3 \;\;;\;\;\; \overline{\Psi^{\pm}_2} \,=\, \Psi^{\pm}_2\,. \end{equation}
This together with (\ref{Cmatrix}) implies that the matrix representation of the operators $\mathcal{C}^+$ and $\mathcal{C}^-$ are Hermitian and independent of each other. So these matrices can be diagonalized and, therefore, the algebraic types $II$, $III$ and $N$ are forbidden. Thus if the signature is Euclidean the Weyl tensor must have one of the following algebraic types \cite{art1}:
\small
\begin{equation}\label{Types-Euclid}
   (O,O)\quad (O,D)\quad \,(O,I)\quad (D,D)\quad(D,I)\quad (I,I)\,.
\end{equation}
\normalsize
An equivalent classification was obtained in \cite{Hacyan} using a mixture of null tetrad and spinorial formalisms. The same classification was also found in \cite{Karlhede} by means of splitting the Weyl tensor as a sum of two 3-dimensional tensors of rank two and using the group $SO(4,\mathbb{R})$ to find canonical forms for such tensors.

\subsection{Lorentzian Signature}
Now assume that the metric $\bl{g}$ is real and Lorentzian. Then the Weyl tensor is real, so that equations (\ref{Real.Loren}) and (\ref{WeylScalars4}) immediately imply the following reality conditions:
$$ \overline{\Psi^{+}_0} \,=\, \Psi^{-}_0 \;;\;\; \overline{\Psi^{+}_1} \,=\, \Psi^{-}_1 \;;\;\; \overline{\Psi^{+}_2} \,=\, \Psi^{-}_2 \;;\;\; \overline{\Psi^{+}_3} \,=\, \Psi^{-}_3 \;;\;\; \overline{\Psi^{+}_4} \,=\, \Psi^{-}_4 \,.$$
Which along with equation (\ref{Cmatrix}) guarantees that the matrix representation of $\mathcal{C}^-$ is the complex conjugate of the matrix representation of $\mathcal{C}^+$. Therefore, in this signature $\mathcal{C}^+$ and $\mathcal{C}^-$ must have the same algebraic type. So from the 21 types listed on eq. (\ref{PossibleTypes}) just the following six types are allowed in the Lorentzian case \cite{art1}:
\small
\begin{equation}\label{Types-Loren}
 (O,O)\quad (D,D) \quad (N,N) \quad (III,III)\quad (II,II) \quad (I,I) \,.
\end{equation}
\normalsize
These types correspond respectively to the Petrov types $O$, $D$, $N$, $III$, $II$ and $I$, retrieving the Petrov classification.  In particular, note that in this signature if $\mathcal{C}^-$ identically zero then $\mathcal{C}^+$ must also vanish, so that non-trivial self-dual Weyl tensors do not exist on the Lorentzian case.

\subsection{Split Signature}
Suppose that $\bl{g}$ is a real metric with split signature. In this case it is possible to find a real null frame, as shown in (\ref{Real.Split}). Thus the Weyl scalars, defined on (\ref{WeylScalars4}), are all real and generally independent of each other. So the matrix representations of $\mathcal{C}^+$ and $\mathcal{C}^-$ are real and generally independent of each other. Therefore, in this case there is no algebraic restriction on the matrices that represent $\mathcal{C}^\pm$, which implies that all the 21 types of eq. (\ref{PossibleTypes}) are allowed \cite{art1}. A classification deeply related to this one was obtained in \cite{Law2} using spinorial calculus. Other, inequivalent, classifications for the Weyl tensor in manifolds of split signature were defined in \cite{Petrov_Book,Law1,Nurwoski2}.

\subsection{Annihilating Weyl Scalars}

In section \ref{Sec._Annihil_Scalars} it was proved that when the signature is Lorentzian each Petrov type can be characterized by the possibility of annihilating some of the Weyl scalars. It turns out that the same thing happens on the generalized classification presented in this chapter, as we shall prove.

Every transformation that maps a null frame into a null frame can be written as a composition of the following three kinds of transformations:
\normalsize
\\
\small
\\
\textbf{(i) Lorentz Boosts}
$$ \bl{E}_1\mapsto \lambda_+\lambda_-\,\bl{E}_1 \;;\;\; \bl{E}_2\mapsto  \lambda_+\lambda_-^{-1}\,\bl{E}_2 \;;\;\;  \bl{E}_3\mapsto \lambda_+^{-1}\lambda_-^{-1} \,\bl{E}_3 \;;\;\; \bl{E}_4\mapsto \lambda_+^{-1}\lambda_-\,\bl{E}_4 $$
\textbf{(ii) Null rotation around} $\bl{E}_1$
$$ \bl{E}_1\mapsto\bl{E}_1\,;\;   \bl{E}_2\mapsto\bl{E}_2 + w_- \bl{E}_1   \,;\; \bl{E}_3\mapsto\bl{E}_3 + w_+ \bl{E}_2 + w_- \bl{E}_4 + w_+w_- \bl{E}_1  \,;\; \bl{E}_4\mapsto\bl{E}_4 + w_+ \bl{E}_1 $$
\textbf{(iii) Null rotation around} $\bl{E}_3$
$$  \bl{E}_1\mapsto\bl{E}_1 + z_- \bl{E}_2 + z_+ \bl{E}_4 + z_+z_- \bl{E}_3  \,;\; \bl{E}_2\mapsto\bl{E}_2 + z_+ \bl{E}_3   \,;\; \bl{E}_3\mapsto\bl{E}_3\,;\;   \bl{E}_4\mapsto\bl{E}_4 + z_- \bl{E}_3 $$
\normalsize
\\
Where $\lambda_\pm$, $w_\pm$ and $z_\pm$ are complex numbers, the six parameters of the group $SO(4;\mathbb{C})$\index{SO(3,1)}. It is interesting to note that under these transformations the Weyl scalars change as:
$$ \Psi_A^+ \,\longmapsto\, F_A(\lambda_+,w_+,z_+,\Psi_B^+) \quad;\quad   \Psi_A^- \,\longmapsto\, F_A(\lambda_-,w_-,z_-,\Psi_B^-)\,.$$
So the parameters $\lambda_-$, $w_-$ and $z_-$ do not appear on the transformation of the operator $\mathcal{C}^+$ while the transformation of $\mathcal{C}^-$ does not depend on $\lambda_+$, $w_+$ and $z_+$. Thanks to this property, the same argument used in section \ref{Sec._Annihil_Scalars} in order to show which Weyl scalars could be made to vanish by a suitable choice of null tetrad remains valid here for both operators $\mathcal{C}^+$ and $\mathcal{C}^-$ individually. Table \ref{Tab._vanishScalars4} summarizes this analysis. Thus, for example, if the Weyl tensor is type $(I,II)$ then it is possible to choose a null frame in which the Weyl scalars $\Psi_0^+$, $\Psi_4^+$, $\Psi_0^-$ , $\Psi_1^-$ and $\Psi_4^-$ vanish simultaneously.
\begin{table}
\begin{center}
\begin{tabular}{l||l}
  \hline
  \hline
  $\mathcal{C}^\pm$ type $I$ $\rightarrow$ $\Psi_0^\pm,\Psi_4^\pm$   &  $\mathcal{C}^\pm$ type $II$ $\rightarrow$ $\Psi_0^\pm,\Psi_1^\pm,\Psi_4^\pm$ \\
    $\mathcal{C}^\pm$ type $D$ $\rightarrow$ $\Psi_0^\pm,\Psi_1^\pm,\Psi_3^\pm,\Psi_4^\pm$ & $\mathcal{C}^\pm$ type $III$ $\rightarrow$ $\Psi_0^\pm,\Psi_1^\pm,\Psi_2^\pm, \Psi_4^\pm$ \\
      $\mathcal{C}^\pm$ type $N$ $\rightarrow$ $\Psi_0^\pm,\Psi_1^\pm,\Psi_2^\pm,\Psi_3^\pm$ & $\mathcal{C}^\pm$ type $O$ $\rightarrow$ $\Psi_0^\pm,\Psi_1^\pm,\Psi_2^\pm,\Psi_3^\pm,\Psi_4^\pm$ \\
  \hline
  \hline
\end{tabular}\caption{\footnotesize{Weyl scalars that can be made to vanish, by a suitable choice of null frame, depending on the algebraic type of the operators $\mathcal{C}^\pm$.}}\label{Tab._vanishScalars4}
\end{center}
\end{table}

In this generalized classification the Weyl tensor shall be called algebraically special when its type is different from $(I,I)$. In such a case one can conveniently choose the signal of the volume-form so that $\mathcal{C}^+$ is not type $I$. Therefore, table \ref{Tab._vanishScalars4} implies that the Weyl tensor is algebraically special if, and only if, it is possible to find a null frame in which  $\Psi_0^+= \Psi_1^+= 0$. This along with eq. (\ref{Cmatrix}) yield that the Weyl tensor is algebraically special if, and only if, $\bl{Z}^{1+}$ is an eigenbivector of the Weyl operator. Since every self-dual null bivector can be written as $\bl{E}^4\wedge\bl{E}^3= \bl{Z}^{1+}$ on a suitable null frame, we arrive at the following theorem \cite{art2}:
\begin{theorem}\label{Theo. EigenBiv}
The Weyl tensor of a 4-dimensional manifold is algebraically special if, and only if, the Weyl operator admits a null eigenbivector.
\end{theorem}

\section{Generalized Goldberg-Sachs Theorem}\label{Sec. GS4D}\index{Goldberg-Sachs theorem}
In this section it will be presented a beautiful generalization of the Goldberg-Sachs (GS) theorem valid in 4-dimensional vacuum\footnote{Throughout this thesis the expressions vacuum manifold and Ricci-flat manifold will be interchanged, they both mean a manifold with vanishing Ricci tensor.} manifolds of arbitrary signature, a result first proved by Pleba\'{n}ski and Hacyan in \cite{Plebanski2}. To this end the notation introduced in section \ref{Sec. Cartan} will be used. In particular, let us recall the following important equations:
\begin{equation}\label{Remember}
  V^\mu\nabla_\mu \bl{E}^a  \equiv - \bl{\omega}^a_{\phantom{a}b}(\bl{V})\,\bl{E}^b\;;\quad \omega_{ab}^{\phantom{ab}c}\equiv \bl{\omega}^c_{\phantom{c}b}(\bl{E}_a) \;;  \quad \nabla_a \bl{E}_b = \omega_{ab}^{\phantom{ab}c}\bl{E}_c \,.
\end{equation}
Where $\bl{\omega}^a_{\phantom{a}b}$ are the so-called connection 1-forms. Since for a null frame the matrix $g_{ab}= \bl{g}(\bl{E}_a,\bl{E}_b)$ is constant it follows that $\bl{\omega}_{ab}= -\bl{\omega}_{ba}$ and $\omega_{abc}= - \omega_{acb}$, where $\bl{\omega}_{ab}\equiv g_{ac}\bl{\omega}^c_{\phantom{c}b}$ and $\omega_{abc}\equiv \omega_{ab}^{\phantom{ab}d}g_{dc}$. Using this notation, the generalized Goldberg-Sachs theorem is given by \cite{Plebanski2}:
\begin{theorem}\label{Theo. GSgen4}
Let $(M,\bl{g})$ be a 4-dimensional manifold with vanishing Ricci tensor. If $\omega_{112}= \omega_{221}= 0$ then $\Psi_0^+= \Psi_1^+= 0$. Conversely, if  $\Psi_0^+= \Psi_1^+= 0$ then it is possible to find a null frame in which the scalars $\Psi_0^+$, $\Psi_1^+$, $\omega_{112}$ and $\omega_{221}$ all vanish.
\end{theorem}

Before proceeding, let us prove that this theorem is equivalent to the Goldberg-Sachs theorem when the signature is Lorentzian. Indeed, using equations (\ref{NullFrame-Tetrad}) and (\ref{Remember}) along with the definition of the shear\index{Shear} parameter, eq. (\ref{Shear}), we find:
\begin{gather*}
  \bl{l}^\mu\nabla_\mu\bl{l} =  \nabla_1\bl{E}_1 = \omega_{11}^{\ph{11}a}\bl{E}_a = \omega_{113}\,\bl{l} - \omega_{114}\,\bl{m} -\omega_{112}\, \overline{\bl{m}}  \\
  \sigma = \bl{g}(m^\mu\,\nabla_{\mu}\bl{l},\bl{m}) =  \bl{g}(\nabla_{2}\bl{E}_1,\bl{E}_2) = -\omega_{21}^{\ph{21}4} = \omega_{212} = -\omega_{221}\,.
\end{gather*}
From which we conclude that the congruence generated by the null vector field $\bl{l}= \bl{E}_1$ is geodesic and shear-free if, and only if, the connection components $\omega_{114}$, $\omega_{112}$ and $\omega_{221}$ all vanish. But equation (\ref{Real.Loren}) implies that on the Lorentzian signature $\omega_{114}$ is the complex conjugate of $\omega_{112}$. Thus $\bl{l}$ will be geodesic and shear-free if, and only if, $\omega_{112}= \omega_{221}= 0$, proving that theorem \ref{Theo. GSgen4} reduces to the usual GS theorem on the Lorentzian signature, see theorem \ref{Theo_Gold-Sachs}.

\begin{figure}[h]
	\centering
		\includegraphics[width=14.8cm]{{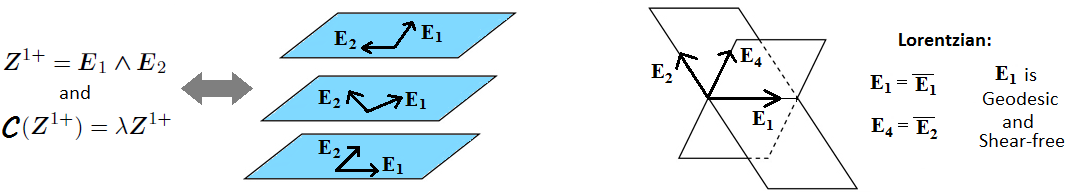}}
	\caption{\footnotesize{In vacuum, the Weyl tensor admits a null eigenbivector if, and only if, the isotropic distribution generated by such bivector is integrable, as depicted on the left hand side of the picture. The vector fields $\bl{E}_1$ and $\bl{E}_2$ are null and orthogonal to each other, generating isotropic planes. On the right hand side of this figure we have the Lorentzian case, where the intersection of a totally null plane and its complex conjugate gives a real null direction $\bl{E}_1$. In this signature if the distribution generated by $\bl{E}_1\wedge\bl{E}_2$ is integrable so will be the distribution generated by $\bl{E}_1\wedge\bl{E}_4$. Moreover, $\bl{E}_1$ will be geodesic and shear-free.}}
	\label{Fig. IsotropicPlanes}
\end{figure}

The condition $\omega_{112}= \omega_{221}= 0$ has a nice geometric interpretation, it is equivalent to say that the complexified manifold can be foliated by totally null leafs. Indeed, using eq. (\ref{Remember}) we find that the Lie bracket of $\bl{E}_1$ and $\bl{E}_2$ is:
\begin{align}
  \nonumber[\bl{E}_1,\bl{E}_2]\,=\,&  \nabla_1\bl{E}_2 - \nabla_2\bl{E}_1 = (\omega_{12}^{\ph{12}a} - \omega_{21}^{\ph{21}a})\bl{E}_a  \\
  \,=\,& (\omega_{123} - \omega_{213})\bl{E}_1 -  (\omega_{124} - \omega_{214})\bl{E}_2 - \omega_{112}\bl{E}_3 - \omega_{221}\bl{E}_4\,.\label{[E1,E2]}
\end{align}
Thus the condition $\omega_{112}= \omega_{221}= 0$ is equivalent to say that the Lie bracket $[\bl{E}_1,\bl{E}_2]$ is a linear combination of $\bl{E}_1$ and $\bl{E}_2$. Since $[\bl{E}_1,\bl{E}_1]$ and $[\bl{E}_2,\bl{E}_2]$ are trivially zero this, in turn, is equivalent to the integrability of the distribution\index{Distribution} generated by $\{\bl{E}_1,\bl{E}_2\}$, see section \ref{Sec. Integrab}. Since such vector fields are null and orthogonal to each other it follows that the vectors tangent to this distribution are all null, this kind of distribution is named isotropic\index{Isotropic}. Therefore, theorem \ref{Theo. GSgen4} guarantees that a vacuum manifold admits an integrable distribution of isotropic planes if, and only if, the Weyl tensor is algebraically special \cite{Plebanski2}. Since $Z^{1+\,\mu\nu}= 2 E_1^{\,[\mu} E_2^{\,\nu]}$ we shall write $\bl{Z}^{1+}= \bl{E}_1\bl{\wedge}\bl{E}_2$ and say that $\bl{Z}^{1+}$ generates the distribution of isotropic planes spanned by the vector fields $\bl{E}_1$ and $\bl{E}_2$. As noticed on the paragraph before theorem \ref{Theo. EigenBiv}, $\bl{Z}^{1+}$ is an eigenbivector of the Weyl operator if, and only if, $\Psi_0^+= \Psi_1^+= 0$, which lead us to the following result \cite{art2}:
\begin{corollary}\label{Cor.EigenPlane4D}\index{Null bivector}
A distribution of isotropic planes in a Ricci-flat 4-dimensional manifold is integrable if, and only if, the null bivector that generates such distribution is an eigenbivector of the Weyl operator.
\end{corollary}\label{Coro. Integ4}
This fact is illustrated in figure \ref{Fig. IsotropicPlanes}. If the metric is real then whenever a distribution is integrable the complex conjugate of such distribution will also be integrable. Particularly, on the Lorentzian signature if $\Delta$ is an integrable distribution of isotropic planes then $\overline{\Delta}$ will also be integrable and $\Delta\cap\overline{\Delta} = Span\{\bl{l}\}$, where $\bl{l}$ is a real null vector field generating a geodesic and shear-free congruence, see figure \ref{Fig. IsotropicPlanes}.

One can also express such integrability result using the dual form of the Frobenius theorem, seen in section \ref{Sec. Integrab}.  In this language the corollary \ref{Coro. Integ4} is equivalent to the claim that given a null bivector $\bl{Z}$, it is possible to find some scalar function $f\neq 0$ such that $d(f\bl{Z})= 0$  if, and only if, $\bl{Z}$ is an eigenbivector of the Weyl operator. Let us state this as a corollary:
\begin{corollary}\label{Coro. dZ=0}
In a Ricci-flat manifold, the Weyl scalars $\Psi_0^+$ and $\Psi_1^+$ vanish if, and only if, it is possible to find a scalar function $f\neq 0$ such that  $d(f\bl{Z}^{1+})= 0$ in a suitable null frame.
\end{corollary}


On the Lorentzian signature a real null vector field $\bl{l}$ is said to be a principal null direction (PND)\index{Principal null directions(PND)} of the Weyl tensor when it is possible to find a null tetrad $\{\bl{l},\bl{n},\bl{m},\overline{\bl{m}}\}$ such that $\Psi_0\equiv \Psi_0^+$ vanishes, in general there exists 4 distinct PNDs. Moreover, this vector field is said to be a repeated PND when, in addition to $\Psi_0$, the Weyl scalar $\Psi_1\equiv \Psi_1^+$ do also vanish. On the general formalism presented in this chapter the concept of privileged null directions might be substituted by privileged null bivectors \cite{art2}. Looking at the definition of $\Psi_0^+$ on eq. (\ref{Tracefree-Bianchi}) it is natural to define a null bivector $\bl{Z}$ to be a principal null bivector (PNB) when $\lan\bl{Z},\mathcal{C}(\bl{Z})\ran= 0$. Furthermore, because of eq. (\ref{Cmatrix}) and theorem \ref{Theo. EigenBiv}, $\bl{Z}$ shall be called a repeated PNB if $\bl{Z}$ is null and $\mathcal{C}(\bl{Z})\propto \bl{Z}$. In general the Weyl tensor will admit 4 self-dual PNBs and 4 anti-self-dual PNBs, as can be verified using the group $SO(4;\mathbb{C})$. In the Lorentzian case  $\bl{l}$ is a PND if, and only if, $\bl{Z}_1= \bl{l}\wedge\bl{m}$ is a self-dual PNB, which, in turn, is equivalent to say that $\overline{\bl{Z}_1}= \bl{l}\wedge\overline{\bl{m}}$ is an anti-self-dual PNB.

As a last comment it is worth pointing out that the generalized GS theorem is also valid in less restrictive situations than the Ricci-flat case. Indeed, on the original article of Pleba\'{n}ski and Hacyan \cite{Plebanski2} it was observed that such theorem remains valid for Einstein manifolds, the ones such that the Ricci tensor is proportional to the metric. Furthermore, in \cite{Nurwoski2} it was worked out the least restrictive version of the generalized GS theorem.

\section{Geometric Consequences of the Generalized\\ Goldberg-Sachs Theorem}
The goal of this section is to use the generalized Goldberg-Sachs theorem in order to prove that certain algebraic types of the Weyl tensor are characterized by the existence of important geometric structures on the manifold. Here it will be assumed that the Ricci tensor of the manifold is identically zero. The results obtained in the present section are based on the article \cite{art2}. Important attempts on the same line can also be found in \cite{Nurwoski2,Robinson Manifolds}. Before proceeding some  definitions and tools of complex differential geometry shall be introduced.

\subsection{Complex Manifolds}\index{Complex manifold}

Let $(M,\bl{g})$ be an even-dimensional manifold, then  an almost complex structure on this manifold is an endomorphism of the tangent bundle, $\mathcal{J}:TM\rightarrow TM$, whose square is minus the identity map, $\mathcal{J}^2=-\mathbf{1}$. Note that the almost complex structure can be seen as a tensor of rank two, $\mathcal{J}^\mu_{\ph{\mu}\nu}$, defined by the following relation:
$$ \mathcal{J}(\bl{V})\,=\, \bl{X} \quad\Longleftrightarrow\quad X^\mu \,=\,  \mathcal{J}^\mu_{\ph{\mu}\nu}\,V^\nu \,.$$
If $\bl{V}$ is some vector field then defining $\bl{V}^\pm\equiv [\bl{V} \mp i \mathcal{J}(\bl{V})]$ we find that $\bl{V}^\pm$ is an eigenvector of $\mathcal{J}$ with eigenvalue $\pm i$. Thus $\mathcal{J}$ splits the tangent bundle as follows:
$$  TM= TM^+ \oplus TM^- \;,  \;\quad TM^\pm\equiv\{\bl{V}\in TM\, | \,\mathcal{J}(\bl{V})=\pm i\,\bl{V}\}\, .$$
The almost complex structure is said to be integrable when the distributions $TM^+$ and $TM^-$ are both integrable, in which case $\mathcal{J}$ is called a complex structure. By means of $\mathcal{J}$ we can define a tensor $\bl{N}$, called the Nijenhuis tensor, whose action on two vector fields yields another vector field as follows:
$$ \bl{N}(\bl{V},\bl{X})=[\bl{V},\bl{X}]-[\mathcal{J}(\bl{V}),\mathcal{J}(\bl{X})]+\mathcal{J}\left([\mathcal{J}(\bl{V}),\bl{X}]\right) + \mathcal{J}\lef[\bl{V},\mathcal{J}(\bl{X})]\rig. $$
It can be proved that $\mathcal{J}$ is integrable if, and only if,  $\bl{N}$ vanishes \cite{Nakahara}. When the almost complex structure leaves the inner products invariant, $\bl{g}\lef\mathcal{J}(\bl{V}),\mathcal{J}(\bl{X})\rig= \bl{g}(\bl{V},\bl{X})$ for all vector fields $\bl{V}$ and $\bl{X}$, the metric is said to be Hermitian with respect to $\mathcal{J}$. In this case one can introduce a 2-form, called the K\"{a}hler form, defined by $\Omega_{\mu\nu}= g_{\rho\nu}\mathcal{J}^\rho_{\ph{\rho}\mu}$. Note that if the metric is Hermitian with respect to $\mathcal{J}$ then the subbundles $TM^+$ and $TM^-$ are both isotropic\index{Isotropic}.

On the chapter \ref{Chap. GR} a manifold of dimension $n$ was defined to be a topological set such that the neighborhood of each point can be smoothly mapped by a coordinate system into a patch of $\mathbb{R}^n$. In addition, it must be required that the transition functions between the coordinate systems of overlapping neighborhoods are smooth. An $n$-dimensional complex manifold\footnote{Do not confuse with a complexified manifold, which is just a regular manifold with all its tensor bundles complexified.} is, likewise, defined as a topological set such that the neighborhood of each point can be smoothly mapped by a coordinate system into a patch of $\mathbb{C}^n$ and such that the transition functions between the coordinates systems of overlapping neighborhoods are not only smooth but also analytic \cite{Nakahara}. This last requirement is more restrictive than it sounds. Indeed, a celebrated theorem on differential geometry, the Newlander-Nirenberg theorem \cite{Newlander}, states that a manifold admits an integrable and real almost complex structure if, and only if, it is a complex manifold. When a complex manifold is endowed with a metric that is invariant by the action of the almost complex structure on the vector fields the manifold is called Hermitian. In this case one can define a 2-form $\bl{\Omega}$, called the K\"{a}hler form of the Hermitian manifold, as defined in the preceding paragraph. If the exterior derivative of the K\"{a}hler form vanishes, $d\bl{\Omega}= 0$, the manifold is said to be a K\"{a}hler manifold. If in addition the Ricci tensor vanishes, as assumed in this chapter, the manifold is called a Calabi-Yau\index{Calabi-Yau manifold} manifold\footnote{Actually, a Calabi-Yau manifold is defined to be a K\"{a}hler manifold with vanishing first Chern class, which is less restrictive than the Ricci-flat condition.}. The Calabi-Yau manifolds are of great relevance for string theory compactifications \cite{BeckerB}.

\subsection{General Results}

Now let $(M,\bl{g})$ be a complexified 4-dimensional manifold of arbitrary signature and $\{\bl{E}_a\}$ a null frame. Then we can define the following almost complex structure \cite{art2}:
\begin{equation}\label{J-def}
   \bl{J} \,\equiv\, i\,\lef \bl{E}_1 \otimes \bl{E}^1 + \bl{E}_2\otimes \bl{E}^2 \rig -i\, \lef \bl{E}_3\otimes \bl{E}^3 + \bl{E}_4\otimes \bl{E}^4 \rig\,.
\end{equation}
Note that the metric $\bl{g}$ is Hermitian with respect to this almost complex structure. For example,
$$\bl{g}\lef\bl{J}(\bl{E}_1),\bl{J}(\bl{E}_3)\rig\,=\, \bl{g}(i\bl{E}_1,-i\bl{E}_3) \,=\,  \bl{g}(\bl{E}_1,\bl{E}_3)\,.$$
It is also immediate to see that $\bl{E}_1$ and $\bl{E}_2$ are eigenvectors of $\bl{J}$ with eigenvalue $i$, while $\bl{E}_3$ and $\bl{E}_4$ are eigenvectors with eigenvalue $-i$. This means that $TM^+= Span\{\bl{E}_1,\bl{E}_2\}$ and $TM^-= Span\{\bl{E}_3,\bl{E}_4\}$. So, using equation (\ref{[E1,E2]}), we conclude that $TM^+$ is integrable if, and only if, $\omega_{112}= \omega_{221}= 0$. Analogously, $TM^-$ is integrable if, and only if, $\omega_{334}= \omega_{443}= 0$. Therefore, we can state:
\begin{equation}\label{Jintegrable}
   \bl{J} \;\,\textrm{is Integrable}  \quad\Longleftrightarrow\quad   \omega_{112}= \omega_{221}= \omega_{334}= \omega_{443}= 0 \,.
 \end{equation}
But theorem \ref{Theo. GSgen4} and equation (\ref{WeylScalars4}) imply that the right hand side of (\ref{Jintegrable}) holds if, and only if, the Weyl scalars $\Psi^{+}_0$, $\Psi^{+}_1$, $\Psi^{+}_3$ and $\Psi^{+}_4$ vanish. Equation (\ref{Cmatrix}), in turn, guarantees that the annihilation of these Weyl scalars is equivalent to say that $\mathcal{C}^+$ is type $D$ or type $O$. So  $\bl{J}$ is integrable if, and only if, $\mathcal{C}^+$ is type $D$ or type $O$. In the same vein, it can be proved that if $(M,\bl{g})$ admits an integrable almost complex structure such that $\bl{g}$ is Hermitian with respect to it then the Weyl tensor must be type $(D,\lozenge)$ or type $(O,\lozenge)$, where $\lozenge$ represents an arbitrary Petrov type \cite{art2}. So the following theorem holds:
\begin{theorem}\label{Theo. Jinteg}
A Ricci-flat 4-dimensional manifold $(M,\bl{g})$ admits an integrable almost complex structure with $\bl{g}$ being Hermitian with respect to it if, and only if, the algebraic type of the Weyl tensor is $(D,\lozenge)$ or $(O,\lozenge)$. Moreover, if such complex structure exists we can always manage to find a null frame in which it takes the form shown on eq. (\ref{J-def}).

\end{theorem}
The K\"{a}hler form is the 2-form $\bl{\Omega}$ such that $\bl{X}\lrcorner\bl{V}\lrcorner\bl{\Omega}=  \bl{g}(\bl{J}(\bl{V}),\bl{X})$ for all vector fields $\bl{V}$ and $\bl{X}$. It is simple matter to prove that it is given by:
\begin{equation}\label{Kahler4}
  \bl{\Omega} \,=\, i \lef \bl{E}^1\wedge \bl{E}^3 \,+\, \bl{E}^4\wedge \bl{E}^2 \rig  \,=\, i\,\sqrt{2}\, \bl{Z}^{3+}\,.
\end{equation}
We can calculate the exterior derivative of this 2-form by means of the first Cartan's structure equation, $d\bl{E}^a+   \bl{\omega}^a_{\ph{a}b}\wedge\bl{E}^b= 0$. The bottom line is:
\begin{equation*}
   d \bl{\Omega} \,=\, -2i\, \bl{\omega}_{12}\wedge\bl{E}^1\wedge \bl{E}^2 \,+\, 2i\, \bl{\omega}_{34}\wedge\bl{E}^3\wedge \bl{E}^4\,.
\end{equation*}
Since $\bl{\omega}_{ab}= \omega_{cba}\bl{E}^c$, it follows that $d \bl{\Omega}= 0$ if, and only if, the connection components $\omega_{321}$, $\omega_{421}$, $\omega_{143}$ and $\omega_{243}$ all vanish. This along with equation (\ref{Jintegrable}) yields:
\begin{equation}\label{dW=0}
  \bl{J} \;\,\textrm{is integrable and} \;\, d \bl{\Omega} =0  \quad\Longleftrightarrow\quad   \bl{\omega}_{12} =  \bl{\omega}_{34} = 0\,.
\end{equation}
Furthermore, let us calculate the covariant derivative of the K\"{a}hler form. Using the identity $\nabla_a\bl{E}^b= \omega_{a\ph{b}c}^{\ph{a}b}\bl{E}^c$ and eq. (\ref{Kahler4}) it is straightforward to prove that:
$$ \nabla_a \bl{\Omega} \,=\, -2i\,\omega_{a21}\,\bl{E}^1\wedge \bl{E}^2   \,+\, 2i\, \omega_{a43}\, \bl{E}^3\wedge \bl{E}^4\,.$$
Thus $\bl{\Omega}$ is covariantly constant if, and only if, $\bl{\omega}_{12}=  \bl{\omega}_{34}= 0$. This together with (\ref{dW=0}) then imply the following useful equivalences:
\begin{equation}\label{Jint-dW-DW}
  \bl{J} \;\,\textrm{Integrable,} \;\, d \bl{\Omega} =0 \quad\Leftrightarrow\quad \bl{\omega}_{12} =  \bl{\omega}_{34} = 0 \quad\Leftrightarrow\quad \nabla_a \bl{\Omega} \,=\,0\,.
\end{equation}

In order to make a connection between these results and the algebraic classification of the Weyl tensor we need to use the second Cartan's structure equation, which in vacuum is:
$$ \frac{1}{2}\, C_{abcd} \, \boldsymbol{E}^c\wedge\boldsymbol{E}^d \,=\, d\boldsymbol{\omega}_{ab} + \boldsymbol{\omega}_{ac}\wedge \boldsymbol{\omega}^c_{\phantom{c}b}\,. $$
Using the definition of the Weyl scalars, this equation can be proved to be equivalent to the following ones:
\begin{equation*}
\begin{cases}
\begin{array}{cl}
    \;\;\,d\bl{\omega}_{12} + \bl{\omega}_{12}\wedge(\bl{\omega}_{24} -\bl{\omega}_{13}) &=\; \Psi_2^+\,\bl{Z}^{1+} + \Psi_0^+\,\bl{Z}^{2+} + \sqrt{2}\Psi_1^+\,\bl{Z}^{3+}\quad\;\\
   -d\bl{\omega}_{34} + \bl{\omega}_{34}\wedge(\bl{\omega}_{24} -\bl{\omega}_{13}) &= \; \Psi_4^+\,\bl{Z}^{1+} + \Psi_2^+\,\bl{Z}^{2+} + \sqrt{2}\Psi_3^+\,\bl{Z}^{3+}\quad\;
   \\
 -\frac{1}{2}d(\bl{\omega}_{24}-\bl{\omega}_{13}  ) +
   \bl{\omega}_{12}\wedge\bl{\omega}_{34}   &=\;  \Psi_3^+\,\bl{Z}^{1+} + \Psi_1^+\,\bl{Z}^{2+} +\sqrt{2}\Psi_2^+\,\bl{Z}^{3+} \quad\;
\end{array}
\\
\\
\begin{array}{cl}
 \nonumber   \;\; \,d\bl{\omega}_{14} + \bl{\omega}_{14}\wedge(\bl{\omega}_{42} -\bl{\omega}_{13}) &=\; \Psi_2^-\,\bl{Z}^{1-} + \Psi_0^-\,\bl{Z}^{2-} + \sqrt{2}\Psi_1^-\,\bl{Z}^{3-}\quad\;\\
   -d\bl{\omega}_{32} + \bl{\omega}_{32}\wedge(\bl{\omega}_{42} -\bl{\omega}_{13}) &=\;  \Psi_4^-\,\bl{Z}^{1-} + \Psi_2^-\,\bl{Z}^{2-} + \sqrt{2}\Psi_3^-\,\bl{Z}^{3-}\quad\;
   \\
 \nonumber -\frac{1}{2}d(\bl{\omega}_{42}-\bl{\omega}_{13}  ) +
   \bl{\omega}_{14}\wedge\bl{\omega}_{32}   &= \; \Psi_3^-\,\bl{Z}^{1-} + \Psi_1^-\,\bl{Z}^{2-} +\sqrt{2}\Psi_2^-\,\bl{Z}^{3-} \quad\;
\end{array}
\end{cases}
\end{equation*}
These two sets of three equations are the self-dual and anti-self-dual parts of the second structure equation respectively. The first important thing to note is that if $\bl{\omega}_{12}= \bl{\omega}_{34}= 0$ then $\Psi_A^+= 0$, so that $\mathcal{C}^+$ is type $O$. Conversely, if $\mathcal{C}^+= 0$ then we can find a null frame such that $\bl{\omega}_{12}= \bl{\omega}_{34}= 0$ and $\bl{\omega}_{24}- \bl{\omega}_{13}= 0$. Thus we can state:
\begin{equation}\label{C^+=0}
  \mathcal{C}^+ \,=\, 0 \quad\Longleftrightarrow\quad \bl{\omega}_{12}= \bl{\omega}_{34}= 0 \;\textrm{ in some null frame}\,.
\end{equation}
A manifold such that $\mathcal{C}^+$ vanishes is dubbed an anti-self-dual manifold\index{Self-dual manifold}. Consider now the isotropic distribution\index{Isotropic} $Span\{\bl{e}_{_{1,\lambda\kappa}}, \,\bl{e}_{_{2,\lambda\kappa}}\}$ with
$$ \bl{e}_{_{1,\lambda\kappa}} \,\equiv\, \lambda\bl{E}_{1}\,+\,\kappa\bl{E}_{4} \quad \textrm{and} \quad  \bl{e}_{_{2,\lambda\kappa}} \,\equiv\, \lambda\bl{E}_{2}\,+\,\kappa\bl{E}_{3}\,,$$
where $\lambda$ and $\kappa$ are constant scalars. Then this distribution will be integrable if, and only if, the Lie bracket of $\bl{e}_{_{1,\lambda\kappa}}$ and $\bl{e}_{_{2,\lambda\kappa}}$ is of the form $f\bl{e}_{_{1,\lambda\kappa}}+ h\bl{e}_{_{2,\lambda\kappa}}$ for some functions $f$ and $h$. Working out such Lie bracket explicitly it is straightforward to prove that this distribution will be integrable for all $\lambda$ and $\kappa$ if, and only if, the following conditions hold:
\begin{align}\label{Integ-Infinity}
 \nonumber \omega_{112} \,=\, \omega_{221}  \,=\, 0 \;\,;\;\;\,\omega_{312}&= \omega_{224}- \omega_{213} \;;\;\; \omega_{412}= \omega_{124}- \omega_{113} \\
 \omega_{334}  \,=\, \omega_{443}  \,=\, 0 \;\,;\;\;\,  \omega_{143}&= \omega_{424}- \omega_{413} \;;\;\; \omega_{243}= \omega_{324}- \omega_{313} \,.
\end{align}
In particular, note that if $\bl{\omega}_{12}= \bl{\omega}_{34}= (\bl{\omega}_{24}- \bl{\omega}_{13}) = 0$ then this infinite family of distributions is integrable. Conversely, if $Span\{\bl{e}_{_{1,\lambda\kappa}}, \,\bl{e}_{_{2,\lambda\kappa}}\}$ is integrable for all $\lambda$ and $\kappa$ then equation (\ref{Integ-Infinity}) holds, so that theorem \ref{Theo. GSgen4} implies that $\Psi_0^+$, $\Psi_1^+$, $\Psi_3^+$ and $\Psi_4^+$ all vanish. Then inserting this and eq. (\ref{Integ-Infinity}) on the self-dual part of the second structure equation we find, after some algebra, that $\Psi_2^+$ must also vanish, so that $\mathcal{C}^+= 0$. Using this result as well as equations (\ref{Jint-dW-DW}) and (\ref{C^+=0}) we arrive at the following theorem:
\begin{theorem}\label{Theo. C^+=0}
In a Ricci-flat 4-dimensional manifold the following conditions are equivalent:\\
(1) The Weyl tensor is type $(O, \lozenge)$, so that $\mathcal{C}^+= 0$\\
(2) There exists a null frame in which $\bl{\omega}_{12}= \bl{\omega}_{34}= 0$\\
(3) $\bl{J}$ is integrable and $d\bl{\Omega}= 0$\\
(4) The K\"{a}hler form, $\bl{\Omega}$, is covariantly constant\\
(5) There exists some null frame in which the isotropic distributions $Span\{\lambda\bl{E}_{1}+\kappa\bl{E}_{4}, \,\lambda\bl{E}_{2}+\kappa\bl{E}_{3}\}$ are integrable for all $\lambda$ and $\kappa$ constants.
\end{theorem}

As shown in chapter \ref{Chap. GR}, in general relativity the gravitational field is represented by a metric $\bl{g}$ of a 4-dimensional manifold while the electromagnetic field is represented by a 2-form $\bl{F}$ on this manifold, with the field equations of this system in the absence of sources being:
$$ R_{\mu\nu}- \frac{1}{2}R\,g_{\mu\nu}= 2G\lef F_{\mu\sigma} F_{\nu}^{\phantom{\nu}\sigma} - \frac{1}{4}\, g_{\mu\nu} F^{\rho\sigma}F_{\rho\sigma}  \rig \;; \quad d\bl{F}= 0 \;;\quad d(\star\bl{F})= 0 \,.$$
Thus if $\bl{Z}$ is a 2-form such that its energy-momentum tensor vanishes, $4Z_{\mu\sigma} Z_{\nu}^{\phantom{\nu}\sigma}= g_{\mu\nu} Z^{\rho\sigma}Z_{\rho\sigma}$, and $\star\bl{Z}\propto \bl{Z}$ then the above field equations become just $R_{\mu\nu}= 0$ and $d\bl{Z}= 0$. Note that the 2-forms $\bl{Z}^{1+}$, $\bl{Z}^{2+}$ and $\bl{\Omega}$ are all self-dual and have vanishing energy-momentum tensor. Furthermore, by what was seen in section \ref{Sec. GS4D}, when the Weyl tensor of a Ricci-flat 4-dimensional manifold is algebraically special we can find a null frame in which $d(f\bl{Z}^{1+})= 0$ for some function $f\neq 0$. In addition, if $\mathcal{C}^+$ is type $D$ then $\bl{Z}^{2+}$ also generates an integrable distribution and, therefore, we can find a function $h\neq 0$ such that $d(h\bl{Z}^{2+})= 0$. Moreover, theorem \ref{Theo. C^+=0} guarantees that if $\mathcal{C}^+= 0$ then $d\bl{\Omega}= 0$. Thus the following theorem holds:
\begin{theorem}\label{Theo. MaxwellFields}
If the Ricci tensor of a 4-dimensional manifold vanishes then depending on the algebraic type of the Weyl tensor it is possible to find a null frame and non-zero functions $f$ and $h$ such that the following 2-forms are solutions to the Einstein-Maxwell equations without sources:\\
$\bullet$ $\mathcal{C}^+$ type $II$, $III$ or $N$: \;$\bl{F}_1= f\bl{Z}^{1+}$\\
$\bullet$ $\mathcal{C}^+$ type $D$: \;$\bl{F}_1= f\bl{Z}^{1+}$ and $\bl{F}_2= h\bl{Z}^{2+}$\\
$\bullet$ $\mathcal{C}^+$ type $O$: \;$\bl{F}_1= f\bl{Z}^{1+}$, $\bl{F}_2= h\bl{Z}^{2+}$ and $\bl{F}_3= \bl{\Omega}= i\sqrt{2}\bl{Z}^{3+}$\,.
\end{theorem}


In the present subsection no assumption was made about the signature of the manifold, nor even it was assumed that the metric is real. In the forthcoming subsections the general results obtained here will be specialized to the case of a real metric for each possible signature.

\subsection{Euclidean Signature}
When the metric is real and Euclidean the vectors of a null frame obey the reality conditions shown on eq. (\ref{Real.Eucli}). Particularly, this implies that the almost complex structure $\bl{J}$ and the K\"{a}hler form $\bl{\Omega}$ are both real. In addition, for this signature just the six algebraic types shown on equation (\ref{Types-Euclid}) are allowed. Thus if the Weyl tensor is not type $(I,I)$ then it must be type $(D,\lozenge)$ or $(O,\lozenge)$, which according to theorem \ref{Theo. Jinteg} is equivalent to say that $\bl{J}$ is integrable on some null frame. Since $\bl{J}$ is real, the Newlander-Nirenberg theorem guarantees that if $\bl{J}$ is integrable then the manifold over the complex field is a complex manifold, more precisely an Hermitian manifold. Therefore we can state the following theorem \cite{art2,Nurwoski2}:
\begin{theorem}
In a 4-dimensional Euclidean manifold with vanishing Ricci tensor, the Weyl tensor is algebraically special if, and only if, the manifold over the complex field is Hermitian.
\end{theorem}
Furthermore, if the type of the Weyl tensor is $(O,\lozenge)$ then theorem \ref{Theo. C^+=0} guarantees that $\bl{\Omega}$ is covariantly constant, $\nabla_a\bl{\Omega}= 0$. In particular the K\"{a}hler form is closed, $d\bl{\Omega}= 0$, which implies that the manifold is a Calabi-Yau manifold\index{Calabi-Yau manifold}. So the following theorem holds:
\begin{theorem}
An Euclidean 4-dimensional Ricci-flat manifold over the complex field is a Calabi-Yau manifold if, and only if, the Weyl tensor is either self-dual, $\mathcal{C}^-= 0$, or anti-self-dual, $\mathcal{C}^+= 0$.
\end{theorem}
This result was first proved in \cite{Broda} using spinorial language and later in \cite{art2} using vectorial formalism.

\subsection{Lorentzian Signature}
If the metric is real and Lorentzian a special phenomenon arises, the self-dual and anti-self-dual parts of the Weyl tensor are complex conjugates of each other, $\mathcal{C}^+ = \overline{\mathcal{C}^-}$. In particular, if a null bivector generates an integrable distribution of isotropic self-dual planes then its complex conjugate generates an integrable distribution of isotropic anti-self-dual planes. Using (\ref{Real.Loren}) we easily find that in this signature $\bl{Z}^{i+}$ is the complex conjugate of $\bl{Z}^{i-}$. Thus if $d\bl{Z}^{1+}= 0$ then the bivector $\bl{F}= \bl{Z}^{1+}+ \bl{Z}^{1-}$ is real and $d\bl{F}= d(\star\bl{F})= 0$. Note also that $\bl{F}$ has the form $\bl{F}= \bl{l}\wedge\bl{e}$ with $\bl{l}$ being a null vector field whereas $\bl{e}$ is space-like and orthogonal to $\bl{l}$, so that $\bl{F}$ is a bivector representing electromagnetic radiation, see section \ref{Sec._Mariot4D}. Theorem \ref{Theo. MaxwellFields} then guarantees that if the Weyl tensor is algebraically special then the space-time admits a real solution for the Maxwell's equations without sources\footnote{$\bl{F}$ is not a solution for the Einstein-Maxwell equations, since its energy-momentum tensor is different from zero. In other words, $\bl{F}$ is just a test field.} corresponding to electromagnetic radiation. This is a classical result obtained by Robinson in \cite{Robinson}, see section \ref{Sec._Mariot4D}. As a last comment note that theorem \ref{Theo. C^+=0} is trivial on the Lorentzian signature, since whenever $\mathcal{C}^+= 0$ the whole Weyl tensor must be identically zero, so that if the Ricci tensor is assumed to vanish then space-time is flat.

\subsection{Split Signature}
Now let us assume that the metric is real and has split signature. As explicitly shown in section \ref{Sec. NullFrame4}, in this case we have two kinds of null frame \cite{Trautman}: (1) a real null frame $\{\bl{E}'_a\}$, so that $\bl{E}'_a= \overline{\bl{E}'_a}$; (2) a complex null frame $\{\bl{E}_a\}$ such that $\bl{E}_3= \overline{\bl{E}_1}$ and $\bl{E}_4= \overline{\bl{E}_2}$. As shown on table \ref{Tab._vanishScalars4}, if  $\mathcal{C}^+$ is algebraically special then we can find a null frame in which $\Psi_0^+= \Psi_1^+= 0$, this null frame can then be either real or complex. Let us work out these two cases separately.

Suppose that the frame in which the Weyl scalars $\Psi_0^+$ and $\Psi_1^+$ vanish is real, then theorem \ref{Theo. GSgen4} implies that the real isotropic distribution generated by $\{\bl{E}'_1,\bl{E}'_2\}$ is integrable. Moreover, if $\mathcal{C}^+$ is type $D$ then the real isotropic distribution $\{\bl{E}'_3,\bl{E}'_4\}$ will also be integrable, so that $\bl{J}$ is integrable. Since in this case $\bl{J}$ and $\bl{\Omega}$ are complex it is useful to define the real tensors $\bl{J}'\equiv -i\bl{J}$ and $\bl{\Omega}'\equiv -i\bl{\Omega}$. Note that, seen as an operator on the tangent bundle, $\bl{J}'$ is such that $\bl{J}'\bl{J}'= \bl{1}$ and $\bl{g}\lef \bl{J}'(\bl{V}), \bl{J}'(\bl{X})\rig= -\bl{g}(\bl{V},\bl{X})$ for all vector fields $\bl{V}$ and $\bl{X}$. Hence the tensor $\bl{J}'$ is called a paracomplex structure, more details about this kind of tensor in this context is available in \cite{parastruct}.

Now let $\mathcal{C}^+$ be algebraically special and assume that the null frame in which $\Psi_0^+= \Psi_1^+= 0$ is not real. Then besides to the isotropic distribution $\{\bl{E}_1,\bl{E}_2\}$, the complex conjugate distribution $\{\bl{E}_3,\bl{E}_4\}$ will also be integrable, so that the almost complex structure $\bl{J}$ is integrable. Note also that, since $\bl{E}_3= \overline{\bl{E}_1}$ and $\bl{E}_4= \overline{\bl{E}_2}$, the complex structure $\bl{J}$ is real. Therefore, the Newlander-Nirenberg theorem guarantees that the manifold over the complex field is an Hermitian manifold. Moreover, theorem \ref{Theo. C^+=0} implies that if $\mathcal{C}^+= 0$ then $\bl{\Omega}$ is covariantly constant. Particularly, in this case the K\"{a}hler form is closed, $d\bl{\Omega}= 0$, so that over the real field the manifold is symplectic\footnote{A symplectic manifold is an even-dimensional manifold endowed with a non-degenerate closed 2-form. In the present case $\bl{\Omega}$ plays the role of a symplectic form.}, while over the complex field it is a Calabi-Yau manifold\index{Calabi-Yau manifold}. In general, the following theorem can be stated \cite{art2}:
\begin{theorem}
Let $(M,\bl{g})$ be a Ricci-flat manifold of split signature. Then it admits an integrable distribution of non-real isotropic planes if, and only if, over the complex field such manifold is Hermitian. In addition, over the complex field such manifold will be Calabi-Yau if, and only if, $\mathcal{C}^+$(or $\mathcal{C}^-$) vanishes.
\end{theorem}

\chapter{Six Dimensions Using Spinors}\label{Chap._Spin6D}
In the previous chapters it has been shown that the Petrov classification and the Goldberg-Sachs (GS) theorem have played a prominent role in the development of general relativity in 4 dimensions. With the increasing interest on higher-dimensional manifolds, see section \ref{Sec.HighD}, it is quite natural to try to develop an algebraic classification for the Weyl tensor valid in dimensions greater than 4, as well as searching for a suitable generalization of the GS theorem. As emphasized in chapter \ref{Chap. Pet}, there are several distinct but equivalent paths to attain the Petrov classification, so one might be tempted to arbitrarily choose one of these methods in order to define an algebraic classification for the Weyl tensor in higher dimensions. However, it turns out that such different approaches lead to inequivalent classifications when the dimension is different from 4. Hence it is important to take a wise path.

Undoubtedly the most neat an elegant route toward Petrov classification in 4 dimensions is the spinorial approach. Therefore, in this chapter the spinors will be used in order to define an algebraic classification for the Weyl tensor valid in 6 dimensions. Furthermore, it will be shown that a generalization of the GS theorem proved in \cite{HigherGSisotropic1,HigherGSisotropic2} can be nicely expressed by means of the spinorial language. The material presented here is based on the article \cite{Spin6D}. After this paper the same issues were explored in \cite{TC-PureSpin} using spinorial formalism in manifolds of arbitrary dimension.
Some previous work on spinors in six dimensions are available in \cite{Spinors in 6D}, where the formalism has been applied to quantum field theory. General aspects of spinors in even-dimensional space-times were also used in \cite{HughstMason}.


Over the last decade there have been several attempts to provide suitable higher-dimensional versions of the Petrov classification and GS theorem. In \cite{5D class.} it was defined an algebraic classification for the Weyl tensor in 5 dimensions using spinors and some applications were made. An algebraic classification for tensors in Lorentzian spaces of arbitrary dimension was defined in \cite{CMPP}, the so-called CMPP\index{CMPP classification} classification. Posterior work then attempted, with partial success, to generalize the GS theorem using the CMPP classification
\cite{M. Ortaggio-Robinson-Trautman,GS-GeodesicPart,GS-5D,GS-HighD}. Further, in \cite{HigherGSisotropic1,HigherGSisotropic2} it was put forward an algebraic classification for the Weyl tensor based on maximally isotropic structures\index{Maximally isotropic}. There it was also proved a higher-dimensional version of the Goldberg-Sachs theorem stating that if the Weyl tensor obeys to certain algebraic restrictions then the manifold admits an integrable maximally isotropic distribution. Here it will be taken advantage of the spinorial formalism in order to express such theorem in an elegant form. Finally, in \cite{art4} it was defined a classification for the Weyl tensor valid in any dimension that naturally generalizes the 4-dimensional bivector approach, there it was also proved a generalization of the GS theorem.

\section{From Vectors to Spinors}

In this section it will be shown how the low-rank tensors of a 6-dimensional vector space are represented in the spinorial formalism. Particularly, the isotropic subspaces will  prove to be elegantly expressed in terms of spinors. The reader is assumed to be familiar with the basics of spinorial formalism and group representation theory, if this is not the case see appendices \ref{App._Cliff&Spinors} and \ref{App._Group} respectively.

Let us first start with the Euclidean vector space $\mathbb{R}^6$, later the results of this case will be extrapolated to the space $\mathbb{C}^6$ in order to obtain results valid in any signature. As explained on appendix \ref{App._Cliff&Spinors}, the universal covering group of $SO(n)$ is $SPin(\mathbb{R}^n)$. More precisely, the latter group is a double covering of the former, $SPin(\mathbb{R}^n)\sim SO(n)\otimes \mathbb{Z}_2$. In particular, it can be proved that $SPin(\mathbb{R}^6)\sim SU(4)$ \cite{Lounesto}. Thus every tensor transforming on a representation of $SO(6)$ can be said to be on a certain representation of $SU(4)$, called the spinorial representation of this tensor. In order to determine the spinorial equivalents for some $SO(6)$ tensors we first need to study the irreducible representations of the group $SU(4)$\index{SU(4)}. Following the notation adopted on appendix \ref{App._Group}, the basic representations of $SU(4)$ are:
\begin{equation}\label{SU(4)-fund.}
  \textbf{4}:\;\; \chi^{A}\,\stackrel{U}{\longrightarrow} \,U^A_{\phantom{A}B} \,\chi^{B}\;\;\;\;\; ;\, \;\;\;\;\; \overline{\textbf{4}}:\;\; \gamma_{A}\,\stackrel{U}{\longrightarrow}\, \overline{U}_A^{\phantom{A}B}\, \gamma_{B}\;.
\end{equation}
Where the indices $A,B,\ldots$ run from 1 to 4 and $U^A_{\phantom{A}B}$ is a $4\times4$ unitary matrix of unit determinant, with $\overline{U}_A^{\phantom{A}B}$ being its complex conjugate. Since a unitary matrix $U$ obeys to $(U^{-1})^t= \overline{U}$, it follows that the representation $\overline{\bl{4}}$ is the inverse of the representation $\textbf{4}$, see eq. (\ref{L(g)v_a}). In particular, this implies that $\chi^{A}\gamma_{A}$ is invariant under the action of $SU(4)$. From now on we shall call the objects transforming on the representation $\textbf{4}$  the spinors of positive chirality, while an object transforming on the representation $\overline{\textbf{4}}$ is a spinor of negative chirality. Taking the complex conjugate of eq. (\ref{SU(4)-fund.}) we find that if $\chi^{A}$ is a spinor\index{Spinor} of positive chirality then its complex conjugate, $\overline{\chi^{A}}$, will be a spinor of negative chirality. Therefore we conclude that the complex conjugation lowers the upper spinorial indices and raises the lower indices, $\overline{\chi^{A}}= \overline{\chi}_{A}$ and $\overline{\gamma_{A}}= \overline{\gamma}^{A}$. A list of the low-dimensional irreducible representations of $SU(4)$ is shown on table \ref{Tab.SU4}. Since all representations of this group can be constructed by means of the direct products of the representation $\textbf{4}$ and its inverse, $\overline{\textbf{4}}$, we say that the fundamental representation of $SU(4)$ is $\textbf{4}$\index{SU(4)}\index{Irreducible representations}.
\begin{table}
\begin{center}
\begin{tabular}{cccc||cl}
\hline
  \hline
  $\bl{1}$ & $\bl{4}$ & $\overline{\bl{4}}$ & $\bl{6}$
  &$\bl{4}\otimes \bl{4}$& $= \,\bl{6} \oplus \bl{10}$    \\
  $\bl{10}$ &$\overline{\bl{10}}$ & $\bl{15}$ & $\bl{20}$
  &$\bl{4}\otimes \overline{\bl{4}}$& $=\,\bl{1} \oplus \bl{15}$    \\
  $\overline{\bl{20}}$ & $\bl{20'}$ & $\bl{20''}$ &$\overline{\bl{20''}}$
  &$\bl{6}\otimes \bl{6}$& $=\,\bl{1} \oplus \bl{15} \oplus \bl{20'}$    \\
  $\bl{35}$ & $\overline{\bl{35}}$ & $\bl{36}$ & $\overline{\bl{36}}$\,
  &$\bl{10}\otimes \bl{6}$& $=\,\bl{15} \oplus \bl{45}$     \\
  $\bl{45}$ &$\overline{\bl{45}}$ & $\bl{50}$ & $\bl{56}$
  &$\bl{10}\otimes \overline{\bl{10}}$& $= \,\bl{1} \oplus \bl{15} \oplus \bl{84}$   \\
  $\overline{\bl{56}}$ & $\bl{60}$ & $\overline{\bl{60}}$ & $\bl{64}$
  &$\bl{15}\otimes \bl{6}$& $=\,\bl{6} \oplus \bl{10} \oplus \overline{\bl{10}}\oplus \bl{64}$    \\
  $\bl{70}$ & $\overline{\bl{70}}$ & $\bl{84}$ &$\bl{84'}$
  &$\bl{20'}\otimes \bl{6}$ & $=\,\bl{6} \oplus \bl{50} \oplus \bl{64}$     \\
  $\overline{\bl{84'}}$ & $\bl{84''}$ & $\overline{\bl{84''}}$ &$\bl{105}$
  &$\bl{15}\otimes \bl{15}$& $=\bl{1} \oplus \bl{15} \oplus \bl{15} \oplus \bl{20'} \oplus \bl{45} \oplus \overline{\bl{45}} \oplus \bl{84}$     \\
     \hline
    \hline
          \end{tabular}\caption{\footnotesize{On the left hand side of this table we have a list of all irreducible representations of the group $SU(4)$ with dimension less than $120$. In this list the inequivalent representations of the same dimension are distinguished by primes. Note that the representations $\bl{1}$, $\bl{6}$, $\bl{15}$, $\bl{20'}$, $\bl{50}$, $\bl{64}$, $\bl{84}$ and $\bl{105}$ are real. Thus, for example, $\overline{\bl{15}}= \bl{15}$. On the right hand side of this table we have the decomposition in irreducible parts of some direct products of the irreducible representations \cite{Slansky}.}}\label{Tab.SU4}
\end{center}
\end{table}

Now let us see how the tensors of $SO(6)$ transform under $SU(4)$. A vector of $\mathbb{R}^6$, $V^\mu$, has six degrees of freedom and, therefore, might be on a non-trivial six-dimensional and  real representation of  $SU(4)$, which according to table \ref{Tab.SU4} is unique, $\bl{6}$. The same table says that this representation can be obtained by decomposing the direct product  $\bl{4}\otimes\bl{4}$ in irreducible parts. Indeed, if $D^{AB}$ is on the representation $\bl{4}\otimes\bl{4}$ then we can split it in two irreducible parts (see appendix \ref{App._Group}):
\begin{equation}\label{4x4=6+10}
  \underbrace{D^{AB}}_{\bl{4}\otimes\bl{4}} \;\longrightarrow\;\;  \underbrace{D^{[AB]}}_{\bl{6}}  \quad+\quad \underbrace{D^{(AB)}}_{\bl{10}}\,.
\end{equation}
Thus a vector $V^\mu$ transforms as an object of the form $V^{AB}= V^{[AB]}$. Another representation of dimension 6 could be provided by $V_{AB}= V_{[AB]}$, let us denote such representation by $\overline{\bl{6}}$. However, it is not hard to verify this representation is, actually, equivalent to the representation $\bl{6}$. Indeed, let $\varepsilon_{ABCD}= \varepsilon_{[ABCD]}$ be the unique completely anti-symmetric symbol such that $\varepsilon_{1234}= 1$. Then its contraction with four arbitrary spinors, $\zeta^{A},\chi^{A},\varphi^{A}$ and $\xi^{A}$, is invariant under $SU(4)$:
\begin{equation}\label{Eps.Contraction}
    \varepsilon_{ABCD}\zeta^{A}\chi^{B}\varphi^{C}\xi^{D}\,\stackrel{U}{\longrightarrow}\, \det(U)\, \varepsilon_{EFGH}\zeta^{E}\chi^{F}\varphi^{G}\xi^{H} = \varepsilon_{ABCD}\zeta^{A}\chi^{B}\varphi^{C}\xi^{D}.
\end{equation}
In the same fashion we can define the object $\varepsilon^{ABCD}= \varepsilon^{[ABCD]}$ with $\varepsilon^{1234}= 1$ and verify that an analogous relation holds for  spinors of negative chirality. Thus if $V^{AB}$ is on the representation $\bl{6}$ then, in order for the combination $V^{AB}\varepsilon_{ABCD}V^{CD}$ be invariant under $SU(4)$, the object  $\varepsilon_{ABCD}V^{CD}$ must be on the inverse representation, $\overline{\bl{6}}$. So that we can define:
\begin{equation}\label{V^AB-V_AB}
   V_{AB} \,\equiv\, \frac{1}{2} \varepsilon_{ABCD}V^{CD}\;\,;\,\; V^{AB} \,\equiv\, \frac{1}{2} \varepsilon^{ABCD}V_{CD} \,.
\end{equation}
Since the representation $\overline{\bl{6}}$ can be obtained from the representation $\bl{6}$ by a simple algebraic operation not involving complex conjugation it follows that these representations are actually equivalent, $\bl{6}= \overline{\bl{6}}$. Because of this we might say that this representation is real. Thus in six dimensions we can raise or low a skew-symmetric pair of indices without changing the representation.

A bivector $B_{\mu\nu}= -B_{\nu\mu}$ in 6 dimensions has 15 degrees of freedom and, therefore, must be in a $15$-dimensional and real representation of $SU(4)$.  According to table \ref{Tab.SU4} both criteria are satisfied by the representation $\bl{15}$. The identity $\bl{4}\otimes\overline{\bl{4}}= \bl{1} \oplus \bl{15}$ says that this representation is given by the objects of the form $B^A_{\ph{A}B}$ with vanishing trace, $B^A_{\ph{A}A}= 0$. The reality of this representation can be understood by the fact that when we take the complex conjugate of $B^A_{\ph{A}B}$ we obtain another trace-less object with one index up and one down.

If $S_{\mu\nu}= S_{(\mu\nu)}$ is a trace-less symmetric tensor on $\mathbb{R}^6$ then it has $20$ independent components. Since it has two indices, it follows that from the $SO(6)$ point of view this tensor is obtained by the direct product of two vectorial representations. Therefore its spinorial equivalent might be contained on the direct product $\bl{6}\otimes\bl{6}$. Table \ref{Tab.SU4} furnish that $\bl{6}\otimes \bl{6}= \bl{1} \oplus \bl{15} \oplus \bl{20'}$, so that the spinorial equivalent of $S_{\mu\nu}$ might be on the representation $\bl{20'}$, which has the form $S^{AB}_{\phantom{AB}CD}=S^{[AB]}_{\phantom{AB}[CD]}$ with vanishing trace, $S^{AB}_{\phantom{AB}CB}=0$. Note that this representation is real.

In six dimensions a 3-vector $T_{\mu\nu\rho}= T_{[\mu\nu\rho]}$ has 20 degrees of freedom and can be obtained by the anti-symmetrization of the direct product of a bivector and a vector. Therefore its spinorial equivalent must be contained on the direct product $\bl{15}\otimes\bl{6}$. By means of  table \ref{Tab.SU4} we have $\bl{15}\otimes \bl{6}= \bl{6} \oplus \bl{10} \oplus \overline{\bl{10}}\oplus \bl{64}$. Thus we conclude that the 3-vectors are on the representation $\bl{10} \oplus \overline{\bl{10}}$ of $SU(4)$\label{3Vector6D}. From the eq. (\ref{4x4=6+10}) we see that the representation $\bl{10}$ is given by $T^{AB}= T^{(AB)}$. So in the spinorial language a 3-vector $T_{\mu\nu\rho}$ is represented by a pair $(T^{AB}, \tilde{T}_{AB})$ of symmetric objects. It is possible to prove that if $\tilde{T}_{AB}= 0$ then the 3-vector is self-dual\index{Self-dual form}, $\star\bl{T}= \bl{T}$. Analogously, whenever $T^{AB}= 0$ the 3-vector is anti-self-dual, $\star\bl{T}= -\bl{T}$.

The Weyl tensor\index{Weyl tensor}\label{Weyl6D} $C_{\mu\nu\rho\sigma}$ is a trace-less object with the symmetries $C_{\mu\nu\rho\sigma}= C_{[\mu\nu][\rho\sigma]}$ and $C_{\mu[\nu\rho\sigma]}= 0$. It can be proved that in 6 dimensions it has 84 independent components. From the first symmetry we see that this tensor can be obtained by a linear combination of the direct product of bivectors, so that its spinorial representation must be contained in $\bl{15}\otimes\bl{15}$. Looking at the expansion of this direct product on table \ref{Tab.SU4} we see that $C_{\mu\nu\rho\sigma}$ must be on the representation $\bl{84}$ of $SU(4)$. Because of the equation $\bl{10}\otimes \overline{\bl{10}}= \bl{1} \oplus \bl{15} \oplus \bl{84}$ we conclude that an object in this representation have the form $\Psi^{AB}_{\ph{AB}CD}$ with $\Psi^{AB}_{\ph{AB}CD}= \Psi^{(AB)}_{\ph{AB}(CD)}$ and $\Psi^{AB}_{\ph{AB}CB}= 0$. The results obtained so far are summarized on table \ref{Tab. Spinors equiv.} \cite{Spin6D}.
\begin{table}[!htbp]
\begin{center}
\begin{tabular}{c|l|c}
  \hline
  \hline
  $SO(6)$ Tensor &   Spinorial Representation & Symmetries   \\ \hline
  $V^\mu$ & $\bl{6}\rightarrow$ $V^{AB}$ & $V^{AB}=-V^{BA}$ \\
  $B_{\mu\nu}$ & $\bl{15}\rightarrow$ $B^A_{\phantom{A}B}$ & $B^A_{\phantom{A}A}=0$ \\
  $S_{\mu\nu}$ & $\bl{20'}\rightarrow$ $S^{AB}_{\phantom{AB}CD}$ & $S^{AB}_{\phantom{AB}CD}=S^{[AB]}_{\phantom{AB}[CD]}, \, S^{AB}_{\phantom{AB}CB}=0$ \\
  $T_{\mu\nu\rho}$ &$\bl{10}\oplus\overline{\bl{10}}\rightarrow$ $(T^{AB},\tilde{T}_{AB})$& $T^{AB}=T^{BA}, \,\tilde{T}_{AB}=\tilde{T}_{BA}$  \\
  $C_{\mu\nu\rho\sigma}$ & $\bl{84}\rightarrow$ $\Psi^{AB}_{\phantom{AB}CD}$ & $\Psi^{AB}_{\phantom{AB}CD}=\Psi^{(AB)}_{\phantom{AB}(CD)}, \, \Psi^{AB}_{\phantom{AB}CB} = 0$ \\
  \hline
  \hline
\end{tabular}
\caption{\footnotesize{Spinorial equivalent of some low rank $SO(6;\mathbb{R})$ tensors. $V^\mu$ is a vector, $S_{\mu\nu}= S_{(\mu\nu)}$ is a trace-less symmetric tensor, $B_{\mu\nu}= B_{[\mu\nu]}$ is a bivector,\, $T_{\mu\nu\rho}= T_{[\mu\nu\rho]}$ is a 3-vector and $C_{\mu\nu\rho\sigma}$ is a tensor with the symmetries of a Weyl tensor. Note that all these representations are real.}}\label{Tab. Spinors equiv.}
\end{center}
\end{table}

\subsection{A Null Frame}
Let $V^\mu$ and $K^\mu$ be two vectors of $\mathbb{R}^6$, then the inner product $\bl{g}(\bl{V},\bl{K})= V^\mu K_{\mu}$ is the only scalar, up to a multiplicative factor, that is invariant under $SO(6)$ and is linear on both vectors. Denoting by $V^{AB}$ and $K^{AB}$ the spinorial equivalents of these vectors then it follows from equation (\ref{Eps.Contraction}) that the scalar $\varepsilon_{ABCD}V^{AB}K^{CD}$ is invariant under $SU(4)$ and, hence, invariant under $SO(6)$. Since such scalar is also linear in $\bl{V}$ and $\bl{K}$ it follows that it must be a multiple of the inner product $V^\mu K_{\mu}$. Because of equation (\ref{V^AB-V_AB}) one conclude that this multiplicative factor might be $2$:
\begin{equation}\label{InnerProd6D}
  V^{\mu}\,K_{\mu} \,=\, \frac{1}{2}\, \varepsilon_{ABCD}V^{AB}K^{CD} \,=\, V^{AB}K_{AB}\,.
\end{equation}

Now let $\{\chi_1^{\,A}, \chi_2^{\,A}, \chi_3^{\,A}, \chi_4^{\,A} \}$ be a basis for the space of positive chirality spinors obeying to the following normalization condition:
\begin{equation}\label{NormalizationSpin6D}
    \varepsilon_{ABCD}\,\chi_1^{\,A}\chi_2^{\,B}\chi_3^{\,C}\chi_4^{\,D} \,=\,1\,.
\end{equation}
Note, in particular, that the choice $\chi_p^{\,A}= \delta_p^{\,A}$ satisfy this constraint. We can use the basis $\{\chi_p^{\,A}\}$ in order to define a dual basis for the space of spinors with negative chirality:
\begin{align*}
   \gamma^1_{\,A} = \varepsilon_{ABCD}\,\chi_2^{\,B}\chi_3^{\,C}\chi_4^{\,D}\;\;;&\;\;
    \gamma^2_{\,A} = -\, \varepsilon_{ABCD}\,\chi_1^{\,B}\chi_3^{\,C}\chi_4^{\,D}\\
  \gamma^3_{\,A} = \varepsilon_{ABCD}\,\chi_1^{\,B}\chi_2^{\,C}\chi_4^{\,D}\;\;;&\;\;
    \gamma^4_{\,A} = -\, \varepsilon_{ABCD}\,\chi_1^{\,B}\chi_2^{\,C}\chi_3^{\,D}
\end{align*}
It is simple matter to verify that the relation $\chi_p^{\,A}\gamma^q_{\,A} = \delta^{\,q}_p$ holds. Then we can define the following frame of vectors, objects on the representation $\bl{6}$:
\begin{gather}
 \nonumber e_1^{\,AB} = \chi_1^{\,[A}\chi_2^{\,B]}\;;\;\; e_2^{\,AB} = \chi_1^{\,[A}\chi_3^{\,B]}\;;\;\; e_3^{\,AB} = \chi_1^{\,[A}\chi_4^{\,B]}\\ \theta^{1\,AB} = \chi_3^{\,[A}\chi_4^{\,B]}\;;\;\; \theta^{2\,AB} = \chi_4^{\,[A}\chi_2^{\,B]}\;;\;\; \theta^{3\,AB} = \chi_2^{\,[A}\chi_3^{\,B]}\,.\label{e^AB}
\end{gather}
By means of equation (\ref{V^AB-V_AB}) one can lower these pairs of skew-symmetric indices yielding:
\begin{gather}
  \nonumber e_{1\,AB} = \gamma^3_{\,[A}\gamma^4_{\,B]}\;;\;\;
    e_{2\,AB} = \gamma^4_{\,[A}\gamma^2_{\,B]}\;;\;\;
     e_{3\,AB} = \gamma^2_{\,[A}\gamma^3_{\,B]}\\
      \theta^1_{\,AB} = \gamma^1_{\,[A}\gamma^2_{\,B]}\;;\;\;
      \theta^2_{\,AB} = \gamma^1_{\,[A}\gamma^3_{\,B]}\;;\;\;
      \theta^3_{\,AB} = \gamma^1_{\,[A}\gamma^4_{\,B]}\,.\label{e_AB}
\end{gather}
Thus using equations (\ref{InnerProd6D}), (\ref{e^AB}) and (\ref{e_AB}) we easily find that the inner products of the frame vectors are:
\begin{equation}\label{NullFrame6D}
 e_{a'}^{\,\,\mu}\,\,e_{b'\,\mu}\,=\, \theta^{a'\,\mu}\,\theta^{b'}_{\,\,\mu}\,=\, 0 \quad;\quad e_{a'}^{\,\,\mu}\,\theta^{b'}_{\,\,\mu}\,=\, \frac{1}{2}\,\delta^{\,\,b'}_{a'}\,.
\end{equation}
In particular, all vectors of the frame $\{\bl{e}_{a'}, \bl{\theta}^{b'}\}$ are null\footnote{Throughout this chapter the following index conventions will be adopted: $A,B,C,\ldots$ are the spinorial indices and pertain to $\{1,2,3,4\}$; $\mu,\nu,\rho,\ldots$ are coordinate indices of $\mathbb{R}^6$, pertaining to $\{1,2,\ldots,6\}$; $a,b,c,\ldots$ are labels for a null frame of $\mathbb{C}^6$ and take the values $\{1,2,\ldots,6\}$; $a',b',c'$ pertain to $\{1,2,3\}$; $p,q$ label a basis of Weyl spinors and pertain to $\{1,2,3,4\}$; $r,s$ label a basis of (anti-)self-dual 3-vectors, running from 1 to 10.}. For later convenience we shall denote such frame by $\{\bl{e}_a\}$ with $\bl{e}_{4}= \bl{\theta^1}$, $\bl{e}_{5}= \bl{\theta^2}$ and $\bl{e}_{6}= \bl{\theta^3}$, or shortly $\bl{e}_{a'+3}= \bl{\theta^{a'}}$. From now on, a frame of vectors $\{\bl{e}_a\}$ in a 6-dimensional space obeying to eq. (\ref{NullFrame6D}) will be called a null frame\index{Null frame}. Defining $g_{ab}\equiv \bl{g}(\bl{e}_a,\bl{e}_b)$ we have $g_{14}= g_{25}= g_{36}= \frac{1}{2}$ while the other components vanish. Using equations (\ref{e^AB}) and (\ref{e_AB}) it is straightforward to prove the following relation:
\begin{equation}\label{ee6D}
  e_a^{\,AB}\,e_{b\,CB} \,+\, e_b^{\,AB}\,e_{a\,CB} \,=\, \frac{1}{2} \,g_{ab}\, \delta_C^{\,A}\,.
\end{equation}

Equation (\ref{e^AB}) enables us to find explicitly the spinorial equivalent of any vector in a vector space of 6 dimensions. More precisely, if $\{\bl{e}_a\}$ is a null frame on this space and $\bl{V}$ is a vector then:
\begin{equation}\label{VectorConvertion}
  \bl{V} \,=\, V^a\,\bl{e}_a   \quad \Longleftrightarrow \quad  V^{AB} \,=\, V^a\,e_a^{\,AB}\,.
\end{equation}
Where $V^a$ are the components of the vector $\bl{V}$ on this null frame and $e_a^{\,AB}$ are the objects defined on equation (\ref{e^AB}). Actually, equation (\ref{VectorConvertion}) teaches us how to convert any tensor to the spinorial language. For example, if $\bl{F}$ is a tensor of rank 2 then its spinorial image will be contained on the representation $\bl{6}\otimes\bl{6}$ and can be written in this formalism as $F^{AB\,CD}= F^{[AB]\,[CD]}$ defined by:
\begin{equation}\label{F-spinRep.}
 F^{AB\,CD} \,=\,  F^{ab} \,e_a^{\,AB}e_b^{\,CD}  \quad \Longleftrightarrow \quad   \bl{F}\,=\, F^{ab}\,\bl{e}_a\otimes \bl{e}_b   \,.
\end{equation}
In particular, if $S_{\mu\nu}$ is a symmetric and trace-less tensor then its spinorial equivalent can be written as:
$$ S^{AB}_{\ph{AB}CD} \,=\, S^{ab}\,e_a^{\,AB}\,e_{b\,CD}\,. $$
Note that using equation (\ref{ee6D}) one can easily see that $S^{AB}_{\ph{AB}CB}= 0$, which agrees with table \ref{Tab. Spinors equiv.}. In the same vein, if $B_{ab}$ is a bivector then its spinorial equivalent is:
$$  B^{AB\,CD} \,=\, B^{ab}\,e_a^{\,AB}\,e_b^{\,CD}\,. $$
However, this does not seem to agree with table \ref{Tab. Spinors equiv.}, since there the bivector is said to be represented by an object of the form $B^A_{\ph{A}B}$ with vanishing trace. But after some algebra it can be proved that the following relation holds:
\begin{equation}\label{Bivetor6D}
\left\{
  \begin{array}{cl}
    B^{AB\, CD} &=\, B^{[A}_{\phantom{A}E}\,\,\varepsilon^{B]ECD}- B^{[C}_{\phantom{C}E}\,\,\varepsilon^{D]EAB} \\
    B^A_{\ph{A}B} &\equiv\, \frac{1}{4}\, B^{AC\,DE} \, \varepsilon_{CDEB}\,=\,\frac{1}{2}\, B^{AC}_{\ph{AC}CB} \,.
  \end{array}
\right.
\end{equation}
So all degrees of freedom of $B^{AB\,CD}$ are contained on the trace-less object  $B^A_{\ph{A}B}$. That is the beauty of representation theory, by means of it one can anticipate how the degrees of freedom of a tensor are stored. Following the same reasoning, if $T_{abc}= T_{[abc]}$ is a 3-vector then its spinorial equivalent will be of the form $T^{AB\,CD\,EF}$,  analogously to eq. (\ref{F-spinRep.}). Nonetheless, according to table \ref{Tab. Spinors equiv.} the degrees of freedom of this tensor must be contained on a pair $(T^{AB},\tilde{T}_{AB})$ such that $T^{AB}= T^{(AB)}$ and $\tilde{T}_{AB}= \tilde{T}_{(AB)}$. By lack of any other option one can assure that $T^{AB}\propto T^{AC\,BD}_{\ph{AC\,BD}CD}$ and $\tilde{T}_{AB}\propto T_{AC\,BD}^{\ph{AC\,BD}CD}$. In order to agree with the notation of \cite{Spin6D} we might choose the proportionality constants to be $1/9$ and $-1/9$ respectively. So we can schematically write \cite{Spin6D}:
$$  T_{abc}= T_{[abc]} \;\Leftrightarrow\; T^{AB\,CD\,EF} \;\Leftrightarrow\; (T^{AB},\tilde{T}_{AB}) \equiv \frac{1}{9} (T^{AC\,BD}_{\ph{AC\,BD}CD}, -T_{AC\,BD}^{\ph{AC\,BD}CD})\,. $$
In a similar fashion, if $C_{abcd}$ is a tensor with the symmetries of a Weyl tensor then table \ref{Tab. Spinors equiv.} says that its degrees of freedom are stored in an object $\Psi^{AB}_{\ph{AB}CD}= \Psi^{(AB)}_{\ph{AB}(CD)}$ with vanishing trace. By lack of any other possibility this object must be a multiple of $C^{AF\ph{CF\,GD}BG}_{\ph{AF}CF\,GD}$, so that one can write \cite{Spin6D}:
\begin{equation}\label{Psi-from-C}
  C_{abcd} \;\;\Leftrightarrow\;\;  C^{AB\,CD\,EF\,GH}  \;\;\Leftrightarrow\;\; \Psi^{AB}_{\ph{AB}CD} \equiv \frac{1}{16}\, C^{AF\ph{CF\,GD}BG}_{\ph{AF}CF\,GD}\,.
\end{equation}

Let $\{\bl{e}_a\}$ be a null frame, then using equations (\ref{e^AB}), (\ref{e_AB}) and (\ref{Bivetor6D}) it is straightforward to find the spinorial equivalents of the bivectors $\bl{e}_a\wedge\bl{e}_b\equiv ({\bl{e}_a\otimes\bl{e}_b} - \bl{e}_b\otimes\bl{e}_a)$, this is summarized on table \ref{Tab. Biv6D}. Analogously, the relation between the Weyl tensor components on a null frame and the components of the object $\Psi^{AB}_{\ph{AB}CD}$ can be obtained, after a lot of algebra, by means of equations (\ref{e^AB}), (\ref{e_AB}) and (\ref{Psi-from-C}), the bottom line is shown on table \ref{Tab. WeylComp6D}.

\begin{table}
\begin{center}
\begin{tabular}{lll}
  \hline
  \hline
  $(e_1\wedge e_2)^A_{\phantom{A}B} = -\frac{1}{4} \chi_1^{\,A}\gamma^4_{\,B}$ & $(e_1\wedge e_3)^A_{\phantom{A}B} = \frac{1}{4} \chi_1^{\,A}\gamma^3_{\,B}$ & $(e_1\wedge\theta^2)^A_{\phantom{A}B} = -\frac{1}{4} \chi_2^{\,A}\gamma^3_{\,B}$  \\
   $(e_1\wedge\theta^3)^A_{\phantom{A}B} = -\frac{1}{4} \chi_2^{\,A}\gamma^4_{\,B}$ &
  $(e_2\wedge e_3)^A_{\phantom{A}B} = -\frac{1}{4} \chi_1^{\,A}\gamma^2_{\,B}$ & $(e_2\wedge\theta^1)^A_{\phantom{A}B} = -\frac{1}{4} \chi_3^{\,A}\gamma^2_{\,B}$ \\
   $(e_2\wedge\theta^3)^A_{\phantom{A}B} = -\frac{1}{4} \chi_3^{\,A}\gamma^4_{\,B}$  &$(e_3\wedge\theta^1)^A_{\phantom{A}B} = -\frac{1}{4} \chi_4^{\,A}\gamma^2_{\,B}$ & $(e_3\wedge\theta^2)^A_{\phantom{A}B} = -\frac{1}{4} \chi_4^{\,A}\gamma^3_{\,B}$  \\
        \multicolumn{2}{ c }{$ (e_1\wedge\theta^1)^A_{\phantom{A}B} = \frac{1}{8} [- \chi_1^{\,A}\gamma^1_{\,B} - \chi_2^{\,A}\gamma^2_{\,B} + \chi_3^{\,A}\gamma^3_{\,B} + \chi_4^{\,A}\gamma^4_{\,B}]$} & $(\theta^1\wedge\theta^2)^A_{\phantom{A}B} = \frac{1}{4} \chi_4^{\,A}\gamma^1_{\,B}$ \\
  \multicolumn{2}{ c }{$ (e_2\wedge\theta^2)^A_{\phantom{A}B} = \frac{1}{8} [- \chi_1^{\,A}\gamma^1_{\,B} + \chi_2^{\,A}\gamma^2_{\,B} - \chi_3^{\,A}\gamma^3_{\,B} + \chi_4^{\,A}\gamma^4_{\,B}]$} & $(\theta^1\wedge\theta^3)^A_{\phantom{A}B} = -\frac{1}{4} \chi_3^{\,A}\gamma^1_{\,B}$\\
  \multicolumn{2}{ c }{$ (e_3\wedge\theta^3)^A_{\phantom{A}B} = \frac{1}{8} [- \chi_1^{\,A}\gamma^1_{\,B} + \chi_2^{\,A}\gamma^2_{\,B} + \chi_3^{\,A}\gamma^3_{\,B} - \chi_4^{\,A}\gamma^4_{\,B}]$} & $(\theta^2\wedge\theta^3)^A_{\phantom{A}B} = \frac{1}{4} \chi_2^{\,A}\gamma^1_{\,B}$  \\
  \hline
  \hline
\end{tabular}
\caption{\footnotesize{The spinorial representation of a basis of bivectors \cite{Spin6D}.}}\label{Tab. Biv6D}
\end{center}
\end{table}

\small
\begin{table}
\begin{center}
\begin{tabular}{llll}
  \hline
  \hline
  $C_{1212}=4\Psi^{44}_{\phantom{44}11}$ & $C_{1213}=-4\Psi^{34}_{\phantom{34}11}$ &
  $C_{1215}=4\Psi^{34}_{\phantom{34}12}$ & $C_{1216}=4\Psi^{44}_{\phantom{44}12}$ \\
  $C_{1313}=4\Psi^{33}_{\phantom{33}11}$ & $C_{1315}=-4\Psi^{33}_{\phantom{33}12}$ &
  $C_{1316}=-4\Psi^{34}_{\phantom{34}12}$ &$C_{1515}=4\Psi^{33}_{\phantom{33}22}$ \\
  $C_{1516}=4\Psi^{34}_{\phantom{34}32}$ & $C_{1616}=4\Psi^{44}_{\phantom{44}22}$&
  $C_{1645}=-4\Psi^{14}_{\phantom{14}24}$ &  $C_{1646}=4\Psi^{14}_{\phantom{14}23}$\\
  $C_{2323}=4\Psi^{22}_{\phantom{22}11}$ & $C_{2326}=4\Psi^{24}_{\phantom{24}13}$ &
  $C_{2335}=4\Psi^{23}_{\phantom{23}14}$ & $C_{2356}=-4\Psi^{12}_{\phantom{12}12}$ \\
  $C_{5656}=4\Psi^{11}_{\phantom{11}22}$ & $C_{2626}=4\Psi^{44}_{\phantom{44}33}$ &
  $C_{2635}=4\Psi^{34}_{\phantom{34}34}$ &$C_{2656}=-4\Psi^{14}_{\phantom{14}23}$ \\
  $C_{3535}=4\Psi^{33}_{\phantom{33}44}$ & $C_{3556}=-4\Psi^{13}_{\phantom{13}24}$ &
  $C_{1242}=-4\Psi^{24}_{\phantom{24}13}$ & $C_{1243}=-4\Psi^{24}_{\phantom{24}14}$ \\
  $C_{1245}=-4\Psi^{14}_{\phantom{14}14}$ & $C_{1246}=4\Psi^{14}_{\phantom{14}13}$ &
  $C_{1342}=4\Psi^{23}_{\phantom{23}13}$ & $C_{1343}=4\Psi^{23}_{\phantom{23}14}$ \\
  $C_{1345}=4\Psi^{13}_{\phantom{13}14}$ & $C_{1346}=-4\Psi^{13}_{\phantom{13}13}$ &
  $C_{1542}=-4\Psi^{23}_{\phantom{23}23}$ & $C_{1543}=-4\Psi^{23}_{\phantom{23}24}$\\
  $C_{1545}=-4\Psi^{13}_{\phantom{13}24}$ & $C_{1546}=4\Psi^{13}_{\phantom{13}23}$ &
  $C_{1642}=-4\Psi^{24}_{\phantom{24}23}$ & $C_{1643}=-4\Psi^{24}_{\phantom{24}24}$ \\
  \end{tabular}

  \begin{tabular}{llll}
  $C_{1223}=4\Psi^{24}_{\phantom{24}11}$ & $C_{1226}=4\Psi^{44}_{\phantom{44}13}$ &
  $C_{1235}=4\Psi^{34}_{\phantom{34}14}$ &$C_{1256}=-4\Psi^{14}_{\phantom{14}12}$ \\
  $C_{1323}=-4\Psi^{23}_{\phantom{23}11}$ &$C_{1326}=-4\Psi^{43}_{\phantom{43}13}$ &
  $C_{1335}=-4\Psi^{33}_{\phantom{33}14}$ &$C_{1356}=4\Psi^{13}_{\phantom{13}12}$ \\
  $C_{1523}=4\Psi^{32}_{\phantom{32}12}$ &$C_{1526}=4\Psi^{34}_{\phantom{34}23}$ &
  $C_{1535}=4\Psi^{33}_{\phantom{33}24}$ &$C_{1556}=-4\Psi^{13}_{\phantom{13}22}$ \\
  $C_{1623}=4\Psi^{24}_{\phantom{24}21}$ &$C_{1626}=4\Psi^{44}_{\phantom{44}23}$ &
  $C_{1635}=4\Psi^{34}_{\phantom{34}24}$ &$C_{1656}=-4\Psi^{14}_{\phantom{14}22}$ \\
  \end{tabular}

  \begin{tabular}{ll}
  $C_{1414}=4(\Psi^{11}_{\phantom{11}11}+\Psi^{22}_{\phantom{22}22}+2\Psi^{12}_{\phantom{12}12})$ & $C_{1425}=4(\Psi^{23}_{\phantom{23}23}-\Psi^{14}_{\phantom{14}14})$\\
  $C_{2525}=4(\Psi^{11}_{\phantom{11}11}+\Psi^{33}_{\phantom{33}33}+2\Psi^{13}_{\phantom{13}13})$ & $C_{1436}=4(\Psi^{24}_{\phantom{24}24}-\Psi^{13}_{\phantom{13}13})$\\
  $C_{3636}=4(\Psi^{11}_{\phantom{11}11}+\Psi^{44}_{\phantom{44}44}+2\Psi^{14}_{\phantom{14}14})$ & $C_{2536}=4(\Psi^{34}_{\phantom{34}34}-\Psi^{12}_{\phantom{12}12})$\\
    \end{tabular}

   \begin{tabular}{lll}
  $C_{1225}=4(\Psi^{14}_{\phantom{14}11}+\Psi^{34}_{\phantom{34}31})$ & $C_{1236}=4(\Psi^{14}_{\phantom{14}11}+\Psi^{44}_{\phantom{44}41})$&
  $C_{1325}=4(\Psi^{23}_{\phantom{23}21}+\Psi^{43}_{\phantom{43}41})$ \\
  $C_{1336}=4(\Psi^{23}_{\phantom{23}21}+\Psi^{33}_{\phantom{33}31})$ & $C_{1525}=4(\Psi^{13}_{\phantom{13}12}+\Psi^{33}_{\phantom{33}32})$ & $C_{1536}=4(\Psi^{13}_{\phantom{13}12}+\Psi^{43}_{\phantom{43}42})$\\
  $C_{1625}=4(\Psi^{14}_{\phantom{14}12}+\Psi^{34}_{\phantom{34}32})$ & $C_{1636}=4(\Psi^{14}_{\phantom{14}12}+\Psi^{44}_{\phantom{44}42})$&
  $C_{1412}=4(\Psi^{14}_{\phantom{14}11}+\Psi^{24}_{\phantom{24}21})$\\
  $C_{1413}=4(\Psi^{33}_{\phantom{33}31}+\Psi^{43}_{\phantom{43}41})$ & $C_{1415}=4(\Psi^{13}_{\phantom{13}12}+\Psi^{23}_{\phantom{23}22})$ & $C_{1416}=4(\Psi^{14}_{\phantom{14}12}+\Psi^{24}_{\phantom{24}22})$\\
  \end{tabular}

  \begin{tabular}{lll}
  $C_{1423}=4(\Psi^{12}_{\phantom{12}11}+\Psi^{22}_{\phantom{22}21})$ & $C_{1426}=4(\Psi^{14}_{\phantom{14}13}+\Psi^{24}_{\phantom{24}23})$ &
  $C_{1435}=4(\Psi^{13}_{\phantom{13}14}+\Psi^{23}_{\phantom{23}24})$ \\
  $C_{1456}=4(\Psi^{31}_{\phantom{31}32}+\Psi^{41}_{\phantom{41}42})$ & $C_{2523}=4(\Psi^{12}_{\phantom{12}11}+\Psi^{23}_{\phantom{23}13})$& $C_{3623}=4(\Psi^{12}_{\phantom{12}11}+\Psi^{24}_{\phantom{24}14})$ \\
  $C_{2526}=4(\Psi^{14}_{\phantom{14}13}+\Psi^{34}_{\phantom{34}33})$ & $C_{3626}=4(\Psi^{14}_{\phantom{14}13}+\Psi^{44}_{\phantom{44}34})$ &
  $C_{2535}=4(\Psi^{13}_{\phantom{13}14}+\Psi^{33}_{\phantom{33}34})$\\
  $C_{3635}=4(\Psi^{13}_{\phantom{13}14}+\Psi^{34}_{\phantom{34}44})$ & $C_{2556}=4(\Psi^{12}_{\phantom{12}22}+\Psi^{14}_{\phantom{14}24})$ &
  $C_{3656}=4(\Psi^{12}_{\phantom{12}22}+\Psi^{13}_{\phantom{13}23})$ \\
      \hline
  \hline
       \end{tabular}
    \caption{\footnotesize{This table displays the relation between Weyl tensor's components in a null frame and its spinorial equivalents \cite{Spin6D}. The missing components of the Weyl tensor can be obtained by making the changes $1\leftrightarrow4$, $2\leftrightarrow5$ and $3\leftrightarrow6$ on the vectorial indices while performing the transformation $\Psi^{AB}_{\;\;\;\;CD}\mapsto \Psi^{CD}_{\;\;\;\;AB}$. Thus, for example, the relation $C_{1212}=4\Psi^{44}_{\;\;\;11}$ implies $C_{4545}= 4\Psi^{11}_{\;\;\;44}$}.}\label{Tab. WeylComp6D}
\end{center}
\end{table}
\normalsize

\subsection{Clifford Algebra in 6 Dimensions}\index{Clifford algebra}

The aim of this subsection is to provide a connection between the spinorial calculus introduced so far and the abstract formalism presented on appendix \ref{App._Cliff&Spinors}. Let us denote the 4-dimensional vector space spanned by the spinors $\{\chi_1^{\,A},\chi_2^{\,A},\chi_3^{\,A},\chi_4^{\,A}\}$ by $S^+$ and call it the space of positive chirality Weyl spinors\index{Weyl spinor}. Analogously, the 4-dimensional space spanned by $\{\gamma^1_{\,A},\gamma^2_{\,A},\gamma^3_{\,A},\gamma^4_{\,A}\}$ will be denoted by $S^-$ and called the space of Weyl spinors with negative chirality. The vector space $S= S^+\oplus S^-$ is then named the space of Dirac spinors, so that a Dirac spinor\index{Dirac spinor} $\bl{\psi}\in S$ is generally written as $\bl{\psi}= \psi^A + \tilde{\psi}_A$. Let us define the inner product of two Dirac spinors by:
\begin{equation}\label{InnerProd.Spinors6D}
  (\bl{\psi}, \bl{\phi}) \,=\, \psi^A\,\tilde{\phi}_A - \phi^A\,\tilde{\psi}_A \,.
\end{equation}
Note that this inner product is skew-symmetric and vanishes if the two spinors $\bl{\psi}$ and $\bl{\phi}$ have the same chirality, as said on appendix \ref{App._Cliff&Spinors}.

On the Clifford algebra formalism the vectors of $V= \mathbb{R}^6$ are linear operators that act on the space of spinors. Therefore, to each vector $\bl{e}_a$ it is associated a linear operator $\check{\bl{e}}_a: S\rightarrow S$ acting on the space of Dirac spinors. The action of this operator is defined by:
\begin{equation}\label{Clifford6D}
  \check{\bl{e}}_a(\bl{\psi}) \,=\, \bl{\phi}\,  \quad\Longleftrightarrow \quad    \phi^A= 2\,e_a^{\,AB}\,\tilde{\psi}_B  \;\,\textrm{ and }\,\;   \tilde{\phi}_A=-2 \,e_{a\,AB}\,\psi^B\,.
\end{equation}
In order to verify that this action is correct note that using equations (\ref{ee6D}) and (\ref{Clifford6D}) we arrive at the following important relation:
$$  \check{\bl{e}}_a\,\check{\bl{e}}_b \,+\, \check{\bl{e}}_b\,\check{\bl{e}}_a  \,=\, 2\,g_{ab}\,\bl{1}\,,$$
where $\bl{1}$ is the identity operator on $S$. Such relation is the analogous of equation \ref{Sym_Clif. Prod} on appendix \ref{App._Cliff&Spinors}. Note also that the inner product defined on (\ref{InnerProd.Spinors6D}) is such that $(\check{\bl{e}}_a(\bl{\psi}),\bl{\phi})= (\bl{\psi},\check{\bl{e}}_a(\bl{\phi}))$, which also agrees with appendix \ref{App._Cliff&Spinors}\footnote{Although the symmetric inner product $\langle\bl{\psi}| \bl{\phi}\rangle\equiv \psi^A\tilde{\phi}_A + \phi^A\tilde{\psi}_A$ is also invariant under $SPin(\mathbb{R}^6)\sim SU(4)$, it does not obey to the property $\langle\check{\bl{e}}_a(\bl{\psi})|\bl{\phi}\rangle= \langle\bl{\psi}|\check{\bl{e}}_a(\bl{\phi})\rangle$. Instead, the identity $\langle\check{\bl{e}}_a(\bl{\psi})|\bl{\phi}\rangle= -\langle\bl{\psi}|\check{\bl{e}}_a(\bl{\phi})\rangle$ holds, so that this inner product is not invariant under the group $Pin(\mathbb{R}^6)$.}. One can define the pseudo-scalar\index{Pseudo-scalar} $\bl{I}$ to be the linear operator on $S$ given by:
$$ \bl{I} \,\equiv\, 2^3(\check{\bl{e}}_1\wedge\check{\bl{\theta}}^1)(\check{\bl{e}}_2\wedge\check{\bl{\theta}}^2)(\check{\bl{e}}_3\wedge\check{\bl{\theta}}^3) \,\equiv\, (\check{\bl{e}}_1\check{\bl{\theta}}^1- \check{\bl{\theta}}^1\check{\bl{e}}_1)
(\check{\bl{e}}_2\check{\bl{\theta}}^2- \check{\bl{\theta}}^2\check{\bl{e}}_2)
(\check{\bl{e}}_3\check{\bl{\theta}}^3- \check{\bl{\theta}}^3\check{\bl{e}}_3)\,. $$
Using (\ref{Clifford6D}) along with equations (\ref{e^AB}) and (\ref{e_AB}) it is possible to prove that $\bl{I}(\bl{\chi})= \bl{\chi}$ for every spinor $\bl{\chi}\in S^+$ and $\bl{I}(\bl{\gamma})= -\bl{\gamma}$ for all $\bl{\gamma}\in S^-$. This justifies calling the spinors of $S^\pm$ the spinors of positive and negative chirality.

\subsection{Isotropic Subspaces}\label{Sec.Iso6D}\index{Isotropic}

Recall that a subspace of $N\subset\mathbb{C}\otimes\mathbb{R}^6$ is said to be isotropic when every vector $n^\mu\in N$ has zero norm, $n^\mu n_\mu= 0$. In particular, a null vector $V^\mu$,  $V^\mu V_\mu= 0$, is said to generate the 1-dimensional isotropic subspace $N_1$ defined by $N_1= \{\lambda V^\mu| \lambda\in\mathbb{C}\}$. In the same vein, a simple bivector $\bl{B}= \bl{V}_1\wedge\bl{V}_2$ is said to generate the subspace $N_2= Span\{\bl{V}_1,\bl{V}_2\}$. Moreover, this bivector $\bl{B}$ is said to be null when $N_2$ is an isotropic subspace, which means that $V_1^{\,\mu}V_{1\,\mu}= V_2^{\,\mu}V_{2\,\mu}=  V_1^{\,\mu}V_{2\,\mu}= 0$. Analogously, a simple 3-vector $\bl{T}= \bl{V}_1\wedge\bl{V}_2\wedge\bl{V}_3$ is said to generate the 3-dimensional subspace $N_3= Span\{\bl{V}_1,\bl{V}_2,\bl{V}_3\}$. Such 3-vector will then be called null\index{Null form} whenever $N_3$ is an isotropic subspace. In 6 dimensions the maximum dimension that an isotropic subspace can have is 3, because of this the 3-dimensional isotropic subspaces are called maximally
isotropic\index{Maximally isotropic} subspaces. In this subsection it will be shown that the null vectors, bivectors and 3-vectors are elegantly expressed in the spinorial language.

Let $V^{AB}= \chi^{[A}\eta^{B]}$ be the spinorial image of the vector $V^\mu$. Then by means of equation (\ref{InnerProd6D}) it is immediate to verify that $V^\mu$ is a null vector. Conversely, if  $V^\mu$ is null it is always possible to find two spinors $\chi^A$ and $\eta^A$ such that the spinorial image of such vector is $V^{AB}= \chi^{[A}\eta^{B]}$ \cite{Spin6D}. Indeed, this can be grasped from the fact that if $\bl{V}$ is null then we can arrange a null frame such that $\bl{V}= \bl{e}_1$, in which case $V^{AB}= \chi_1^{\,[A}\chi_2^{\,B]}$.

In a similar fashion, $\bl{B}$ is a null bivector\index{Null bivector} if, and only if, its spinorial image is $B^A_{\ph{A}B}= \chi^A\gamma_B$ for some spinors $\chi^A$ and $\gamma_A$ such that $\chi^A\gamma_A= 0$ \cite{Spin6D}. In this case isotropic subspace generated by $\bl{B}$ is the one spanned by the vectors $V^{AB}= \chi^{[A}\eta^{B]}$ for all $\eta^A$ such that $\eta^A\gamma_A= 0$. For instance, if $\{\bl{e}_a\}$ is a null frame then $\bl{B}= \bl{e}_1\wedge\bl{e}_2$ is a null bivector such that $B^A_{\ph{A}B}\propto \chi_1^{\,A}\gamma^4_{\,B}$, see table \ref{Tab. Biv6D}.

Finally, a 3-vector $\bl{T}$ is a null 3-vector\index{Null form} if, and only if, its spinorial image $(T^{AB},\tilde{T}_{AB})$ is either of the form $(\chi^A\chi^B,0)$ or $(0,\gamma_A\gamma_B)$. In the former case the isotropic subspace generated by $\bl{T}$ is $N_3^+=\{V^{AB}= \chi^{[A}\eta^{B]}\,|\,\eta^{A}\in S^+\}$, while on the latter case the isotropic subspace is $N_3^-=\{V_{AB}= \gamma_{[A}\zeta_{B]}\,|\,\zeta_{A}\in S^-\}$. Using equations (\ref{InnerProd6D}) and (\ref{Clifford6D}) one can easily see that if $\bl{n}\in N_3^+$ then $\check{\bl{n}}(\bl{\chi})= 0$. In the jargon introduced in appendix \ref{App._Cliff&Spinors} this means that the spinor $\bl{\chi}$ is the pure spinor\index{Pure spinor} associated with the maximally isotropic subspace $N_3^+$. Analogously, one can prove that if $\bl{m}\in N_3^-$ then $\check{\bl{m}}(\bl{\gamma})= 0$, which means that the $\bl{\gamma}$ is the pure spinor associated with the maximally isotropic subspace $N_3^-$. The results of this subsection are summarized on the table \ref{Tab. Isotropic6D}.
\begin{table}[!htbp]
\begin{center}
\begin{tabular}{lcl}
  \hline
  \hline
  Null Vector & \;$V^{AB}= \chi^{[A}\eta^{B]}$ & \;\,$Span\{\;\chi^{[A}\eta^{B]}\;\}$ \\
  Null Bivector & \;$B^A_{\ph{A}B}= \chi^A\gamma_B$,\;$\chi^A\gamma_A= 0$ & \;\,$Span\{\,\chi^{[A}\eta^{B]}\;|\;\eta^A\gamma_A= 0\,\}$ \\
  \multicolumn{2}{l}{ Null 3-vector $\left\{
  \begin{array}{l}
    \;T^{AB}= \chi^A\chi^B\,,\; \tilde{T}_{AB}=0 \\
    \;T^{AB}= 0\,, \;\tilde{T}_{AB}= \gamma_A\gamma_B\\
    \end{array}
\right.$}& $\begin{array}{l}
    Span\{\,\chi^{[A}\eta^{B]}\;|\;\eta^{A}\in S^+\,\}\\
    Span\{\,\gamma_{[A}\zeta_{B]}\;|\;\zeta_{A}\in S^-\,\}\\
    \end{array}$ \\
    \hline\hline
      \end{tabular}
      \caption{\footnotesize{On the central column we have the spinorial form of a null $p$-vector. The column on the right shows the isotropic subspaces generated by the respective null $p$-vectors.}}\label{Tab. Isotropic6D}
\end{center}
\end{table}


\section{Other Signatures}

So far we dealt only with the Euclidean space $\mathbb{R}^6$, now it is time to investigate the other signatures. In the previous chapter it was shown that in four dimensions one can grasp the distinct signatures as different reality conditions on the complexified space, see section \ref{Sec. NullFrame4}. The same thing is valid in any dimension. Particularly, in 6 dimensions if $\{\bl{e}_a\}$ is a null frame then we can have the following reality conditions according to the signature\index{Signature} \cite{Trautman}:
\small
\begin{equation}\label{Reality Cond 6D}
    \begin{cases}
    \mathbb{R}^6 \;\textrm{(Euclidean)}\rightarrow\;\;  \overline{\bl{e}_1}=\bl{\theta}^1\;,\; \overline{\bl{e}_2}=\bl{\theta}^2\;,\;\overline{\bl{e}_3}=\bl{\theta}^3 \\
    \\
    \mathbb{R}^{5,1} \;\textrm{(Lorentzian)}\rightarrow\;\;  \overline{\bl{e}_1}=\bl{e}_1\;,\; \overline{\bl{\theta}^1}=\bl{\theta}^1\;,\; \overline{\bl{e}_2}= \bl{\theta}^2 \;,\; \overline{\bl{e}_3}= \bl{\theta}^3\\
\\
    \mathbb{R}^{4,2} \rightarrow\;\;
    \begin{cases}
    \overline{\bl{e}_1}=\bl{e}_1\;,\; \overline{\bl{\theta}^1}=\bl{\theta}^1 \;,\; \overline{\bl{e}_2}=\bl{e}_2\;,\; \overline{\bl{\theta}^2}=\bl{\theta}^2 \;,\; \overline{\bl{e}_3}=\bl{\theta}^3 \\
        \overline{\bl{e}_1}=-\bl{\theta}^1 \;,\; \overline{\bl{e}_2}=\bl{\theta}^2 \;,\; \overline{\bl{e}_3}=\bl{\theta}^3
    \end{cases}\\
\\
    \mathbb{R}^{3,3} \;\textrm{(Split)}\rightarrow\;\;
    \begin{cases}
    \textrm{Real Basis}\\
    \overline{\bl{e}_1}=\bl{e}_1 \;,\; \overline{\bl{\theta}^1}=\bl{\theta}^1 \;,\; \overline{\bl{e}_2}=-\bl{\theta}^2 \;,\; \overline{\bl{e}_3}=\bl{\theta}^3\,.
    \end{cases}
        \end{cases}
\end{equation}
\normalsize

Therefore, in order to obtain results valid in any signature we just have to work on the vector space $\mathbb{C}^6$ and then choose the desired reality condition according to eq. (\ref{Reality Cond 6D}). So instead of working with the group $SPin(\mathbb{R}^6)\sim SU(4)$ we shall deal with its complexification, which is the group $SPin(\mathbb{C}^6)\sim SL(4;\mathbb{C})$\footnote{In order to see that the complexification of $SU(4)$ is $SL(4;\mathbb{C})$ remember that on the Lie algebra formalism the elements of $SU(4)$ are of the form $U= e^{i(a^j H_j)}$, where $\{H_j\}$ is a basis of Hermitian trace-less matrices and $a^j$ are real numbers. Then, complexify $SU(4)$ means allow the scalars $a^j$ to assume complex values. This implies that elements of the complexified group are of the form $S= e^{iM}$, with $M$ being the sum of a trace-less Hermitian matrix and a trace-less anti-Hermitian matrix. Thus $M$ can be any trace-less matrix, so that $S$ is a general $4\times4$ matrix with unit determinant.}. The group $SL(4;\mathbb{C})$\index{SL(4;C)} has four inequivalent irreducible representations of dimension 4:
$$ \textbf{4}:\;\; \chi^{A}\stackrel{S}{\longrightarrow} S^A_{\ph{A}B} \,\chi^{B}\;\;\;\; ; \;\;\;\;
         \widetilde{\textbf{4}}:\;\; \gamma_{A}\stackrel{S}{\longrightarrow} S^{-1\,B}_{\phantom{-1\,B}A} \,\gamma_{B}\,  $$
\begin{equation}\label{Represent. SL(4;C)}
            \overline{\textbf{4}}:\;\; \gamma_{\dot{A}} \stackrel{S}{\longrightarrow} \overline{S}_{\dot{A}}^{\phantom{A}\dot{B}}\, \gamma_{\dot{B}} \;\;\;\; ; \;\;\;\;
        \widetilde{\overline{\textbf{4}}}:\;\; \chi^{\dot{A}}\stackrel{S}{\longrightarrow} \overline{S}^{-1\phantom{B}\dot{A}}_{\phantom{-1}\dot{B}} \,\chi^{\dot{B}}\,.
\end{equation}
Where $S^A_{\ph{A}B}$ is a $4\times4$ matrix of unit determinant, $S^{-1\,A}_{\ph{-1\,A}B}$ is its inverse and $\overline{S}_{\dot{A}}^{\ph{A}\dot{B}}$ its complex conjugate. From equation (\ref{Represent. SL(4;C)}) we see that if $\chi^A$ transforms on the representation $\textbf{4}$ then its complex conjugate will be on the representation $\overline{\textbf{4}}$, so that we can write $\overline{\chi^A}= \overline{\chi}_{\dot{A}}$.
Note that if $S$ is unitary then $S^{-1}=\overline{S}^{\,t}$, which implies that in this case the transformations $\widetilde{\textbf{4}}$ and      $\overline{\textbf{4}}$ are equivalent, as well as the transformations $\textbf{4}$ and $\widetilde{\overline{\textbf{4}}}$. This is the reason of why the group $SU(4)$ has just two inequivalent irreducible representations of dimension 4.

Since $SPin(\mathbb{C}^6)\sim SL(4;\mathbb{C})$ is a double cover for the group $SO(6;\mathbb{C})$ it follows that every tensor transforming on a representation of the latter group can be seen as an object transforming on some representation of the former. Furthermore, since $\textbf{4}$ is the fundamental representation of $SL(4;\mathbb{C})$ then, as long as we do not take complex conjugates, every tensor of $SO(6;\mathbb{C})$ can be said to be on a composition of the representations $\textbf{4}$ and $\widetilde{\textbf{4}}$. Thus, almost all the results obtained for the Euclidean space $\mathbb{R}^6$ can be carried for the complex space $\mathbb{C}^6$. In particular, except for the table \ref{Tab.SU4}, all the above tables remain valid on the complex case. Note also that, since $\det(S)= 1$, equation (\ref{InnerProd6D}) is still valid.

The differences between the Euclidean case and the other signatures shows up only when the operation of complex conjugation is performed. As explained before, on the Euclidean case the complex conjugation of an object on the representation $\textbf{4}$ turns out to be on the representation $\overline{\textbf{4}}= \widetilde{\textbf{4}}$, while on  the other signatures the complex conjugate will be on the representation $\overline{\textbf{4}}\neq \widetilde{\textbf{4}}$. Thus on the Euclidean case one can easily verify whether a tensor is real using the spinorial language. For example, in this signature a vector $V^{AB}$ is real when $\overline{V^{AB}}\equiv \overline{V}_{AB}= V_{AB}$, while a bivector is real if $\overline{B}_{A}^{\ph{A}B}= B^B_{\ph{B}A}$. In the other signatures one cannot directly compare $V^{AB}$ to its complex conjugate, since the latter is on the representation $\overline{\bl{4}}$ and the equation $\overline{V}_{\dot{A}\dot{B}}= V_{AB}$ is non-sense. This kind of comparison can be done only after introducing a charge conjugation
operator\index{Charge Conjugation}, which provides a map between the representations $\overline{\bl{4}}$ and $\widetilde{\textbf{4}}$, see appendix \ref{App._Cliff&Spinors}. If $\bl{\psi}$ is a Dirac spinor then its charge conjugate is the spinor $\bl{\psi}^c$ such that $[\check{\bl{e}}_a(\bl{\psi})]^c= \check{\overline{\bl{e}}}_a(\bl{\psi}^c)$. For instance, one can use equations (\ref{e^AB}), (\ref{e_AB}), (\ref{Clifford6D}) and (\ref{Reality Cond 6D}) to prove that on the Euclidean and Lorentzian cases the charge conjugation can be respectively given by\footnote{Note that the inner product introduced on (\ref{InnerProd.Spinors6D}) is such that $\overline{(\bl{\psi},\bl{\phi})}= (\bl{\psi}^c,\bl{\phi}^c)$.}:
\small
\begin{gather}
  \textbf{\textrm{Euclidean}} \left\{
  \begin{array}{cccc}
       \bl{\chi}_1^c \,=\, \bl{\gamma}^1&   \bl{\chi}_2^c \,=\,\bl{\gamma}^2 &
\bl{\chi}_3^c \,=\,\bl{\gamma}^3 &    \bl{\chi}_4^c \,=\,\bl{\gamma}^4  \,\\
       \bl{\gamma}^{1\,c} \,=\,-\bl{\chi}_1  &   \bl{\gamma}^{2\,c} \,=\, -\bl{\chi}_2&
\bl{\gamma}^{3\,c} \,=\, -\bl{\chi}_3&     \bl{\gamma}^{4\,c} \,=\,-\bl{\chi}_4   \end{array}
\right. \label{CharConj-Euc}\\
\nonumber \\
  \textbf{\textrm{Lorentzian}} \left\{
  \begin{array}{cccc}
        \bl{\chi}_1^c \,=\, \bl{\chi}_2&  \bl{\chi}_2^c \,=\, -\bl{\chi}_1&
\bl{\chi}_3^c \,=\, -\bl{\chi}_4&     \bl{\chi}_4^c \,=\, \bl{\chi}_3 \\
\bl{\gamma}^{1\,c} \,=\,\bl{\gamma}^{2\,c}  &   \bl{\gamma}^{2\,c} \,=\,- \bl{\gamma}^{1\,c}&
\bl{\gamma}^{3\,c} \,=\, -\bl{\gamma}^{4\,c}&     \bl{\gamma}^{4\,c} \,=\, \bl{\gamma}^{3\,c}.
\end{array}
\right. \label{CharConj-Lor}
\end{gather}
\normalsize
But, as far as the $SO(6;\mathbb{C})$ tensors are concerned, one can avoid using the charge conjugation operation by making direct use of equation (\ref{Reality Cond 6D}), which sometimes is profitable.

\section{An Algebraic Classification for the Weyl Tensor}\label{Sec. Alg.Clas6D}

The intent of the present section is to use the spinorial formalism just introduced in order to define a natural algebraic classification for the Weyl tensor. The role played by the spinorial language here is to uncover relations that are hard to guess using the vectorial formalism.

As a warming example let us work out an algebraic classification for bivectors in 6 dimensions. Note that the spinorial form of a bivector, $B^A_{\ph{A}B}$, enables us to associate to each bivector $\bl{B}$ the following map on the space of Dirac spinors \cite{Spin6D}:
$$ \mathcal{B}: S\rightarrow S \;\,,\quad     \bl{\psi} = \psi^A + \tilde{\psi}_A \,\stackrel{\mathcal{B}}{\longmapsto}\, \bl{\phi} =  \underbrace{B^A_{\ph{A}B}\,\psi^B}_{\phi^A} \,+\, \underbrace{\tilde{\psi}_B\,B^B_{\ph{B}A}}_{\tilde{\phi}_A} \,.$$
It is simple matter to verify that this operator is self-adjoint with respect to the inner product defined on (\ref{InnerProd.Spinors6D}), meaning that $(\mathcal{B}(\bl{\psi}_1),\bl{\psi}_2)= (\bl{\psi}_1,\mathcal{B}(\bl{\psi}_2))$.
Note also that it preserves the spaces $S^+$ and $S^-$. Indeed, plugging $\tilde{\psi}_A= 0$ in the above equation we get $\tilde{\phi}_A= 0$. Analogously, if $\psi^A$ vanishes then $\phi^A= 0$. Hence we have $\mathcal{B}= \mathcal{B}^+\oplus \mathcal{B}^-$, where $\mathcal{B}^\pm$ are the restrictions of the operator $\mathcal{B}$ to the spaces $S^\pm$. If $\{\bl{\chi}_p\}$ is a basis for the space of Weyl spinors of positive chirality, $S^+$, then one can define its dual basis $\{\bl{\gamma}^p\}$ for the space $S^-$ as the basis such that $(\bl{\chi}_p,\bl{\gamma}^q)= \delta^{\,q}_p$. The matrix representations of the operators $\mathcal{B}^\pm$ on these bases are then easily seen to be $\mathcal{B}^+_{pq}= (\mathcal{B}(\bl{\chi}_q),\bl{\gamma}^p)$ and $\mathcal{B}^-_{pq}= (\bl{\chi}_p,\mathcal{B}(\bl{\gamma}^q))$. Thus using the fact that $\mathcal{B}$ is self-adjoint we find $\mathcal{B}^+_{pq}= \mathcal{B}^-_{qp}$.

One can use the operator $\mathcal{B}$ to algebraically classify the bivectors in six dimensions according to the Segre type of this operator, see appendix \ref{App._Segre}. But since $\mathcal{B}= \mathcal{B}^+\oplus \mathcal{B}^-$, then classify $\mathcal{B}$ is equivalent to classify $\mathcal{B}^\pm$. Furthermore, once the matrix representation of $\mathcal{B}^-$ is the transpose of the matrix representation of $\mathcal{B}^+$ it follows that the algebraic types of the operators $\mathcal{B}^+$ and $\mathcal{B}^-$  are the same. Thus we just really need to classify $\mathcal{B}^+$. As an example note that if the bivector is null, $B^A_{\ph{A}B}= \chi^A\gamma_B$ with $\chi^A\gamma_A= 0$, then one can always arrange a basis such that $\bl{\chi}_1= \bl{\chi}$ and $\bl{\gamma}_2= \bl{\gamma}$. In this basis we have
$$  \mathcal{B}^+_{pq} \,=\, \diag(\left[
                                     \begin{array}{cc}
                                       0 & 1 \\
                                       0 & 0 \\
                                     \end{array}
                                   \right], 0, 0
 )\,.$$
So that the refined Segre classification\index{Segre classification} of $\mathcal{B}^+$ is $[\,|2,1,1]$. The converse of this result is also true, leading us to the conclusion that a bivector in six dimensions is null if, and only if, its algebraic type is $[\,|2,1,1]$. Note that such algebraic classification for bivectors heavily depends on the spinors and can hardly be attained using just the vectorial formalism.
\\

Now let us try to define an algebraic classification for the Weyl tensor. According to table \ref{Tab. Spinors equiv.}, in six dimensions a tensor with the symmetries of the Weyl tensor is represented by an object of the form $\Psi^{AB}_{\ph{AB}CD}$ that is symmetric on both pairs of indices, $\Psi^{AB}_{\ph{AB}CD}= \Psi^{(AB)}_{\ph{AB}(CD)}$, and trace-less, $\Psi^{AB}_{\ph{AB}CB}= 0$. Then, since the 3-vectors are represented by a pair of symmetric tensors $(T^{AB},\tilde{T}_{AB})$, it follows that the Weyl tensor can be seen as an operator $\mathcal{C}: \Lambda^3\rightarrow \Lambda^3$,  with $\Lambda^3$ denoting the space of 3-vectors, whose action is \cite{Spin6D}:
\begin{equation}\label{C-6D}
  \left(\,T^{AB},\tilde{T}_{AB}\,\right) \,\stackrel{\mathcal{C}}{\longmapsto}\, \left(\,T'^{AB},\tilde{T}'_{AB}\,\right) = \left(\,\Psi^{AB}_{\ph{AB}CD}T^{CD},\tilde{T}_{CD}\Psi^{CD}_{\ph{CD}AB}\,\right) \,.
\end{equation}
Let us denote the space of self-dual 3-vectors,  $\tilde{T}_{AB}= 0$, by $\Lambda^{3+}$ and the space of anti-self-dual 3-vectors, $T^{AB}= 0$, by $\Lambda^{3-}$. Then it is immediate to verify the spaces $\Lambda^{3\pm}$ are preserved by the operator $\mathcal{C}$. Indeed, plugging $\tilde{T}_{AB}= 0$ on equation (\ref{C-6D}) we find that $\tilde{T}'_{AB}= 0$. Analogously, if $T^{AB}= 0$ then $T'^{AB}= 0$. So the operator $\mathcal{C}$ that acts on the 20-dimensional space $\Lambda^3$ can be seen as the direct sum of two operators acting on 10-dimensional spaces, $\mathcal{C}= \mathcal{C}^+ \oplus \mathcal{C}^-$. Where $\mathcal{C}^\pm$ are the restrictions of $\mathcal{C}$ to the spaces $\Lambda^{3\pm}$.

Thus one can classify the Weyl tensor according to the refined Segre types of the operators $\mathcal{C}^\pm$. However, let us see that the algebraic types of $\mathcal{C}^+$ and $\mathcal{C}^-$ always coincide, so that we just need to classify the operator $\mathcal{C}^+$. To this end it is useful to introduce the following basis for the space of 3-vectors:
\small
\begin{equation*}
  \begin{array}{llll}
       T_1^{\;AB}= \chi_1^{\,A}\chi_1^{\,B} & T_2^{\;AB}= \sqrt{2}\,\chi_1^{\,(A}\chi_2^{\,B)} & T_3^{\;AB}= \sqrt{2}\,\chi_1^{\,(A}\chi_3^{\,B)} & T_4^{\;AB}= \sqrt{2}\,\chi_1^{\,(A}\chi_4^{\,B)}  \\
      T_5^{\;AB}=  \chi_2^{\,A}\chi_2^{\,B} & T_6^{\;AB}= \sqrt{2}\,\chi_2^{\,(A}\chi_3^{\,B)} &  T_7^{\;AB}= \sqrt{2}\,\chi_2^{\,(A}\chi_4^{\,B)}  & T_8^{\;AB}= \chi_3^{\,A}\chi_3^{\,B}  \\
      T_9^{\;AB}=  \sqrt{2}\,\chi_3^{\,(A}\chi_4^{\,B)} & T_{10}^{\;AB}= \chi_4^{\,A}\chi_4^{\,B} & \tilde{T}^1_{\;AB}= \gamma^1_{\,A}\gamma^1_{\,B}  & \tilde{T}^2_{\;AB}= \sqrt{2} \gamma^1_{\,(A}\gamma^2_{\,B)} \\
     \tilde{T}^3_{\;AB}= \sqrt{2} \gamma^1_{\,(A}\gamma^3_{\,B)} & \tilde{T}^4_{\;AB}= \sqrt{2} \gamma^1_{\,(A}\gamma^4_{\,B)} & \tilde{T}^5_{\;AB}= \gamma^2_{\,A}\gamma^2_{\,B} & \tilde{T}^6_{\;AB}= \sqrt{2} \gamma^2_{\,(A}\gamma^3_{\,B)} \\
     \tilde{T}^7_{\;AB}= \sqrt{2} \gamma^2_{\,(A}\gamma^4_{\,B)} & \tilde{T}^8_{\;AB}=  \gamma^3_{\,A}\gamma^3_{\,B} & \tilde{T}^9_{\;AB}=\sqrt{2} \gamma^3_{\,(A}\gamma^4_{\,B)}  & \tilde{T}^{10}_{\;AB}= \gamma^4_{\,A}\gamma^4_{\,B}
   \end{array}
\end{equation*}
\normalsize
Abstractly we shall denote by $\bl{T}_r$ the self-dual 3-vector whose spinorial image is $(T_r^{\;AB},0)$ and by $\tilde{\bl{T}}^r$ the anti-self-dual 3-vector  $(0,\tilde{T}^r_{\;AB})$. Then $\{\bl{T}_r\}$ provides a basis for $\Lambda^{3+}$, while $\{\tilde{\bl{T}}^r\}$ provides a basis for $\Lambda^{3-}$. It is simple matter to verify that the following identities hold:
$$ T_r^{\;AB}\,\tilde{T}^s_{\;AB} \,=\, \delta^{\,s}_r \quad ;\quad T_r^{\;AB}\,\tilde{T}^r_{\;CD} \,=\, \delta^{\,(A}_C\delta^{\,B)}_D\,.  $$
Using the first relation above we find that the actions of the operators $\mathcal{C}^\pm$ are given by
\begin{gather*}
  \mathcal{C}^+(\bl{T}_s) \,=\,  \bl{T}_r\, \mathcal{C}^+_{rs} \quad \textrm{with} \quad  \mathcal{C}^+_{rs}\,\equiv\, \tilde{T}^r_{\;AB}\,\Psi^{AB}_{\ph{AB}CD}\,T_s^{\;CD}\, \\
  \mathcal{C}^-(\tilde{\bl{T}}^s) \,=\, \tilde{\bl{T}}^r\,\mathcal{C}^-_{rs} \quad \textrm{with} \quad  \mathcal{C}^-_{rs}\,\equiv\,  \tilde{T}^s_{\;AB}\,\Psi^{AB}_{\ph{AB}CD}\,T_r^{\;CD}\,.
\end{gather*}
Thus we have that $\mathcal{C}^+_{rs}= \mathcal{C}^-_{sr}$ and, therefore, the algebraic types of $\mathcal{C}^+$ and $\mathcal{C}^-$ are always the same. Note also that these operators are trace-less, $\mathcal{C}^+_{rr} = \tilde{T}^r_{\;AB}\,\Psi^{AB}_{\ph{AB}CD}\,T_r^{\;CD} = \Psi^{AB}_{\ph{AB}AB}= 0$. Thus the algebraic classification for the Weyl tensor proposed here amounts to compute the refined Segre type of the trace-less operator $\mathcal{C}^+$ \cite{Spin6D}. As an example let suppose that the Weyl tensor has the form $\Psi^{AB}_{\ph{AB}CD}= f^{AB}h_{CD}$ with $f^{AB}h_{CB}= 0$. Then one can choose a basis $\{\bl{\textsf{T}}_r\}$ for $\Lambda^{3+}$ such that $\textsf{T}_1^{\;AB}= f^{AB}$ and $\textsf{T}_r^{\;AB}h_{AB}= \delta^{\,2}_r$. In this basis the matrix representation of $\mathcal{C}^+$ is given by:
$$  \mathcal{C}^+_{rs}  \,=\,  \diag(\left[
                                     \begin{array}{cc}
                                       0 & 1 \\
                                       0 & 0 \\
                                     \end{array}
                                   \right], 0, 0, 0, 0, 0, 0, 0, 0  )\,.  $$
The refined Segre classification of this matrix is $[\,|2,1,1,1,1,1,1,1,1]$. Thus, in this example, we shall say that the algebraic classification of the Weyl tensor is $[\,|2,1,1,1,1,1,1,1,1]$.

A special phenomenon occurs when the signature is Euclidean. In this case equation (\ref{CharConj-Euc}) enables us to say that the 3-vectors $\bl{T}_r$ are the complex conjugates of the 3-vectors $\tilde{\bl{T}}^r$. Furthermore, if the  Weyl tensor is real then $\overline{\Psi^{AB}_{\ph{AB}CD}}= \Psi^{CD}_{\ph{CD}AB}$, so that we have:
$$ \overline{ \mathcal{C}^+_{rs}} \,\,=\,\, \overline{\tilde{T}^r_{\;AB}}\;\overline{\Psi^{AB}_{\ph{AB}CD}}\;\overline{T_s^{\;CD}} \,\,=\,\,
  T_r^{\;AB}\,\Psi^{CD}_{\ph{CD}AB}\,\tilde{T}^s_{\;CD} \,\,=\,\, \mathcal{C}^+_{sr}\,. $$
Hence, when the signature is Euclidean and the Weyl tensor is real, the matrix representation of $\mathcal{C}^+$ is Hermitian and, therefore, can be diagonalized. This is an enormous constraint for the possible algebraic types of the Weyl tensor, since one can anticipate that all Jordan blocks of $\mathcal{C}^+$ will have dimension one.

In spite of the resemblances, it is worth noting that there is one important difference between the bivector classification and the Weyl tensor classification introduced in the present section. While on the former the operator $\mathcal{B}$ acts on the space of spinors, which has no vectorial corresponding, on the latter the operator $\mathcal{C}$ acts on the space of 3-vectors, which does have a vectorial equivalent. Thus the operator $\mathcal{C}$ must admit an expression without the use of spinors. Indeed, it can be proved that this operator is proportional to the following map:
\begin{equation}\label{C 3-vec6D}
    T_{\mu\nu\alpha}\,\,\longmapsto \,\,T'_{\mu\nu\alpha}\,=\, C^{\rho\sigma}_{\phantom{\rho\sigma}[\mu\nu} \, T_{\alpha]\rho\sigma}\,.
\end{equation}
Then the operator $\mathcal{C}^+$ is proportional to the restriction of the above map to the subspace of self-dual 3-vectors, $\star\bl{T}= \bl{T}$.

As last comment it is worth mentioning that in 6 dimensions one can also classify the Weyl tensor using the fact that this tensor provides an operator on the space of bivectors, $B_{\mu\nu}\mapsto C_{\mu\nu\rho\sigma}B^{\rho\sigma}$. Actually such classification can obviously be done in any dimension, a fact that was exploited in \cite{BivectHighDim} with the aim of refining the CMPP classification. The advantage of the Weyl tensor classification using 3-vectors, introduced in this section, is that it turns out to be nicely related to some integrability properties, as will be shown in what follows.

\section{Generalized Goldberg-Sachs}\label{Sec.GS6D}\index{Goldberg-Sachs theorem}

On reference \cite{HigherGSisotropic2} it was proved a beautiful partial generalization of the Goldberg-Sachs (GS) theorem valid in manifolds of all dimensions greater than 4, as well as in any signature. The goal of the presented section is to prove that in 6 dimensions such theorem can be elegantly expressed and acquires a beautiful geometrical interpretation when the spinorial formalism is used. Moreover, it will be shown that this theorem is nicely related to the algebraic classification of the Weyl tensor introduced in the previous section. In what follows the spinorial objects will be fields over a 6-dimensional manifold $(M,\bl{g})$, so that the vector spaces treated so far are now the tangent spaces of this manifold\footnote{In order for the manifold admit a spinor bundle its topology must be constrained, see \cite{Cliff_Rigo} for example. However, since from the physical point of view we are interested on local phenomena this fact will be ignored.}.

Let be $N$ be a maximally isotropic\index{Maximally isotropic} distribution over a Ricci-flat\footnote{Actually the theorem proved in \cite{HigherGSisotropic2} is more general and remains  valid even if certain components of the Ricci tensor are different from zero. Its original version is expressed in a conformally invariant way in terms of the Cotton-York tensor. But, for simplicity, from now on we shall assume the Ricci tensor to vanish.} manifold of dimension greater than four and arbitrary signature. Then the theorem presented in \cite{HigherGSisotropic2} states that if the Weyl tensor is such that $C_{\mu\nu\rho\sigma}V_1^{\,\mu}V_2^{\,\nu}V_3^{\,\rho}= 0$ for all vector fields $\bl{V}_1$, $\bl{V}_2$ and $\bl{V}_3$ tangent to $N$ and is generic otherwise\footnote{The proof of this theorem requires that some generality conditions are satisfied by the Weyl tensor, so the imposition of ``generic otherwise'' is certainly sufficient, but it is not clear at all what is the necessary requirement. For example, in the section 3.4.2 of reference \cite{HigherGSisotropic1} some cases are shown in which the generality assumption can be relaxed. Also, at section 5.3 of \cite{GS-5D} it is said that in five dimensions there exist many cases such that the generality conditions can be neglected if the Ricci identities are used. As such, we will ignore this requirement in the present discussion.\label{Foot-Generality}} then the maximally isotropic distribution $N$ is locally integrable. Note that this theorem is a partial generalization of the GS theorem to higher dimensions \cite{Plebanski2}. In six dimensions given a maximally isotropic distribution $N$, one can always arrange a null frame $\{\bl{e}_a\}$ such that $N= Span\{\bl{e}_1,\bl{e}_2,\bl{e}_3\}$. Thus, supposing that $(M,\bl{g})$ is Ricci-flat and that the Weyl tensor obeys the generality conditions then:
\begin{equation}\label{GS-vec6D}
 C_{a'b'c'd}\,=\,0 \;\quad \Longrightarrow \;\quad  Span\{\bl{e}_1,\bl{e}_2,\bl{e}_3\} \;\; \textrm{ is Integrable.}
\end{equation}
Where in the above equation the indices $a',b',c'$ pertain to $\{1,2,3\}$, while the index $d$ runs from 1 to 6. A careful look at table \ref{Tab. WeylComp6D} reveals that the algebraic condition on the left hand side of eq. (\ref{GS-vec6D}) has the following equivalent in the spinorial language:
\begin{equation}\label{Cijka-v1}
   C_{a'b'c'd}\,=\,0\;\;\Longleftrightarrow\;\; \begin{cases}
\Psi^{AE}_{\phantom{AE}11}=0\\
\Psi^{AB}_{\phantom{AB}1D}=0
\end{cases} \;\; \forall\;\; A,B\neq1  \,.
\end{equation}
Actually, it is an immediate consequence of the identity $\Psi^{AB}_{\phantom{AB}CB}= 0$ that the first constraint on the right side of eq. (\ref{Cijka-v1}) is contained on the second constraint. Thus one can say that the condition $C_{a'b'c'd}= 0$ is tantamount to $\Psi^{AB}_{\phantom{AB}1D}=0$ for all $A,B\neq1$. This last constraint, in turn, can be reexpressed as:
$$  C_{a'b'c'd}=0 \;\Leftrightarrow\; (\,\varepsilon_{AEFG}\,\varepsilon_{BHIJ}\,\Psi^{GJ}_{\phantom{GJ}CD}\,)\,\chi_1^{\,A}\, \chi_1^{\,B} \, \chi_1^{\,C}=0\;\Leftrightarrow\; \chi_1^{\,[E}\Psi^{A][B}_{\phantom{A][B}CD}\,\chi_1^{\,F]}\chi_1^{\,C}= 0 \,. $$
But note that the spinor $\bl{\chi}_1$ is just the pure spinor associated to the maximally isotropic distribution $Span\{\bl{e}_1,\bl{e}_2,\bl{e}_3\}$, see subsection \ref{Sec.Iso6D} and appendix \ref{App._Cliff&Spinors}. Thus the theorem of reference \cite{HigherGSisotropic2} can be elegantly expressed in terms of spinors as follows:
\begin{theorem}\label{Theo. GS-spin-6D}
Let $(M,\bl{g})$ be a Ricci-flat 6-dimensional manifold whose Weyl tensor obeys the constraint $\chi^{[E}\Psi^{A][B}_{\phantom{A][B}CD}\,\chi^{F]}\chi^{C}= 0$ for some spinor $\bl{\chi}\in S^+$ and is generic otherwise (see \cite{HigherGSisotropic2}). Then the maximally isotropic distribution associated to the pure spinor $\bl{\chi}$ is integrable.
\end{theorem}
For completeness, let us remark that by means of table \ref{Tab. WeylComp6D} one can also prove that the following equivalences hold:
$$ \begin{array}{lll}
     C_{a'b'cd}\,=\,0\;\;\Leftrightarrow & (\,\varepsilon_{AEFG}\,\Psi^{GB}_{\phantom{GB}CD}\,)\chi_1^{\,A}\, \chi_1^{\,C} =0 &\Leftrightarrow\;\; \chi_1^{\,[E}\Psi^{A]B}_{\phantom{A]B}CD}\chi_1^{\,C}= 0 \\
     C_{a'bcd}\,=\,0\;\;\Leftrightarrow & (\,\varepsilon_{AEFG}\,\Psi^{GB}_{\phantom{GB}CD}\,)\chi_1^{\,A}\, =0   &\Leftrightarrow\;\; \chi_1^{\,[E}\Psi^{A]B}_{\phantom{A]B}CD}= 0\,.\\
   \end{array} $$

In the previous paragraph we oriented the null frame in such a way that the maximally isotropic distribution was spanned by $\{\bl{e}_1,\bl{e}_2,\bl{e}_3\}$. This is a self-dual distribution, meaning that the 3-vector $\bl{T}= \bl{e}_1\wedge\bl{e}_2\wedge\bl{e}_3$ is self-dual. But we could also have assumed that the distribution was generated by $\{\bl{\theta}^1,\bl{\theta}^2,\bl{\theta}^3\}$, which is an anti-self-dual distribution. In such a case the associated pure spinor is $\bl{\gamma}^1$, which has negative chirality. In this circumstance the integrability condition of theorem \ref{Theo. GS-spin-6D} might be replaced by $\gamma^1_{[E}\Psi^{AB}_{\phantom{AB}C][D}\gamma^1_{F]}\gamma^1_{A}= 0$.
\\

Now let us see that theorem \ref{Theo. GS-spin-6D} can be expressed in terms of the map $\mathcal{C}$ defined in section \ref{Sec. Alg.Clas6D}. Indeed, using equations (\ref{C-6D}) and (\ref{Cijka-v1}) we immediately find:
$$ C_{a'b'c'd}=0 \;\Rightarrow\; \Psi^{AB}_{\phantom{AB}11}= 0 \;\textrm{ if}\;A\neq1 \;\Rightarrow\; \Psi^{AB}_{\phantom{AB}11}\propto \chi_1^{\,A}\chi_1^{\,B} \;\Rightarrow\; \mathcal{C}^+(\bl{T}_1)\propto \bl{T}_1\,.$$
Where in the above equation the 3-vector $\bl{T}_1$ is the one whose spinorial equivalent is $(\chi_1^{\,A}\chi_1^{\,B},0)$. In the vectorial language this 3-vector is proportional to $\bl{e}_1\wedge\bl{e}_2\wedge\bl{e}_3$. Thus we proved that if the integrability condition for a maximally isotropic distribution is satisfied then the null 3-vector that generates it is an eigen-3-vector of the operator $\mathcal{C}^+$. This is a partial generalization of the corollary \ref{Cor.EigenPlane4D} of chapter \ref{Chap. Gen4D}. Furthermore, using the above results we have:
$$ C_{a'b'c'd}=0 \;\Leftrightarrow\; \Psi^{AB}_{\phantom{AB}1C}= 0 \;\textrm{ if}\;A,B\neq1 \;\Leftrightarrow\; \Psi^{AB}_{\phantom{AB}CD}\chi_1^{\,C}\chi_p^{\,D} = \chi_1^{\,(A}\eta_p^{\,B)}\,. $$
Where $\{\eta_p^{\,B}\}$ is some set of four spinors. The above equation means that if the integrability condition $C_{a'b'c'd}=0$ is satisfied then the subspace formed by the 3-vectors of the form $(\chi_1^{\,(A}\eta^{B)},0)$ for all $\bl{\eta}\in S^+$ is invariant by the action of $\mathcal{C}^+$. Using the 3-vector basis introduced in section \ref{Sec. Alg.Clas6D} this is the subspace spanned by\footnote{On the vectorial formalism the referred subspace is the one spanned by the 3-vectors $\bl{e}_1\wedge\bl{e}_2\wedge\bl{e}_3$, $\bl{e}_1\wedge(\bl{e}_2\wedge\bl{\theta}^2+\bl{e}_3\wedge\bl{\theta}^3)$, $\bl{e}_2\wedge(\bl{e}_1\wedge\bl{\theta}^1+\bl{e}_3\wedge\bl{\theta}^3)$ and $\bl{e}_3\wedge(\bl{e}_1\wedge\bl{\theta}^1+\bl{e}_2\wedge\bl{\theta}^2)$.} $\{\bl{T}_1,\bl{T}_2,\bl{T}_3,\bl{T}_4\}$. The results of this paragraph enables us to rephrase theorem \ref{Theo. GS-spin-6D} as follows:
\begin{theorem}\label{Theo. GS-AlgClas-6D}\index{Pure spinor}\index{Maximally isotropic}
Let $(M,\bl{g})$ be a Ricci-flat 6-dimensional manifold whose Weyl operator $\mathcal{C}^+$ keeps invariant the subspace spanned by the 3-vectors of the form $T^{AB}= \chi^{(A}\eta^{B)}$ for all $\eta^{A}\in S^+$, with $\mathcal{C}^+$ being generic otherwise. Then the maximally isotropic distribution associated to the pure spinor $\bl{\chi}$ is integrable and the 3-vector $T^{AB}= \chi^{A}\chi^{B}$ is an eigen-3-vector of $\mathcal{C}^+$.
\end{theorem}


\subsection{Lorentzian Signature}

Now let us assume that $(M,\bl{g})$ is a manifold whose metric $\bl{g}$ is real and has Lorentzian signature. If the Weyl tensor satisfies the integrability condition $C_{a'b'c'd}= 0$ then, by the previous results, we know that $\mathcal{C}^+(\bl{T}_1)\propto \bl{T}_1$. Furthermore, the subspace $\mathcal{A}\equiv Span\{\bl{T}_1,\bl{T}_2,\bl{T}_3,\bl{T}_4\}$ is invariant under $\mathcal{C}^+$, where
$$  T_1^{\,AB}= \chi_1^{\,A}\chi_1^{\,B}\quad  \textrm{and} \quad \mathcal{A} = \{\,T^{AB}= \chi_1^{\,(A}\eta^{B)}\,|\; \eta^A\in\,S^+\,\}\,. $$
Since the metric is assumed to be real it follows that the Weyl tensor is also real, as well as the operator $\mathcal{C}^+$. Thus the complex conjugate of these constraints are likewise valid, leading us to the conclusion that $\mathcal{C}^+(\overline{\bl{T}_1})\propto \overline{\bl{T}_1}$ and that the subspace $\overline{\mathcal{A}}$ is also invariant by the action of $\mathcal{C}^+$. By means of equation (\ref{CharConj-Lor}) we have that
$$  \overline{T_1^{\,AB}} = T_5^{\,AB}= \chi_2^{\,A}\chi_2^{\,B}\quad  \textrm{and} \quad \overline{\mathcal{A}} = \{\,T^{AB}= \chi_2^{\,(A}\eta^{B)}\,|\; \eta^A\in\,S^+\,\}\,. $$
Note that since the subspaces $\mathcal{A}$ and $\overline{\mathcal{A}}$ are invariant under $\mathcal{C}^+$ so will be $\mathcal{A}\cap\overline{\mathcal{A}}= Span\{T^{AB}= \chi_1^{\,(A}\chi_2^{\,B)}\}$. From which we conclude that the 3-vector $\bl{T}_2$ is an eigen-3-vector of the operator $\mathcal{C}^+$. These results along with theorem \ref{Theo. GS-AlgClas-6D} lead us to the following corollary \cite{Spin6D}:
\begin{corollary}
Let $(M,\bl{g})$ be a Ricci-flat Lorentzian manifold, then the integrability conditions for the maximally isotropic distribution generated by $\{\bl{e}_1,\bl{e}_2,\bl{e}_3\}$ are:\\
(1) The 3-vectors $\bl{T}_1$, $\bl{T}_2$ and $\bl{T}_5$ are eigen-3-vectors of the Weyl operator $\mathcal{C}^+$\\
(2) The subspaces $\mathcal{A}= Span\{\bl{T}_1,\bl{T}_2,\bl{T}_3,\bl{T}_4\}$ and $\overline{\mathcal{A}} =Span\{\bl{T}_2,\bl{T}_5,\bl{T}_6,\bl{T}_7\}$ are invariant by the action of $\mathcal{C}^+$.
\end{corollary}

If the metric is real then whenever a distribution is integrable the complex conjugate of this distribution will also be integrable, that is the geometrical origin of the above corollary. Using eq. (\ref{Reality Cond 6D}) we conclude that the complex conjugate of the distribution $Span\{\bl{e}_1,\bl{e}_2,\bl{e}_3\}$ is the distribution spanned by $\{\bl{e}_1,\bl{\theta}^2,\bl{\theta}^3\}$. The pure spinor associated to the latter maximally isotropic distribution is $\bl{\chi}_2$. Note that the intersection of the distributions $Span\{\bl{e}_1,\bl{e}_2,\bl{e}_3\}$ and $Span\{\bl{e}_1,\bl{\theta}^2,\bl{\theta}^3\}$ is the 1-dimensional distribution tangent to the real and null vector field $\bl{e}_1$. Since the leafs of an integrable maximally isotropic distribution are totally geodesic \cite{Mason_KY}, it follows that if these two distributions are integrable then the vector field $\bl{e}_1$ is geodesic. But, differently from the 4-dimensional case, the congruence generated by $\bl{e}_1$ generally is not shear-free. Finally, it is easy to verify that if $C_{a'b'c'd}= 0$ then the vector field $\bl{e}_1$ turns out to be a multiple Weyl aligned null direction, meaning that the components $C_{1\alpha1\beta}$, $C_{1\alpha\beta\kappa}$ and $C_{141\alpha}$ vanish for all $\alpha, \beta, \kappa\neq 1,4$.

\section{Example, Schwarzschild in 6 Dimensions}

In this section it will be used the spinorial formalism in order to analyze the 6-dimensional Schwarzschild space-time, the unique spherically symmetric vacuum solution in 6 dimensions. In a suitable coordinate system the metric of this manifold is given by:
$$ \textrm{ds}^2=-h^2\textrm{dt}^2 + h^{-2}\textrm{dr}^2 + r^2\left\{\textrm{d}\phi_1^2 + \sin^2\phi_1 \left[\textrm{d}\phi_2^2 + \sin^2\phi_2\, ( \textrm{d}\phi_3^2 + \sin^2\phi_3\,\textrm{d}\phi_4^2 ) \right]\right\}\,,$$
where $h^2= (1-\alpha \,r^{-3})$. The Schwarzschild metric in higher dimensions is sometimes also called the Tangherlini metric \cite{Tangherlini}.  A convenient null frame on this space-time is defined by:
\begin{gather*}
  \bl{e}_1=\frac{1}{2}\left(h\partial_r + h^{-1}\partial_t\right) \;\;;\;\; \bl{e}_2=\frac{1}{2}\left(\frac{1}{r}\partial_{\phi_1} +\frac{i}{r\sin\phi_1}\partial_{\phi_2}\right) \\
   \bl{e}_3=\frac{1}{2}\left(\frac{1}{r\sin\phi_1\sin\phi_2}\partial_{\phi_3} +\frac{i}{r\sin\phi_1\sin\phi_2\sin\phi_3}\partial_{\phi_4}\right)  \;\;;\nonumber \\
  \bl{e}_4=\frac{1}{2}\left(h\partial_r - h^{-1}\partial_t\right) \;\;;\;\; \bl{e}_5=\frac{1}{2}\left(\frac{1}{r}\partial_{\phi_1} -\frac{i}{r\sin\phi_1}\partial_{\phi_2}\right) \\ \bl{e}_6=\frac{1}{2}\left(\frac{1}{r\sin\phi_1\sin\phi_2}\partial_{\phi_3} -\frac{i}{r\sin\phi_1\sin\phi_2\sin\phi_3}\partial_{\phi_4}\right)\,.
\end{gather*}
Since this space-time is a vacuum solution its Ricci tensor vanishes, so that the Riemann tensor is equal to the Weyl tensor. Up to the trivial symmetries, $C_{abcd}= C_{[ab][cd]}= C_{cdab}$, the non-vanishing components of the Weyl tensor are:
\begin{gather*}
  C_{1414}=-\frac{3\alpha}{2r^5}\;;\;\; C_{1245}=C_{1346}=C_{1542}=C_{1643}=-\frac{3\alpha}{8r^5} \;; \\
  C_{2356}=C_{2552}=C_{2653}=C_{3636}=\frac{\alpha}{4r^5}\,.
\end{gather*}
This reveals that such tensor is of type $D$ on the CMPP classification\index{CMPP classification}, with $\bl{e}_1$ and $\bl{e}_4$ being multiple WANDs \cite{CMPP}. One can then use table \ref{Tab. WeylComp6D} to prove that the spinorial equivalent of this Weyl tensor is:
\begin{gather}
 \nonumber \Psi^{AB}_{\phantom{AB}CD} \,=\, -\frac{\alpha}{8r^5}[\chi_1^{\,A}\chi_1^{\,B}\gamma^1_{\,C}\gamma^1_{\,D} +\chi_2^{\,A}\chi_2^{\,B}\gamma^2_{\,C}\gamma^2_{\,D}
  + \chi_3^{\,A}\chi_3^{\,B}\gamma^3_{\,C}\gamma^3_{\,D}+
  \chi_4^{\,A}\chi_4^{\,B}\gamma^4_{\,C}\gamma^4_{\,D} ]\;+ \\
  - 2\frac{\alpha}{8r^5}[ \chi_1^{\,(A}\chi_2^{\,B)}\gamma^1_{\,(C}\gamma^2_{\,D)} +
  \chi_3^{\,(A}\chi_4^{\,B)}\gamma^3_{\,(C}\gamma^4_{\,D)}   ] \;\label{Psi-Schwarz6D}+ \\ \nonumber
  +3\frac{\alpha}{8r^5}[ \chi_1^{\,(A}\chi_3^{\,B)}\gamma^1_{\,(C}\gamma^3_{\,D)} +
  \chi_1^{\,(A}\chi_4^{\,B)}\gamma^1_{\,(C}\gamma^4_{\,D)} +
  \chi_2^{\,(A}\chi_3^{\,B)}\gamma^2_{\,(C}\gamma^3_{\,D)} +
  \chi_2^{\,(A}\chi_4^{\,B)}\gamma^2_{\,(C}\gamma^4_{\,D)} ]\,.
\end{gather}
It is then immediate to verify that the matrix representation of the operator $\mathcal{C}^+$ on the basis $\{\bl{T}_r\}$, defined in section \ref{Sec. Alg.Clas6D}, is given by:
$$ \mathcal{C}^+_{rs}\,=\, -\frac{\alpha}{16r^5}\, \textrm{diag}(2,2,-3,-3,2,-3,-3,2,2,2)\,.$$
Leading us to the conclusion that the algebraic type of the Weyl tensor of the 6-dimensional Schwarzschild space-time\index{Segre classification} is $[(1,1,1,1,1,1),(1,1,1,1)|\,]$.

Using the expressions for the null frame $\{\bl{e}_a\}$ defined above, it is straightforward to compute the following Lie brackets:
\begin{gather*}
  [\bl{e}_1,\bl{e}_2]=-\frac{h}{2r}\bl{e}_2 \;\;;\;\; [\bl{e}_1,\bl{e}_3]=-\frac{h}{2r}\bl{e}_3 \;\;;\;\; [\bl{e}_1,\bl{e}_4]=\frac{3\alpha}{4r^4}h^{-1}(\bl{e}_1-\bl{e}_4) \;; \\
  [\bl{e}_2,\bl{e}_3]=-\frac{1}{2r}(\cot\phi_1+i\frac{\cot\phi_2}{\sin\phi_1})\bl{e}_3 \;\;;\;\; [\bl{e}_2,\bl{e}_4]=\frac{h}{2r}\bl{e}_2 \;; \\ [\bl{e}_2,\bl{e}_5]=\frac{\cot\phi_1}{2r}(\bl{e}_2-\bl{e}_5)  \;\;;\;\; [\bl{e}_2,\bl{e}_6]=-\frac{1}{2r}(\cot\phi_1+i\frac{\cot\phi_2}{\sin\phi_1})\bl{e}_6 \;;\\ [\bl{e}_3,\bl{e}_4]=\frac{h}{2r}\bl{e}_3 \;\;;\;\; [\bl{e}_3,\bl{e}_6]=\frac{\cot\phi_3}{2r\sin\phi_1\sin\phi_2}(\bl{e}_3-\bl{e}_6)\,.
\end{gather*}
The missing commutators can be obtained by taking the complex conjugate of these relations and using eq. (\ref{Reality Cond 6D}). From these commutation relations one conclude that the distributions spanned by $\{\bl{e}_1,\bl{e}_2,\bl{e}_3\}$, $\{\bl{e}_1,\bl{e}_5,\bl{e}_6\}$, $\{\bl{e}_4,\bl{e}_2,\bl{e}_6\}$, $\{\bl{e}_4,\bl{e}_5,\bl{e}_3\}$, $\{\bl{e}_4,\bl{e}_5,\bl{e}_6\}$, $\{\bl{e}_4,\bl{e}_2,\bl{e}_3\}$,  $\{\bl{e}_1,\bl{e}_5,\bl{e}_3\}$ and $\{\bl{e}_1,\bl{e}_2,\bl{e}_6\}$ are all integrable. Since the pure spinors\index{Pure spinor} associated to these maximally isotropic distributions are respectively $\bl{\chi}_1$, $\bl{\chi}_2$, $\bl{\chi}_3$, $\bl{\chi}_4$, $\bl{\gamma}_1$, $\bl{\gamma}_2$, $\bl{\gamma}_3$ and $\bl{\gamma}_4$, it is natural to wonder whether such spinors obey the algebraic condition of theorem \ref{Theo. GS-spin-6D}. Using eq. (\ref{Psi-Schwarz6D}) it is simple matter to verify that the integrability constraints
$$  \chi_p^{\,[E}\Psi^{A][B}_{\phantom{A][B}CD}\,\chi_p^{\,F]}\chi_p^{\,C}= 0\quad \textrm{and}  \quad\gamma^p_{[E}\Psi^{AB}_{\phantom{AB}C][D}\gamma^p_{F]}\gamma^p_{A}= 0$$
are, indeed, valid for all $p\in\{1,2,3,4\}$. In addition to these eight distributions, there exist infinitely many independent maximally isotropic integrable distributions on this manifold\footnote{The author thanks Marcello Ortaggio for pointing out this fact. Comments in the same lines can also be found in section 8.3 of \cite{GS-HighD}, where it was argued that Robinson-Trautman space-times with transverse spaces of constant curvature admit infinitely many isotropic structures. See also the footnote in the section 5.2 of reference \cite{GS-5D}.}. Since the 4-sphere is conformally flat, it follows that one can manage to find a coordinate system in which the metric of this space-time takes the form
$$ \textrm{ds}^2=-h^2\,\textrm{dt}^2 \,+\, h^{-2}\,\textrm{dr}^2 \,+\, r^2g(y_p)\,\left[dy_1^2+ dy_2^2+ dy_3^2 + dy_4^2\right] \,.$$
Defining $\bl{k}_1=a^p\partial_{y_p}$ and $\bl{k}_2=b^p\partial_{y_p}$ with $a^p$ and $b^p$ being complex constants such that $\delta_{pq}a^pa^q= \delta_{pq}b^pb^q= \delta_{pq}a^pb^q= 0$, then it is immediate to verify that the maximally isotropic distributions $\{\bl{e}_1,\bl{k}_1,\bl{k}_2\}$ and $\{\bl{e}_4,\bl{k}_1,\bl{k}_2\}$ are integrable for all $a^p,b^p$ \cite{Spin6D}. As a final comment it is worth remarking that there exist some pure spinors that obey the integrability condition while the associated maximally isotropic distributions are not integrable, which is possible because the Weyl tensor of the Schwarzschild space-time does not satisfy the generality condition assumed on ref. \cite{HigherGSisotropic2}. For instance, although the pure spinor $\bl{\eta}= \bl{\chi}_1+f\bl{\chi}_2$ obeys the constraint $\eta^{[E}\Psi^{A][B}_{\phantom{A][B}CD}\,\eta^{F]}\eta^{C}= 0$ for all functions $f$, its associated distribution, $Span\{\bl{e}_1, (\bl{e}_2+f\bl{e}_6), (\bl{e}_3-f\bl{e}_5)\}$, is not integrable if $f\neq 0$.

\chapter{Integrability and Weyl Tensor Classification in All Dimensions}\label{Chap. AllD}

Throughout this thesis it has been repeatedly advocated that, since the Petrov classification and the Goldberg-Sachs (GS) theorem have played a prominent role in the development of general relativity in 4 dimensions, it is worth looking for higher-dimensional generalizations of these results. Hopefully this could be helpful in the search of new exact solutions to Einstein's equation in higher dimensions, as it proved to be in 4 dimensions \cite{Kerr,typeD}. It is also worth mentioning that recently it was made a connection between Navier-Stokes' and Einstein's equations \cite{Navier-Stokes} in which the algebraic classification of the Weyl tensor plays an important role, which gives a further motivation for a investigation on these subjects.

In the previous chapter it was taken advantage of the spinorial language in order to define an algebraic classification for the Weyl tensor. Such classification proved to be valuable because it is connected to a generalization of the GS theorem in 6 dimensions. Given the success of the spinorial formalism in 4 and 6 dimensions it seems reasonable trying to use this language in higher-dimensional spaces. However, it is hard to deal with spinors in arbitrary dimensions since some important details can heavily depend on the specific dimension. Moreover, in dimensions greater than 6 not all Weyl spinors are pure, which represents a further drawback. In spite of these difficulties this path was adopted in \cite{TC-PureSpin}.

The aim of the present chapter is to define an algebraic classification for the Weyl tensor valid in arbitrary dimension and associate such classification with integrability properties using the vectorial formalism. Here the Weyl tensor will be used to define operators acting on the bundle of differential forms, so that the refined Segre classification of these operators provides an algebraic classification for the Weyl tensor. In this approach the Petrov classification and the spinorial classification defined in chapter \ref{Chap._Spin6D} emerge as special cases. The material presented here is based in the article \cite{art4}.

As in the previous chapters it will be assumed that the manifold is complexified, so that the results can be carried to any signature by a suitable choice of reality condition. For simplicity the metric is supposed to be real, so that the Weyl tensor is real. All calculations here are local, therefore global issues shall be neglected.

\section{Algebraic Classification for the Weyl Tensor}

In what follows the reader is assumed to be familiar with the formalism of differential forms, for a quick review see section \ref{Sec. DiffForms} of chapter \ref{Chap. GR}. Let $(M,\bl{g})$ be an $n$-dimensional manifold of signature $s$. Since we are interested on local results we can always assume that such manifold is endowed with a volume-form\index{Volume-form} $\epsilon_{\mu_1\ldots\mu_n}$. By means of this tensor one can define the Hodge dual of a $p$-form as in equation (\ref{HodgeDual-Def}). For clearness on the notation we shall abstractly denote the Hodge dual map by $\mathcal{H}_p$\index{Hodge dual}\label{HodgeOperator}:
$$ \left\{
  \begin{array}{ll}
    \mathcal{H}_p: \Gamma(\wedge^pM)\rightarrow \Gamma(\wedge^pM)   \\
   \bl{F} \;\mapsto\; \mathcal{H}_p(\bl{F}) \,=\, \star\bl{F}\,.\\
    \end{array} \right. $$
Where $\Gamma(\wedge^pM)$ is the space of $p$-forms\footnote{Actually this operator is defined just locally. So that, formally, its domain should be written as $\Gamma(\wedge^pM)|_{N_x}$, where $N_x\subset M$ is the neighborhood of some point $x\in M$.}. Denote the identity operator on $\Gamma(\wedge^pM)$ by $\bl{1}_p$. Then using the complete skew-symmetry of the volume-form along with equation (\ref{EE_DELTA}) it is immediate to see that the following identity holds:
\begin{equation}\label{HpHn-p}
   \mathcal{H}_{n-p}\, \mathcal{H}_p \,=\, (-1)^{[(n-p)p+\frac{n-s}{2}]}\,\,\bl{1}_p  \,.
\end{equation}

The Weyl tensor\index{Weyl tensor} $C_{\mu\nu\rho\sigma}$ is the trace-less part of the Riemann tensor and, therefore, has the following symmetries:
\begin{equation*}
  C_{\mu\nu\rho\sigma}=C_{[\mu\nu][\rho\sigma]}= C_{\rho\sigma\mu\nu}\;\;;\;\; C_{\mu[\nu\rho\sigma]}=0\;\;;\;\; C^\mu_{\phantom{\mu}\nu\mu\sigma}=0\,.
\end{equation*}
Inspired by equation (\ref{C 3-vec6D}) one can use this tensor to introduce an operator $\mathcal{C}_p$ acting on the bundle of $p$-forms, with $p\geq2$, whose definition is \cite{art4}:
\begin{equation}\label{Cp-All}
\left\{
  \begin{array}{ll}
    \mathcal{C}_p: \Gamma(\wedge^pM)\rightarrow \Gamma(\wedge^pM)   \\
   \bl{F} \;\mapsto\; \mathcal{C}_p(\bl{F}) \,=\, \frac{1}{p!} \,\lef C^{\rho\sigma}_{\phantom{\rho\sigma}\nu_1\nu_2}F_{\nu_3\ldots\nu_{p} \,\rho\sigma}\rig\,dx^{\nu_1}\wedge dx^{\nu_2}\wedge\ldots\wedge dx^{\nu_p}\,.
    \end{array} \right.
\end{equation}
Note that for $p=2$ this operator reduces to the well-known bivector operator, $B_{\mu\nu}\mapsto C_{\mu\nu\rho\sigma}B^{\rho\sigma}$, whose properties in arbitrary dimension were explored in \cite{BivectHighDim}. Furthermore, in 6 dimensions when $p=3$ such operator is proportional to the Weyl operator defined in the previous chapter using spinors, see eq. (\ref{C 3-vec6D}). Now let us prove that $\mathcal{C}_p$ commutes with the Hodge dual map.
\begin{align*}
  \left[ \mathcal{H}_p\right.&\left.\mathcal{C}_p(F) \right]^{\nu_1\ldots\nu_{n-p}}\; =\; \frac{1}{p!}\, \epsilon^{\mu_1\ldots\mu_{p} \, \nu_1\ldots\nu_{n-p}} C^{\alpha\beta}_{\phantom{\alpha\beta}\mu_1\mu_2}\, F_{\mu_3\ldots\mu_{p} \alpha\beta}\,  \\
  &=\,\frac{1}{p!}\, \epsilon^{\mu_1\ldots\mu_{p} \, \nu_1\ldots\nu_{n-p}} C^{\alpha\beta}_{\phantom{\alpha\beta}\mu_1\mu_2}\, \left[ \mathcal{H}_{n-p}\mathcal{H}_p(F)\right]_{\mu_3\ldots\mu_{p} \alpha\beta}\,(-1)^{[(n-p)p+ \frac{n-s}{2}]}  \\
  &=\,  \frac{(-1)^{[(n-p)p+ \frac{n-s}{2}]}}{p!\,(n-p)!}\, \epsilon^{\mu_1\ldots\mu_{p} \, \nu_1\ldots\nu_{n-p}} C^{\alpha\beta}_{\phantom{\alpha\beta}\mu_1\mu_2}\,  \,\epsilon_{ \sigma_1 \ldots \sigma_{n-p} \mu_3\ldots\mu_{p} \alpha\beta}
   \left[ \mathcal{H}_p(F)\right]^{\sigma_1 \ldots \sigma_{n-p}} \\
  &=\, \frac{(p-2)!\,(n-p+2)!}{p!\,(n-p)!} \, \delta_{\alpha}^{\;[\mu_1}\delta_\beta^{\;\mu_2}\delta_{\sigma_1}^{\;\nu_1}\ldots\delta_{\sigma_{n-p}}^{\;\nu_{n-p}]} \, C^{\alpha\beta}_{\phantom{\alpha\beta}\mu_1\mu_2} \left[ \mathcal{H}_p(F)\right]^{\sigma_1 \ldots \sigma_{n-p}}\\
  &=\,  C^{\phantom{\mu_1\mu_2}[\nu_1\nu_2}_{\mu_1\mu_2} \,\left[ \mathcal{H}_p(F)\right]^{\nu_3 \ldots \nu_{n-p}]\mu_1\mu_2} \;=\;
    \left[ \mathcal{C}_{n-p}\,\mathcal{H}_p(F) \right]^{\nu_1\ldots\nu_{n-p}} \,.
 \end{align*}
Where equations (\ref{EE_DELTA}) and (\ref{HpHn-p}) were used. This proves that the following important relation holds:
\begin{equation}\label{HC=CH -All}
    \mathcal{H}_{p}\,\mathcal{C}_p \;=\; \mathcal{C}_{n-p}\,\mathcal{H}_p  \,.
\end{equation}
In particular, since the operator $\mathcal{H}_p$ is invertible, see eq. (\ref{HpHn-p}), the above relation implies that $\mathcal{C}_{n-p}= \mathcal{H}_{p}\mathcal{C}_p\mathcal{H}_{p}^{-1}$. So the operators $\mathcal{C}_{n-p}$ and $\mathcal{C}_{p}$ are connected by a similarity transformation. Recall that on equation (\ref{Cp-All}) the operator $\mathcal{C}_{p}$  was not defined for $p=0$ and $p=1$. However, we can use equation (\ref{HC=CH -All}) in order to define these operators in terms of $\mathcal{C}_{n}$ and $\mathcal{C}_{n-1}$. For instance,
\begin{align*}
 \left[\, \mathcal{C}_1(F)\,\right]^\mu =&  \left[\, \mathcal{H}_1^{-1}\,\mathcal{C}_{n-1}\,\mathcal{H}_1(F)\,\right]^\mu \propto \,\left[ \, \mathcal{H}_{n-1}\,\mathcal{C}_{n-1}\,\mathcal{H}_1(F)\,\right]^\mu  \\
 \propto& \;\,\epsilon^{\nu_1\ldots\nu_{n-1}\mu}\,C^{\alpha\beta}_{\ph{\alpha\beta}\nu_1\nu_2}\,\epsilon_{\sigma\nu_3\ldots\nu_{n-1}\alpha\beta}\,F^\sigma \\
 \propto& \;\, \delta_{\alpha}^{\;[\nu_1}\delta_\beta^{\;\nu_2}\delta_{\sigma}^{\;\mu]} \, C^{\alpha\beta}_{\ph{\alpha\beta}\nu_1\nu_2}\, F^\sigma
     = C^{[\nu_1\nu_2}_{\ph{[\nu_1\nu_2}\nu_1\nu_2} \,F^{\mu]} = 0\,.
\end{align*}
Where equation (\ref{EE_DELTA}) and the trace-less property of the Weyl tensor were used. In the same fashion one can prove that the operator $\mathcal{C}_0$ is identically zero. Therefore, using these results along with eq. (\ref{HC=CH -All}), we conclude that in a manifold of dimension $n$ we have:
\begin{equation}\label{C=0-All}
  \mathcal{C}_0 \equiv 0 \quad;\quad \mathcal{C}_1 \equiv 0 \quad;\quad \mathcal{C}_{n-1} = 0 \quad;\quad \mathcal{C}_n = 0 \,.
\end{equation}

The refined Segre types of the operators $\mathcal{C}_p$, for all possible values of $p$, provide an algebraic classification for the Weyl tensor. But, because of equation (\ref{C=0-All}), we do not need to worry about the cases $p=0$, $p=1$, $p=n-1$ and $p=n$. Moreover, since $\mathcal{C}_p$ and $\mathcal{C}_{n-p}$ are connected by a similarity transformation they have the same algebraic type according to the refined Segre classification. Therefore, we just need to consider the values of $p$ between 2 and $n/2$. \emph{So the algebraic classification for the Weyl tensor established here amounts to gathering the refined Segre types of the operators $\mathcal{C}_p$ for the integer values of $p$ contained on the interval $2\leq p \leq n/2$} \cite{art4}.

\subsection{Inner Product of $p$-forms}

It will prove to be valuable introducing the following symmetric inner product on the space of $p$-forms:
\begin{equation}\label{InnerProd-All}
   \langle\bl{F}, \bl{K}\rangle \,\equiv\, F^{\nu_1\nu_2\ldots\nu_p}\,K_{\nu_1\nu_2\ldots\nu_p} \,.
\end{equation}
Where in the above equation $\bl{F}$ and $\bl{K}$ are $p$-forms. Since the metric $\bl{g}$ is non-degenerate it follows that the inner product $\langle\,,\rangle$ is also non-degenerate. Moreover, using the Weyl tensor symmetry $C_{\mu\nu\rho\sigma}= C_{\rho\sigma\mu\nu}$ it is trivial verifying that the operator $\mathcal{C}_p$ is self-adjoint with respect to such inner product:
$$  \langle\bl{F}, \mathcal{C}_p(\bl{K})\rangle \,=\,  \langle\mathcal{C}_p(\bl{F}), \bl{K}\rangle \,.$$

Now let $\{\bl{F}_r\}$ be some basis for the space of $p$-forms\footnote{Actually, because of topological obstructions, generally we can define such basis just locally. Therefore, we have $\bl{F}_r\in \Gamma(\wedge^pM)|_{N_x}$. Where, formally, $\Gamma(\wedge^p M)|_{N_x}$ is the restriction of the space of sections of the $p$-form bundle to the neighborhood $N_x$ of some point $x\in M$. Roughly speaking, $\Gamma(\wedge^p M)|_{N_x}$ is the space spanned by the $p$-form fields in the neighborhood $N_x$.}\label{P-FormSections2}, with\footnote{The indices $r,s,\ldots$ run from 1 to $\frac{n!}{p!(n-p)!}$.} $\langle\bl{F}_r, \bl{F}_s\rangle= f_{rs}$. Since this inner product is non-degenerate it follows that the matrix $f_{rs}$ is invertible, let us denote its inverse by $f^{rs}$. Thus defining the $p$-forms $\bl{F}^r\equiv f^{rs}\bl{F}_s$ we find that $\langle\bl{F}_r, \bl{F}^s\rangle= \delta_r^{\,s}$. So if $\bl{F}$ is some $p$-form then its expansion on the basis $\{\bl{F}_r\}$ is given by $\bl{F}= \langle\bl{F}^r, \bl{F}\rangle\,\bl{F}_r$. Using index notation, the latter equation is  tantamount to:
\begin{equation}\label{FF-All}
   \lef F^r \rig_{\nu_1\nu_2\ldots\nu_p} \, \lef F_r \rig^{\mu_1\mu_2\ldots\mu_p} \,=\,  \delta_{\nu_{1}}^{\;[\mu_{1}}\delta_{\nu_{2}}^{\;\mu_{2}} \ldots\delta_{\nu_{p}}^{\;\mu_{p}]} \,.
\end{equation}
The action of the operator $\mathcal{C}_p$ on this basis is given by:
$$ \mathcal{C}_p(\bl{F}_s)\,\equiv\, \bl{F}_r\,\mathcal{C}_{rs}\;, \;\;\textrm{where }\; \mathcal{C}_{rs}\,=\, \langle \bl{F}^r ,\mathcal{C}_p(\bl{F}_s)\rangle \,. $$
Using this one can easily prove that the trace of $\mathcal{C}_p$ is zero. Indeed, by means of (\ref{FF-All}) we have
\begin{align}
  \nonumber \textrm{tr}(\mathcal{C}_p)&= \mathcal{C}_{rr}= \lef F^r\rig^{\mu_1\mu_2\ldots\mu_p} C^{\alpha\beta}_{\phantom{\alpha\beta}\mu_1\mu_2} \lef F_r \rig_{\mu_3\mu_4\ldots\mu_p\alpha\beta} \\
  &= C^{\alpha\beta}_{\phantom{\alpha\beta}\mu_1\mu_2} \delta_{\alpha}^{\;[\mu_{1}}\delta_{\beta}^{\;\mu_{2}}\delta_{\mu_{3}}^{\;\mu_{3}}\ldots\delta_{\mu_{p}}^{\;\mu_{p}]} \propto C^{\alpha\beta}_{\phantom{\alpha\beta}\alpha\beta}=0\,.\label{TrCp-All}
\end{align}

The signature of the inner product $\langle\,,\rangle$ depends on the signature of the metric $\bl{g}$. In particular, if the metric is Euclidean then it is immediate to verify that the inner product defined in (\ref{InnerProd-All}) is positive-definite. Therefore, since the operator $\mathcal{C}_p$  is real and self-dual with respect to $\langle\,,\rangle$, it follows that on the Euclidean signature $\mathcal{C}_p$ can be diagonalized. More explicitly, if the metric $\bl{g}$ is positive-definite then so will be $\langle\,,\rangle$, which means that, locally (in a neighborhood $N_x$), one can find a real basis $\{\hat{\bl{F}}_r\}$ for $\Gamma(\wedge^pM)$ such that $\langle\hat{\bl{F}}_r, \hat{\bl{F}}_s\rangle= \delta_{rs}$. The matrix representation of $\mathcal{C}_p$ in this basis is then real and symmetric and, therefore, can be diagonalized. This represents a huge limitation on the possible algebraic types that the operator $\mathcal{C}_p$ can have. Let us state this as a theorem \cite{art4}:
\begin{theorem}\label{Theo.C-Eucli-All}
When the signature of $\bl{g}$ is Euclidean the operator $\mathcal{C}_p$ admits a trace-less diagonal matrix representation with real eigenvalues. Particularly, this guarantees that on the refined Segre classification of this operator all numbers inside the square bracket are equal to 1.
\end{theorem}

\subsection{Even Dimensions}

In this subsection it will be proved that a particularly interesting simplification occurs when the dimension of the manifold is even. If the dimension of $(M,\bl{g})$ is $n=2m$, with $m$ being an integer, then equation (\ref{HpHn-p}) implies that
\begin{equation*}
  \mathcal{H}_{m}\, \mathcal{H}_{m} \,=\,  (-1)^{\frac{s}{2}}\,\,\bl{1}_m \;\Longrightarrow\; \mathcal{H}_{m}\, \mathcal{H}_{m} \,=\, \varrho^2\,\bl{1}_m \; \;
  \left\{
  \begin{array}{ll}
    \varrho = 1 \;\textrm{ if $\frac{s}{2}$ is even}\\
     \varrho = i \;\textrm{ if $\frac{s}{2}$ is odd.}\\
    \end{array} \right.
\end{equation*}
So locally the space of $m$-forms can be split into the direct sum of two subspaces of the same dimension, the eigenspaces of $\mathcal{H}_{m}$\label{SelfDualSpace}:
$$ \Gamma(\wedge^mM) \,=\, \Lambda^{m+} \oplus \Lambda^{m-}\,, \quad  \Lambda^{m\pm}=\{\,\bl{F}\in \Gamma(\wedge^mM) \,|\,\, \mathcal{H}_{m}(\bl{F}) = \pm\varrho\,\bl{F}\,\} \,.$$
An element of $\Lambda^{m+}$ is said to be a self-dual $m$-form\index{Self-dual form}, while an element of $\Lambda^{m-}$ is called an anti-self-dual $m$-form. Note that these spaces are interchanged when we multiply the volume-form by $-1$. The subspaces $\Lambda^{m\pm}$ can equivalently be defined as follows:
$$ \Lambda^{m\pm}\,=\, \left\{\; \lef\bl{F} \,\pm\, \frac{1}{\varrho}\,\mathcal{H}_{m}(\bl{F}) \rig  \;|\;\, \bl{F}\in \Gamma(\wedge^mM)  \;\right\}\;;\quad  \left\{
  \begin{array}{ll}
    \varrho = 1 \;\textrm{ if $\frac{s}{2}$ is even}\\
     \varrho = i \;\textrm{ if $\frac{s}{2}$ is odd.}\\
    \end{array} \right. $$
From which we see that if $\frac{s}{2}$ is even then the spaces $\Lambda^{m\pm}$ are real, while if $\frac{s}{2}$ is odd then the elements of $\Lambda^{m\pm}$ must be complex. Furthermore, since the operator $\mathcal{H}_{m}$ is real, if $\frac{s}{2}$ is odd then the complex conjugate of a self-dual $m$-form is anti-self-dual. Note also that the operator $\mathcal{H}_{m}$ can be self-adjoint or anti-self-adjoint with respect to the inner product $\langle\,,\rangle$ depending on the dimension of the manifold:
\begin{align*}
  \langle \bl{F}, \mathcal{H}_{m}(\bl{K}) \rangle \,&=\, \frac{1}{m!}\,\epsilon_{\nu_1\ldots\nu_m\mu_1\ldots\mu_m} \,F^{\mu_1\ldots\mu_m}\, K^{\nu_1\ldots\nu_m}\\
  &=\, \frac{(-1)^{m^2}}{m!}\,\epsilon_{\mu_1\ldots\mu_m\nu_1\ldots\nu_m} \,F^{\mu_1\ldots\mu_m}\, K^{\nu_1\ldots\nu_m} \,=\, (-1)^{m}\,\langle  \mathcal{H}_{m}(\bl{F}), \bl{K}\rangle
\end{align*}
Using the above equation one can easily see that if  $m$ is even then the inner product $\langle \bl{F}^+, \bl{K}^- \rangle$ vanishes whenever $\bl{F}^+\in \Lambda^{m+}$ and $\bl{K}^-\in \Lambda^{m-}$. Analogously, if $m$ is odd then the inner products $\langle \bl{F}^+, \bl{K}^+ \rangle$ and $\langle \bl{F}^-, \bl{K}^- \rangle$ vanish for all $\bl{F}^+,\bl{K}^+ \in \Lambda^{m+}$ and $\bl{F}^-,\bl{K}^-\in \Lambda^{m-}$. These results are summarized by the below theorem \cite{art4}.
\begin{theorem}\label{Theo. -SplitAll}
Let $(M,\bl{g})$ be a manifold of signature $s$ and dimension $n=2m$, with $m$ being an integer. Then the Hodge dual map splits the space of $m$-forms into a direct sum of its eigenspaces, $\Gamma(\wedge^mM)= \Lambda^{m+} \oplus \Lambda^{m-}$. When $s$ is a multiple of 4 the spaces $\Lambda^{m+}$ and $\Lambda^{m-}$ are both real, otherwise they must be complex conjugates of each other. Furthermore, if $m$ is even then the spaces $\Lambda^{m+}$ and $\Lambda^{m-}$ are orthogonal to each other, while if $m$ is odd both spaces $\Lambda^{m\pm}$ are isotropic with respect to the inner product $\langle\,,\rangle$.
\end{theorem}

Now plugging $n=2m$ and $p=m$ on equation (\ref{HC=CH -All}) yields that the operators $\mathcal{C}_{m}$ and $\mathcal{H}_{m}$ commute. Thus the spaces $\Lambda^{m+}$ and $\Lambda^{m-}$ are both preserved by the action of $\mathcal{C}_{m}$. So, the latter operator can be written as the direct sum of its restrictions to the spaces $\Lambda^{m\pm}$:
\begin{equation}\label{C^+- All}
\mathcal{C}_{m} \,=\, \mathcal{C}^+ \oplus \mathcal{C}^-\,, \quad  \mathcal{C}^{\pm}\,\equiv\,\frac{1}{2}\lef \mathcal{C}_{m} \,\pm\, \frac{1}{\varrho}\,\mathcal{C}_{m}\mathcal{H}_{m} \rig\,.
\end{equation}
Note that the action of $\mathcal{C}^+$ on an element of $\Lambda^{m-}$ gives zero, as well as the restriction of $\mathcal{C}^-$ to $\Lambda^{m+}$ is identically zero. Therefore, generally it is useful to assume that the domains of the operators $\mathcal{C}^\pm$ are the spaces $\Lambda^{m\pm}$, instead of the whole bundle of $m$-forms. It is worth remarking that eq. (\ref{C^+- All}) imposes huge restrictions on the possible algebraic types of the operator $\mathcal{C}_m$.

A special phenomenon happens when $m$ is odd. In this case, because of theorem \ref{Theo. -SplitAll}, one can always introduce a basis\footnote{Now the indices $r,s$ and $t$ run from 1 to $\frac{1}{2}\cdot\frac{(2m)!}{m!m!}$\,.} $\{\bl{F}^+_r\}$ for $\Lambda^{m+}$ and a basis $\{\bl{F}^-_r\}$ for $\Lambda^{m-}$ such that $\langle \bl{F}^+_r, \bl{F}^-_s \rangle= \delta_{rs}$. Indeed, since $\langle\,,\rangle$ is non-degenerate we just need to start with a basis  for $\Lambda^{m+}$  and a basis for $\Lambda^{m-}$ and then use the Gram-Schmidt process in order to redefine the latter. Thus when $m$ is odd the operators have the following matrix representations:
$$  \left.
  \begin{array}{ll}
    \mathcal{C}_{rs}^+ \,=\, \langle \,\bl{F}^-_r, \mathcal{C}^+(\bl{F}^+_s) \,\rangle \,=\, \langle \,\bl{F}^-_r, \mathcal{C}_m(\bl{F}^+_s) \,\rangle\\
     \mathcal{C}_{rs}^- \,=\, \langle \,\bl{F}^+_r, \mathcal{C}^-(\bl{F}^-_s) \,\rangle \,=\, \langle\, \bl{F}^+_r, \mathcal{C}_m(\bl{F}^-_s) \,\rangle\\
    \end{array}\; \right\} \;\Longrightarrow \; \mathcal{C}_{rs}^+ \,=\, \mathcal{C}_{sr}^-  \,.$$
Where on the last step it was used the fact that $\mathcal{C}_m$ is self-adjoint. Thus, when $m$ is odd the matrix representation of $\mathcal{C}^+$ is the transpose of the matrix representation of $\mathcal{C}^-$ and, therefore, these operators have the same algebraic type. So if the dimension $n$ is even but not a multiple of four, classify $\mathcal{C}_{\frac{n}{2}}$ is tantamount to classify $\mathcal{C}^+$.

In the same vein, if the signature $s$ is not a multiple of 4 then the spaces $\Lambda^{m+}$ and $\Lambda^{m-}$ are connected by complex conjugation, see theorem \ref{Theo. -SplitAll}. Therefore, in this case the degrees of freedom of the operators $\mathcal{C}^+$ and $\mathcal{C}^-$ are connected by a reality condition. More precisely, the operator $\mathcal{C}^+$ is the complex conjugate of $\mathcal{C}^-$, which can be easily seen from equation (\ref{C^+- All}) along with the fact that the operators $\mathcal{C}_m$ and $\mathcal{H}_m$ are both real:
$$ \frac{s}{2} \;\textrm{ is odd}\quad \Longrightarrow \quad \mathcal{C}^\pm\,=\, \frac{1}{2}\,\lef\,\mathcal{C}_m \,\mp\, i\,\mathcal{C}_m\,\mathcal{H}_m\,\rig \quad \Longrightarrow \quad  \mathcal{C}^+\,=\, \overline{\mathcal{C}^-}\,.$$
Thus, in such a case $\mathcal{C}^+$ and $\mathcal{C}^-$ have the same refined Segre type. So that in order to classify $\mathcal{C}_m$ we just need to compute the algebraic type of $\mathcal{C}^+$.

Since there is no scalar that can be constructed using just the Weyl tensor and the volume-form linearly, it is reasonable to expect that  both operators $\mathcal{C}^+$ and $\mathcal{C}^-$ have vanishing trace. Indeed, using (\ref{C^+- All}) along with the fact that $\mathcal{C}_{m}$ is trace-less, see eq. (\ref{TrCp-All}), it follows that:
\begin{align*}
 \tr(\mathcal{C}^\pm) \,&=\, \frac{\pm1}{2\varrho}\tr(\mathcal{C}_{m}\mathcal{H}_{m}) \,\propto\, (F^r)^{\;\mu_1\ldots\mu_m}\, C^{\alpha\beta}_{\phantom{\alpha\beta}\mu_1\mu_2} \,\epsilon^{\nu_1\ldots\nu_m}_{\phantom{\nu_1\ldots\nu_m}\mu_3\ldots\mu_m\alpha\beta} \,(F_r)_{\;\nu_1\ldots\nu_m} \\
  &\propto\, C^{\alpha\beta}_{\phantom{\alpha\beta}\mu_1\mu_2} \,\epsilon^{\nu_1\ldots\nu_m}_{\phantom{\nu_1\ldots\nu_m}\mu_3\ldots\mu_m\alpha\beta}\,
\delta_{[\nu_1}^{\;\mu_{1}}\ldots\delta_{\nu_{m}]}^{\;\mu_{m}}\,=\, C_{\alpha\beta\mu_1\mu_2} \, \epsilon^{\alpha\beta\mu_1\ldots\mu_m}_{\phantom{\alpha\beta\mu_1\ldots\mu_m}\mu_3\ldots\mu_m} \,=\, 0\,.
\end{align*}
Where on the last step it was used the Bianchi identity, $C_{[\mu\nu\rho]\sigma}= 0$. The previous results then lead us to the following theorem \cite{art4}.
\begin{theorem}\label{Theo.CSplit-All}
In a manifold of even dimension $n=2m$ the operator $\mathcal{C}_{m}$ is the direct sum of its restrictions to the spaces $\Lambda^{m\pm}$, $\mathcal{C}_{m}= \mathcal{C}^+\oplus\mathcal{C}^-$. The operators $\mathcal{C}^+$ and $\mathcal{C}^-$ have vanishing trace. Moreover, they carry the same degrees of freedom both when $m$ is odd and when the signature of the manifold is not a multiple of 4, more precisely the following relations hold:\\
(1) $m$ is odd $\;\Rightarrow\;$ $\mathcal{C}^+$ is the adjoint of $\mathcal{C}^-$, $\langle \bl{F}, \mathcal{C}^+(\bl{K})\rangle= \langle  \mathcal{C}^-(\bl{F}), \bl{K}\rangle$\\
(2) $\frac{s}{2}$ is odd $\;\Rightarrow\;$ $\mathcal{C}^+$ is the complex conjugate of $\mathcal{C}^-$, $\mathcal{C}^+= \overline{\mathcal{C}^-}$.\\
On the other hand, if $m$ and $\frac{s}{2}$ are both even then the operators $\mathcal{C}^+$ and $\mathcal{C}^-$ generally carry different degrees of freedom. In particular, on the latter case the reality condition relates $\mathcal{C}^+$  with itself as well as $\mathcal{C}^-$ with itself, so that both operators are real.
\end{theorem}
An immediate consequence of this theorem is that whenever $m$ or $\frac{s}{2}$ are odd the refined Segre type of the operators $\mathcal{C}^+$ and $\mathcal{C}^-$ coincide. Thus, is such cases in order to classify $\mathcal{C}_m$ we just need to compute the refined Segre type of $\mathcal{C}^+$.

Note that the chapters \ref{Chap. Gen4D} and \ref{Chap._Spin6D} provide explicit examples for the theorems proved in the present chapter, let us perform few comparisons. In the previous chapters it was proved that respectively in 4 and 6 dimensions the operators $\mathcal{C}_2$ and $\mathcal{C}_3$ can be diagonalized when the signature is Euclidean, which agrees with theorem \ref{Theo.C-Eucli-All}. In 4 dimensions we proved that the operator $\mathcal{C}^+$ is the complex conjugate $\mathcal{C}^-$ if the signature is Lorentzian, which endorses theorem \ref{Theo.CSplit-All}, since in this case $\frac{s}{2}=1$. In 6 dimensions it was proved, using the spinorial formalism, that in a suitable basis $\mathcal{C}^-$ is the transpose of $\mathcal{C}^+$, since in such case $m=3$ this agrees with theorem \ref{Theo.CSplit-All}. Finally, recall that in chapter \ref{Chap. Gen4D} it was shown that in a 4-dimensional manifold of split signature the operators $\mathcal{C}^+$ and $\mathcal{C}^-$ are both real and independent of each other. Since on the latter case $m=2$ and $\frac{s}{2}=0$ are both even, this again supports theorem \ref{Theo.CSplit-All}.

In 4 dimensions a manifold\index{Self-dual manifold} is said to be self-dual if $\mathcal{C}^-= 0$ and $\mathcal{C}^+\neq 0$, see chapter \ref{Chap. Gen4D}. Such manifolds have been widely studied in the past \cite{SD-Manifolds,Broda}, in particular it has been shown that Einstein's vacuum equation on self-dual manifolds reduces to a single second-order differential equation \cite{SD-Manifolds}. Now it is natural wondering whether the notion of self-dual manifolds can be extended to higher dimensions. According to theorem \ref{Theo.CSplit-All} this is not possible neither if $\frac{n}{2}$ is odd nor if $\frac{s}{2}$ is odd, since in these cases the constraint $\mathcal{C}^-= 0$ implies $\mathcal{C}^+= 0$. However, if the dimension and the signature are both multiples of four then the self-dual manifolds could, in principle, be defined. Nevertheless, it turns out that laborious calculations reveal that in 8 dimensions if $\mathcal{C}^-$ vanishes then $\mathcal{C}^+= 0$, irrespective of the signature being a multiple of four. Although the present author has worked out only the 8-dimensional case, such result seems to indicate that the self-dual manifolds cannot be defined if the dimension is different from 4.

\subsection{An Elegant Notation}\label{Sec.Notation-All}

In this subsection it will be introduced an elegant and useful notation to manage the operators $\mathcal{C}_p$. To this end the formalism presented in section (\ref{Sec. Cartan}) will be extensively used. Let $\{\bl{e}_a\}$ be a frame of vector fields on the manifold $(M,\bl{g})$, with $\{\bl{e}^a\}$ being the dual frame of 1-forms such that $\bl{e}^a(\bl{e}_b)= \delta^a_{\phantom{a}b}$. Assuming that the Ricci tensor vanishes, so that the Riemann tensor is equal to the  Weyl tensor, the curvature 2-form is then defined by
\begin{equation}\label{Curv2form-All}
 \mathbb{C}^a_{\phantom{a}b}\, \equiv \, \frac{1}{2}\,C^a_{\phantom{a}bcd}\,\bl{e}^{c}\wedge \bl{e}^{d}\,.
\end{equation}
Now let $\bl{F}$ be a $p$-form, with $p\geq 2$, then we can associate to it a set of $(p-2)$-forms defined by
\begin{equation}\label{F-elegant-All}
\mathbb{F}^{\ph{a}b}_{a}\, \equiv \,\frac{2}{p!}\, F^{\ph{a}b}_{a\ph{ab}c_1c_2\ldots c_{p-2}}\, \bl{e}^{c_1}\wedge \bl{e}^{c_2}\wedge \ldots \wedge \bl{e}^{c_{p-2}}\,.
\end{equation}
In particular, note that $\bl{F}=\frac{1}{2}\mathbb{F}_{ab}\wedge\bl{e}^{a}\wedge \bl{e}^{b}$, where $\mathbb{F}_{ab}\equiv \mathbb{F}^{\ph{a}c}_{a}g_{cb}$. Then using equations (\ref{Curv2form-All}) and (\ref{F-elegant-All}) we have
\begin{align*}
  \mathbb{C}^a_{\phantom{a}b}\wedge \mathbb{F}^{\phantom{a}b}_{a} \,&=\, \frac{1}{p!} \,C^a_{\phantom{a}bc_1c_2}\, F^{\phantom{a}b}_{a\phantom{b}c_3c_4\ldots c_{p}}\, \bl{e}^{c_1}\wedge \bl{e}^{c_2}\wedge\bl{e}^{c_3}\wedge \ldots \wedge \bl{e}^{c_{p}} \\
  \,&=\,\frac{1}{p!} \,C^{ab}_{\phantom{ab}c_1c_2} \,F_{c_3c_4\ldots c_{p}ab}\, \bl{e}^{c_1}\wedge \ldots \wedge \bl{e}^{c_{p}} \,=\, \mathcal{C}_p(\bl{F})\,\,\Rightarrow
\end{align*}
\begin{equation}\label{Cp-Eleg-All}
   \mathcal{C}_p(\bl{F}) \,=\,  \mathbb{C}^a_{\phantom{a}b}\wedge \mathbb{F}^{\phantom{a}b}_{a}\,.
\end{equation}

Now let us define the $(p-1)$-form $\textbf{D}\mathbb{F}^{\phantom{a}b}_{a}\equiv d\mathbb{F}^{\phantom{a}b}_{a} +  \bl{\omega}^b_{\phantom{b}c}\wedge \mathbb{F}_{a}^{\phantom{a}c} - \bl{\omega}^c_{\phantom{c}a}\wedge \mathbb{F}_{c}^{\phantom{c}b}$, where $\bl{\omega}^a_{\phantom{a}b}$ are the connection 1-forms defined on eq. (\ref{w-connect-1-form}). Then taking the exterior derivative of the identity $\bl{F}=\frac{1}{2}\mathbb{F}_{ab}\wedge\bl{e}^{a}\wedge \bl{e}^{b}$ and using the first Cartan structure equation we find that $d\bl{F}= \frac{1}{2} \bl{e}^{a}\wedge \bl{e}^{b}\wedge \textbf{D}\mathbb{F}_{ab}$, where $\textbf{D}\mathbb{F}_{ab}\equiv g_{bc}\textbf{D}\mathbb{F}^{\phantom{a}c}_{a}$. When the Ricci tensor vanishes, as assumed here, the second Cartan structure equation is $\mathbb{C}^a_{\ph{a}b}= d\bl{\omega}^a_{\ph{a}b} + \boldsymbol{\omega}^a_{\phantom{a}c}\wedge \boldsymbol{\omega}^c_{\phantom{c}b}$. Taking the exterior derivative of this relation we easily find that $d\mathbb{C}^a_{\phantom{a}b}= \mathbb{C}^a_{\phantom{a}c}\wedge \bl{\omega}^c_{\phantom{c}b} - \bl{\omega}^a_{\phantom{a}c} \wedge \mathbb{C}^c_{\phantom{c}b}$. Then using this result while computing the exterior derivative of equation (\ref{Cp-Eleg-All}) lead us to the identity $d\left[\mathcal{C}_p(\bl{F}) \right] =\mathbb{C}^{ab}\wedge \textbf{D}\mathbb{F}_{ab}$. The results of this paragraph are summarized by the following equations:
\begin{gather}
  d\bl{F} \,=\,\frac{1}{2}\,   \bl{e}^{a}\wedge \bl{e}^{b}\wedge \textbf{D}\mathbb{F}_{ab} \quad\quad ; \quad\quad d\left[\mathcal{C}_p(\bl{F}) \right] \,=\, \mathbb{C}^{ab}\wedge \textbf{D}\mathbb{F}_{ab} \label{dCp-All}\\
  \nonumber \textbf{D}\mathbb{F}_{ab}\,\equiv\, g_{bc}\lef d\,\mathbb{F}^{\phantom{a}c}_{a} +  \bl{\omega}^c_{\phantom{c}d}\wedge \mathbb{F}_{a}^{\phantom{a}d} - \bl{\omega}^d_{\phantom{d}a}\wedge \mathbb{F}_{d}^{\phantom{d}c}\rig\,.
\end{gather}

As a simple application of this notation, suppose that $\bl{F}$ is a $p$-form such that $\textbf{D}\mathbb{F}_a^{\phantom{a}b}=\bl{\varphi}\wedge\mathbb{F}_a^{\phantom{a}b}$ for some 1-form $\bl{\varphi}$. Then equation (\ref{dCp-All}) immediately implies that:
\begin{equation}\label{dF-All}
 d\,\bl{F} \,=\, \bl{\varphi}\wedge\bl{F} \quad \textrm{and} \quad  d\left[\mathcal{C}_p(\bl{F}) \right] \,=\, \bl{\varphi}\wedge\mathcal{C}_p(\bl{F})\,.
\end{equation}
This, in turn, implies that if $\bl{F}$ is a simple form then, according to the Frobenius theorem, the vector distribution annihilated by $\bl{F}$ is integrable, see section \ref{Sec. Integrab}. Analogously, if $\mathcal{C}_p(\bl{F})$ is a simple $p$-form then eq. (\ref{dF-All}) guarantees that the vector distribution annihilated by $\mathcal{C}_p(\bl{F})$ is integrable.

\section{Integrability of Maximally Isotropic Distributions}\index{Isotropic}

Let $\{\bl{e}_1,\bl{e}_2\}$ be a vector distribution generating isotropic planes on a Ricci-flat 4-dimensional manifold, then the celebrated Goldberg-Sachs theorem\index{Goldberg-Sachs theorem} states that such distribution is integrable if, and only if, the 2-form $\bl{B}= \bl{e}_1\wedge\bl{e}_2$ is such that $\mathcal{C}_2(\bl{B})\propto \bl{B}$, see chapter \ref{Chap. Gen4D}. A partial generalization of this theorem was proved in chapter \ref{Chap._Spin6D} with the help of a theorem of Taghavi-Chabert \cite{HigherGSisotropic2}. More precisely, it was shown that in 6 dimensions if the operator $\mathcal{C}_3$ obeys to certain algebraic constraints then the manifold admits an integrable maximally isotropic distribution. The aim of the present section is to generalize this result to all even dimensions, \textit{i.e.}, express the integrability condition for a maximally isotropic distribution in terms of algebraic constraints on the operator $\mathcal{C}_m$. From now on in this chapter, we shall assume that the manifold $(M,\bl{g})$ has dimension $n= 2m$, with $m$ being an integer.

Before proceeding let us set few conventions and recall some important definitions. Up to a multiplicative factor there exists a one-to-one relation between vector field distributions\index{Distribution} and simple forms. More explicitly, if  $Span\{\bl{V}_1,\bl{V}_2,\ldots,\bl{V}_p\}$ is a $p$-dimensional distribution of vector fields then any non-zero $p$-form proportional to $F^{\nu_1\ldots\nu_p}= p!\,V_1^{[\nu_1}V_2^{\nu_2}\ldots V_p^{\nu_p]}$ is said to generate such distribution. In abstract notation we shall right $\bl{F}= \bl{V}_1\wedge\bl{V}_2\wedge\ldots\wedge\bl{V}_p$.  A distribution of vector fields is called \emph{isotropic} if every vector field $\bl{V}$ tangent to such distribution has zero norm, $\bl{g}(\bl{V},\bl{V})= 0$. In particular all vector fields tangent to an isotropic distribution are orthogonal to each other. A simple form $\bl{F}$ is then said to be \emph{null}\index{Null form} if its associated distribution is isotropic. Following the convention adopted in the previous chapter, a frame $\{\bl{e}_a\}= \{\bl{e}_{a'}, \bl{e}_{a'+m}=\bl{\theta}^{a'}\}$ of vectors fields is called a \emph{null frame}\index{Null frame} whenever the inner products between the frame vectors are:
$$ \bl{g}(\bl{e}_{a'}, \bl{e}_{b'})\,=\, 0 \,=\, \bl{g}(\bl{\theta}^{a'}, \bl{\theta}^{b'}) \quad;\quad  \bl{g}(\bl{e}_{a'}, \bl{\theta}^{b'})= \frac{1}{2}\,\delta^{\,b'}_{a'}\,,$$
where the indices $a,b,c,\ldots$ run from 1 to $2m$, while the indices $a', b', c', \ldots$ pertain to the set $\{1,2,\ldots,m\}$. In $n= 2m$ dimensions, the maximum dimension that an isotropic distribution can have is $m$. Therefore, an $m$-dimensional isotropic distribution is called
\emph{maximally isotropic}\index{Maximally isotropic}. In particular, note that if $\{\bl{e}_a\}$ is a null frame then $\bl{e}_1\wedge\bl{e}_2\wedge\ldots\wedge\bl{e}_m$ is a null $m$-form and its associated distribution is maximally isotropic.

As commented in section \ref{Sec.GS6D}, in reference \cite{HigherGSisotropic2} it was proved a theorem that partially generalizes the GS theorem to higher dimensions. Using the notation adopted here, such theorem can be conveniently stated as follows: \emph{If the Weyl tensor of a Ricci-flat manifold is such that $C_{a'b'c'd}= 0$, and is generic otherwise\footnote{See footnote \ref{Foot-Generality} of chapter \ref{Chap._Spin6D} for comments on this generality condition.}, then the maximally isotropic distribution $Span\{\bl{e}_{a'}\}$ is integrable.} The intent of the present section is to express the algebraic condition $C_{a'b'c'd}= 0$ in terms of the operator $\mathcal{C}_m$. With this aim it is of particular help to define the subspaces $\mathcal{A}_q\subset \Gamma(\wedge^mM)$ as follows:
\begin{equation}\label{Aq-All}
 \mathcal{A}_q \,\equiv\,\{\,\bl{F} \in \Gamma(\wedge^mM) \;|\; \bl{e}_{a'_q}\lrcorner \ldots \bl{e}_{a'_2}\lrcorner \bl{e}_{a'_1}\lrcorner \bl{F} \,=\,0 \;\; \forall\; a'_1, \ldots, a'_p\in (1,\ldots, m) \,\}\,.
\end{equation}
Where $\bl{e}\lrcorner \bl{F}$ means the interior product of the vector field $\bl{e}$ on the differential form $\bl{F}$ (see section \ref{Sec. DiffForms}). These subspaces can be equivalently defined by:
$$ \mathcal{A}_q \,=\, \textsf{A}_1\oplus \textsf{A}_2\oplus\cdots\oplus \textsf{A}_q\;; \quad \textsf{A}_q\,\equiv\,Span\{\bl{\theta}^{a'_1}\wedge\cdots\wedge\bl{\theta}^{a'_{q-1}}\wedge\bl{e}_{a'_q}\wedge\cdots\wedge\bl{e}_{a'_m}\}\,.  $$

Now let us use the notation of section \ref{Sec.Notation-All} in order to express the invariance of the subbundle $\mathcal{A}_1$ under the action of $\mathcal{C}_m$ in terms of the Weyl tensor components. If $\bl{F}$ is an $m$-form pertaining to $\mathcal{A}_1$ then $\bl{F}\propto \bl{e}_1\wedge\bl{e}_2\wedge\ldots\wedge\bl{e}_m$. In particular it follows that $\mathbb{F}_{a'b}= 0$ and $\bl{e}_{a'}\lrcorner\mathbb{F}_{bc}= 0$,  so that eq. (\ref{Cp-Eleg-All}) implies:
\begin{align}
 \nonumber \bl{e}_{c'}\lrcorner\,\mathcal{C}_m(\bl{F}) \,&=\, \lef \bl{e}_{c'}\lrcorner\,\mathbb{C}_{ab} \rig \wedge \mathbb{F}^{ab} \,+\, \mathbb{C}_{ab}\wedge\lef\bl{e}_{c'}\lrcorner\,\mathbb{F}^{ab}\rig   \,=\,  \lef\bl{e}_{c'}\lrcorner\,\mathbb{C}_{ab}\rig\wedge \mathbb{F}^{ab} \\
   &=\, C_{abc'd}\,\bl{e}^d\wedge \mathbb{F}^{ab} \,=\,
  C_{a'b'c'd}\,\bl{e}^d\wedge \mathbb{F}^{a'b'}\label{InvA1-All}
\end{align}
From this equation we easily see that if $C_{a'b'c'd}= 0$ then $\bl{e}_{a'}\lrcorner\,\mathcal{C}_m(\bl{F})= 0$, which means that $\mathcal{C}_m(\bl{F})$ pertain to $\mathcal{A}_1$. Thus the integrability condition for the distribution generated by $\bl{e}_1\wedge\bl{e}_2\wedge\ldots\wedge\bl{e}_m$ implies that such $m$-vector is an eigen-$m$-vector of the operator $\mathcal{C}_m$. On the other hand, equation (\ref{InvA1-All}) guarantees that if $\mathcal{C}_m(\bl{F})\in \mathcal{A}_1$ then $C_{a'b'c'd'}= 0$ for all $a',b',c',d'$ and $C_{a'b'c'}^{\ph{a'b'c'}d'}= 0$ if either $d'= a'$ or $d'= b'$. Particularly, in 4 dimensions these two constraints imply that the whole integrability condition $C_{a'b'c'd}= 0$ is satisfied, while in higher dimensions this is not true anymore. Similar manipulations lead to the following interesting theorem \cite{art4}:
\begin{theorem}\label{Theo.GSTC-All}
The three statements below are equivalent:\\
(1) The Weyl tensor obeys the integrability condition $C_{a'b'c'd}=0$\\
(2) The subbundles $\mathcal{A}_1$ and $\mathcal{A}_2$ are invariant under the action of $\mathcal{C}_m$\\
(3) All subbundles $\mathcal{A}_q$, $q\in \{1,2,\ldots, m\}$, are invariant by the action of $\mathcal{C}_m$.
\end{theorem}
This theorem along with the theorem of reference \cite{HigherGSisotropic2} immediately imply the following corollary:
\begin{corollary}\label{Coro-C_inv-Integ}
In a Ricci-flat manifold of dimension $n=2m$, if the operator $\mathcal{C}_m$  preserves the spaces $\mathcal{A}_1$ and $\mathcal{A}_2$, with $\mathcal{C}_m$ being generic otherwise, then the maximally isotropic distribution generated by $\bl{e}_1\wedge \bl{e}_2\wedge\ldots\wedge \bl{e}_m$ is integrable.
\end{corollary}
In 4 dimensions these results recover part of the corollary \ref{Cor.EigenPlane4D} obtained in chapter \ref{Chap. Gen4D}, while in 6 dimensions we retrieve  theorem \ref{Theo. GS-AlgClas-6D} of chapter \ref{Chap._Spin6D}. For the details see \cite{art4}.

Since the operator $\mathcal{C}_m$ preserves the spaces $\Lambda^{m\pm}$ then it follows that if $\mathcal{A}_q$ is an eigenspace of $\mathcal{C}_m$ so will be the subbundles $\mathcal{A}_q^\pm\equiv \mathcal{A}_q\cap\Lambda^{m\pm}$. In 4 dimensions we have that $\mathcal{A}_1^-= 0$ and  $\mathcal{A}_2^-= \Lambda^{m-}$. Since these spaces are trivially preserved by the action of $\mathcal{C}_2$ it follows that the invariance of the subbundles $\mathcal{A}_q$ under  $\mathcal{C}_2$ imposes no constraint over $\mathcal{C}^-$. Differently, in higher dimensions, $m>2$, we have $\dim(\mathcal{A}^-_2)= \frac{1}{2}(m+m^2)< \frac{1}{2} \frac{(2m)!}{m!\,m!}= \dim(\Lambda^{m-})$. So, in these cases, if $\mathcal{A}_2$ is invariant by $\mathcal{C}_m$ then the operator $\mathcal{C}^-$ must admit a non-trivial eigenspace, leading us to the following theorem:
\begin{theorem}\index{Maximally isotropic}
While in 4 dimensions the integrability condition for the self-dual planes generated by $\bl{e}_1\wedge \bl{e}_2$ imposes restrictions only over $\mathcal{C}^+$, with $\mathcal{C}^-$ being arbitrary; in higher dimensions the integrability condition for the self-dual maximally isotropic distribution generated by $\bl{e}_1\wedge \bl{e}_2\wedge\ldots\wedge \bl{e}_m$ constrains both operators, $\mathcal{C}^+$ and $\mathcal{C}^-$.
\end{theorem}


\section{Optical Scalars and Harmonic Forms}

In this section the 4-dimensional concept of optical scalars introduced in chapter \ref{Chap. Theorems4} will be generalized to higher dimensional manifolds.  Moreover, it will be shown that the existence of certain harmonic forms imposes constraints on these scalars. To this end, and in order to match the standard notation \cite{Rev. Ortaggio}, let us define a semi-null frame $\{\bl{l}, \bl{n}, \bl{m}_i\}$ to be a frame of vector fields whose inner products are\footnote{The indices $i,j,k,\ldots$ run from 2 to $n-1$, where $n$ is the dimension of the manifold.}:
$$ \bl{g}(\bl{l},\bl{l}) = \bl{g}(\bl{n},\bl{n}) =\bl{g}(\bl{l},\bl{m}_i)=\bl{g}(\bl{n},\bl{m}_i)= 0\;;\quad  \bl{g}(\bl{l},\bl{n})= 1\;; \quad \bl{g}(\bl{m}_i,\bl{m}_j)=\delta_{ij} \,. $$
Then the optical scalars\index{Optical scalars} associated to the null congruence generated by $\bl{l}$ are defined by:
\begin{equation*}\label{OpticalSca-All}
 M_0\,=\, l^\nu n^\mu \, \nabla_\nu l_\mu\ \;\;\;;\;\; M_i\,=\, l^\nu m_i^\mu \, \nabla_\nu l_\mu\ \;\;\;;\;\; M_{ij}\,=\, m_j^\nu m_i^\mu \, \nabla_\nu l_\mu\ \,.
\end{equation*}
It is simple matter to prove that $\bl{l}$ is geodesic if, and only if, $M_i= 0$, the parametrization being affine when $M_0= 0$. Furthermore, the congruence generated by $\bl{l}$ is hyper-surface-orthogonal, $l_{[\mu} \nabla_\nu l_{\rho]}= 0$, if, and only if, $M_i$ and $M_{[ij]}$ both vanish. In the Lorentzian signature the vector fields of a semi-null frame can be chosen to be real, so that in such a case the optical scalars are real. The $(n-2)\times(n-2)$ matrix $M_{ij}$ is dubbed the optical matrix of the null congruence generated by $\bl{l}$. Analogously to what was done in chapter \ref{Chap. Theorems4} it is useful to split this matrix as a sum of a symmetric and trace-less matrix, a skew-symmetric matrix and a term proportional to the identity:
$$   M_{ij} \,=\, \sigma_{ij} \,+\, A_{ij} \,+\, \theta\,\delta_{ij} \;; \quad  \theta\equiv\frac{1}{n-2}\,\delta^{ij} M_{ij}\;;\;\; \sigma_{ij}\equiv M_{(ij)}- \theta\,\delta_{ij}\;; \;\; A_{ij}\equiv M_{[ij]}\,. $$
The scalar $\theta$ is called the expansion, $\sigma_{ij}$ is named the shear\index{Shear} matrix, while $A_{ij}$ is  called the twist matrix. In particular, if $\sigma_{ij}= 0$ we shall say that the congruence is shear-free.

Before proceeding let us introduce some jargon. A $p$-form $\bl{K}$ is called harmonic\index{Harmonic form} if it is closed, $d\bl{K}= 0$, and co-closed, $d(\star\bl{K})= 0$. In terms of components this means that the following differential equations hold:
\begin{equation}\label{Harmonic-All}
\nabla_{[\alpha}\, K_{\mu_1\mu_2 \ldots \mu_p]} \,=\,0 \quad \textrm{and} \quad \nabla^{\alpha}\, K_{\alpha\mu_2 \ldots \mu_p}\,=\,0  \,.
\end{equation}
Note, in particular, that if $\bl{L}$ is a closed 1-form then, by the Poincar\'{e} lemma \cite{Nakahara}, it follows that locally there exists some scalar function $f$ such that $L_\mu= \nabla_\mu f$. Thus the 1-form $\bl{L}$ will be harmonic if $\nabla^\mu\nabla_\mu f= 0$, which is the well-known equation satisfied by a harmonic function. In the CMPP\index{CMPP classification} classification \cite{CMPP} we say that a $p$-form $\bl{K}$ is type $N$ with $\bl{l}$ being a multiple aligned null direction if $\bl{K}$ admits the following expansion:
\begin{equation}\label{TypeN-All}
  K^{\mu_1\mu_2 \ldots \mu_p}= p!\, f_{j_2j_3\ldots j_p} \, l^{[\mu_1}m_{j_2}^{\mu_2}m_{j_3}^{\mu_3}\ldots m_{j_p}^{\mu_p]}\,.
\end{equation}
Where $f_{j_2j_3\ldots j_p}= f_{[j_2j_3\ldots j_p]}$ are scalars and it is being assumed a sum over the indices $j_2,\ldots,j_p$. In what follows it will be proved that if a manifold admits a harmonic form that is type $N$ then the optical scalars of its multiple aligned null direction are constrained.

Let $\bl{K}\neq 0$ be a harmonic $p$-form of type $N$ on the CMPP classification with $\bl{l}$ being its multiple aligned null direction, which means that the equations (\ref{Harmonic-All}) and (\ref{TypeN-All}) hold. Since $K^{\alpha\beta\mu_3 \ldots \mu_p}l_\beta= 0$ it follows that:
\begin{gather*}
  0 = \nabla_{\alpha}\lef K^{\alpha\beta\mu_3 \ldots \mu_p}\, l_\beta  \rig = K^{\alpha\beta\mu_3 \ldots \mu_p}\, \nabla_{\alpha}l_\beta =
 p!\, f_{j_2j_3\ldots j_p} \, l^{[\alpha}m_{j_2}^{\beta}m_{j_3}^{\mu_3}\ldots m_{j_p}^{\mu_p]} \,\nabla_{\alpha}l_\beta \\
  = h_1\, f_{j_2j_3\ldots j_p} \,m_{j_2}^{[\beta}m_{j_3}^{\mu_3}\ldots m_{j_p}^{\mu_p]}\,l^\alpha \,\nabla_{\alpha}l_\beta \,+\,
  h_2 \,l^\beta \,\nabla_{\alpha}l_\beta \,(\,\cdots\,) \,+\\ +\,h_3\, f_{j_2j_3\ldots j_p} \,l^{[\mu_3}m_{j_4}^{\mu_4}\ldots m_{j_p}^{\mu_p]}\, m_{j_2}^{\alpha}m_{j_3}^{\beta}\,\nabla_{\alpha}l_\beta  \\ =\,h_4\, f_{j_2j_3\ldots j_p}
  \, m_{j_3}^{\mu_3}\ldots m_{j_p}^{\mu_p}\,M_{j_2} \,+\, 0 \,+\, h_5 \,f_{j_2j_3\ldots j_p}l^{[\mu_3}m_{j_4}^{\mu_4}\ldots m_{j_p}^{\mu_p]}\, M_{j_2j_3}\,.
\end{gather*}
Where in the above equation the $h$'s are non-zero unimportant constants. We, thus, arrive at the following constraints:
\begin{equation}\label{Mf-All}
   M_i\, f_{ij_3\ldots j_p} \,=\,0 \quad; \quad  A_{ij}f_{ij k_4\ldots k_p}\,=\,0\,.
\end{equation}
In a similar fashion, expanding the equation $\nabla_{[\alpha}\, K_{\mu_1\mu_2 \ldots \mu_p]} l^\alpha m_{j_1}^{\phantom{j_1} \mu_1}\ldots m_{j_p}^{\phantom{j_p} \mu_p}= 0$ we arrive, after some careful algebra, at the following relation:
\begin{equation*}
  M_{[j_1}\,f_{j_2 \ldots j_p]} \,=\, 0\,.
\end{equation*}
In particular, the contraction of this identity with $M_{j_1}$ along with equation (\ref{Mf-All}) lead us to the relation $M_iM_i= 0$. Analogously, working out the equality $0=(\nabla^{\alpha}\, K_{\alpha\mu_2 \ldots \mu_p})m_{j_2}^{\phantom{j_2} \mu_2}\ldots m_{j_p}^{\phantom{j_p} \mu_p}$ it easily follows that:
\begin{equation}\label{Alg-All}
  K_{\alpha\mu_2 \ldots \mu_p} \nabla^\alpha (m_{j_2}^{\phantom{j_2} \mu_2}\ldots m_{j_p}^{\phantom{j_p} \mu_p})\,=\, (p-1)!\,\, l^\alpha \nabla_\alpha f_{j_2\ldots j_p} \,+\, (p-1)!\, f_{j_2\ldots j_p} \nabla^\alpha l_\alpha\,.
\end{equation}
Now expanding the relation $\nabla_{[\alpha}\, K_{\mu_1\mu_2 \ldots \mu_p]}l^\alpha n^{\mu_1}m_{j_2}^{\phantom{j_2} \mu_2}\ldots m_{j_p}^{\phantom{j_p} \mu_p}= 0$ and using the identity $\nabla^\alpha l_\alpha = M_0+ (n-2)\theta$ along with equation (\ref{Alg-All}) it follows that:
$$  2 (p-1) \,f_{i[j_3 \ldots j_p} \, \sigma_{j_2] i} \;=\; (n-2p)\, \theta \, f_{j_2 \ldots j_p}\,. $$
These results are summarized by the following theorem \cite{art4}:
\begin{theorem}\label{Theo. OptScal-All}\index{Harmonic form}
If  $K^{\mu_1\mu_2\ldots \mu_p}= p!\, f_{j_2\ldots j_p}\, l^{[\mu_1} m_{j_2}^{\mu_2}\ldots m_{j_p}^{\mu_p]}$ is a non-zero $p$-form such that $d\bl{K}= 0$ and $d(\star\bl{K})= 0$ then the following relations hold:\\
(1) $M_i\, f_{ij_3\ldots j_p} \,=\,0$\\
(2) $ M_{[j_1}\, f_{j_2 \ldots j_p]} \,=\, 0$\\
(3) $2(p-1) \,f_{i[j_3 \ldots j_p} \, \sigma_{j_2] i} \;=\; (n-2p) \,\theta\, f_{j_2 \ldots j_p}$\\
(4) $M_i M_i \,=\, 0$\\
(5) $A_{ij}f_{ij k_4\ldots k_p}\,=\,0$.
\end{theorem}
On the Lorentzian signature it is possible to introduce a real semi-null frame, so that the optical scalars are real in such frame. In this case the equation $M_iM_i= 0$ implies that $M_i= 0$, which means that the real vector field $\bl{l}$ is geodesic. The particular case $p= 2$ of the above theorem in Lorentzian manifolds was obtained before on ref. \cite{M. Ortaggio-Robinson-Trautman}. Similar results for arbitrary $p$ on the Lorentzian signature were also obtained, by means of the so-called GHP formalism, in ref. \cite{GHP}, where the identities $\emph{(1)}$, $\emph{(2)}$ and $\emph{(3)}$ can be explicitly found on the proof of the Lemma 3 of \cite{GHP}\footnote{The author thanks Harvey S. Reall for pointing out this reference.}.

\section{Generalizing Mariot-Robinson and Goldberg-Sachs Theorems}

As explained in section \ref{Sec._Mariot4D}, the Mariot-Robinson\index{Mariot-Robinson theorem} theorem guarantees that a 4-dimensional Lorentzian manifold admits a null bivector\index{Null bivector} $\bl{F}\propto \bl{l}\wedge\bl{m}$ obeying to the source-free Maxwell's equations, $d\bl{F}= 0$ and $d\star\bl{F}= 0$, if, and only if, the real null vector field $\bl{l}$ is geodesic and shear-free. But in 4 dimensions the proper geometric generalization to arbitrary signature of a geodesic and shear-free null congruence is the existence of an integrable distribution of isotropic planes, see section \ref{Sec. GS4D}. Then it follows that the Mariot-Robinson theorem provides a connection between the existence of null solutions for Maxwell's equations and the existence of an integrable maximally isotropic distribution in 4 dimensions. By means of the results presented in section \ref{Sec. Integrab} it is not so hard to generalize this theorem to arbitrary even dimensions. Let $\bl{F}= \bl{e}_1\wedge\ldots\wedge\bl{e}_m$ be a null $m$-form on a $2m$-dimensional manifold, so that it generates the maximally isotropic distribution $Span\{\bl{e}_{a'}\}$. Note that since $\bl{e}_{a'}\lrcorner\bl{F}= 0$, this distribution coincides with the distribution annihilated by $\bl{F}$. Now from the results of section \ref{Sec. Integrab} it follows that the latter distribution is integrable if, and only if, there exists some function $h\neq 0$ such that $d(h\bl{F})= 0$. But a null $m$-form must always be self-dual or anti-self-dual, $\star\bl{F}= \pm\varrho\bl{F}$ with $\varrho$ equal to 1 or $i$, which can be grasped from the discussion below equation \ref{eo_psi} on appendix \ref{App._Cliff&Spinors}. Thus we conclude that if $d(h\bl{F})= 0$ then $d\star(h\bl{F})= \pm\varrho d(h\bl{F})= 0$, leading us to the following generalized version of the Mariot-Robinson theorem \cite{HughstMason,art4}:
\begin{theorem}\label{Theo.Mariot-All}\index{Null form}
In a $2m$-dimensional manifold a null $m$-form $\bl{F}'$ generates an integrable maximally isotropic distribution if, and only if, there exists some function $h\neq 0$ such that $\bl{F}= h\bl{F}'$ obeys the equations $d\bl{F}= 0$ and $d(\star\bl{F})= 0$.
\end{theorem}

Now let $\{\bl{e}_a\}= \{\bl{e}_{a'},\bl{\theta}^{b'}\}$ be a null frame on a $2m$-dimensional manifold. Then we can use it in order to define the following semi-null frame:
\begin{equation*}
 \bl{l}= \bl{e}_1\;,\;\; \bl{n}= 2\bl{\theta}^1 \;,\;\; \bl{m}_j= (\bl{e}_j + \bl{\theta}^j) \;,\;\;
 \bl{m}_{j+m-1}= -i(\bl{e}_j- \bl{\theta}^j)\;;\;\; j\,\in\,\{2,3,\ldots,m\}\,.
\end{equation*}
In such a basis the null $m$-form $\bl{F}= \bl{e}_1\wedge\bl{e}_2\wedge\ldots\wedge\bl{e}_m$ can be written as follows\footnote{For example, in 6 dimensions, $m= 3$, we have the following expression: $$  \widehat{f}_{j_2j_3}\,\equiv\, \frac{2!}{4}\, \lef \delta_{[j_2}^{2} + i \delta_{[j_2}^{4} \rig
\lef \delta_{j_3]}^{3} + i \delta_{j_3]}^{5}\rig\,=\,
 \frac{1}{2}\lef  \delta_{[j_2}^{2}\delta_{j_3]}^{3}\,+\, i\,\delta_{[j_2}^{2}\delta_{j_3]}^{5} \,+\, i\,\delta_{[j_2}^{4}\delta_{j_3]}^{3} \,-\, \delta_{[j_{2}}^{4}\delta_{j_{3} ]}^{5}\rig $$.}:
\begin{equation}\label{F-semiNull}
\left\{
  \begin{array}{l}
    F^{\mu_1\mu_2\ldots\mu_m} \, \equiv\, m!\, e_1^{\,[\mu_1}\ldots e_m^{\,\mu_m]} \,=\, m! \, \widehat{f}_{j_2j_3\ldots j_m}\, l^{[\mu_1}m_{j_2}^{\,\mu_2}\ldots m_{j_m}^{\,\mu_m]}  \\
    \\
    \widehat{f}_{j_2j_3\ldots j_m}\,\equiv\, \frac{(m-1)!}{2^{m-1}}\, \lef \delta_{[j_2}^{2} + i \delta_{[j_2}^{m+1} \rig \lef \delta_{j_3}^{3} + i \delta_{j_3}^{m+2} \rig\cdots \lef \delta_{j_m]}^{m} + i \delta_{j_m]}^{2m-1} \rig  \\
    \end{array}
\right.
\end{equation}
Thus the $m$-form $\bl{F}$ is type $N$ on the CMPP classification with $\bl{l}= \bl{e}_1$ being a multiple aligned null direction. It is worth noting that the definition $\bl{l}\equiv \bl{e}_1$ was quite arbitrary, since we could have chosen $\bl{l}$ to be any non-zero vector field tangent to the distribution generated by the null form $\bl{F}$. A special phenomenon happens when the signature is Lorentzian, in this case the real part of a maximally isotropic distribution is always 1-dimensional \cite{Trautman}. Thus on the Lorentzian case we shall choose $\bl{l}$ to be tangent to the unique real null direction on the distribution generated by $\bl{F}$. Now the successive combination of theorem \ref{Theo.Mariot-All}, then equation (\ref{F-semiNull}) and finally theorem \ref{Theo. OptScal-All} immediately lead us to the following corollary:
\begin{corollary}\label{Coro-Integ-OptSca}
If $Span\{\bl{e}_1,\bl{e}_2,\ldots, \bl{e}_m\}$ is an integrable maximally isotropic distribution on a manifold of dimension $n= 2m$ then the optical scalars of the null congruences generated by vector fields tangent to such distribution are constrained as follows:\\
(1) $M_i\, \widehat{f}_{ij_3\ldots j_p} \,=\,0$\\
(2) $ M_{[j_1}\, \widehat{f}_{j_2 \ldots j_m]} \,=\, 0$\\
(3) $ \widehat{f}_{i[j_3 \ldots j_m} \, \sigma_{j_2] i} \,=\,0$\\
(4) $M_i M_i \,=\, 0$\\
(5) $A_{ij}\widehat{f}_{ij k_4\ldots k_m}\,=\,0$.
\end{corollary}
Particularly, on the Lorentzian signature if $\bl{l}$ is a real vector field tangent to such distribution then the item $\emph{(4)}$ implies that $\bl{l}$ is geodesic. It is worth mentioning that in appendix C of ref. \cite{GS-HighD} the integrability of a maximally isotropic distribution is expressed in terms of the Ricci rotation coefficients of a null frame. Note that in the above corollary no condition is assumed over the Ricci tensor. A simple application of this result on 6-dimensional manifolds has been worked out on \cite{art4}.

The original version of the Goldberg-Sachs\index{Goldberg-Sachs theorem} theorem establish an equivalence between algebraic restrictions on the Weyl operator $\mathcal{C}_2$ and the existence of a null congruence whose optical scalars are constrained in Ricci-flat 4-dimensional space-times, see theorem \ref{Theo_Gold-Sachs} in chapter \ref{Chap. Theorems4}. Now by a simple merger of corollaries \ref{Coro-C_inv-Integ} and \ref{Coro-Integ-OptSca} one can state an analogous result valid in even-dimensional manifolds of arbitrary signature \cite{art4}:
\begin{theorem}\label{Theo. GS-OptSca-All}
In a Ricci-flat manifold of dimension $n=2m$ if the operator $\mathcal{C}_m$  preserves the spaces $\mathcal{A}_1$ and $\mathcal{A}_2$, with $\mathcal{C}_m$ being generic otherwise, then the optical scalars of the null congruences generated by vectors fields tangent to the maximally isotropic distribution $Span\{\bl{e}_1, \bl{e}_2, \ldots,\bl{e}_m\}$ are constrained as follows:\\
(1) $M_i\, \widehat{f}_{ij_3\ldots j_p} \,=\,0$\\
(2) $ M_{[j_1}\, \widehat{f}_{j_2 \ldots j_m]} \,=\, 0$\\
(3) $\widehat{f}_{i[j_3 \ldots j_m} \, \sigma_{j_2] i} \,=\,0$\\
(4) $M_i M_i \,=\, 0$\\
(5) $A_{ij}\widehat{f}_{ij k_4\ldots k_m}\,=\,0$.
\end{theorem}
Where the subbundles $\mathcal{A}_q$ were defined in (\ref{Aq-All}), while the object $\widehat{f}_{j_2j_3\ldots j_p}$ was defined in equation (\ref{F-semiNull}). Again, in the particular case of the Lorentzian signature if $\bl{l}$ is a real vector field tangent to such distribution then the equation $M_iM_i= 0$ guarantees that $\bl{l}$ is geodesic.

Theorem \ref{Theo. GS-OptSca-All} is a partial generalization of the Goldberg-Sachs theorem to even-dimensional manifolds. Note, however, that while in 4 dimensions the GS theorem is an equivalence relation, the theorem presented here goes just in one direction, stating that algebraic restrictions on the Weyl tensor imply the existence of constrained null congruences, but not the converse. Furthermore, while in 4-dimensional manifolds of Lorentzian signature the item $\emph{(3)}$ of theorem \ref{Theo. GS-OptSca-All} implies that the null congruence is shear-free, in higher dimensions this is not true anymore. Indeed, a simple count of degrees of freedom reveals that the higher the dimension the more restrictive the shear-free condition becomes. Indeed, in $n$ dimensions the object $\nabla_\mu l_\nu$ has $D= n(n-1)$ non-trivial components, since $l^\nu\nabla_\mu l_\nu$ is automatically zero. On the other hand, the shear matrix  $\sigma_{ij}$ has $S= \frac{1}{2}(n-1)(n-2)-1$ independent components. Note that the rate $S/D$ becomes higher and higher as the dimension increases, approaching the limit $S/D\rightarrow\frac{1}{2}$ as the dimension goes to infinity. This gives a hint that in dimensions greater than four the Goldberg-Sachs theorem cannot be trivially generalized stating that a simple algebraic restriction on the Weyl tensor is equivalent to the existence of a null congruence that is geodesic and shear-free, since the latter condition is too strong.

\chapter{Conclusion and Perspectives}\label{Chap. Conclusion}

As demonstrated in chapter \ref{Chap. Pet}, there exist several ways to approach the Petrov classification in 4-dimensional Lorentzian manifolds.
One of these methods is the bivector line of attack, which treats the Weyl tensor as an operator, $\mathcal{C}_2$, on the space of bivectors. Although this was the original path taken for defining such classification, during the past decades it has been overlooked in favor of other methods like the spinorial approach. However, in this thesis it was proved that the bivector method can be quite fruitful and full of geometric significance. Indeed, in chapter \ref{Chap. Gen4D} this approach was used in order to generalize the Petrov classification to 4-dimensional manifolds of arbitrary signature in a unified way. Furthermore, it was proved that the null eigen-bivectors of $\mathcal{C}_2$ generate integrable isotropic planes, providing a convenient way to state the Goldberg-Sachs (GS) theorem. In particular, this form of interpreting the GS theorem yielded connections between the algebraic type of the Weyl tensor and the existence of geometric structures as symplectic forms and complex structures.

In chapter \ref{Chap. AllD} it was shown that the bivector operator $\mathcal{C}_2$ is just a single member of an infinite class of linear operators $\mathcal{C}_p$ sending $p$-forms into $p$-forms that can be constructed out of the Weyl tensor in arbitrary dimension. It was proved that such operators have nice properties as commuting with the Hodge dual map and being self-dual with respect to a convenient inner product. Particularly, when the signature is Euclidean these operators can be diagonalized, which makes the algebraic classification rather simple in this case. Moreover, when the dimension is even, $n= 2m$, the operator $\mathcal{C}_m$ plays a prominent role, as it can be nicely used to express the integrability condition of maximally isotropic distributions. In chapter \ref{Chap. AllD} it was also proved a generalized version of the Goldberg-Sachs theorem, valid in even-dimensional spaces of arbitrary signature, stating that certain algebraic constraints on the operator $\mathcal{C}_m$ imply the existence of null congruences with restricted optical scalars. These results teaches us   that while in 4 dimensions the bivectors are featured objects, in $n= 2m$ dimensions this role is played by the $m$-forms.

Since the most elegant approach to the Petrov classification and its associated theorems uses spinors, it is natural to employ such language in order to provide a higher-dimensional generalization of these results. This was the route taken in chapter \ref{Chap._Spin6D}, where the spinorial formalism in 6 dimensions was  developed \textit{ab initio}. There it is shown how to represent the $SO(6;\mathbb{C})$ tensors in terms of spinors, which reveals the possibility of classifying the bivectors and the Weyl tensor in a simple way. In particular, this Weyl tensor classification coincides with the one attained by means of the operator $\mathcal{C}_3$. An important feature of spinors is that they constitute the most suitable tool to describe isotropic subspaces, as explicitly illustrated on subsection \ref{Sec.Iso6D}. Particularly, the maximally isotropic distributions are represented by the so-called pure spinors. Because of this property, the spinorial formalism was shown to provide a simple and elegant form to express the integrability condition of a maximally isotropic distribution.\\

The work presented in this thesis can be enhanced in multiple forms. For example, the operators $\mathcal{C}_p$ and their relation with integrability properties
deserve further investigation. Since in $2m$ dimensions the operator $\mathcal{C}_m$ is connected to the integrability of $m$-dimensional isotropic distributions, a natural question to be posed is whether the operators $\mathcal{C}_p$ are, likewise, associated to the integrability of $p$-dimensional isotropic distributions irrespective of the manifold dimension. Another interesting quest is trying to provide links between the algebraic type of the Weyl tensor and the existence of hidden symmetries on the manifold. A more ambitious project would be to study which algebraic conditions might be imposed to the operator $\mathcal{C}_m$ in order for Einstein's vacuum equation to be analytically integrable, just as in 4 dimensions the type $D$ condition allows the complete integration of Einstein's equation. Concerning the 6-dimensional spinorial formalism introduced here, certainly further progress can be accomplished as soon as a connection is introduced on the spinor bundle. In particular, the generality condition referred to on the footnote \ref{Foot-Generality} of chapter \ref{Chap._Spin6D} can, probably, be better understood by means of the spinorial language. In addition, once such connection is introduced the 6-dimensional twistors can be investigated.

The main goal behind the research shown on the present thesis was to give a better understanding of general relativity in higher dimensions, particularly to provide further tools to study geometrical properties of higher-dimensional black holes. But, besides general relativity, this piece of work can, hopefully, be applied to other branches of physics and mathematics. For instance, higher-dimensional manifolds are of great relevance in string theory and supergravity, so that the results obtained here could be useful. More broadly, this work can be applied to physical systems whose degrees of freedom form a differentiable manifold with dimension greater than 3. In particular, by means of Caratheodory's formalism, it follows that integrable distributions are of interest to thermodynamics (see section \ref{Sec. Integrab}), which suggests a possible application for the results presented here. Finally, since spinors are acquiring increasing significance in physics it follows that the 6-dimensional spinorial language developed here can have multiple utility. For instance, in order to retrieve our 4-dimensional space-time out of a 10-dimensional manifold of string theory one generally need to compactify 6 dimensions, so that 6-dimensional manifolds are of particular relevance.

\appendix


\chapter{Segre Classification and its Refinement}\label{App._Segre}

Segre classification\index{Segre classification} is a well-known form to classify square matrices (or linear operators) over the complex field. Essentially this classification amounts to specify the eigenvalue structure of the matrix in a compact code. In this appendix such classification will be explained and a refinement will be presented.

It is a standard result of linear algebra that given a square matrix $M$ over the complex field it is always possible to find a basis in which such matrix acquires the so-called Jordan canonical form \cite{Hof-Kunze}. This means that it is always possible to find an invertible matrix $B$ such that $M'=BMB^{-1}$ assumes the following block-diagonal form:
\begin{equation}\label{M'}
  M' \,=\, \diag(J_1,J_2,\ldots,J_q)\;, \textrm{ where }\; J_i = \left[
                                                                   \begin{array}{ccccc}
                                                                     \lambda_i& 1 & 0    &\ldots& 0 \\
                                                                     0         &\lambda_i& 1    &  & \vdots \\
                                                                     0   & 0 & \ddots &  & 0 \\
                                                                    \vdots  & \vdots &  & \lambda_i & 1 \\
                                                                   0        & 0 & \ldots & 0 & \lambda_i \\
                                                                   \end{array}
                                                                 \right]\,, \;\lambda_i\in\mathbb{C}\,.
\end{equation}
Note that $J_i$ can also be the $1\times1$ matrix $J_i=\lambda_i$. The blocks $J_i$ are called the Jordan blocks of the matrix $M$. Each block $J_i$ admits just one eigenvector and its eigenvalue is $\lambda_i$. Thus, for example, if we manage to put the $5\times5$ matrix $G$ on the Jordan canonical form
$$ G' \,=\, \left[
              \begin{array}{ccccc}
                2 & 1 & 0 & 0 & 0\\
                0 & 2 & 0 & 0 & 0 \\
                0 & 0 & 3 & 0 & 0 \\
                0 & 0 & 0 & 5 & 1 \\
                0 & 0 & 0 & 0 & 5 \\
              \end{array}
            \right]\,, \textrm{ then } \; J_1 = \left[
                                                  \begin{array}{cc}
                                                    2 & 1 \\
                                                    0 & 2 \\
                                                  \end{array}
                                                \right]\,,\; J_2 = 3 \,\textrm{ and } \; J_3 = \left[
                                                               \begin{array}{cc}
                                                                 5 & 1 \\
                                                                 0 & 5 \\
                                                               \end{array}
                                                             \right].
 $$
In particular this canonical form implies that the matrix $G$ admits just three different eigenvectors (apart from a multiplicative scale). The eigenvalues of these eigenvectors are $\lambda_1=2$, $\lambda_2=3$ and $\lambda_3=5$.

The Jordan canonical form of a matrix is unique up to the ordering of the Jordan blocks $J_i$. In particular, the dimensions of the Jordan Blocks are invariant under the change of basis, which opens up the possibility of introducing an invariant classification. The Segre classification of a matrix amounts to \emph{list the dimensions of all the Jordan blocks and bound together, inside round brackets, the dimensions of the blocks with the same eigenvalue.} This classification can be refined if we separate the dimensions of the blocks with eigenvalue zero putting them on the right of the dimensions of the other blocks, using a vertical bar to separate \cite{Spin6D}. As a pedagogical example, let us work out the Segre type (ST) and the refined Segre type (RST) of the matrix $F$:
\begin{equation}\label{MatrixApp}
 F\,=\,\left[
    \begin{array}{cccccc}
      \kappa& 1 & 0 & 0 & 0 &0 \\
      0 &\kappa& 1 & 0 & 0 & 0  \\
      0 & 0 &\kappa & 0 & 0 & 0 \\
      0 & 0&0 & \alpha & 0 & 0 \\
      0 & 0&0 & 0 & \beta & 1 \\
      0 & 0&0 & 0 & 0 & \beta \\
    \end{array}
  \right]\,.
\end{equation}
The types depend on the values of $\kappa$, $\alpha$ and $\beta$. Some of the possibilities are:
\begin{align*}
  \kappa,\alpha,\beta\neq0 \, \textrm{ and all different} &\Rightarrow \; \textrm{ST: } [3,2,1] \;\;;\;\;\; \textrm{RST: } [3,2,1|\,]\\
  \alpha,\beta\neq0=\kappa \, \textrm{ and } \alpha\neq\beta&\Rightarrow \;\textrm{ST: } [3,2,1] \;\;;\;\;\; \textrm{RST: } [2,1|3]\\
  \alpha=\beta\neq0\,,\;\kappa=0 \,  &\Rightarrow\; \textrm{ST: } [3,(2,1)] \;;\;\; \textrm{RST: } [(2,1)|3]\\
  \alpha=\beta=0\,,\;\kappa\neq0 \,  &\Rightarrow\; \textrm{ST: } [3,(2,1)] \;;\;\; \textrm{RST: } [3|2,1]\,.
\end{align*}
Note that the order of the numbers between the square bracket and the vertical bar does not matter. As a final example it is displayed below all the possible refined Segre types that a trace-less $3\times3$ matrix can have. This result will be used in chapter \ref{Chap. Pet}.
$$
(A):\quad\left[
  \begin{array}{ccc}
    \lambda_1& 0         & 0 \\
    0        & \lambda_2 & 0\\
    0        & 0         & \lambda_3 \\
  \end{array}
\right] \longrightarrow\left\{
  \begin{array}{cl}
  \lambda_i\neq0 \textrm{ and } \lambda_i\neq\lambda_j\;\;\forall\;i,j  &\rightarrow\,  [1,1,1|\,]     \\
  \lambda_1=0 \textrm{ and } \lambda_i\neq\lambda_j\;\;\forall\;i,j &\rightarrow\, [1,1|1] \\
  \lambda_1=\lambda_2\neq0,\; \lambda_3= -2\lambda_1 &\rightarrow\, [(1,1),1|\,]\\
 \lambda_1=\lambda_2= \lambda_3= 0 &\rightarrow\, [\,|1,1,1]\\
         \end{array}
\right.
$$

$$
(B):\quad\left[
  \begin{array}{ccc}
    \lambda& 1         & 0 \\
    0        & \lambda & 0\\
    0        & 0         & -2\lambda \\
  \end{array}
\right] \longrightarrow\left\{
  \begin{array}{cl}
    \lambda \neq 0 &\rightarrow \; [2,1|\,]  \\
    \lambda =0 &\rightarrow \;[\,|2,1] \\
    \end{array}
\right.
$$

$$
(C):\quad\left[
  \begin{array}{ccc}
    0& 1         & 0 \\
    0        & 0 & 1\\
    0        & 0         & 0 \\
  \end{array}
\right] \longrightarrow\; [\,|3]
  $$
It is worth noting that the trace-less condition restricted enormously the number of possible algebraic types. For instance, the types $[(1,1)|1]$, $[2|1]$ and $[3|\,]$ are some examples of types that are incompatible with the trace-less assumption.

\newpage
%
%
%

\chapter{Null Tetrad Frame}\label{App._NullTetrad}

In 1962 E. T. Newman and R. Penrose introduced a tetrad frame formalism in which all basis vectors are null \cite{NewPen-Tetrad}, which can be accomplished only if complex vectors are used. This was a novelty at the time and since then this kind of basis has proved to be useful in many general relativity calculations.
According to \cite{chandrasekhar} the reason that led Penrose to introduce a null basis was his faith that the fundamental structures of general relativity are the light-cones.

If $(M,\bl{g})$ is a 4-dimensional Lorentzian manifold then a null tetrad frame is a set of four null vector fields $\{\bl{l},\bl{n},\bl{m},\overline{\bl{m}}\}$ that span the tangent space at every point. The vector fields $\bl{l}$ and $\bl{n}$ are real, while $\bl{m}$ and $\overline{\bl{m}}$ are complex and conjugates to each other. In a null tetrad frame the only non-zero inner products are assumed to be:
$$  \bl{g}(\bl{l},\bl{n})\,=\, 1 \quad \textrm{ and } \quad  \bl{g}(\bl{m},\overline{\bl{m}})\,=\, -1\,.$$
Therefore the metric can be written as follows:
$$  g_{\mu\nu} \,=\, 2\,l_{(\mu}n_{\nu)} \,-\, 2\,m_{(\mu}\overline{m}_{\nu)}\,. $$
Which can be easily verified by contracting this metric with the basis vectors. Given an orthonormal frame $\{\hat{\bl{e}}_0,\hat{\bl{e}}_1,\hat{\bl{e}}_2,\hat{\bl{e}}_3\}$, with $\bl{g}(\hat{\bl{e}}_a,\hat{\bl{e}}_b)= \eta_{ab} = \diag(1,-1,-1,-1)$, then we can easily construct a null tetrad by defining:
$$ \bl{l}=\frac{1}{\sqrt{2}}(\hat{\bl{e}}_0+\hat{\bl{e}}_1) \; ; \; \bl{n} =\frac{1}{\sqrt{2}}(\hat{\bl{e}}_0-\hat{\bl{e}}_1) \; ; \; \bl{m} =\frac{1}{\sqrt{2}}(\hat{\bl{e}}_2+i\hat{\bl{e}}_3) \; ; \; \overline{\bl{m}} =\frac{1}{\sqrt{2}}(\hat{\bl{e}}_2-i\hat{\bl{e}}_3)  $$
The null tetrads can be elegantly expressed in terms of spinors. Let $\{\bl{o},\bl{\iota}\}$ be a spinor frame, \textit{i.e.}, spinors such that $o_{_A}\iota^A= 1$ (see section \ref{Sec.Spinor4d}), then it can be easily shown that the following vectors form a null tetrad:
\begin{equation}\label{NulltetradSPIN}
    l^\mu \,\sim\, o^A \overline{o}^{\dot{A}} \; ;\;\; n^\mu \,\sim\, \iota^A\overline{\iota}^{\dot{A}} \; ;\;\; m^\mu \,\sim\, o^A\overline{\iota}^{\dot{A}} \; ;\;\; \overline{m}^\mu \,\sim\, \iota^A\overline{o}^{\dot{A}} \,.
\end{equation}

\newpage

\chapter{Clifford Algebra and Spinors}\label{App._Cliff&Spinors}\index{Clifford algebra}
The Clifford Algebra, also called geometric algebra, was created by the English mathematician William Kingdon Clifford around 1880. His intent was to unify Hamilton's work on quaternions and Grassmann's work about exterior algebra. Since the first paper of Clifford on the subject has been published in an obscure journal at the time, it went unnoticed until the beginning of the XX century, when \'{E}lie Cartan discovered the spinors \cite{Cartan}, objects related to unknown representations of the $SO(n)$ group. Actually, it seems that R. Brauer and H. Weyl have been the first ones to connect Cartan's spinors with the geometric algebra \cite{Weyl_35}.

An algebra is, essentially, a vector space in which an associative multiplication between the vectors is defined. Clifford algebra is a special kind of algebra defined on vector spaces endowed with inner products. Let $V$ be an $n$-dimensional vector space endowed with the non-degenerate inner product $<\,,>$, then the Clifford product of two vectors $\bl{a},\bl{b}\in V$ is defined to be such that its symmetric part gives the inner product:
\begin{equation}\label{Sym_Clif. Prod}
  \bl{a}\bl{b}\,+\, \bl{b}\bl{a} \,=\, 2<\bl{a},\bl{b}>\,.
\end{equation}
If $\{\hat{\bl{e}}_1,\hat{\bl{e}}_2,\ldots,\hat{\bl{e}}_n\}$ is an orthonormal basis for $V$, $<\hat{\bl{e}}_i,\hat{\bl{e}}_j>= \pm\delta_{ij}$, then it follows from (\ref{Sym_Clif. Prod}) that $\hat{\bl{e}}_i\hat{\bl{e}}_j=-\hat{\bl{e}}_j\hat{\bl{e}}_i$ if $i\neq j$. Analogously, $\hat{\bl{e}}_i\hat{\bl{e}}_j\hat{\bl{e}}_k$ is totally skew-symmetric if $i\neq j\neq k\neq i$. Thus we conclude that a general element of $\mathcal{C}l(V)$, the Clifford algebra of $V$, can always be put in the following form:
$$\bl{\omega} \,=\, w + w^i\,\hat{\bl{e}}_i + w^{ij} \hat{\bl{e}}_i\hat{\bl{e}}_j + \ldots + w^{i_1\ldots i_n} \hat{\bl{e}}_{i_1} \ldots \hat{\bl{e}}_{i_n}\, ,  $$
where $w$ is a real (or complex) number and $w^{i_1\ldots i_p}$ are skew-symmetric tensors with values on the real (or complex) field. Thus we conclude that the exterior algebra of $V$, $\wedge V$, provides a basis for $\mathcal{C}l(V)$. In other words, the vector space of the Clifford algebra associated to $V$ is $\wedge V$. By what was just seen it is natural to define the wedge product of vectors to be the totally anti-symmetric part of the Clifford product:
\begin{equation}\label{a^b}
  \bl{a}_1\wedge \bl{a}_2 \wedge\ldots\wedge \bl{a}_p \,=\, \frac{1}{p!}\,\sum_\sigma \,  (-1)^{\epsilon_\sigma}\, \bl{a}_{\sigma(1)} \bl{a}_{\sigma(2)} \ldots \bl{a}_{\sigma(p)}\,,
\end{equation}
where the sum runs over all permutations of $\{1,2,\ldots,p\}$ and $\epsilon_\sigma$ is even or odd depending on the parity of the permutation $\sigma$. In particular, note that $\hat{\bl{e}}_1 \hat{\bl{e}}_2 \ldots \hat{\bl{e}}_p = \hat{\bl{e}}_1\wedge \hat{\bl{e}}_2 \wedge\ldots\wedge \hat{\bl{e}}_p$. With this definition we find that given the vectors $\bl{a},\bl{b}\in V$ then $\bl{a}\bl{b}-\bl{b}\bl{a}=2 \bl{a}\wedge \bl{b}$. Using this and eq. (\ref{Sym_Clif. Prod}) we arrive at the following formula for the Clifford product of two vectors:
\begin{equation}\label{Cliff. Product}
  \bl{a}\bl{b} \,=\, <\bl{a},\bl{b}> \,+\, \bl{a}\wedge \bl{b}\,.
\end{equation}
Using equations (\ref{a^b}) and (\ref{Cliff. Product}) it can be proved, for instance, that
\begin{equation}\label{abc}
  \bl{a}\bl{b}\bl{c} \,=\, <\bl{b},\bl{c}> \bl{a} \,+\, <\bl{a},\bl{b}>\bl{c} \, \,-\, <\bl{a},\bl{c}> \bl{b} \,+\, \bl{a}\wedge \bl{b}\wedge \bl{c}\,.
\end{equation}
A non-zero linear combination of the wedge product of $p$ vectors, $\bl{a}_1\wedge \bl{a}_2\wedge \ldots\wedge \bl{a}_p$, is called a $p$-vector or an element of order $p$. Since the Clifford product of two elements of even order yields another even order element, it follows that the set of all elements of $\mathcal{C}l(V)$ with even order forms a subalgebra, denoted $\mathcal{C}l(V)^+$.
\\
\\
\textbf{Example:}
\newline
As a simple example let us work out the Clifford algebra of the vector space $\mathbb{R}^{0,2}$. $\mathcal{C}l(\mathbb{R}^{0,2})$ is generated by $\{1,\, \hat{\bl{e}}_1,\, \hat{\bl{e}}_2,\, \hat{\bl{e}}_1\wedge \hat{\bl{e}}_2\}$, where $\hat{\bl{e}}_1\hat{\bl{e}}_1 = -1 = \hat{\bl{e}}_2\hat{\bl{e}}_2$ and $\hat{\bl{e}}_1\hat{\bl{e}}_2 = \hat{\bl{e}}_1\wedge \hat{\bl{e}}_2$. Note also that
$$(\hat{\bl{e}}_1\wedge \hat{\bl{e}}_2)(\hat{\bl{e}}_1\wedge \hat{\bl{e}}_2)= \hat{\bl{e}}_1\hat{\bl{e}}_2 \hat{\bl{e}}_1\hat{\bl{e}}_2 = -\hat{\bl{e}}_1\hat{\bl{e}}_1\hat{\bl{e}}_2\hat{\bl{e}}_2 = -1\,.$$
Thus defining $\boldsymbol{i}=\hat{\bl{e}}_1$, $\boldsymbol{j}=\hat{\bl{e}}_2$ and $\boldsymbol{k}=\hat{\bl{e}}_1\wedge \hat{\bl{e}}_2$, we find that $\boldsymbol{i}^2=\boldsymbol{j}^2=\boldsymbol{k}^2=\boldsymbol{i}\boldsymbol{j}\boldsymbol{k}=-1$, \textit{i.e.}, $\mathcal{C}l(\mathbb{R}^{0,2})$ is the quaternion algebra. In particular note that it admits the following matrix representation:
$$1 \sim \left[
           \begin{array}{cc}
             1 & 0 \\
             0 & 1 \\
           \end{array}
         \right] \;; \, \boldsymbol{i} \sim \left[
                                 \begin{array}{cc}
                                   0 & i \\
                                   i & 0 \\
                                 \end{array}
                               \right]  \;; \, \boldsymbol{j} \sim \left[
                                 \begin{array}{cc}
                                   0 & -1 \\
                                   1 & 0 \\
                                 \end{array}
                               \right] \;; \,\boldsymbol{k} \sim \left[
                                 \begin{array}{cc}
                                   i & 0 \\
                                   0 & -i \\
                                 \end{array}
                               \right]\,.
  $$
\hfill\(\Box\)
\\

An important element of $\mathcal{C}l(V)$ is the so-called pseudo-scalar\index{Pseudo-scalar}, $\bl{I}=\hat{\bl{e}}_1\hat{\bl{e}}_2\ldots \hat{\bl{e}}_n$. If $s$ is the signature of the inner product, it is not difficult to prove that the Clifford product of $\bl{I}$ with itself is given by
\begin{equation}\label{I^2}
 \bl{I}^2 \,=\, (-1)^{\frac{1}{2}[n(n-1)+(n-s)]} \,\,.
\end{equation}
Defining the reversion operation by $(\bl{a}_1\bl{a}_2\ldots \bl{a}_p)^t\equiv \bl{a}_p\ldots \bl{a}_2 \bl{a}_1$ it follows that the Hodge dual\index{Hodge dual} of an element of $\wedge V$ can be easily expressed in terms of the Clifford algebra, more precisely we have that
\begin{equation}\label{Hodge_Cliff}
 \star\bl{\omega} = (-1)^{\frac{1}{2}[n(n-1)+(n-s)]} \, (\bl{I} \bl{\omega})^t\,.
\end{equation}

Now let us see the deep connection between geometric algebra and rotations. Let $\bl{n}\in V$ be a normalized vector, $\bl{n}^2 = <\bl{n},\bl{n}>=\pm1$, and $\bl{a}\in V$ be an arbitrary vector. Then by means of (\ref{Sym_Clif. Prod}) it easily follows that:
\begin{equation}\label{Reflection}
  -\bl{n}\,\bl{a}\,\bl{n}^{-1} \,=\, -(-\bl{a}\bl{n} + 2<\bl{n},\bl{a}>)\bl{n}^{-1} \,=\, \bl{a}- 2<\bl{n},\bl{a}>\bl{n}^{-1} \,.
\end{equation}
Where $\bl{n}^{-1}=\pm \bl{n}$ when $\bl{n}^2=\pm1$. The combination $\bl{a} - 2  <\bl{n},  \bl{a}>\bl{n}^{-1}$ is the exactly the reflection of the vector $\bl{a}$ with respect to the plane orthogonal to $\bl{n}$. Indeed, if $\bl{a}$ is orthogonal to $\bl{n}$ then it gives $\bl{a}$, while if $\bl{a}$ is parallel to $\bl{n}$ such combination yields $-\bl{a}$. It can be proved that in $n$ dimensions any rotation can be decomposed as a product of at most $n$ reflections \cite{Cartan}. Thus is natural to define the following groups contained on the Clifford algebra:
\begin{align}\label{Pin}
 \nonumber Pin(V) \,&=\,\{\bl{\varphi}\in \mathcal{C}l(V) \,|\, \bl{\varphi}= \bl{n}_p\ldots \bl{n}_2\bl{n}_1,\, \bl{n}_i\in V \textrm{ and } \bl{n}_i^2=\pm1\}  \\
  SPin(V) \,&=\,\{\bl{\varphi}\in \mathcal{C}l(V) \,|\, \bl{\varphi}= \bl{n}_{2p}\ldots \bl{n}_2\bl{n}_1,\, \bl{n}_i\in V \textrm{ and } \bl{n}_i^2=\pm1 \}
\end{align}
Note that $SPin(V)=Pin(V)\cap\mathcal{C}l(V)^+$, \textit{i.e}, $SPin(V)$ is the subgroup of $Pin(V)$ formed by the elements of even order. It is simple matter to verify that $Pin(V)$ and $SPin(V)$ are indeed groups under the Clifford multiplication. Then, by what was seen above, we conclude that the elements of these groups can be used to implement reflections and pure rotations on an arbitrary vector $\bl{a}\in V$.
\begin{align*}
  \textbf{\textrm{Rotation $\boldsymbol{+}$ Reflection}} \,&:\,\;    (-1)^p \,\bl{\varphi} \,\bl{a}\,\bl{\varphi}^{-1} \;,\; \bl{\varphi}\in Pin(V)\\
  \textbf{\textrm{Pure Rotation}} \,&:\,\;     \bl{\varphi} \,\bl{a}\,\bl{\varphi}^{-1} \;,\; \bl{\varphi}\in SPin(V)
\end{align*}
Indeed, these transformations are just a composition of the reflections seen on eq. (\ref{Reflection}). In particular, it is immediate to verify that the norm of $\bl{a}$ is preserved. Note that $\bl{\varphi}$ and $-\bl{\varphi}$ accomplish the same transformation on a vector, which results on the following important relations:
$$ O(V) = Pin(V)/\mathbb{Z}_2 \quad;\quad  SO(V) = SPin(V)/\mathbb{Z}_2 \,.$$
Moreover, it can be proved that $Pin(V)$ and $SPin(V)$ are the universal covering groups of the orthogonal groups $O(V)$ and $SO(V)$ respectively. We can also define the group $SPin_+(V)$ as being the subgroup of $SPin(V)$ formed by the elements $\varphi_+\in SPin(V)$ such that $\varphi_+^t\varphi_+=1$. Note that the action of the groups $Pin(V)$ and $Spin(V)$ on $V$ yield elements on $V$, thus the vector space $V$ provides a representation for these groups. But this representation is quadratic and therefore it is not faithful, since $\varphi$ and $-\varphi$ are represented by the same operation on $V$. In what follows we will see that the space of spinors\index{Spinor} gives a linear and faithful representation for these groups, actually for the whole Clifford algebra. But before proceeding let us see an explicit example of how the rotations shows up on the geometric algebra formalism.
\\
\\
\textbf{Example:}\\
Let $\{\hat{\bl{e}}_1,\hat{\bl{e}}_2,\ldots,\hat{\bl{e}}_n\}$ be an orthonormal basis for the Euclidian vector space $\mathbb{R}^n$, $<\hat{\bl{e}}_i,\hat{\bl{e}}_j>=\delta_{ij}$. Now defining $\bl{n}_1=\hat{\bl{e}}_1$, $\bl{n}_2= \cos\theta\, \hat{\bl{e}}_1 + \sin\theta\, \hat{\bl{e}}_2$ and $\bl{\varphi}_{_\theta}=\bl{n}_2\bl{n}_1$, it is simple matter to prove the following relations:
\begin{align*}
  \bl{\varphi}_{_\theta}\, \hat{\bl{e}}_1\,\bl{\varphi}_{_\theta}^{-1} \,&=\, \bl{n}_2\bl{n}_1 \,\hat{\bl{e}}_1\, \bl{n}_1\bl{n}_2 \,=\, \cos(2\theta)\, \hat{\bl{e}}_1 + \sin(2\theta)\, \hat{\bl{e}}_2  \\
 \bl{\varphi}_{_\theta} \, \hat{\bl{e}}_2\,\bl{\varphi}_{_\theta}^{-1} \,&=\, \bl{n}_2\bl{n}_1 \,\hat{\bl{e}}_2\, \bl{n}_1\bl{n}_2 \,=\, - \sin(2\theta)\, \hat{\bl{e}}_1 + \cos(2\theta)\, \hat{\bl{e}}_2 \\
 \bl{\varphi}_{_\theta} \, \hat{\bl{e}}_j\,\bl{\varphi}_{_\theta}^{-1} \,&=\, \bl{n}_2\bl{n}_1 \,\hat{\bl{e}}_j\, \bl{n}_1\bl{n}_2 \,=\, \hat{\bl{e}}_j \textrm{ if } j\geq3
\end{align*}
Thus $\bl{\varphi}_{_\theta}\in SPin(\mathbb{R}^n)$ accomplish a rotation of $2\theta$ on the plane generated by $\{\hat{\bl{e}}_1,\hat{\bl{e}}_2\}$. As a final remark note that $\bl{\varphi}_{_\theta}=\bl{n}_2\bl{n}_1 = (\cos\theta -\sin\theta\, \hat{\bl{e}}_1\hat{\bl{e}}_2)$ can be formally represented by $\bl{\varphi}_{_\theta}=e^{-\theta\, \hat{\bl{e}}_1\hat{\bl{e}}_2}$, as can be easily verified expanding the exponential in series. Thus, in general, the element $\bl{\varphi} = e^{-\theta\, \hat{\bl{e}}_i\wedge \hat{\bl{e}}_j}$ undertakes a rotation of $2\theta$ on the plane generated by $\{\hat{\bl{e}}_i,\hat{\bl{e}}_j\}$.
\hfill\(\Box\)
\\

Spinors\index{Spinor} can be roughly defined as the elements of a vector space on which the less-dimensional faithful representation of the Clifford algebra acts. In order to be more precise we shall define what a minimal left ideal is. In what follows it will be assumed, for simplicity, that the dimension of $V$ is even, $n=2r$ with $r\in\mathbb{N}$. We call $L\subset\mathcal{C}l(V)$ a left ideal of the algebra $\mathcal{C}l(V)$ when $L$ is invariant under the action on the left of the whole algebra:
$$ \textrm{$L$ is a left ideal} \;\Leftrightarrow\;  \bl{\omega}\, \bl{\zeta} \,=\, \bl{\zeta}'\in L \quad \forall \quad \bl{\zeta}\in L \,\textrm{ and } \,\bl{\omega}\in\mathcal{C}l(V) \,. $$
In particular, note that a left ideal is a subalgebra. A minimal left ideal is a left ideal that as an algebra admits no proper left ideal, \textit{i.e}, is a left ideal that admits no left ideal other than itself and the zero element.

Note that a left ideal $L\subset\mathcal{C}l(V)$ provides a representation of the Clifford algebra, sice $L$ is a vector space and, by definition, this algebra maps $L$ into $L$. A minimal left ideal $S\subset\mathcal{C}l(V)$ furnish the less-dimensional faithful representation of $\mathcal{C}l(V)$, the so-called spinorial representation of the Clifford algebra. Therefore the elements of $S$ are called spinors. It can be proved that if $n=2r$ is the dimension of the vector space $V$ then the dimension of the spinor space is $2^r$ \cite{Spinor-thesis,Cliff_Rigo}. Particularly, this implies that the algebra $\mathcal{C}l(V)$ and the groups $Pin(V)$, $SPin(V)$, $O(V)$ and $SO(V)$ can all be faithfully represented by $2^r\times2^r$ matrices.

Although the pseudo-scalar\index{Pseudo-scalar} $\bl{I}$ always commutes with the elements of even order, when the dimension is even it does not commute with the elements of odd order, so in this case the spinorial representation of $\bl{I}$ is not a multiple of the identity. From equation (\ref{I^2}) we see that $\bl{I}^2=\varepsilon^2$, with $\varepsilon=1$ or $\varepsilon=i$ depending on the dimension and on the signature. Thus when acting on $S$ the pseudo-scalar $\bl{I}$ splits this space into a direct sum of two subspaces of dimension $2^{r-1}$.
$$ S\,=\, S^+ \oplus S^-\;;\quad S^{\pm}=\{\bl{\psi}\in S \,|\, \bl{I}\bl{\psi}=\pm\varepsilon \bl{\psi}\} $$
The elements of $S^{\pm}$ are called Weyl spinors (or semi-spinors) of positive and negative chirality\index{Weyl spinor}. Since $\bl{I}$ commutes with $\mathcal{C}l(V)^+$ it follows that if $\bl{\psi}^{\pm}\in S^{\pm}$ and $\bl{\omega}_+\in\mathcal{C}l(V)^+$ then  $\bl{\omega}_+\bl{\psi}^{\pm}$  will also pertain to $S^{\pm}$. This means that in even dimensions the spinorial representation of $\mathcal{C}l(V)^+$ splits in two blocks of dimension $2^{r-1}\times2^{r-1}$.
$$  \mathcal{C}l(V)^+ \,\sim\, \left[
                                 \begin{array}{cc}
                                   R_+ & 0 \\
                                   0 & R_- \\
                                 \end{array}
                               \right]
$$
Where $R_{\pm}$ is the restriction of the spinorial representation of $\mathcal{C}l(V)^+$ to $S^{\pm}$. The representations $R_{\pm}$ are generally faithful and independent of each other. Since the group $SPin(V)$ is formed just by elements of even order it then follows that it generally admits representations of dimension $2^{r-1}$ and, consequently, the same is valid for the group $SO(V)$. For instance, the following relations are valid \cite{Lounesto}:
$$ SPin(\mathbb{R}^2) \sim U(1) \quad; \; SPin(\mathbb{R}^{3,1}) \sim Sl(2,\mathbb{C}) \quad; \; SPin(\mathbb{R}^6) \sim SU(4)\,.$$
In order to make clear the concepts introduced so far, let us work out a simple example.
\\
\\
\textbf{Example:}\\
Let $\{\hat{\bl{e}}_1,\hat{\bl{e}}_2\}$ be an orthonormal basis for the space $V=\mathbb{R}^2$, so that $\hat{\bl{e}}_1\hat{\bl{e}}_1=\hat{\bl{e}}_2\hat{\bl{e}}_2=1$ and $\hat{\bl{e}}_1\hat{\bl{e}}_2= \hat{\bl{e}}_1\wedge \hat{\bl{e}}_2$. In particular $\{1,\, \hat{\bl{e}}_1,\, \hat{\bl{e}}_2,\,  \bl{I}=\hat{\bl{e}}_1\hat{\bl{e}}_2\}$ forms a basis for $\mathcal{C}l(\mathbb{R}^2)$. A general element of $SPin(\mathbb{R}^2)$ has the following form:
$$ \Phi = [\cos(\phi_2)\hat{\bl{e}}_1+\sin(\phi_2)\hat{\bl{e}}_2][\cos(\phi_1)\hat{\bl{e}}_1+\sin(\phi_1)\hat{\bl{e}}_2]= \cos\theta-\sin\theta\, \hat{\bl{e}}_1\wedge \hat{\bl{e}}_2\,,$$
where $\theta=\phi_1-\phi_2$. Hence the elements of $SPin(\mathbb{R}^2)$ are labeled by a single real number $\theta\in[0,2\pi)$. Moreover, since
\begin{align*}
  \Phi_{\theta_{1}}\,\Phi_{\theta_{2}} \,&=\, (\cos\theta_1-\sin\theta_1\, \hat{\bl{e}}_1\wedge \hat{\bl{e}}_2)(\cos\theta_2-\sin\theta_2\, \hat{\bl{e}}_1\wedge \hat{\bl{e}}_2) \\
  \,&=\, \cos(\theta_1+\theta_2)-\sin(\theta_1+\theta_2)\, \hat{\bl{e}}_1\wedge \hat{\bl{e}}_2 \,=\, \Phi_{(\theta_{1} + \theta_{2})}\,,
\end{align*}
it follows that $SPin(\mathbb{R}^2)\sim U(1)$. The rotation implemented by $\Phi_{\theta}$ is the following:
\begin{align*}
  \hat{\bl{e}}_1 \,\rightarrow\, \hat{\bl{e}}'_1=\Phi_{\theta} \,\hat{\bl{e}}_1\,\Phi_{\theta}^{-1} \,&=\, \cos(2\theta)\hat{\bl{e}}_1 + \sin(2\theta)\hat{\bl{e}}_2   \\
 \hat{\bl{e}}_2 \,\rightarrow\, \hat{\bl{e}}'_2=\Phi_{\theta} \,\hat{\bl{e}}_2\,\Phi_{\theta}^{-1} \,&=\, -\sin(2\theta)\hat{\bl{e}}_1 + \cos(2\theta)\hat{\bl{e}}_2\,.
\end{align*}
Now let us see that $S=\{\bl{\psi}\in\mathcal{C}l(\mathbb{R}^2)\,|\,\bl{\psi} = \alpha(1+\hat{\bl{e}}_1)+\beta \hat{\bl{e}}_2(1+\hat{\bl{e}}_1) \;\forall\, \alpha,\beta\in\mathbb{C}\}$ is a minimal left ideal of this Clifford algebra. Indeed, defining $\bl{\psi}_1\equiv(1+\hat{\bl{e}}_1)$ and $\bl{\psi}_2\equiv\hat{\bl{e}}_2(1+\hat{\bl{e}}_1)$ we easily prove that
$$\hat{\bl{e}}_1\,(\alpha\bl{\psi}_1+\beta\bl{\psi}_2)= \alpha\bl{\psi}_1 -  \beta\bl{\psi}_2 \;\textrm{ and }\; \hat{\bl{e}}_2\,(\alpha\bl{\psi}_1+\beta\bl{\psi}_2)= \beta\bl{\psi}_1 +\alpha\bl{\psi}_2\,,$$
which implies that $S$ is invariant by the left action of $\mathcal{C}l(\mathbb{R}^2)$. It is also simple matter to verify that $S$ admits no proper left ideal, which implies that $\{\bl{\psi}_1,\bl{\psi}_2\}$ can be seen as a basis for the spinor space. The spinors $\bl{\psi}^{\pm}= \bl{\psi}_1\pm i\bl{\psi}_2$ are Weyl spinors\index{Weyl spinor}, since they obey the relation $\bl{I}\bl{\psi}^{\pm}=\pm i \bl{\psi}^{\pm}$. The action of the group $SPin(\mathbb{R}^2)$ on the semi-spinors is the following:
$$ \Phi_{\theta}\,\bl{\psi}^+ \,=\, e^{-i\theta}\bl{\psi}^+\quad;\;  \Phi_{\theta}\,\bl{\psi}^- \,=\, e^{i\theta}\bl{\psi}^- \,.$$
Particularly, note that taking $\theta=\pi$ the vectors remain unchanged by the action of the group $SPin(\mathbb{R}^2)$ while the spinors change the sign. This is an example of a well-known property of spinors, they are multiplied by $-1$ when a rotation of $2\pi$ is executed on the space.
\hfill\(\Box\)
\\

Given a spinor\index{Spinor} $\bl{\psi}\in S$ we can associate to it a vector subspace $N_{\bl{\psi}}\subset V$ called the null subspace of $\bl{\psi}$ and defined by $ N_{\bl{\psi}} \,=\, \{\bl{a}\in V \,|\, \bl{a}\,\bl{\psi} \,=\,0  \}  $. This vector subspace has the property of being totally null (isotropic)\index{Isotropic}, \textit{i.e.}, all vectors of $N_{\bl{\psi}}$ are orthogonal to each other. Indeed, assuming that $\bl{\psi}\neq 0$
it follows that
$$ 2<\bl{a},\bl{b}>\,\bl{\psi} \,=\, (\bl{a}\bl{b}+\bl{b}\bl{a}) \,\bl{\psi} \,=\,0 \quad\forall\; \bl{a},\bl{b}\in N_{\bl{\psi}} \;\Rightarrow\; <\bl{a},\bl{b}>\,=\,0\,.$$
In a vector space of complex dimension $n=2r$, the maximal dimension that an isotropic subspace can have is $r$. Therefore a totally null subspace with this dimension is dubbed maximally isotropic\index{Maximally isotropic}. When the subspace $N_{\bl{\psi}}$ is maximally isotropic the spinor $\bl{\psi}$ is said to be a pure spinor\index{Pure spinor}. Apart from a multiplicative constant, the association between pure spinors and maximally isotropic subspaces is one-to-one. It is worth noting that in general the sum of two pure spinors is not a pure spinor, indeed the purity condition is a quadratic constraint on the spinor \cite{Spinor-thesis}.

Now let us prove that every pure spinor must be a Weyl spinor\index{Weyl spinor}. Let $V$ be a complexified vector space and $\{\bl{e}_1,\bl{e}_2,\ldots,\bl{e}_r\}$ be the basis of a maximally isotropic subspace $N_{\bl{\psi}}$, thus $<\bl{e}_a,\bl{e}_b>=0$. We can complete this basis with $r$ other vectors $\{\bl{\theta}^a\}$ in order to form a basis for the whole vector space $V$ such that $<\bl{e}_a,\bl{\theta}^b>=\frac{1}{2}\delta_a^{\,b}$ and $<\bl{\theta}^a,\bl{\theta}^b>=0$. Then we have that
\begin{equation}\label{I_nullframe}
\bl{I} \,\propto\, (\bl{e}_1\wedge\bl{\theta}^1)(\bl{e}_2\wedge\bl{\theta}^2)\ldots(\bl{e}_r\wedge\bl{\theta}^r)\,.
\end{equation}
By definition $\bl{e}_a\,\bl{\psi}=0$, therefore
\begin{gather}
  \nonumber (\bl{e}_a\bl{\theta}^b)\,\bl{\psi} \,=\, (\bl{e}_a\bl{\theta}^b+ \bl{\theta}^b\bl{e}_a)\,\bl{\psi} \,=\,2<\bl{e}_a,\bl{\theta}^b>\bl{\psi} \,=\,\delta_a^{\,b}\, \bl{\psi} \;\Rightarrow \\
  (\bl{e}_a\wedge\bl{\theta}^b)\,\bl{\psi} = \frac{1}{2}(\bl{e}_a\bl{\theta}^b- \bl{\theta}^b\bl{e}_a)\,\bl{\psi} = \frac{1}{2}(\bl{e}_a\bl{\theta}^b)\,\bl{\psi} = \frac{1}{2}\delta_a^{\,b} \, \bl{\psi}\,.\label{eo_psi}
\end{gather}
Then equations (\ref{I_nullframe}) and (\ref{eo_psi}) imply that $\bl{I}\,\bl{\psi}\propto\bl{\psi}$. This, in turn, guarantees that the pure spinor $\bl{\psi}$ must be a Weyl spinor. Conversely, if $n=2,4,6$ then all Weyl spinors are pure, but in higher dimensions this is not true \cite{Spinor-thesis}. Using (\ref{Hodge_Cliff}) it is also simple matter to prove that the Hodge dual of the $r$-vector $\bl{e}_1\wedge \bl{e}_2\wedge\ldots\wedge \bl{e}_r$ is a multiple of this $r$-vector.

The space of spinors, $S$, can be endowed with an operation called charge conjugation\index{Charge Conjugation}, $c:S\rightarrow S$. This is an anti-linear operation whose action, $\bl{\psi}\mapsto\bl{\psi}^c$, is such that the following property holds:
$$ (\bl{\omega}\,\bl{\psi})^c \,=\, \overline{\bl{\omega}} \, \bl{\psi}^c \quad \forall\; \bl{\omega} \,\in\,\mathcal{C}l(V) \textrm{ and }  \; \bl{\psi}\,\in\,S\, ,$$
where $\overline{\bl{\omega}}$ is the complex conjugate of $\bl{\omega}$. The charge conjugation has different features depending on the signature and on the dimension of the vector space, see \cite{Spinor-thesis} for example. For instance, on the Minkowski space, $\mathbb{R}^{1,3}$, such operation changes the chirality of a Weyl spinor and its square gives the identity, while for $\mathbb{R}^{5,1}$ the spaces $S^{\pm}$ are invariant and $(\bl{\psi}^c)^c=-\bl{\psi}$.

Another important property of the spinor space is that it is always possible to introduce a non-degenerate bilinear inner product, $(\,,): S\times S \rightarrow \mathbb{C}$, that is invariant by the group $SPin_+(V)$. Indeed, defining $(\bl{\psi},\bl{\chi})=f(\bl{\psi}^t\bl{\chi})$ for some function $f:\mathcal{C}l(V)\rightarrow \mathbb{C}$ we find that $(\bl{\omega}\bl{\psi},\bl{\chi})=(\bl{\psi},\bl{\omega}^t\bl{\chi})$. Hence making a $SPin_+(V)$ transformation on the spinors, $S\mapsto \bl{\varphi}_+S$, we find that $(\bl{\psi},\bl{\chi}) \mapsto (\bl{\varphi}_+\bl{\psi},\bl{\varphi}_+\bl{\chi}) = (\bl{\psi},\bl{\varphi}_+^t\bl{\varphi}_+\bl{\chi}) = (\bl{\psi},\bl{\chi})$, since $\bl{\varphi}_+\in SPin_+(V)$. A particularly simple choice for $f$ would be $f(\bl{\omega})=[\bl{\omega}]_0$, where $[\bl{\omega}]_0$ is the scalar part (zero order term) of $\bl{\omega}\in \mathcal{C}l(V)$. But in order for the inner product to be non-degenerate we must judiciously choose the function $f$, as $f=[\,]_0$ may not obey to this criterium. The general formalism for the choice of an adequate $f$ is very tricky and more details can be found in \cite{Cliff_Rigo,Lounesto}.

The inner product $(\,,\,)$ can be symmetric or skew-symmetric depending on the dimension of $V$. For example, in two dimensions it is symmetric,  while in four and six dimensions it is skew-symmetric \cite{Spinor-thesis}. Furthermore, in four dimensions the inner product of two semi-spinors of \emph{opposite} chirality vanishes, while in two and six dimensions the inner product of Weyl spinors of the \emph{same} chirality vanish \cite{Spinor-thesis}.

In Physics, the Clifford algebra and the spinor formalism is usually used in a less abstract way, making use of the so-called Dirac matrices \cite{Polch.2}. If the metric of a $2r$-dimensional vector space $V$ is $g_{ab}$ then the Dirac matrices, $\gamma_a$, are defined to be $2^r\times2^r$ matrices such that $\{\gamma_a,\gamma_b\} = (\gamma_a\gamma_b +\gamma_b\gamma_a) = 2g_{ab}$. The $2^r$-dimensional vector space on which these matrices act is called the space of spinors. Using the tools presented in this appendix it is not hard to guess the origin this practical approach. The matrices $\gamma_a$ are just a matrix representation of the vectors $\{\hat{\bl{e}}_a\}$ of $\mathcal{C}l(V)$ and the anti-commutation relation $\{\gamma_a,\gamma_b\} = 2g_{ab}$ is the matrix realization of equation (\ref{Sym_Clif. Prod}). Since the Dirac matrices provide a faithful representation of minimal dimension for the vectors of the Clifford algebra they must be $2^r\times2^r$ matrices and the column vectors on which these matrices act should be called spinors.

The material presented in this appendix is just a scratch on the rich field of geometric algebra. There are many nice references on Clifford algebra and spinors. The classical reference that presents the ``modern'' approach to the subject is the book of C. Chevalley \cite{Chevalley}. Introductory texts with applications in Physics can be found in \cite{Lasenby,Hestenes}, while geometric applications and historical notes are available in \cite{Snygg}. More advanced and rigorous treatments are found in \cite{Cliff_Rigo,Lounesto}.

\chapter{Group Representations}\label{App._Group}

In this appendix it will be explained what is a representation of a group and how to construct higher-dimensional representations out of a lower-dimensional one. First let us recall some basic definitions on group theory. Let $G$ be a set endowed with a product $g_1\cdot g_2=g_3$ such that $g_3\in G$ for all $g_1,g_2\in G$. Then $G$ is called a group when the following three properties hold:
(1) There exists an element $e\in G$, called the identity element, such that $e\cdot g= g$ for all $g\in G$;
(2) For every element $g\in G$ there exists an element $g^{-1}\in G$, called the inverse of $g$, such that $g\cdot g^{-1}= e$;
(3) The product is associative, $g_1\cdot (g_2\cdot g_3)= (g_1\cdot g_2)\cdot g_3$ for all $g_1, g_2, g_3\in G$.
A map $H:G\rightarrow G'$ between two groups $G$ and $G'$ is called a homomorphism if $H(g_1)\cdot H(g_2)= H(g_1\cdot g_2)$ for all $g_1,g_2\in G$.

Whenever a physical system has a symmetry the group theory can be used in order to simplify the analysis. Although sometimes it is possible to move on just using the abstract concept of a group, generally it is necessary to use a down-to-earth approach, such as expressing the group elements by matrices. A representation of a group $G$ on the vector space $V$ is a homomorphism $L:G\rightarrow GL(V)$, where $GL(V)$ is the group formed by all invertible linear operators acting on $V$. Since vector spaces are ubiquitous in physics it follows that representation theory is a quite helpful tool in many branches of this science. If $\dim(V)= n$ we say that $L$ is an $n$-dimensional representation. Note that every group admits a trivial representation of dimension $1$ given by $I:G\rightarrow GL(\mathbb{R})= \mathbb{R}^*$ with $I(g)= 1$ for all $g\in G$. Two representations $L_1$ and $L_2$ of the group $G$ on the vector space $V$ are said to be equivalent when there exists some $B\in GL(V)$ such that $L_2(g)= BL_1(g)B^{-1}$ for all $g\in G$.

Let us adopt the index notation and denote a vector of the $n$-dimensional vector space $V$ by $v^a$, with $a\in \{1,2,\ldots,n\}$. Then a representation of the group $G$ on this vector space is an association of a matrix $L^{a}_{\ph{a}b}(g)$ to every $g\in G$. Since this association is, by definition, a homomorphism, the identity $L^{a}_{\ph{a}c}(g_1)L^{c}_{\ph{c}b}(g_2)= L^{a}_{\ph{a}b}(g_1\cdot g_2)$ must hold for all $g_1,g_2\in G$. Once specified a representation $L$ of the group $G$ on the vector space $V$, we then say that the action of a group element $g$ on a vector $v^a$ amounts to the following transformation:
\begin{equation}\label{L(g)v}
  v^a \,\stackrel{g}{\longrightarrow} \, L^{a}_{\ph{a}b}(g)\,v^b\,.
\end{equation}
In abstract notation we can write $\bl{v}\rightarrow L(g)\bl{v}$. Given such representation one can define another representation $P: G\rightarrow GL(V)$ called the inverse representation and defined by $\bl{v}\rightarrow P(g)\bl{v}$, with $P(g)$ being the transpose of $L(g)$ inverse,  $P(g)\equiv (L(g)^{-1})^t$. Let us verify that this is, indeed, a representation:
\begin{align*}
   P(g_1)P(g_2) =&\, (L(g_1)^{-1})^t(L(g_2)^{-1})^t  \,=\, \left(L(g_2)^{-1}L(g_1)^{-1}\right)^t \\
  =&\, [\left( L(g_1)L(g_2)\right)^{-1}]^t \,=\, \left(L(g_1\cdot g_2)^{-1}\right)^t    \,=\,  P(g_1\cdot g_2)\,.
\end{align*}
Note that generally the representations $L$ and $P$ are not equivalent. By definition the representation $P$ acts on the same vector space of the representation $L$, but it is useful to pretend that $P$ acts on a different vector space $V'$ that is isomorphic to $V$ and whose vectors are denoted with an index down, $u_a\in V'$. So the representation $P$ has the following action:
\begin{equation}\label{L(g)v_a}
   u_a \,\stackrel{g}{\longrightarrow} \, P_{a}^{\ph{a}b}(g)\,u_b\quad;\quad P_{a}^{\ph{a}b}(g)\equiv [L(g)^{-1}]^b_{\ph{b}a}\,.
\end{equation}
On the jargon we say that $v^a$ is on the $L$ representation while $u_a$ is on the $P$ representation. Note that in this case the contraction $v^au_a$ is invariant by the action of the group $G$, which is equivalent to say that $v^au_a\in \mathbb{R}$ is on the trivial representation, $I$.

Suppose that the vector space $V$ has a proper subspace $K\subset V$ such that $L(g)\bl{k}\in K$ for all $\bl{k}\in K$ and for all $g\in G$. Then the restriction of $L(g)$ to this subspace provides a representation for the group $G$ on the lower-dimensional vector space $K$. When this happens we say that the representation $L$ is reducible, otherwise it is called irreducible. The irreducible representations\index{Irreducible representations} of a group are the building blocks of a general representation, since every representation of $G$ can be understood as a composition of some irreducible representations of this group. For instance, it is well-known that the irreducible representations of the rotation group on $\mathbb{R}^3$, $SO(3)$, are labeled by $l\in\{0,\,\frac{1}{2},\, 1, \,\frac{3}{2}, \,2,\cdots\}$, the angular momentum quantum number. The dimension of the representation dubbed $l$ is $(2l+1)$. Here we shall label an irreducible representation of a group by its dimension in bold face. Thus the representations $\bl{2}$ and $\bl{3}$ of $SO(3)$ mean the ones with $l=\frac{1}{2}$ and $l=1$ respectively. Moreover, the trivial representation $I$ might be denoted by $\bl{1}$.

Given an irreducible representation $\bl{n}$ of a group $G$, generally it is possible to generate other irreducible representations of $G$ by means of the direct products of the representation $\bl{n}$ with itself. We can understand this as follows, the representation $\bl{n}$ associates to every $g\in G$ an $n\times n$ matrix $L(g)$. Then taking the direct product $L(g)\otimes L(g)$  we obtain an $n^2\times n^2$ matrix for every $g$. These matrices also provide a representation for the group $G$, but generally this representation is not irreducible, since in general such $n^2\times n^2$ matrices will admit proper invariant subspaces. Then looking for the invariant subspaces of these matrices one can split the new representation into its irreducible parts. For example, the direct product of the irreducible representations $l'$ and $l''$ of the group $SO(3)$ is equal to the direct sum of all irreducible representations contained on the interval $|l''-l'|\leq l\leq (l'+l'')$. This is usually written as \cite{GroupWu}:
\begin{equation}\label{WignerEck}
l'\otimes l'' \,=\,  (l'+l'') \,\oplus\, (l'+l''-1)  \,\oplus\, (l'+l''-2) \,\oplus\cdots\oplus\, |l'-l''|\,.
\end{equation}

As an instructive example let us work out the direct product of some irreducible representations of the group $SO(n)$. Let $R: SO(n)\rightarrow GL(\mathbb{R}^n)$ be the usual representation of this group that associates to every element of $SO(n)$ an $n\times n$ orthogonal matrix $R$ with unit determinant, $RR^t= \bl{1}$  and $\det(R)= 1$. This irreducible representation is denoted by $\bl{n}$ and its action on $\mathbb{R}^n$ is given by:
$$ v^a \,\stackrel{R}{\longrightarrow}\, R^a_{\ph{a}b}\,v^b \,. $$
We say that the tensor $T^{ab}$ is on the representation $\bl{n}\otimes\bl{n}$ if its transformation under the group $SO(n)$ is given by:
$$  T^{ab} \,\stackrel{R}{\longrightarrow}\, R^a_{\ph{a}c}\,R^b_{\ph{b}d}\,T^{cd} \,. $$
It is simple matter to verify that this representation is reducible. Indeed, note that the subspace formed by the symmetric tensors $T^{ab}= T^{(ab)}$ is invariant under the action of the representation $\bl{n}\otimes\bl{n}$. Suppose that $S^{ab}$ is symmetric, then
\begin{align*}
  S^{ab} \,\stackrel{R}{\longrightarrow}\,&\; R^a_{\ph{a}c}\,R^b_{\ph{b}d}\,S^{cd} = R^a_{\ph{a}c}\,R^b_{\ph{b}d}\,\frac{1}{2}\,[S^{cd} + S^{dc} ] \\
   &= \frac{1}{2}\,[R^a_{\ph{a}c}\,R^b_{\ph{b}d} + R^a_{\ph{a}d}\,R^b_{\ph{b}c} ] \,S^{cd} = R^{(a}_{\ph{a}c}\,R^{b)}_{\ph{b}d}\,S^{cd}\,,
\end{align*}
which is also symmetric. In the same vein, the space of skew-symmetric tensors $T^{ab}= T^{[ab]}$ is, likewise, invariant under the action of the representation $\bl{n}\otimes\bl{n}$. Moreover, we can easily convince ourselves that the restriction of the representation $\bl{n}\otimes\bl{n}$ to the space of skew-symmetric tensors is irreducible. Differently, the representation provided by the symmetric tensors can be split in two irreducible representations. Indeed, note that the symmetric tensors of the form $T^{ab}= \lambda\,\delta^{ab}$ are invariant by $SO(n)$:
$$ \lambda\,\delta^{ab} \stackrel{R}{\longrightarrow}\, \lambda\,R^a_{\ph{a}c}\,R^b_{\ph{b}d}\,\delta^{cd} = \lambda\,\delta^{ab}\,,  $$
where it was used that fact that $R$ is an orthogonal matrix. Note that the inverse of the representation $\bl{n}$ for the group $SO(n)$ is the representation $\bl{n}$ itself, which can be verified using equation (\ref{L(g)v_a}) and the identity $(R^{-1})^t= R$ valid for orthogonal matrices. Thus a general tensor $T^{ab}$ on the representation $\bl{n}\otimes\bl{n}$ of the group $SO(n)$ can be written as the following sum of irreducible parts:
$$ T^{ab} \,=\, \lef T^{(ab)} - \lambda \, \delta^{ab}\rig  \,+\, T^{[ab]}  \,+\, \lambda\, \delta^{ab}  \;\;;\quad \lambda\equiv\frac{1}{n}\,\delta_{cd}T^{cd}\,. $$
These irreducible parts are respectively called the symmetric trace-less part, the skew-symmetric part and the trace of the representation $\bl{n}\otimes\bl{n}$. In terms of dimensions this is written as:
\begin{equation}\label{Dim.T^ab}
  \bl{n}\otimes\bl{n}  \,=\,  \left[\bl{\frac{1}{2}n(n+1)-1}\right]  \,\,\oplus \,\,  \bl{\frac{1}{2}n(n-1)}\,\,\oplus \,\, \bl{1}     \,.
\end{equation}
Where $[\frac{1}{2}n(n+1)-1]$ is the number of components of a symmetric tensor with vanishing trace, $S^{ab}= S^{ba}$ and $\delta_{ab}S^{ab}= 0$, $\frac{1}{2}n(n-1)$ is the number of independent components of a skew-symmetric tensor, $A^{ab}= -A^{ba}$, and $1$ represents the single degree of freedom contained in $\lambda$, the trace of $T^{ab}$. Note that for $n=3$ this is consistent with the formula (\ref{WignerEck}) valid for the group $SO(3)$:
$$  [\,l'=1\,]\otimes[\,l''=1\,] \,\,\bl{=}\, \,[\,l=2\,]  \,\,\oplus \,\, [\,l=1\,]  \,\,\oplus \,\, [\,l=0\,]  \,.$$
Since the dimension of the irreducible representation labeled by $l$ is $(2l+1)$, it follows that the above equation is equivalent to:
$$ \bl{3}\otimes\bl{3} \,=\,  \bl{5} \,\,\oplus \,\,  \bl{3} \,\,\oplus \,\,  \bl{1}  \,,$$
which agrees with equation (\ref{Dim.T^ab}) when $n=3$. As a last example let us look for the irreducible parts of the representation  $\bl{n}\otimes\bl{n}\otimes\bl{n}$ of the group $SO(n)$. An object in this representation is a tensor with three indices, $N^{abc}$, transforming as follows:
\begin{equation}\label{N-transform}
   N^{abc} \,\stackrel{R}{\longrightarrow}\, R^a_{\ph{a}d}\,R^b_{\ph{b}e}\,R^c_{\ph{c}f}\,N^{def} \,.
\end{equation}
Let us try to separate the parts of this tensor that are invariant under this transformation for a general orthogonal matrix $R^a_{\ph{a}b}$. In what follows we shall display the dimension of each representation below the respective invariant terms, with the irreducible representations being denoted by bold face. The first trivial separation of the tensor $N^{abc}$ in parts that are invariant under the transformation (\ref{N-transform}) is given by:
$$ \underbrace{N^{abc}}_{n^3} \;\longrightarrow\;\;  \underbrace{N^{a(bc)}}_{\frac{1}{2}n^2(n+1)}  \quad,\quad \underbrace{N^{a[bc]}}_{\frac{1}{2}n^2(n-1)}\,. $$
Then the first term on the right hand side of the above equation splits on the following invariant parts:
$$ \underbrace{N^{a(bc)}}_{\frac{1}{2}n^2(n+1)} \;\longrightarrow\;\; \underbrace{\delta_{ab}N^{a(bc)}}_{\bl{n}} \quad,\quad \underbrace{\delta_{bc}N^{a(bc)}}_{\bl{n}} \quad,\quad \underbrace{\hat{N}^{a(bc)}}_{\frac{1}{2}n^2(n+1)-2n} \,.$$
Where $\hat{N}^{a(bc)}$ is a tensor such that $\delta_{ab}\hat{N}^{a(bc)}= 0$ and $\delta_{bc}\hat{N}^{a(bc)}= 0$. This tensor, in turn, gives rise to the following irreducible parts:
$$ \underbrace{\hat{N}^{a(bc)}}_{\frac{1}{2}n^2(n+1)-2n}  \;\longrightarrow\;\; \underbrace{\hat{N}^{(abc)}}_{\bl{\frac{1}{3!}n(n+1)(n+2)-n}} \quad,\quad  \underbrace{\tilde{N}^{a(bc)}}_{\bl{\frac{1}{3}n(n^2-4)}} \,.$$
Where $\tilde{N}^{a(bc)}$ is a tensor obeying to the following constraints $\tilde{N}^{(abc)}= 0$, $\delta_{ab}\tilde{N}^{a(bc)}= 0$ and $\delta_{bc}\tilde{N}^{a(bc)}= 0$. In the same vein, the tensor $N^{a[bc]}$ splits on the following invariant parts:
$$ \underbrace{N^{a[bc]}}_{\frac{1}{2}n^2(n-1)} \;\longrightarrow\;\; \underbrace{\delta_{ab}N^{a[bc]}}_{\bl{n}} \quad,\quad \underbrace{\hat{N}^{a[bc]}}_{\frac{1}{2}n^2(n-1)-n} \,.$$
With $\hat{N}^{a[bc]}$ being a trace-less tensor, $\delta_{ab}\hat{N}^{a[bc]}= 0$. This tensor, in turn, lead to the following irreducible parts:
$$ \underbrace{\hat{N}^{a[bc]}}_{\frac{1}{2}n^2(n-1)-n}  \;\longrightarrow\;\; \underbrace{\hat{N}^{[abc]}}_{\bl{\frac{1}{3!}n(n-1)(n-2)}} \quad,\quad  \underbrace{\tilde{N}^{a[bc]}}_{\bl{\frac{1}{3}n(n^2-4)}} \,.$$
Where $\tilde{N}^{a[bc]}$ is a tensor such that $\tilde{N}^{[abc]}= 0$ and $\delta_{ab}\tilde{N}^{a[bc]}= 0$. Therefore, the representation $\bl{n}\otimes\bl{n}\otimes\bl{n}$ splits on the following irreducible parts:
\begin{align}
  \nonumber\bl{n}\otimes\bl{n}\otimes\bl{n} \,=\,\,& \bl{n} \,\,\oplus\,\, \bl{n} \,\,\oplus\,\, \bl{n} \,\,\oplus\,\, \bl{\frac{1}{3}n(n^2-4)} \,\,\oplus\,\, \bl{\frac{1}{3}n(n^2-4)} \\
  & \oplus\,\,  \bl{\frac{n(n-1)(n-2)}{6}} \,\,\oplus\,\,  \bl{\left[\frac{n(n+1)(n+2)}{6}-n\right]} \,.\label{Dim-N^abc}
\end{align}
In particular, for the group $SO(3)$ we have:
\begin{equation}\label{N^abc-3}
  \bl{3}\otimes\bl{3}\otimes\bl{3} \,=\, \bl{3} \,\,\oplus\,\, \bl{3} \,\,\oplus\,\, \bl{3} \,\,\oplus\,\,  \bl{5} \,\,\oplus\,\, \bl{5} \,\,\oplus\,\, \bl{1} \,\,\oplus\,\,  \bl{7}\,.
\end{equation}
One can easily use equation (\ref{WignerEck}) in order to verify that this result is correct:
\begin{align*}
   \bl{3}\otimes\bl{3}\otimes\bl{3} \,=\,& \,\bl{3} \otimes \left[ \,\bl{5} \,\,\oplus \,\,  \bl{3} \,\,\oplus \,\,  \bl{1}\,\right] =\left[\,\bl{3} \otimes \bl{5} \,\right] \,\,\oplus \,\, \left[\,\bl{3} \otimes \bl{3}\,\right] \,\,\oplus \,\, \left[\,\bl{3} \otimes \bl{1}\,\right] \\
  =\,& \left[ \,  \bl{7} \,\,\oplus \,\,  \bl{5} \,\,\oplus \,\,  \bl{3} \, \right]  \,\,\oplus \,\, \left[ \,\bl{5} \,\,\oplus \,\,  \bl{3} \,\,\oplus \,\,  \bl{1}\,\right] \,\,\oplus \,\, \left[\,\bl{3} \,\right]\,,
\end{align*}
which agrees with equation (\ref{N^abc-3}). Equations (\ref{Dim.T^ab}) and (\ref{Dim-N^abc}) show us that starting with the irreducible representation $\bl{n}$ of the group $SO(n)$ we can take direct products in order to construct other irreducible representations. In general this kind of procedure can be used for any group. If a group $G$ admits an irreducible representation $\bl{f}$ such that all irreducible representations of $G$ can be constructed using the direct products of this representation, its inverse and its complex conjugate, then $\bl{f}$ is called the fundamental representation of $G$. For instance, the fundamental representation of the group $SO(3)$ is the one with $l=\frac{1}{2}$, in which the rotations of $\mathbb{R}^3$ are represented by $2\times2$ unitary matrices of unit determinant (spinorial representation).

\printindex

\newpage

\addcontentsline{toc}{chapter}{List of Symbols}

\begin{flushleft}
\huge{\textbf{List of Symbols}}
\end{flushleft}
\vspace{0.5cm}

\begin{align*}
\partial_\mu      \quad\quad& \textrm{ Partial derivative } \frac{\partial}{\partial x^\mu}:  \small{\textsf{  Page \pageref{PartialD}.}}  \\
C_{\mu\nu\rho\sigma} \quad\quad& \textrm{ Weyl Tensor:} \small{\textsf{  Page \pageref{WeylTensor}.}}  \\
 T_{[a_1a_2\ldots a_p]} \quad\quad& \textrm{ Skew-symmetric part of the tensor } T_{a_1a_2\ldots a_p}:  \small{\textsf{  Page \pageref{Symmetrization}.}}   \\
 T_{(a_1a_2\ldots a_p)} \quad\quad& \textrm{ Symmetric part of the tensor } T_{a_1a_2\ldots a_p}:  \small{\textsf{ Page  \pageref{Symmetrization}.}}   \\
 \epsilon_{\mu_1\mu_2\ldots\mu_n} \quad\quad&  \textrm{ Volume-form of the $n$-dimensional manifold:}   \small{\textsf{  Page \pageref{VolumeForm}.}}\quad   \\
 \bl{g} \quad\quad&   \textrm{ The metric of the manifold:}   \small{\textsf{  Page \pageref{Metric}.}}  \\
 \lrcorner \;,\;\; \bl{V}\lrcorner\bl{F} \quad\quad&  \textrm{ Interior product:}   \small{\textsf{  Page \pageref{InteriorProduct}.}}   \\
  \star\bl{F} \quad\quad&   \textrm{ Hodge dual of a differential form:}   \small{\textsf{  Page \pageref{HodgeDual}.}}   \\
  \bl{\omega}^a_{\ph{a}b}\,,\; \omega_{ab}^{\ph{ab}c}\,,\; \omega_{abc}  \quad\quad&    \textrm{ Connection 1-form and its components:} \small{\textsf{  Page \pageref{Connection}.}} \\
  \Psi_0\,,\;\Psi_1\,,\;\ldots,\Psi_4 \quad\quad& \textrm{ Weyl scalars in 4 dimensions: } \small{\textsf{  Pages \pageref{WeylScalars1} and \pageref{WeylScalars2}.}} \\
   \sigma \quad\quad& \textrm{ Shear of a null congruence:} \small{\textsf{  Pages \pageref{Shear1} and \pageref{Shear}.}} \\
    \Gamma(\wedge^p M)\quad\quad&    \textrm{ Space of local sections of the $p$-form bundle:} \small{\textsf{  Pages \pageref{P-FormSections} and \pageref{P-FormSections2}.}}\quad\quad\\
     \Lambda^{m+}\quad\quad&  \textrm{ Space of self-dual $m$-forms in $2m$ dimensions:}   \small{\textsf{  Page \pageref{SelfDualSpace}.}}\\
    \mathcal{H}_p \quad\quad& \textrm{ Hodge dual operator on $p$-forms:}   \small{\textsf{  Page \pageref{HodgeOperator}.}} \\
      \mathcal{C}_p\quad\quad& \textrm{ Weyl operator on $p$-forms:}   \small{\textsf{  Page \pageref{Cp-All}.}} \\
     \mathcal{C}^\pm  \quad\quad&  \textrm{ Restriction of the Weyl operator to $\Lambda^{m\pm}$:}   \small{\textsf{  Page \pageref{C^+- All}.}} \\
   \mathcal{A}_q \quad\quad& \textrm{ Particular subbundle of $\Gamma(\wedge^m M)$:}   \small{\textsf{  Page \pageref{Aq-All}.}} \\
         M_i\,,\;\sigma_{ij}\,,\;A_{ij}\,,\;\theta\quad\quad&  \textrm{ Optical scalars of a null congruence:}   \small{\textsf{  Page \pageref{OpticalSca-All}.}}\\
          \Psi^{AB}_{\ph{AB}CD} \quad\quad&  \textrm{ Spinorial representation of the Weyl tensor in 6D:}   \small{\textsf{  Page \pageref{Weyl6D}.}} \\
   (T^{AB}\,,\, \tilde{T}_{AB}) \quad\quad& \textrm{ Spinorial representation of a 3-vector in 6D:}  \small{\textsf{  Page \pageref{3Vector6D}.}}\\
   Span\{\bl{V}_i\}\quad\quad&    \textrm{ Vector distribution generated by the vector fields $\bl{V}_i$:} \small{\textsf{  Page \pageref{Span}.}}\\
\end{align*}

\end{document}